\documentclass[10pt,twoside,openright]{report}
\usepackage[a4paper,centering,top=28mm,bottom=22mm,right=20mm]{geometry}
\usepackage[titles]{tocloft}
\usepackage[utf8]{inputenc}
\usepackage{acronym}
\usepackage[usenames,dvipsnames,svgnames,x11names,table]{xcolor}
\usepackage{graphicx}
\usepackage{fancyhdr}
\usepackage{adjustbox}
\usepackage{emptypage}
\usepackage{mfirstuc}
\usepackage{amssymb}
\usepackage{tikz}
\usepackage{bera} %
\usepackage[toc,page]{appendix}
\usepackage[style=ieee,citestyle=numeric,backend=bibtex,natbib=true,sorting=none,doi=true,isbn=false,hyperref=true,alldates=year]{biblatex}
\usepackage{url}
\usepackage{hyperref}
\usepackage{breakurl}
\usepackage{listings}
\lstdefinestyle{lua}{language={[5.0]Lua},basicstyle=\footnotesize\ttfamily,frame=lines,captionpos=b}
\usepackage{xspace}
\usepackage{pifont}
\usepackage{colortbl}
\usepackage{subfig}
\usepackage{booktabs}
\usepackage[most]{tcolorbox}
\usepackage{amsmath}
\usepackage{cleveref}
\usepackage{multirow}
\usepackage[noend]{algpseudocode}
\usepackage[ruled,vlined,linesnumbered]{algorithm2e}
\usepackage[binary-units, per-mode=symbol, detect-weight=true, detect-family=true]{siunitx}
\usepackage[abbreviations]{foreign}
\usepackage{multicol}
\usepackage{needspace}
\usepackage{tabularx}
\usepackage[inline]{enumitem}
\usepackage{etoolbox}
\usepackage{pdfpages}
\usepackage[parfill]{parskip}
\usepackage[french,british]{babel}
\usepackage{titlesec}
\titlespacing*{\section}
{0pt}{4.5mm minus 1mm}{1,2mm}
\titlespacing*{\subsection}
{0pt}{3.3mm minus 1mm}{0mm}
\titlespacing*{\subsubsection}
{0pt}{3mm minus 1mm}{0mm}

\hypersetup{
    colorlinks=true,
    urlcolor={MidnightBlue},
    citecolor={RoyalPurple},
    linkcolor={Black},
}

\bibliography{references}
\renewcommand*{\bibfont}{\small}

\DeclareSIUnit\groupsize{\lvert \mathcal{G} \rvert} %
\DeclareSIUnit\operation{ops}
\DeclareSIUnit\requests{req}
\DeclareSIUnit{\nothing}{\relax}

\Crefname{algocf}{Algorithm}{Algorithms}
\AtBeginEnvironment{algorithm}{%
	\DontPrintSemicolon
	\SetKwProg{Procedure}{Procedure}{:}{end}
	\SetKwInOut{Input}{input}
	\SetKwInOut{Output}{output}
	\fontsize{8}{9.6}
}

\makeatletter
\renewcommand{\algocf@makecaption@ruled}[2]{%
	\global\sbox\algocf@capbox{\hskip\AlCapHSkip%
		\setlength{\hsize}{\columnwidth}%
		\addtolength{\hsize}{-2\AlCapHSkip}%
		\parbox[t]{\hsize}{\algocf@captiontext{#1}{#2}}}%
}%
\makeatother

\graphicspath{ {figures/} }

\renewcommand{\chaptermark}[1]{\markboth{\chaptername\ \thechapter.\ #1}{}}

\fancypagestyle{plain}{
  \fancyhf{} 
  \fancyfoot[LE,RO]{\thepage}
  }

\newboolean{showcomments}
\setboolean{showcomments}{true}
\ifthenelse{\boolean{showcomments}}
{ \newcommand{\mynote}[3]{
        \fbox{\bfseries\sffamily\scriptsize#1}
        {\small$\blacktriangleright$\textsf{\emph{\color{#3}{#2}}}$\blacktriangleleft$}}}
{ \newcommand{\mynote}[3]{}}

\usepackage{etoolbox}
\makeatletter
\patchcmd\@acf{\AC@acl}{\AC@foo}{}{}
\patchcmd\@acf{\AC@acl}{\AC@foo}{}{}
\patchcmd\@acf{\AC@foo}{\hskip\z@\AC@acl}{}{}
\patchcmd\@acf{\AC@foo}{\hskip\z@\AC@acl}{}{}
\makeatother

\def\DocumentTitle{Distributed systems and trusted execution environments: \\
                   Trade-offs and challenges}
\def\AuthorName{Rafael Pereira Pires}
\def\unine{Universit\'e de Neuch\^atel}

\title{\DocumentTitle}
\author{\AuthorName}
\newcommand{\ibbesgx}{\acs{IBBE}-\acs{SGX}\xspace}
\newcommand{\securestreams}{\textsc{SecureStreams}\xspace}
\newcommand{\cyclosa}{\textsc{Cyclosa}\xspace}
\newcommand{\mapreduce}{\textsc{MapReduce}\xspace}
\newcommand{\asky}{A-Sky\xspace}
\newcommand{\tor}{\ac{Tor}\xspace}
\newcommand{\peas}{PEAS\xspace}
\newcommand{\goopir}{GooPIR\xspace}
\newcommand{\tmn}{TrackMeNot\xspace}
\newcommand{\xsearch}{\textsc{X-S\small{earch}}\xspace}

\newcommand{\queryscrambler}{QueryScrambler\xspace}
\newcommand{\tmns}{\textsc{TMN}\xspace}
\newcommand{\luavm}{\mbox{\textsc{Lua}} interpreter\xspace}
\newcommand{\zmq}{\mbox{\textsc{ZeroMQ}}\xspace}
\newcommand{\y}{\textcolor{Green4}{\ding{51}}}
\newcommand{\n}{\textcolor{Red4}{\ding{55}}}
\newcommand{\bbw}{\emph{BBW}\xspace}
\newcommand{\accessmonitor}{\textsc{AccessControl}\xspace}
\newcommand{\writerproxy}{\textsc{WriterShield}\xspace}
\newcommand{\wrt}{with regard to\xspace}
\newcommand{\Wrt}{With regard to\xspace}

\hypersetup{%
    pdftitle=\DocumentTitle,%
    pdfauthor=\AuthorName,%
    pdfsubject={PhD dissertation, IIUN UniNE, Document created on \today},%
    pdfkeywords={thesis, dissertation, cloud computing, security, SGX},%
}

\makeatletter
\renewcommand*\@makechapterhead[1]{%
   \vspace*{30\p@}%
   {\parindent \z@ \raggedright \normalfont
     \ifnum \c@secnumdepth >\m@ne
         \huge\bfseries \@chapapp\space \thechapter
         \par\nobreak
         \vskip 20\p@
     \fi
     \interlinepenalty\@M
     \Huge \bfseries #1\par\nobreak
     \vskip 30\p@
   }}
\renewcommand*\@makeschapterhead[1]{%
   \vspace*{30\p@}%
   {\parindent \z@ \raggedright
     \normalfont
     \interlinepenalty\@M
     \Huge \bfseries  #1\par\nobreak
     \vskip 30\p@
   }}
\makeatother

\begin{document}
\pagenumbering{roman}

\thispagestyle{empty}
{
\centering
\includegraphics[height=2.5cm]{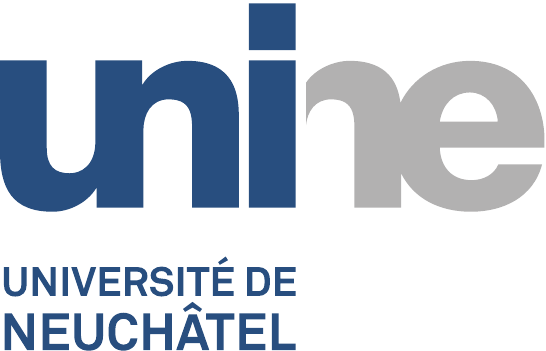}
\vspace{1cm}
{
\large
\\
\unine\\
Faculté des sciences\\
Institut d'informatique\\
}
\vspace{1cm}
\begingroup
\centering
\sffamily\bfseries\fontsize{26}{31.2}\selectfont
\DocumentTitle
\\[1.5cm]
\normalfont\large
par
\\[0.25em]
\sffamily\bfseries\Large
\textbf{\AuthorName}
\\[0.4in]
\normalfont\normalsize
Thèse
\\[0.5em]
présentée à la Faculté des sciences
\\[0.5em]
pour l'obtention du grade de Docteur \`es sciences
\\[1cm]
Acceptée sur proposition du jury:
\\[0.5em]
\textbf{Prof. Pascal Felber}, directeur de th\`ese
\\
\unine, Suisse
\\[0.5cm]
\textbf{Prof. Marcelo Pasin}, codirecteur de th\`ese
\\
Haute école Arc, Suisse
\\[0.5cm]
\textbf{Dr. Valerio Schiavoni}
\\
\unine, Suisse
\\[0.5cm]
\textbf{Prof. Daniel Lucani}
\\
Aarhus University, Danemark
\\[0.5cm]
\textbf{Prof. Ga\"{e}l Thomas}
\\
Telecom SudParis, France
\vfill
Soutenue le 3 d\'ecembre 2019

\endgroup
}
\cleardoublepage

\includepdf{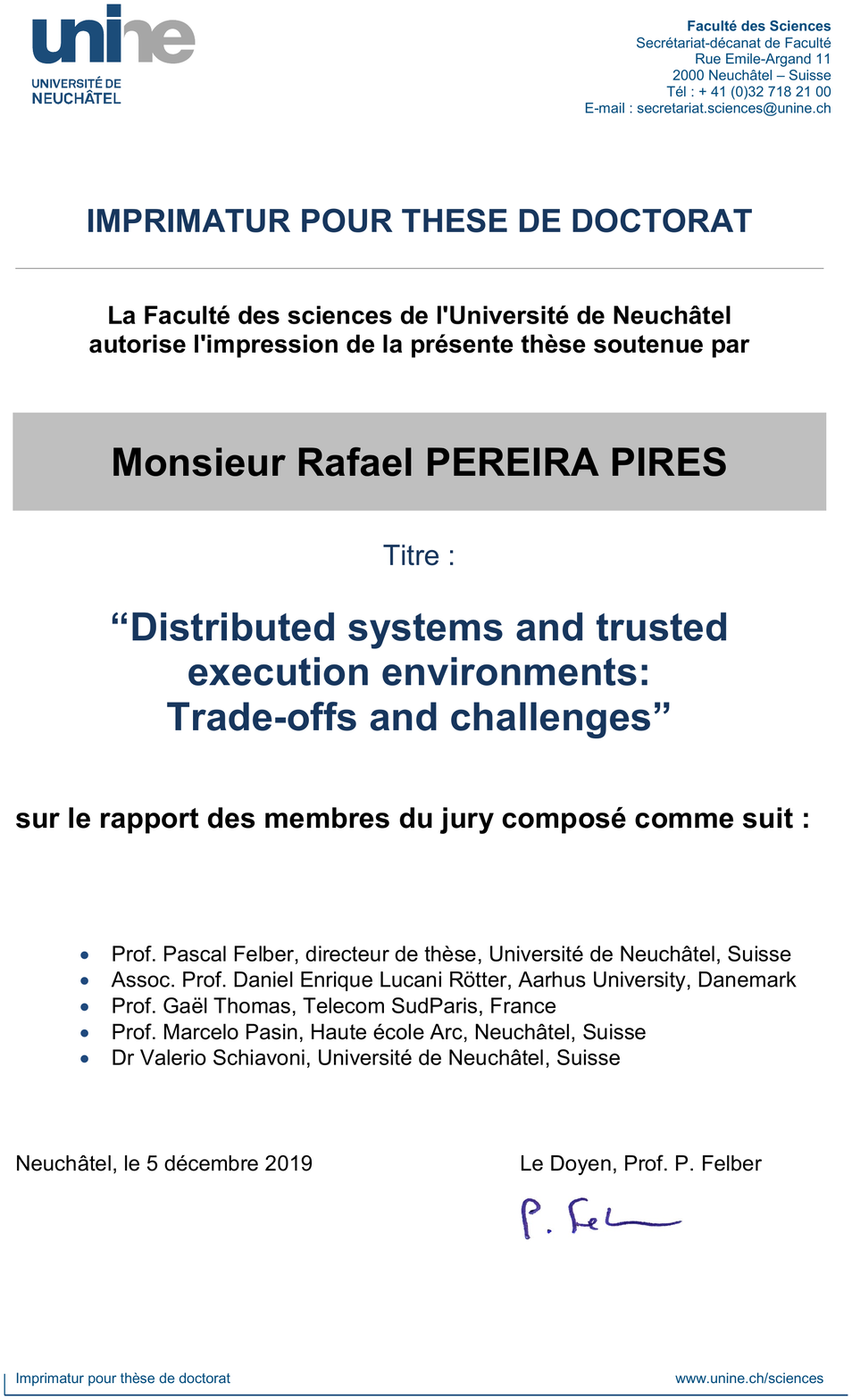}
\cleardoublepage

\chapter*{Abstract}
\addcontentsline{toc}{chapter}{Abstract}

Security and privacy concerns in computer systems have grown in importance with the ubiquity of connected devices.
Additionally, cloud computing boosts such distress as private data is stored and processed in multi-tenant infrastructure providers.
In recent years, \acp{TEE} have caught the attention of scientific and industry communities as they became largely available in user- and server-class machines.

\acp{TEE} provide security guarantees based on cryptographic constructs built in hardware.
Since silicon chips are difficult to probe or reverse engineer, they can offer stronger protection against remote or even physical attacks when compared to their software counterparts.
Intel \ac{SGX}, in particular, implements powerful mechanisms that can shield sensitive data even from privileged users with full control of system software.

Designing secure distributed systems is a notably daunting task, since they involve many coordinated processes running in geographically-distant nodes, therefore having numerous points of attack.
In this work, we essentially explore some of these challenges by using Intel \ac{SGX} as a crucial tool.
We do so by designing and experimentally evaluating several elementary systems ranging from communication and processing middleware to a peer-to-peer privacy-preserving solution.

We start with support systems that naturally fit cloud deployment scenarios, namely content-based routing, batching and stream processing frameworks.
Our communication middleware protects the most critical stage of matching subscriptions against publications inside secure enclaves and achieves substantial performance gains in comparison to traditional software-based equivalents.
The processing platforms, in turn, receive encrypted data and code to be executed within the trusted environment.
Our prototypes are then used to analyse the manifested memory usage issues intrinsic to \ac{SGX}.

Next, we aim at protecting very sensitive data: cryptographic keys.
By leveraging \acp{TEE}, we design protocols for group data sharing that have lower computational complexity than legacy methods.
As a bonus, our proposals allow large savings on metadata volume and processing time of cryptographic operations, all with equivalent security guarantees.

Finally, we focus our attention on privacy-preserving systems.
After all, users cannot modify some existing systems like web-search engines, and the providers of these services may keep individual profiles containing sensitive private information about them.
We aim at achieving indistinguishability and unlinkability properties by employing techniques like sensitivity analysis, query obfuscation and leveraging relay nodes.
Our evaluation shows that we propose the most robust system in comparison to existing solutions \wrt user re-identification rates and results' accuracy in a scalable way.

All in all, this thesis proposes new mechanisms that take advantage of \acp{TEE} for distributed system architectures.
We show through an empirical approach on top of Intel \ac{SGX} what are the trade-offs of distinct designs applied to distributed communication and processing, cryptographic protocols and private web search.

\textbf{Keywords:} security, privacy, \acs{TEE}, \acs{SGX}, distributed systems, communication, processing, cryptographic protocols, web search.

\cleardoublepage
 
\begin{otherlanguage}{french}
\chapter*{R\'esum\'e}
\addcontentsline{toc}{chapter}{R\'esum\'e}

Les problèmes de sécurité et de confidentialité des systèmes informatiques ont pris de l'importance avec l'omniprésence des périphériques connectés.
En outre, l'informatique en nuage accroît cette détresse, car les données privées sont stockées et traitées par des fournisseurs d'infrastructure hébergeant de multiples locataires.
Ces dernières années, les environnements d'exécution de confiance, ou TEE («~trusted execution environments~») ont attiré l'attention des communautés scientifiques et industrielles, car ils sont devenus largement disponibles sur des machines de type utilisateur et serveur.

Les TEE fournissent des garanties de sécurité basées sur des constructions cryptographiques intégrées au matériel.
Les puces de silicium étant difficiles à sonder ou étudier par ingénierie inverse, elles offrent une protection renforcée contre les attaques distantes, voire physiques, par rapport à leurs homologues logicielles.
En particulier, les extensions de protection logicielle Intel SGX («~software guard extensions~») implémentent de puissants mécanismes qui peuvent protéger les données sensibles même contre les utilisateurs privilégiés disposant du contrôle total du logiciel de système.

La conception de systèmes repartis sécurisés est une tâche particulièrement ardue, car ils impliquent de nombreux processus coordonnés exécutés dans des n\oe{}uds géographiquement distants, ce qui entraîne de nombreux points d'attaque.
Dans ce travail, nous explorons essentiellement certains de ces défis en utilisant Intel \ac{SGX} comme pierre angulaire.
Nous le faisons en concevant et en évaluant de manière expérimentale plusieurs systèmes élémentaires allant du logiciel médiateur de communication et de traitement à une solution de protection de la confidentialité pair-à-pair.

Nous commençons par des systèmes de support qui s'adaptent naturellement aux scénarios de déploiement dans le \textit{cloud}, à savoir~: des infrastructures de routage en fonction du contenu, de traitement par lots et de traitement de flux.
Notre logiciel médiateur de communication protège la phase de mise en correspondance d'abonnements avec des publications, qui est la plus critique, en l'effectuant à l'intérieur d'enclaves sécurisées.
Il permet des gains substantiels en performance par rapport aux équivalents traditionnels basés sur des logiciels.
Les plates-formes de traitement reçoivent à leur tour des données chiffrées et du code à exécuter dans l'environnement sécurisé.
Nos prototypes sont ensuite utilisés pour analyser les problèmes d'utilisation de mémoire qui sont inhérents à \ac{SGX}.

Nous visons ensuite à protéger des données très sensibles~: les clés cryptographiques.
En utilisant les \ac{TEE}, nous concevons des protocoles pour le partage de données de groupe présentant une complexité de calcul inférieure à celle des méthodes traditionnelles.
De plus, nos propositions permettent d'importantes économies quant au volume de méta-données produites et au temps de traitement des opérations cryptographiques, le tout avec des garanties de sécurité équivalentes.

Enfin, nous concentrons notre attention sur les systèmes préservant la confidentialité.
Après tout, les utilisateurs ne peuvent pas modifier certains systèmes existants, tels que les moteurs de recherche Web.
Les fournisseurs de ces services peuvent conserver des profils individuels contenant des informations confidentielles les concernant.
Nous visons à obtenir des propriétés d'indiscernabilité et indissociabilité en utilisant des techniques telles que l'analyse de sensibilité, l'obscurcissement de requêtes et l'utilisation de n\oe{}uds relais.
Notre évaluation montre que nous proposons le système le plus robuste par rapport aux solutions existantes en ce qui concerne les taux de ré-identification des utilisateurs et la précision des résultats de manière évolutive.

Dans sa globalité, cette thèse propose de nouveaux mécanismes tirant parti des \ac{TEE} pour les architectures de systèmes repartis.
Nous montrons par approche empirique au-dessus de Intel \ac{SGX} quels sont les compromis entre plusieurs modèles de conception distincts appliqués à la communication et au traitement répartis, aux protocoles cryptographiques et à la recherche privée sur le Web.

\textbf{Mots clés~:} sécurité, confidentialité, \ac {TEE}, \acs {SGX}, systèmes repartis, communication, traitement, protocoles cryptographiques, recherche sur le Web.
\end{otherlanguage}

\cleardoublepage
 
\thispagestyle{empty}
\begin{flushright}
\vspace*{5cm}
\emph{To you, who is reading this.}
\end{flushright}
\cleardoublepage
 
\chapter*{Acknowledgements}
\addcontentsline{toc}{chapter}{Acknowledgements}

Doing a PhD is not trivial, as a good friend might say.
We work hard, \emph{on est schlagés du galetas}, at any time, at weekends and on holidays.
And yet our masterpieces are rejected by evil reviewers.
Nice ones accept them though, and that is amazing. 
We celebrate and we travel.
I personally presented my work in meetings, conferences and seminars that took place in Switzerland, Denmark, Romania, France, Germany, Spain, Italy, Luxembourg and the USA.
It is the joy, but then it is gone.
Like addicts, we react by working hard for the next one.
C'est la vie~!
In spite of pros and cons, a PhD is made by people, many people.
And I dedicate this work to them.

First and foremost, I would like to express my gratitude to my advisors, Prof. Pascal Felber and Prof. Marcelo Pasin.
Pascal provided me with all the support I needed.
His cleverness in conceiving and assigning tasks and his amazing network of partners were crucial for this work's accomplishments.
Apart from that, he excels in every kind of sport.
We had a remarkable work meeting while hiking down a mountain in Sinaia, Romania, and went a few times for indoor climbing during lunch breaks.
Thanks for everything!

Pasin was responsible for my coming to Neuch\^atel. 
Once I arrived, he helped me with administrative stuff, gave me private lectures about computer science subjects (my years in industry made me somewhat rusty) and made excellent Brazilian barbecues at his place.
Along the way, we interacted in several technical discussions, project deliverables and also in career counselling.
Muito obrigado!

Many thanks to the external committee members Prof. Daniel Lucani and Prof. Ga\"el Thomas, who provided valuable comments and suggestions for improving this manuscript.
I am also really grateful to all the co-authors with whom I had the privilege to work.
Special thanks to Dr. Emanuel Onica, Dr. Stefan Contiu and Dr. Sonia Ben Mokhtar.

Dr. Valerio Schiavoni, who was also a committee member, provided me valuable comments.
We worked together in part of the results of this thesis.
He is skilled in subtle task assignment and impressively follows up the work of virtually all PhD students in the group.
But not only: he goes along by working hard on experiments and writing, even if we happen to be in a casino-hotel in Las Vegas at 3 a.m.
Grazie mille!

Thanks to all colleagues in the Department, including but not limited to 
Heverson Ribeiro,
Raluca Halalai,
Aur\'elien Havet,
Sukanya Nath,
Laurent Hayez and
Catherine Ikae.
Many thanks to Roberta Barbi and her discreet laugh,
Andrei Lapin and his rollers,
Yarco Hayduk and his colourful attire, 
Mascha Kurpicz and her language abilities,
Mirco Kocher and his taste for mushrooms, 
Maria Carpen-Amarie and her strong opinions,
Raziel Gómez and his salsa,
Rapha\"el Barazzutti and his washing machine,
Veronica Estrada and her \emph{sangre latino},
Christian G\"ottel and his motorbike,
Lars Nielsen and his denmarkness, 
Nils Schaetti and his silence, 
Isabelly Rocha and her cross-fit, 
Rémi Dulong and his musical instruments, and to
Peterson Yuhala and his kebabs. 
Every lunch, board games evenings and gatherings in bars, street festivals or in someone's place helped to build a friendly atmosphere.
This had the power to attenuate tensions of any kind (including pre-deadline ones) and gave strength and motivation to go to work every day.

\begin{otherlanguage}{french}
Merci au personnel de l'Institut d'informatique, y compris les secrétaires Émilie Auclair et Debora Mendes, les concierges Álvaro et Afrim Sadiku, et aux maîtres-assistants et professeurs Peter Kropf, Jacques Savoy, Alain Sandoz, Hervé Sanglard, Hugues Mercier et Lorenzo Leonini.
C'était toujours agréable de partager avec vous des moments aux soupers de fin d'année, à la torée neuchâteloise en été et dans l'institut.

Un grand merci à Dorian Burihabwa, qui a relu ce texte et suggéré des corrections.
D'ailleurs, malgré ses recommandations pour les mauvais films, j'ai bien apprécié ses visites quotidiennes dans notre bureau (S\'eb et moi) et sa disponibilit\'e en étant toujours pr\^et \`a boire un verre en soirée ou un brunch les week-ends.

Je remercie en particulier mon pote et compagnon de bureau, S\'ebastien Vaucher.
On a bossé ensemble sur quelques projets, et il m'a appris pas mal de choses.
Ses connaissances sur la Suisse et sur tous les r\'eglements de n'importe quoi sont impressionnantes.
Mais c'est surtout sa bonne humeur qui m'a fait prendre plaisir de passer mon temps au bureau. Merci bien~!

S\'ebastien et R\'emi m'ont appris un tas d'expressions francophones.
Quelques-unes restent pour moi un mystère et boule de gomme, mais au moins je sais maintenant qu'il ne faut jamais pousser Mémé dans les orties, malgré le fait que parfois on tape d'abord et ensuite on interroge.
En fait, Dorian, Rémi et S\'ebastien ont été les responsables du d\'eclenchement de mon fran\c{c}ais. Merci les gars~!
\end{otherlanguage}

Meu muito obrigado vai aos amigos do Brasil que vieram me visitar em Neuch\^atel: Edmar Ara\'ujo, Ana e Cesar Guerra; e aos que me acolheram em Porto Alegre (e em Paris) --- Marcelo Neves, Mônica Marcuzzo e Rubens Belusso ---, em Floripa --- Luiza Delpretto e Edmar ---, em Miami --- Roberta e Juliano Reckziegel ---, e em Barcelona/Melbourne --- Bárbara Cardoso e Danillo Santos. 
Agradeço também a Natália Bortolás, que fez parte dessa jornada.

Enfim, agradeço imensamente à minha família.
Meus pais, Jorge e Rejane, irmãos Gaspar e Bárbara, sobrinhos e sobrinha, avós, tias e tios, cunhada e cunhado, primas e primos.
O incentivo, suporte e carinho de vocês são as bases sobre as quais todas as minhas conquistas são construídas.

Well, until here I wrote in 6 languages, cited 11 countries and 57 names (non-exhaustive list), who in turn are citizens of 20 countries.
It is a good indication that this work is the result of a multi-cultural environment and had many, many contributors.
Vielen Dank an alle.

\cleardoublepage
 
\newcommand{\listacronymname}{List of acronyms}
\renewcommand{\listfigurename}{List of figures}
\renewcommand{\listtablename}{List of tables}
\renewcommand{\listalgorithmcfname}{List of algorithms}

\tocloftpagestyle{plain}
\tableofcontents
\cleardoublepage

\chapter*{\listacronymname\markboth{\listacronymname}{}}
\addcontentsline{toc}{chapter}{\listacronymname}
\begin{multicols}{2}
\begin{acronym}[IND-CCA]
\acro{ABE}{attribute-based encryption}
\acro{AE}{authenticated encryption}
\acro{AES}{advanced encryption standard}
\acro{AESM}{application enclave service manager}
\acro{AMD}{advanced micro devices}
\acro{ANOBE}{anonymous broadcast encryption}
\acro{AOL}{America online}
\acro{API}{application programming interface}
\acro{ARM}{advanced RISC machine}
\acro{ASLR}{address space layout randomisation}
\acro{ASPE}{asymmetric scalar-product preserving encryption}
\acro{ASID}{address space identifier}
\acro{BE}{broadcast encryption}
\acro{BIOS}{basic input/output system}
\acro{CA}{certification authority}
\acro{CDF}{cumulative distribution function}
\acro{CBR}{content-based routing}
\acro{CPU}{central processing unit}
\acro{CSV}{comma separated value}
\acro{CTR}{counter mode}
\acro{DAG}{directed acyclic graph}
\acro{DH}{Diffie-Hellman}
\acro{DMA}{direct memory access}
\acro{DoS}{denial of service}
\acro{DRAM}{dynamic random-access memory}
\acro{ecall}{enclave call}
\acro{ECC}{elliptic curve cryptography}
\acro{ECDH}{elliptic-curve Diffie–Hellman}
\acro{ECDSA}{elliptic curve digital signature algorithm}
\acro{EDL}{enclave definition language}
\acro{EK}{endorsement key}
\acro{EPC}{enclave page cache}
\acro{ES}{encrypted state}
\acro{FE}{functional encryption}
\acro{GCM}{Galois counter mode}
\acro{glibc}{GNU C library}
\acro{GMP}{GNU multiple precision arithmetic library}
\acro{GPG}{GNU privacy guard}
\acro{GPU}{graphics processing unit}
\acro{HE}{hybrid encryption}
\acro{HE-PKI}{hybrid encryption with public key}
\acro{HE-IBE}{hybrid encryption with identity-based encryption}
\acro{HIBE}{hierarchical identity-based encryption}
\acro{HMAC}{keyed-hash message authentication code}
\acro{HTTP}{hypertext transfer protocol}
\acro{IAS}{Intel attestation service}
\acro{IND-CCA}{indistinguishability under non-adaptive and adaptive chosen ciphertext attack}
\acro{IoT}{Internet of things}
\acro{IBBE}{identity-based broadcast encryption}
\acro{IBE}{identity-based encryption}
\acro{IO}[I/O]{input and output}
\acro{IP}{Internet protocol}
\acro{IP2}[IP]{intellectual property}
\acrodefplural{IP2}[IPs]{intellectual properties}
\acro{ISA}{instruction set architecture}
\acro{ISV}{independent software vendor}
\acro{IV}{initialisation vector}
\acro{JMH}{Java microbenchmark harness}
\acro{JSON}{JavaScript object notation}
\acro{JVM}{Java virtual machine}
\acro{LDA}{latent Dirichlet allocation}
\acro{LoC}{line of code}
\acrodefplural{LoC}[LoC]{lines of code}
\acro{LLC}{last level cache}
\acro{LTS}{long term support}
\acro{MAC}{message authentication code}
\acro{MA}[M\&A]{merger and acquisition}
\acrodefplural{MA}[M\&A]{mergers and acquisitions}
\acro{MC}{memory controller}
\acro{ME}{management engine}
\acro{MEE}{memory encryption engine}
\acro{MMU}{memory management unit}
\acro{ocall}{outside call}
\acro{OS}{operating system}
\acrodefplural{OS}[OSes]{operating systems}
\acro{PBC}{pairing-based cryptography}
\acro{P2P}{peer-to-peer}
\acro{PGP}{pretty good privacy}
\acro{PIR}{private information retrieval}
\acro{PKE}{public key encryption}
\acro{PKI}{public key infrastructure}
\acro{PMC}{performance monitoring counters}
\acro{PRM}{process reserved memory}
\acro{PSE}{platform services enclave}
\acro{PSW}{platform software}
\acro{pubsub}[pub/sub]{publish/subscribe}
\acro{RAM}{random-access memory}
\acro{REST}{representational state transfer}
\acro{RMW}{read-modify-write}
\acro{ROP}{return-oriented programming}
\acro{RPC}{remote procedure call}
\acro{RSA}{Rivest–Shamir–Adleman}
\acro{RSS}{RDF Site Summary}
\acro{RTT}{round-trip time}
\acro{S3}{simple storage service}
\acro{SCBR}{secure content-based routing}
\acro{SCONE}{secure container environment}
\acro{SDK}{software development kit}
\acro{SEV}{secure encrypted virtualisation}
\acro{SGX}{software guard extensions}
\acro{SHA}{secure hash algorithm}
\acro{SME}{secure memory encryption}
\acro{SoC}{system on chip}
\acro{SRAM}{static random-access memory}
\acro{SSD}{solid-state drive}
\acro{SPI}{serial peripheral interface}
\acro{STL}{standard template library}
\acro{TA}{trusted authority}
\acro{TCB}{trusted computing base}
\acro{TCP}{transmission control protocol}
\acro{TEE}{trusted execution environment}
\acro{TLB}{translation lookaside buffer}
\acro{TLS}{transport layer security}
\acro{Tor}{the onion router}
\acro{TPM}{trusted platform module}
\acro{UUID}{universally unique identifier}
\acro{VDR}{virtual data room}
\acro{VM}{virtual machine}
\acro{YCSB}{Yahoo! cloud serving benchmark}
\end{acronym}
 \end{multicols}
\cleardoublepage

\advance\cftfignumwidth 0.5em\relax
\setlength{\cftfigindent}{0cm}
\listoffigures
\addcontentsline{toc}{chapter}{\listfigurename}
\cleardoublepage

\pagenumbering{arabic}

\chapter{Introduction}
\label{sec:intro}
\acresetall

This work explores design strategies for distributed systems that leverage \acp{TEE}. 
We aim at achieving better security and privacy guarantees while maintaining or improving performance in comparison to existing equivalent approaches.

\section{\label{sec:intro:context}Context and motivation}
Legacy systems perform access control by separating software roles into privileged and user modes.
While privileged system software like hypervisors and \acp{OS} control machine resources, user-mode applications must rely on system software correctness, \ie, that they are free of exploitable bugs, and honesty, \ie, that they are not malicious.

The correctness aspect is already hard to verify.
After all, system software is very complex and comprises millions of \acp{LoC}.
To illustrate, more than \num{10000} computer systems' vulnerabilities are being discovered each year~\cite{cve:2019:vulnerabilities} and the number of software bugs is proportional to its size.
Presuming that an adversary would be able to exploit one of them is not a far-fetched assumption.

When it comes to honesty though, confidence may be even lower.
With the raise of cloud computing, many computer systems are deployed on third-party infrastructure providers.
The reasons for doing so are mostly related to the large costs of acquiring and maintaining private data centres.
In such shared environments, at least the hypervisors~\cite{popek:1974:virtualization} are under the provider's control.
Even if one trusts the infrastructure provider and all of its employees with physical access or administrative credentials to the machines, still they would have to hope that other tenants running on the same platforms are not malicious.

A large code-base consisting of system software on which user applications must trust is therefore not the best approach from a security standpoint.
\acp{TEE}, in turn, invert this logic.
They make it possible for only a small piece of code to be considered safe, \ie, belonging to the \ac{TCB}.
This dramatically reduces the trust surface, as in this case it suffices to believe that the \ac{TEE} hardware implementation is correct and has no backdoors, apart from having confidence on the reduced piece of software that runs in isolation, \ie, within an \emph{enclave}.

In the last quarter of 2015, a few months before this work started, Intel released the first commercially available machines with \ac{SGX}~\cite{snyder:2015:skylake}.
It was the first \ac{TEE} accessible in consumer-grade computers that offered out-of-the-box attestation mechanisms in conjunction with memory encryption, integrity and freshness guarantees.
These assurances were attainable by user applications that could then be protected from higher privileged entities like the hypervisor or the operating system.
Up to this date, it is still unmatched by alternatives (see \Cref{sec:back:tees}).

With processes that coordinate over multiple interconnected machines, distributed systems face attacks on all fronts of their deployment.
Reaching dependability~\cite{avizienis:2004:dependability} in such systems is not a trivial task.
However, \acp{TEE} provide an opportunity to achieve this goal.
This thesis explores these possibilities through the design and evaluation of experimental prototypes representative of fundamental distributed systems relying on \acp{TEE}---materialised with Intel \ac{SGX}---to offer security and privacy.
Doing so, we wish to answer the following questions:
\begin{itemize}
\item How can \acp{TEE} help to achieve security and privacy in distributed systems?
\item What are the drawbacks?
\item What benefits may come from using \acp{TEE}?
\end{itemize}

\subsubsection{Assumptions}
We assume that Intel \ac{SGX} behaves correctly, \ie, there are no bugs or backdoors.
Additionally, we do not deal with side-channel attacks against \ac{SGX}.
We consider such attacks outside the scope of this dissertation and that the research community provides some solutions that could possibly be incorporated (see \Cref{sec:sgx:vulnerabilities}).
Furthermore, we are also aware that \ac{DoS} attacks cannot be prevented, \eg malicious agents might drop requests or refuse to initialise enclaves.
Besides, we do not handle rollback attacks of persistent data. Standard solutions with monotonic counters could be used, although we are aware that they might be ineffective (see \Cref{sec:sgx:platservices}).
Finally, we assume that all the cryptographic primitives and libraries used in enclaves are trusted and cannot be forged.

\section{Contributions}

The contributions of this work are as follows.
\begin{itemize}
\item 
Design, implementation and evaluation of \textbf{\ac{SCBR}}, the first system that demonstrates the practical benefits of \ac{SGX} for privacy-preserving \ac{CBR}.
We protect compute-intensive matching operations in the trusted environment, so that efficient algorithms that operate on plaintext data can be used. 
As a consequence, we reach performance gains of one order of magnitude in comparison to a software-only solution with analogous guarantees.
We also analyse \ac{SGX} overheads when surpassing its memory limits.
This work was done in collaboration with Christof Fetzer, from the Technical University of Dresden, Germany.
\item 
Proposal of \textbf{Lightweight \mapreduce}, a processing framework based on a programming model extensively used for parallel data processing in distributed environments.
We ported a Lua interpreter engine to run inside secure enclaves and leverage it as execution unit that operates on code and data provisioned in encrypted form.
From the usability perspective, a user can just write \mapreduce scripts and let the framework handle data encryption and dissemination.
Besides, we observe the performance influence of going beyond the \ac{LLC} in enclave executions.
Daniel Gravril and Emanuel Onica, from the Alexandru Ioan Cuza University of Ia\c{s}i, Romania, collaborated with this work.
\item 
\textbf{SecureStreams}: a reactive middleware built on top of Lua libraries.
Its architecture relies on Lua \ac{VM} pairs on each node, \ie, one running inside enclaves and another outside.
This way, only sensitive data processing is relayed to trusted environments, while message queuing and the pipeline management is kept outside.
We analyse performance losses when compared to unsafe executions in terms of throughput and scalability.
This was joint work with Aur\'elien Havet and Valerio Schiavoni, from our Institute, and Romain Rouvoy, from the University of Lille, France.
\begin{table}
	\centering
	\caption{\label{tab:overview}Papers related to this dissertation.}
\begin{tabular}{@{}clcccc@{}}
\toprule
\multicolumn{1}{l}{Chapter} & Keyword   & Section & Publication & Source code & Venue \\ \midrule
\multirow{4}{*}{\ref{chap:background}} 
& \acs{SEV} vs. \acs{SGX} & \ref{sec:amdsev} & \cite{gottel:2018:sevsgx} &  & SRDS'18 \\
& Malware  & \ref{sec:sgx:vulnerabilities} & \cite{mogage:2019:malware} &  & SRDS'19  \\
& SecureCloud & \ref{sec:securecloud} & \cite{kelbert:2017:securecloud}  &  & DATE'17 \\
& Scheduling & \ref{sec:securecloud} & \cite{vaucher:2018:sched} & \cite{vaucher:2017:schedulersource} & ICDCS'18 \\
  \cmidrule(l){1-6} 
\multirow{3}{*}{\ref{chap:clouds}} & \acs{SCBR} & \ref{sec:scbr} & \cite{pires:2016:scbr} & \cite{pires:2016:scbrsource} & Middleware'16 \\
                            & MapReduce & \ref{sec:lwmp} & \cite{pires:2017:lwmp} & \cite{pires:2017:lwmrsource} & CCGRID'17 \\
                            & \securestreams & \ref{sec:sstreams} & \cite{havet:2017:securestreams} & \cite{havet:2017:sstreamssource} & DEBS'17 \\ \cmidrule(l){1-6}
\multirow{2}{*}{\ref{chap:sharing}} & \ibbesgx & \ref{sec:ibbe} & \cite{contiu:2018:ibbe} & \cite{pires:2017:ibbesource}  & DSN'18 \\
                            & \asky  & \ref{sec:asky} & \cite{contiu:2019:asky} & \cite{pires:2018:askysource}  & SRDS'19 \\ \cmidrule(l){1-6}
\multirow{2}{*}{\ref{chap:privacy}} & X-Search & \ref{sec:xsearch} & \cite{mokhtar:2017:xsearch} &  & Middleware'17 \\
                            & \cyclosa  & \ref{sec:cyclosa} & \cite{pires:2018:cyclosa} &  & ICDCS'18 \\ \cmidrule(l){1-6}
\end{tabular}

 \end{table}
\item 
Introduction of \textbf{\ibbesgx}, a new cryptographic access control extension for collaborative editing of shared data.
Thanks to \acp{TEE}, we are able to cut part of the computational complexity of an \ac{IBBE} scheme.
Shielding a master key inside the trusted environment allows us to spare considerable computation time by avoiding the usage of an \ac{IBBE} public-key during encryption.
Because of this, we improve performance by orders of magnitude in comparison to \ac{HE}, both in terms of membership changes and produced metadata, consequently also profiting in storage and network usage.
\item
To handle anonymity among group members, \textbf{A-Sky} is presented.
Instead of relying on costly asymmetric cryptography like \ac{PGP}, secure enclaves allow A-Sky to create key envelopes using efficient symmetric operations hence achieving faster execution times and shorter ciphertexts.
In addition, we only require the usage of a \ac{TEE} proxy for writing to the shared storage and leave the dominant data consumption operations directly in charge of rightful readers.
We propose an end-to-end system based on micro-services and \iac{REST} interface before evaluating its performance and scalability.
Both \ibbesgx and \asky resulted from our collaboration with Stefan Contiu and Laurent R\'eveill\`ere, from the University of Bordeaux, France.
Additionally, S\'ebastien Vaucher, from our Institute, contributed in both.
\item
\textbf{\xsearch}, in turn, leverages \acp{TEE} for providing a privacy-preserving solution for Web search.
In order to prevent service providers from keeping accurate user profiles and therefore obstruct privacy breaches, we propose a \ac{SGX} proxy between users and search engines.
From the service provider's perspective, queries originate from another source, thus becoming more difficult to link them back to their issuing users.
Since the proxy operates in the trusted environment, we can safely store past user queries and use them to obfuscate requests, so that the search engine cannot distinguish real from fake queries.
These strategies combined offer stronger privacy guarantees and outperform previous approaches in latency and throughput.
Our conjoint efforts with Sonia Ben Mokhtar and Antoine Boutet, from the University of Lyon, France, led to this work.
\item
Finally, we contribute with the proposal of \textbf{Cyclosa}.
Instead of a centralised proxy, we now spread the load across a \ac{P2P} network of \ac{SGX} relay nodes.
Each one may issue their own queries through the decentralised network and also forward requests to the search engine on behalf of others, always having enclaves as intermediaries.
Obfuscation is done through different paths, thus facilitating the delivery of results by simply discarding those that handle fake queries and therefore achieving perfect results' accuracy. 
Additionally, we propose an adaptive privacy protection solution based on sensitivity analysis that reduces the risk of user re-identification.
With that, we solve the issue of possibly being blacklisted by search engines because of centralised proxies while meeting scalability and accuracy.
In addition to Ben Mokhtar and Boutet, Sara Bouchenak---also from Lyon---contributed in \cyclosa.  Likewise, David Goltzsche and R\"udiger Kapitza, from the Technical University of Braunschweig, Germany, joined the team. Valerio Schiavoni, from our Institute, contributed in both \xsearch and \cyclosa.
\end{itemize}

These contributions were previously published in conference proceedings in papers co-authored by Pascal Felber and Marcelo Pasin, my supervisors, and me. 
\Cref{tab:overview} establishes the relation between publications and the corresponding section in this document where they are mentioned.
Additionally, we indicate the publication venue and a reference to the corresponding source code, when available.

\section{Outline}

This manuscript is organised in six chapters, the first being this introduction.
The remaining chapters are arranged as follows.

\textbf{\Cref{chap:background}} presents the background on \acp{TEE} and particularly on Intel \ac{SGX}.
We focus on their main features and peculiarities, which finally endorse our choice for \ac{SGX}. 
The SecureCloud project is briefly described before we present the state-of-the-art by mentioning related work specific to each of our contributions.

In \textbf{\Cref{chap:clouds}}, we explore the cloud scenario depicted in \Cref{sec:intro:context} by designing and evaluating systems supposed to be deployed in such unreliable environments.
A communication middleware (\Cref{sec:communication}) and two distributed processing frameworks are presented (\Cref{sec:processing}).
\Cref{sec:scbr} presents \ac{SCBR}, where we essentially put a \ac{pubsub} matcher unit inside an enclave, \ie, the component that handles sensitive data.
The processing frameworks, one for batch execution (\Cref{sec:lwmp}) and another for processing stream of events (\Cref{sec:sstreams}), used a Lua script interpreter within enclaves, so that code and data can come on demand.
The goal in this chapter is to quantify the \ac{SGX} performance implications in the context of practical systems, besides learning about the main design concerns involved in building these secure applications.
Our results show that whenever cache and \ac{EPC} limits are surpassed, there are performance penalties.
Although not surprising, these results---along with the designs---contribute towards sensitive data processing in untrusted public clouds.

\acp{TEE} are not only useful for shielding legacy artefacts like script processors.
In fact, simply porting memory-eager systems may not be a good idea as they suffer considerable overheads in the current version of \ac{SGX}.
Instead, in \textbf{\Cref{chap:sharing}}, we capitalise on a trusted environment for protecting cryptographic keys.
Particularly, for group communication and data sharing.
In \Cref{sec:ibbe}, we generate a master secret within an enclave, from where it never leaves in unencrypted form.
Since this key can only be used in protected execution,  we can cut some computational cost of an \ac{IBBE} scheme by using simpler cryptographic primitives.
That brings significant advantages when compared to \ac{HE} in terms of metadata size and performance.
At the same time, it obviates the need for relying on a \ac{PKI}. 
Apart from that, we also design in \Cref{sec:asky} a distributed system that offers anonymity among group members.
For that, we use enclaves to harbour keys and group membership data while attaching some metadata to encrypted files that can then be stored in untrusted clouds.
A user who is member of the group for a given file will be able to retrieve the key from the attached metadata.
The take-away message is that \acp{TEE}, apart from shielding traditional applications, can also have an essential role in the design of efficient cryptographic protocols.

Until this point, we covered the design of distinct server-side systems.
Nevertheless, often times users must rely on established services which are unlikely to change.
That is the case for web search engines, which are able to quietly track user activities in spite of correctly answering their queries.
Because of this, \textbf{\Cref{chap:privacy}} handles privacy-preserving solutions from the user perspective.
Our first approach, described in \Cref{sec:xsearch}, adds a centralised proxy to hide user identities from search engines.
At the same time, we obfuscate requests using past queries that are securely kept within enclaves, so that the introduced noise hinders the service provider's ability to keep meaningful profiles.
But a toll must be paid: the introduced fake queries have consequences in the quality of results, which need to be filtered.
Moreover, its centralised nature allows search engines to easily block the \ac{SGX} proxies, as the proxies potentially issue considerable amounts of traffic.
Besides, with a growing number of users they would become performance bottlenecks.
To counter these disadvantages, we propose in \Cref{sec:cyclosa} a fully decentralised approach.
It consists in a user-side peer-to-peer solution that combines performance, scalability, fault tolerance and accuracy of results, since in this case no result filtering is needed.
	
Finally, \textbf{\Cref{chap:conclusion}} revisits our achievements on the three parts: cloud deployment systems, crypto schemes and client-side privacy protection.
Before concluding, we discuss some avenues for further research.

\chapter{Background and related work}
\label{chap:background}
\acresetall

Integrity and confidentiality of applications are enforced by means of logical isolation.
Virtual address spaces and privileged instructions, for instance, are hardware mechanisms used by \acp{OS} to prevent unauthorized processes from getting access to potentially sensitive pieces of memory (\eg, belonging to other users) or operations (\eg, interacting with input and output peripherals).
Likewise, hardware virtualisation extensions \cite{intel:2019:vmx,amd:2019:amdv} are used by hypervisors to isolate multiple \acp{OS} running on the same physical host.
Such a model requires that both the \ac{OS} and the hypervisor be trusted by a user application, since it uses their services.

In multi-tenant setups, as in cloud environments, at least the hypervisor is under the provider's control.
Despite the legal protection assured by possible contractual agreements between applications and infrastructure providers, security concerns may arise from this separation of administrative domains.
The potential threats are numerous: distinct law jurisdictions, corrupt employees with privileged access, malicious co-located tenants etc.

From a technical perspective, \acp{OS} and hypervisors are composed of millions of lines of source code, resulting in bloated \acp{TCB}.
The number of bugs, which may be reduced by means of \ac{OS} formal proofs \cite{klein:2009:osformalverif}, is proportional to software size~\cite{gaffney:1984:bugs}.
Other tenants running on the same computers can profit from them to gain access to sensitive data.
Moreover, system administrators of the cloud provider have access to all application data.
Hosted applications can further be compromised either by privileged administrators acting maliciously or as a consequence of having their credentials hijacked~\cite{zetter:2016:hacksysadmins}.

\section{\label{sec:back:tees}Trusted execution environments}

Traditional hardware isolation mechanisms protect user applications from one another (memory control), multiple \acp{OS} from one another (virtualisation), and the system software from user applications (privileged instructions).
However, none of them provide isolation to user applications from the system software.
Apart from logical isolation, other mechanisms must also be employed, for instance, when attackers get physical access to platforms.
In such cases, they could wiretap memory buses or read memory content through cold boot attacks~\cite{halderman:2009:coldboot,gruhn:2013:coldboot}.

Cryptography is widely used to protect data in transit, \eg, using \ac{TLS} \cite{dierks:2015:rfc5246}, or at rest, for storage.
When data are processed though, ciphertexts must be deciphered before loaded into system memory.
During processing, data confidentiality is hence threatened by privileged users and physical attackers.
Cryptographic approaches like homomorphic encryption could be employed, although they are not considered practical due to their prohibitive cost~\cite{gottel:2018:sevsgx}.

\begin{table}[t]
	\centering
	\caption{\label{tab:tee:comparison}Comparison of TEEs.}
\begin{tabular*}{\textwidth}{r @{\extracolsep{\fill}} c  c  c  c }
		\toprule
		&\acs{TPM} & TrustZone & \acs{AMD} \acs{SEV} & Intel \acs{SGX} \\
		\midrule
		Released & 2009 & 2005 & 2016 & 2015 \\
		Running mode & coprocessor &  special mode  & hypervisor & user-level \\
		\acs{ISA} & n/a & \acs{ARM} & x86\_64 & x86\_64 \\
		Executes arbitrary code & \n & \y & \y & \y \\
		Secret hardware key  & \y & \n* & \y & \y \\
		Attestation and Sealing& \y & \n*  & \y & \y \\
		Memory encryption & n/a & \n & \y & \y \\
		Memory integrity & n/a & \n  & \n & \y \\
		Resilient to wiretap & \n & \n  & \y & \y \\
		\acs{IO} from \acs{TEE} & \y & \y  & \y & \n \\
		\acs{TEE} usable memory limit & n/a & system \acs{RAM} & system \acs{RAM} &\SI{93.5}{\mebi\byte}\\
		\acs{TCB} & \begin{tabular}{c}Data and funcs \\in the \acs{TPM} chip\end{tabular} & \begin{tabular}{c}\acs{OS} and apps in \\the secure world\end{tabular} & Entire \acsp{VM} & \begin{tabular}{c}Trusted \\app partition\end{tabular} \\
		\bottomrule
\end{tabular*}
{$\!\begin{aligned}
	\textrm{n/a:}&\textrm{~not applicable} \\
	\textrm{*:}&\textrm{~possible with additional hardware}
\end{aligned}$}
 \end{table}

To solve issues like isolation of user applications from system software and providing cryptographic layers of isolation against physical attacks, \acp{TEE} were proposed.
Different implementations vary in terms of features, which we summarize in \Cref{tab:tee:comparison}.
In this overview, we leave aside academic proposals (\eg, Sanctum~\cite{costan:2016:sanctum}, Bastion~\cite{champagne:2010:bastion} and AEGIS~\cite{suh:2003:aegis}) and focus on popular commercial solutions:
\begin{itemize}
\item Trusted platform modules (\acsu{TPM}s) are dedicated coprocessors, and therefore do not allow for the execution of general purpose software.
Their \ac{TCB} boundary is the microcontroller package, which contain the memories they need to operate.
Memory integrity and confidentiality guarantees serve hence no purpose, as memories are inside this package and cannot be tampered with or wiretapped without physical violation.
Wiretapping its \ac{IO} bus, however, could reveal sensitive information, \eg, data unsealed by the \ac{TPM}.
Besides, the \ac{TPM} state can only be reset by physical presence, which makes it suitable for preventing remote attacks.
The \ac{TPM} specification was standardised in 2009 through ISO/IEC 11889.
\item TrustZone is a \ac{TEE} for the \ac{ARM} \ac{ISA} that switches between secure and unsecure modes.
It provides isolation by means of interrupts routing and restrictions on the memory bus and \ac{MMU}. However, there is no built-in cryptographic primitives that could provide a root of trust for persistent sealing and attestation \cite{sabt:2015:dualexec}.
Since \ac{ARM} is a \ac{IP2} core provider and not a chip manufacturer, vendors might implement such additional security, although most often they are not publicly disclosed. 
In 2018, \ac{ARM} announced families of semiconductor \acp{IP2} called CryptoIsland and CryptoCell that can possibly be integrated to \ac{ARM} \acp{CPU} in a single \ac{SoC}. 
Conceptually, that would be like having a \ac{TPM} inside the processor package, allowing the usage of hardware keys as root of trust (like Intel \ac{SGX}).
\item \acs{AMD} \ac{SEV} provides automatic inline encryption and decryption of memory traffic, granting confidentiality for data in use by \acp{VM}.
Cryptographic operations are performed by hardware and are transparent to applications, which do not need to be modified.
Keys are generated at boot time and secured in a coprocessor integrated to the \ac{SoC}.
It was conceived for cloud scenarios, where guest \acp{VM} might not trust the hypervisor.
Apart from including the whole guest \ac{OS} in the \ac{TCB}, it does not provide memory integrity and freshness guarantees. Rowhammer attacks~\cite{kim:2014:rowhammer} might corrupt data and rollback attacks are not detected.
\item Intel \ac{SGX} is primarily conceived for shielding micro-services, so that the \ac{TCB} would be minimised.
Automatic memory encryption and integrity protection are performed by hardware over a reserved memory area fixed at booting time, defined in the \ac{BIOS} and limited to \SI{128}{\mebi\byte}.
Whatever is kept in this area is automatically encrypted and integrity checked by hardware.
The trust boundary is the \ac{CPU} package, which holds hardware keys upon which attestation and sealing services are built.
Applications are partitioned into trusted (shielded in \emph{enclaves}) and untrusted parts.
The \ac{OS} is considered untrusted, which prevents enclaves from directly issuing system calls.
\ac{SGX} enclaves are subject to side-channel attacks and denial of service.

\end{itemize}
We further detail these technologies in the following sections.

\subsection{Trusted platform modules}

\Iac{TPM}~\cite{tcg:2011:tpm} is an independent and specialized tamper-resistant coprocessor.
Its root of trust comes from an integrated asymmetric pair of \acp{EK} physically burnt in the component.
The private endorsement key is never made available to users, irrespective of their system privileges.
Rather than providing general purpose computations, \acp{TPM} perform a small set of security operations such as random number generation, cryptographic hash functions, \eg, \ac{SHA}, public- and symmetric-key cryptographic algorithms, \eg, \ac{AES} and \ac{RSA}. 
Some implementations also provide trusted time and monotonic counters (see \Cref{sec:sgx:platservices}).
Typically, \iac{TPM} securely holds software measurements and cryptographic keys which may be used for trusted boot, remote attestation and data sealing.

\begin{figure}[t]
	\centering
	\includegraphics{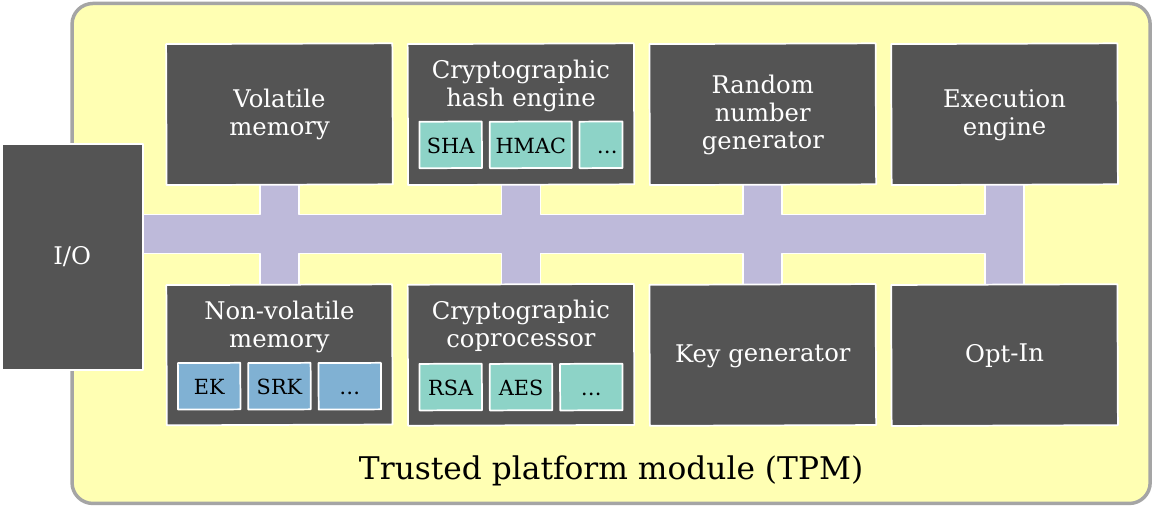}
	\caption{\label{fig:tee:tpm}TPM components.}
\end{figure}

Access to \acp{TPM} is based on the \emph{ownership} granted to whom first sets a shared secret.
Access control is operated by a component called \emph{Opt-In}, which maintains flags associated to the \ac{TPM}'s state.
When someone takes ownership, a storage root key (SRK) is created and can be used for sealing data in persistent storage.
Only the owner is able to remotely use the \ac{TPM} in its full capabilities, although it may be factory reset through the assertion of physical presence.
Nevertheless, other users can perform operations  allowed by the owner, like querying software measurements, storing keys and using crypto primitives.
\Cref{fig:tee:tpm} illustrates internal components of a \ac{TPM}.

Being a separate component, \acp{TPM} are detached from other isolation mechanisms such as memory page tables and privileged instructions.
Security is therefore highly dependant on how \ac{TPM} chips are wired to main processors, what are the privileges users must hold to get access to such chips and whether adversaries might have physical access. Wiretapping the \ac{TPM}'s \ac{IO} lines may compromise sensitive data.

\subsection{\label{sec:back:trustzone}ARM TrustZone}

ARM TrustZone \cite{arm:2009:trustzone,pinto:2019:trustzone} is a popular \ac{TEE}, mostly used in low-energy or mobile devices.
It provides a protection domain called \emph{secure world} that cannot be accessed by the \emph{normal world}.
Each physical processor core is viewed as two virtual ones: secure and non-secure, which are used in a time-sliced manner. 
The context switch is made through a \emph{monitor mode} (ARMv8-A) accessed by a special instruction, or through hardware exception mechanisms (ARMv8-M).
At booting time, a secure firmware initialises the platform and decides what is part of secure and non-secure worlds by configuring the interrupt controller and setting up memory partitions and peripherals.
Then, the processor switches to the normal world, typically yielding control to the normal (or rich) \ac{OS} bootloader.

In the system bus, a supplementary control signal (a $33^{rd}$ bit) called Non-Secure bit (NS) is set by hardware whenever a transaction (read or write) is made by components that belong to the normal world.
Being so, once addresses are decoded, they cannot match any component that is part of the secure world.
Additional mechanisms provide the means for securing memory regions and interrupts, which are also partitioned between the two worlds and can only be configured in secure mode.
Such mechanisms provide security building blocks such as trusted \ac{IO} paths and secure storage.

\begin{figure}
	\centering
	\includegraphics{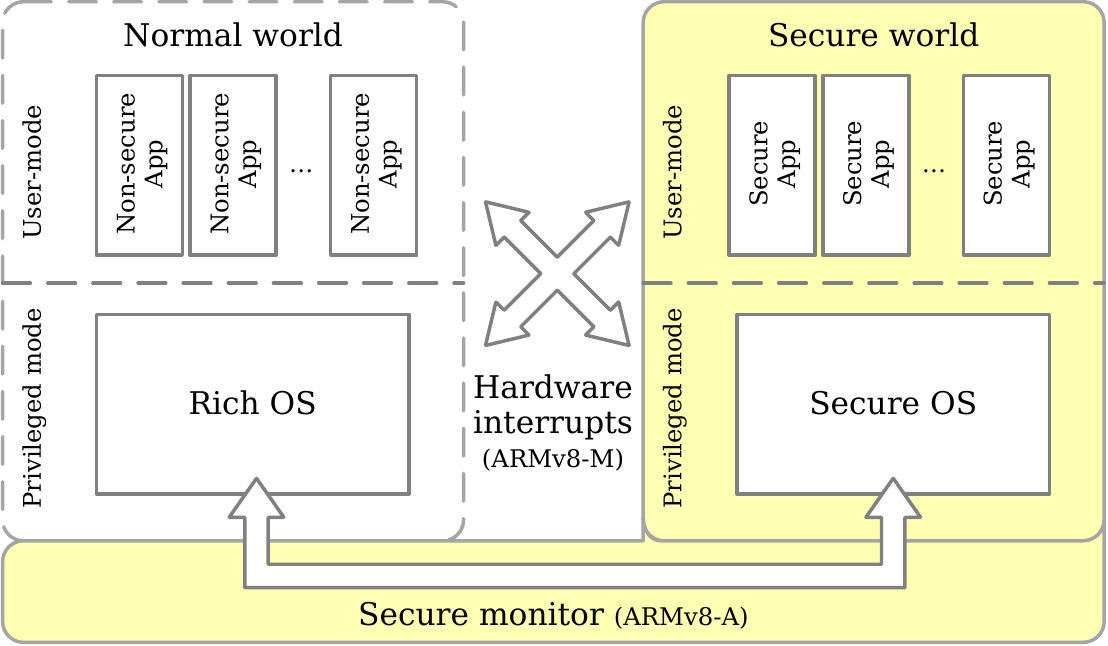}
	\caption{\label{fig:tee:trustzone}TrustZone scheme.}
\end{figure}

Remote attestation and sealing do not come out of the box, although some works simply assume the availability of a secure hardware key in the secure world~\cite{linaro:2019:optee,santos:2014:trustedlanguage}, since \ac{ARM} chip vendors are free to add their own \acp{IP2} to manufactured \acp{SoC}.
In conjunction with a \ac{TPM} that can solely communicate with the secure world, one can envision a root of trust for providing these services~\cite{zhao:2014:roottrust}.
		
TrustZone provides, however, only one secure world.
As a consequence, multiple secure applications need to share it, increasing the chances of dividing the protected environment with potential attackers.
Alternatively, a requirement of at most one application in the secure world can be enforced, although this could be too restrictive in some scenarios.
Moreover, the secure world is under the control of a separate \ac{OS}, which means that it can still be accessed by some (possibly compromised) system administrator.

\subsection{\label{sec:amdsev}AMD \ac{SEV}}
\begin{figure}
	\centering
	\includegraphics{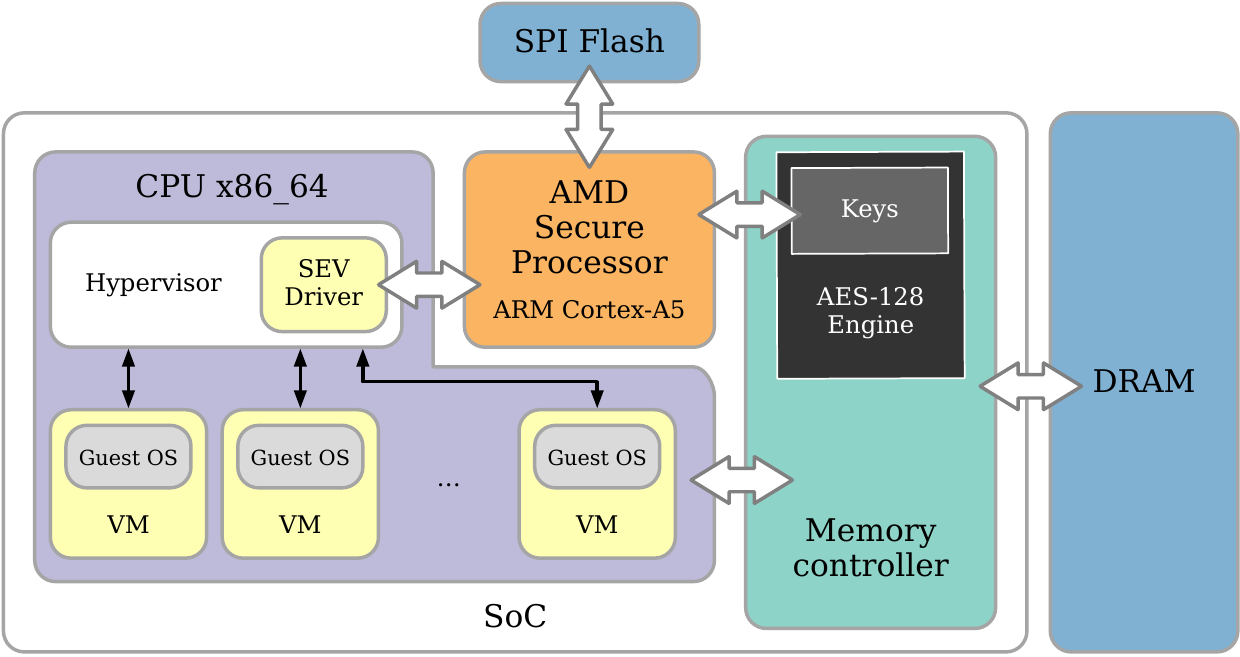}
	\caption{\label{fig:tee:sev}AMD SEV architecture.}
\end{figure}

The \ac{AMD} \ac{SEV} protects data in use by \acp{VM}.
It relies on a hardware encryption engine embedded in the \ac{MC} which automatically performs cryptographic operations, given that appropriate keys are provided to it, and adds minimum performance impact and no requirement for application changes.
Key generation and management are performed in the \ac{AMD} secure processor, a dedicated security subsystem based on an \ac{ARM} Cortex-A5 coprocessor physically isolated from the rest of the \ac{SoC}.
It contains a dedicated \ac{RAM}, non-volatile storage in a \ac{SPI} flash and cryptographic engines.
\Cref{fig:tee:sev} depicts the high level architecture.

While the \ac{SME} feature uses a single key for system-wide memory encryption, \ac{SEV} involves multiple encryption keys, one per \ac{VM}.
When using \ac{SME}, ephemeral keys are randomly generated at boot time by the secure processor and are used to encrypt memory pages that are marked by system software within page tables through the \emph{C-bit}.
Alternatively, all pages may be encrypted if the Transparent \ac{SME}, or T\ac{SME}, is activated during boot time.
T\ac{SME} does not require modifications in the \ac{OS} and prevents physical attacks like cold boot.

Unlike \ac{SME}, \ac{SEV} allows the association of one encryption key per hardware virtual machine in a way that the hypervisor has no longer access to everything within guest \acp{VM}.
The same cryptographic isolation is present in the other direction, \ie, the hypervisor has a separate key that prevents the guest from reading its assigned memory pages even in case of a malicious attack resulting in the break of memory logical isolation.
Despite that, guests and the hypervisor still communicate, \ie, the latter performs scheduling and device emulation for \acp{VM}.

Apart from resources management, the hypervisor also manages keys, although it has no access to them.
This is done by the manipulation of \acp{ASID}, which are specified when \acp{VM} are launched and are used to determine the encryption key. Essentially, \acp{ASID} are used as indexes into a key table in the \ac{MC} that determines the key in use by the \ac{AES} engine.
Besides, such identifier is also employed to tag data in the cache, in a way that only their owner is able to hit on that cache line and get access to the corresponding data.

Memory pages containing instructions and guest page tables are always private, so that attacks such as code injection become more difficult. 
Data pages, on the other hand, may be shared by each guest \ac{OS} through the C-bit manipulation. 
In particular, this is used for \ac{DMA}, which is only allowed in shared memory pages.
From the software perspective, this requires changes in the guest \ac{OS} and hypervisor, so that they can set memory pages access privileges. 
That is done through an open source \ac{AMD} secure processor driver, which communicates with the secure processor firmware.
The \ac{AMD} secure processor only runs signed firmware, which is a closed source package that \ac{AMD} provides to \ac{BIOS} vendors.

Key management allows for platform authentication, virtual machines start-up, migration and snapshot.
Chips are fused with unique keys, including the chip \acl{EK} (C\acs{EK}) derived from chip-unique values and an \ac{AMD} signing key for authentication.
Apart from that, platform owners may set a \ac{CA} key which may be used for allowing migration exclusively to other platforms that had been signed by the same \ac{CA}.
Once combined, the C\acs{EK} and the \acs{CA} key produce the platform \acl{EK} (P\acs{EK}) which is encrypted and stored in the \ac{SPI} flash.
The P\acs{EK} is finally used to derive the platform \acl{DH} key (P\acs{DH}) in conjunction with guest \acp{OS} every time the machine is powered on or reset. 
That way, the hypervisor is not able to intercept communication between guest \acp{OS} and the \ac{AMD} secure processor.

Originally, the processor register state of \acp{VM} could be exploited by hypervisors running under \ac{AMD} \ac{SEV}.
For that reason, \ac{AMD} released the \ac{SEV} \ac{ES} that allows guests to restrict access to the register state, albeit possibly sharing it with hypervisors.
Although \ac{AMD} \ac{SME}/\ac{SEV}/\ac{SEV}-\ac{ES} allow memory and registers encryption and platform attestation to a finer granularity (\acp{VM}), these technologies are vulnerable to memory integrity manipulation (\eg, rowhammer) and rollback attacks, \ie, when a previous state (even if encrypted and signed) is replayed back and considered genuine.

\section{Intel \ac{SGX}}

Intel \ac{SGX} provides a \ac{TEE} by employing distinguished approaches, particularly with respect to running privileges (user-level) and security guarantees (memory integrity and freshness).
Limiting the secure execution to user-level processes allows to purge the \ac{OS} out of the \ac{TCB}, therefore dramatically reducing the size of the latter.
Memory integrity and freshness in conjunction with confidentiality guarantees, on the other hand, make the secure environment robust against all kinds of memory tampering.
These features, however, are not costless.
A unit of code protected by \ac{SGX}, or \emph{enclave}, often needs to use \ac{OS} services, and such interactions become more expensive.
Likewise, the bookkeeping that enables memory safety is made by hardware, hence consuming silicon area and memory at runtime.
Because of that, enclaves are limited to use a small amount of protected memory.
Whenever they need to use more than that, considerable overheads arise.

\begin{figure}[t]
	\centering
	\includegraphics{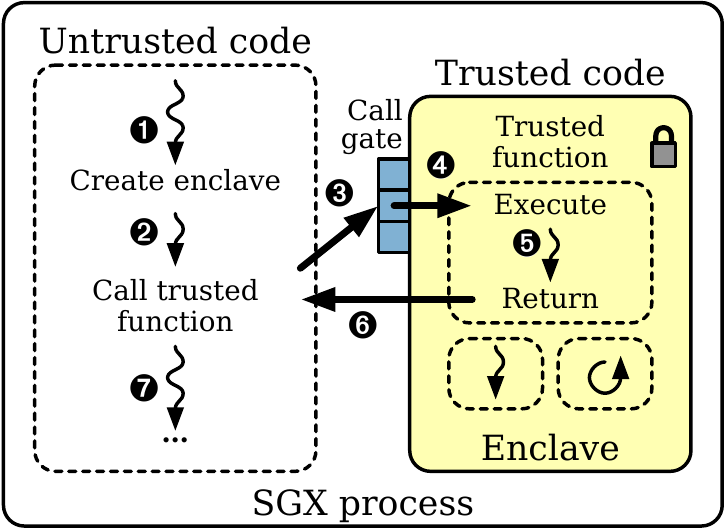}
	\caption{SGX execution flow.}
	\label{fig:sgx}
\end{figure}

\ac{SGX} aims to shield code execution against inspection and tampering from privileged code (\eg infected \ac{OS}) and certain physical attacks.
It does so by offering an instruction set extension in x86\_64 \acp{CPU} manufactured by Intel.
\ac{SGX} seamlessly encrypts and tracks integrity tags of memory in use by enclaves.
In this way, it is guaranteed that at any time during enclave execution the most recent and untampered plaintext data is present inside the \ac{CPU} package.
As a consequence, computations done in the enclave cannot be seen from the outside and any modification attempts are detected, including replay of previously authenticated values.
The \ac{TCB} of an \ac{SGX} enclave is composed of the \ac{CPU} itself, and the code running within.
The assumption is that opening the \ac{CPU} package is difficult for an attacker, and leaves clear evidence of the breach.

To enable an application to use enclaves, the developer must provide a signed shared library (\texttt{.so} or \texttt{.dll}) that will execute inside an enclave.
The library itself is not encrypted and can be inspected before being started. 
Due to this, no secret should be stored inside the code.
An enclave is provided with secrets, like certificates and keys, with the help of a remote attestation protocol.
This protocol can prove that an enclave runs on a genuine Intel processor with \ac{SGX} and can verify that its identity matches that of the code that is expected to run \cite{gueron:2016:sgxmemory}.
As a result of the attestation, a shared secret that enables a secure communication channel is established and permits the remote entity to provide the enclave with encrypted secrets.

\ac{SGX} applications are partitioned between trusted and untrusted segments.
Typically, the trusted part is supposed to handle sensitive data and computations, whereas the untrusted is responsible for performing system calls and handling non-sensitive backing operations.
While trusted code has access to the whole process memory space, the untrusted code is limited to its own memory pages.
Transferring the control from one running mode to the other happens in a similar fashion to \acp{RPC} in the sense that arguments are serialised and replicated in another memory space before switching modes.
Calls made from the untrusted code to the trusted one are called \acp{ecall}, whereas the ones in the opposite direction are called \acp{ocall}.
\Cref{fig:sgx} depicts the basic execution flow of \ac{SGX}.
First, an enclave is created (\ding{202}).
As soon as a program needs to execute a trusted function (\ding{203}), it performs an \ac{ecall} (\ding{204}).
The call goes through the \ac{SGX} call gate to bring the execution flow inside the enclave (\ding{205}).
Once the trusted function is executed by one of the enclave's threads (\ding{206}), its result is encrypted and sent back (\ding{207}) before giving the control to the main processing thread (\ding{208}).

\begin{figure}[b]
	\centering
	\includegraphics{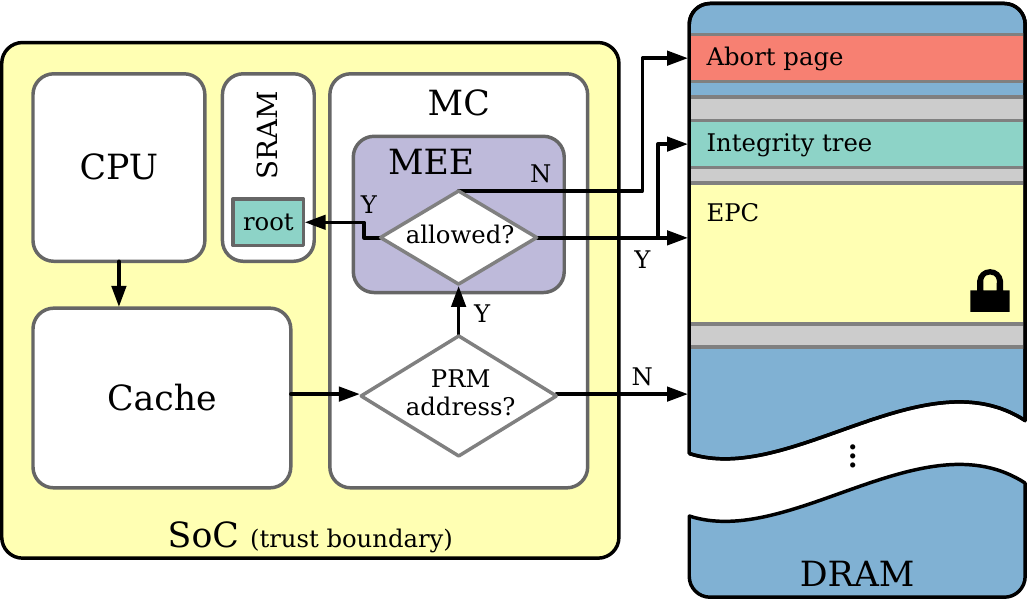}
	\caption{SGX memory management.}
	\label{fig:sgx:inegritytree}
\end{figure}

Intel provides a set of tools to aid the coding and fulfilment of \ac{SGX} requirements.
The \ac{SGX} \ac{SDK}~\cite{intel:2019:sgxsdk} comprises a generator for proxies and stubs written in C that are supposed to be compiled and linked to both trusted and untrusted code.
This generation is based on a text-based configuration file that follows the syntax of the \ac{EDL}, which basically defines the interface of edge routines.
Since system calls and \ac{IO} instructions are not allowed inside enclaves, Intel also provides libraries that are guaranteed to comply with these limitations.
It also contains wrappers for specific functions, like accessing the true random number generator by the \texttt{RDRAND} instruction, which uses hardware noise as source of entropy.
Moreover, it provides functionalities to improve performance like the switchless calls~\cite{tian:2018:switchless}, which basically implement asynchronous handling of \acp{ocall} to reduce the number of transitions between trusted and untrusted modes.
The \ac{SDK} also includes the \emph{enclave signing tool} responsible for measuring and signing the shared library to be loaded as an enclave.
In the following sections, we further detail some of the \ac{SGX} features, their implications and limitations.

\vspace{-0.5em}
\subsection{\label{sec:sgx:memory}Memory protection}

Encrypted memory is provided in a reserved memory area predefined at boot time.
It is called \ac{PRM} and is limited to \SI{128}{\mebi\byte} in the first version of \ac{SGX}, although future releases might relax this limitation~\cite{mckeen:2016:sgxdynmem}.
Within the \ac{PRM}, an area of at most \SI{93.5}{\mebi\byte}~\cite{vaucher:2018:sched} called \ac{EPC} can be used by application’s memory pages, while the remaining area is used to maintain \ac{SGX} metadata.
If this limit is surpassed, enclave pages are subject to a swapping mechanism implemented in the Intel \ac{SGX} driver, resulting in severe performance penalties~\cite{pires:2016:scbr,brenner:2016:securekeeper,arnautov:2016:scone}.

Normally, memory transactions that miss the cache are handled by the processor's \ac{MC}.
If they correspond to \ac{PRM} addresses though, a component of the \ac{MC} called \ac{MEE} takes over~\cite{gueron:2016:sgxmemory}.
\Cref{fig:sgx:inegritytree} illustrates this mechanism.
The \ac{MEE} is responsible for providing cryptographic operations, tamper-resistance and replay protection.
It generates random keys at every boot, one for cryptographic operations and another for \acp{MAC}.
The safety guarantees are achieved by
\begin{enumerate*}[label={(\roman*)}]
	\item encrypting data before sending them to \ac{DRAM};
	\item updating an integrity tree;
	\item decrypting data fetched from \ac{DRAM}; and
	\item verifying the integrity tree.
\end{enumerate*}

An integrity tree is typically implemented as a Merkle tree~\cite{elbaz:2009:integritytrees}, where each node holds a hash digest of its children.
The root is the digest of the entire data, and each leaf the hash of one data unit.
Being so, an update requires to re-compute only the implicated leaf and its ancestors.
Instead of simple hashes, \ac{SGX} \ac{MEE} uses stateful \acp{MAC}. 
The state is a \emph{nonce} (non repeating number, coined to be used only once) called by Intel as the \emph{version} of each data unit, which has the size of a cache line, \ie, \SI{64}{\byte}.
The tree is stored in untrusted memory, except for its root that is kept in an internal on-die \ac{SRAM} and inaccessible from outside.
It reflects the integrity of the most recently written protected area at any given time.
Any mismatch during a verification causes a \ac{MC} lock, which ultimately requires a machine reboot, so that the \ac{MEE} restarts with fresh keys~\cite{gueron:2016:sgxmemory}.

\begin{figure}
	\centering
	\includegraphics{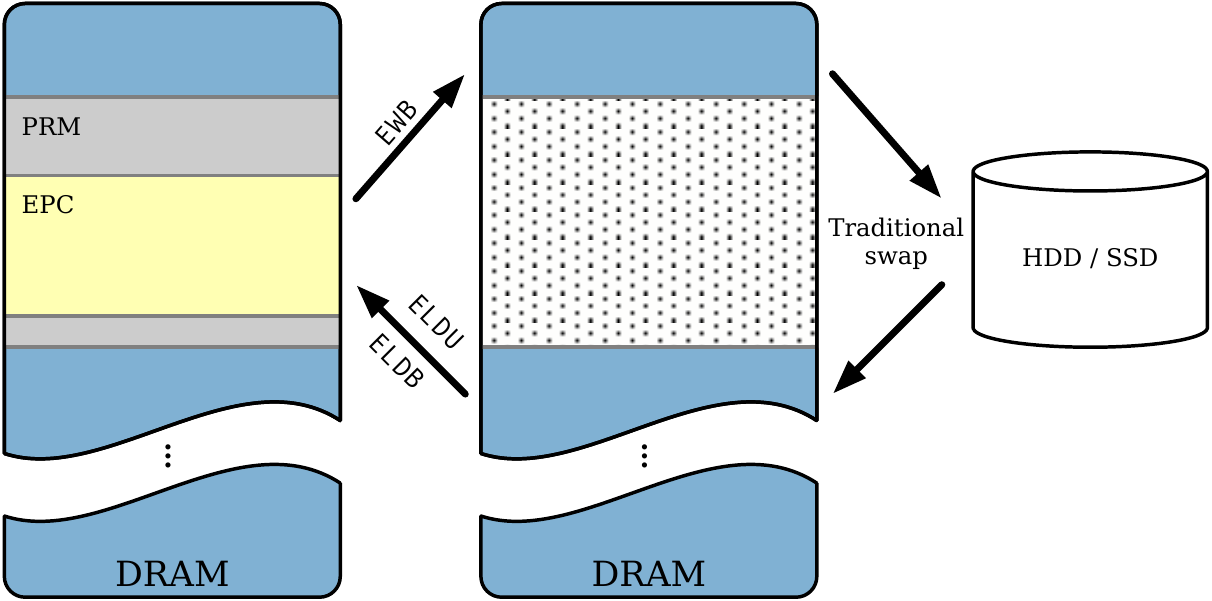}
	\caption{SGX paging.}
	\label{fig:sgx:swap}
\end{figure}

Accesses to enclave pages that do not reside in the \ac{EPC} trigger page faults.
The \ac{SGX} driver interacts with the \ac{CPU} to choose which pages to evict.
Such pages of \SI{4}{\kibi\byte} are moved to main memory through the privileged instruction \texttt{EWB}, which also computes a cryptographic \ac{MAC} of the page and stores it in unprotected memory.
Additionally, \texttt{EWB} generates a nonce of \SI{8}{\byte}, or \emph{page version}, that ensures the freshness of the evicted page once it is loaded back in. These nonces are stored in special \ac{EPC} pages called \emph{Version Arrays}, which are not assigned to any enclave and therefore cannot be read by software.
After a page is evicted, the \ac{SGX} driver loads the requested page, \ie, the one that triggered the fault, from main memory through the instructions \texttt{ELDB} or \texttt{ELDU}. These, in turn, decrypt the page and perform integrity and freshness checks.
\Cref{fig:sgx:swap} illustrates memory swaps.

Despite the close interaction between the \ac{SGX} driver and the \ac{CPU} for paging, one does not need to trust the driver since it cannot violate the confidentiality nor the integrity of enclave pages, which are maintained and checked by hardware.
External snooping, such as eavesdropping the bus or the \ac{EPC} memory content, will only reveal ciphertext undistinguishable from random data.
Likewise, data modifications or feeding integral old data will be detected through mismatching authentication codes.

\vspace{-0.5em}
\subsection{Limitations and performance implications}
\label{sec:sgx:perfissues}

The main performance bottlenecks when using Intel \ac{SGX} appear during transitions between trusted and untrusted modes (inside/outside the enclaves) and under intensive memory usage, notably in two stages:
\begin{enumerate*}[label={(\roman*)}]
\item when exceeding the processor's \ac{LLC}, which requires cache evictions and \ac{DRAM} fetches, along with cryptographic and integrity operations; and 
\item when exceeding the \ac{EPC} size, which triggers memory swaps serviced by the underlying \ac{OS}.
\end{enumerate*}

\begin{figure}[b]
\centering
\includegraphics{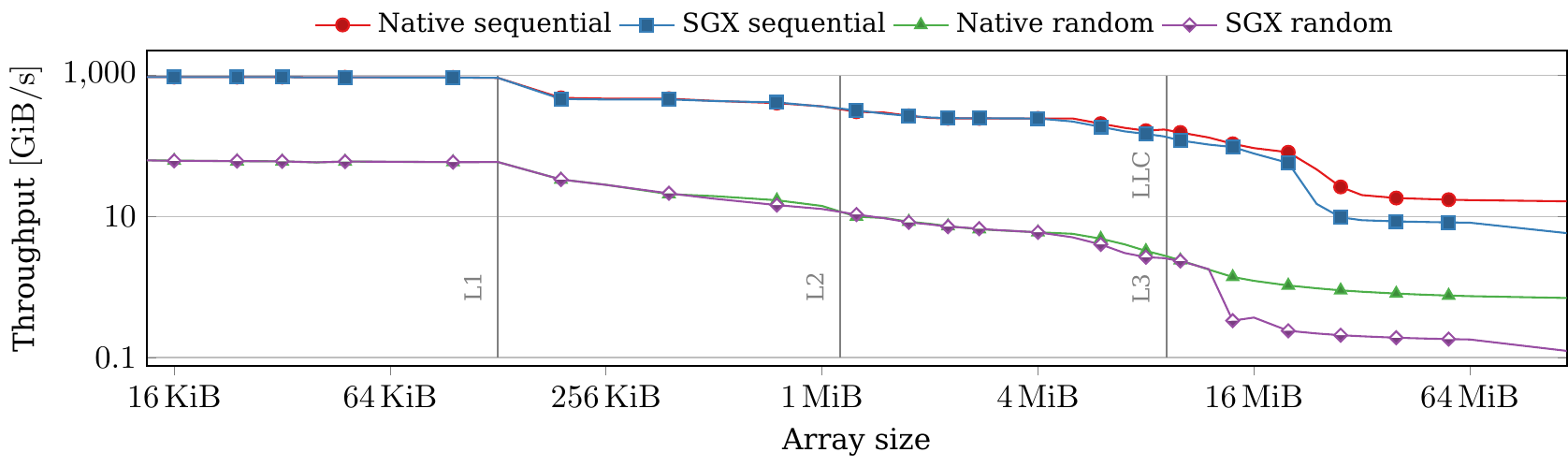}
\vskip 5mm
\caption[SGX caching effects.]{\label{fig:sgxcache}SGX caching effects according to sequential and random reads.}
\end{figure}

Transitions are essential as enclaves cannot perform system calls.
This is one of the main design concerns when porting legacy applications to \ac{SGX} enclaves.
Every disk or network access, for instance, must be handled by untrusted code.
While switches between user and privileged modes when executing system calls are completed in \num{150} \ac{CPU} cycles~\cite{soares:2010:flexsc, weisse:2017:hotcalls}, enclave transitions (\texttt{EENTER}, \texttt{ERESUME} or \texttt{EEXIT} instructions) take more than \num{3000} cycles each~\cite{orenbach:2017:eleos}, \ie, an \ac{ocall}, where the control leaves and enters back in the enclave, is more than \num{40}$\times$ slower in comparison to system call transitions.
This does not take into account the extra work performed in edge routines (\eg, checking memory bounds) and collateral costs of enclave transitions like the ones caused by flushing the \ac{TLB} for security reasons.

To mitigate this issue, several systems~\cite{arnautov:2016:scone, orenbach:2017:eleos, weisse:2017:hotcalls} proposed the asynchronous treatment of transitions, which was later integrated to the Intel \ac{SGX} \ac{SDK} 2.2 as a feature named \emph{switchless calls}~\cite{tian:2018:switchless}.
This is essentially done by
\begin{enumerate*}[label={(\roman*)}]
	\item having at least one thread inside and another outside the enclave;
	\item establishing a communication channel between them (shared memory in untrusted area); and
	\item using synchronisation primitives to signal events (spin locks).
\end{enumerate*}
If we take an \ac{ocall} as an example, an enclave thread would enqueue a request in a data structure residing in untrusted memory. The request is then acquired and processed by an untrusted worker thread which puts the result back in the shared memory. At this point, an enclave thread can collect and consume the result.
This approach, although simple, brings along several practical matters.

While considerable performance improvements of using switchless calls can be observed in heavy workloads, gains are diminished when asynchronous calls are sparse.
In idle workloads, a lot of time and energy is spent in busy waiting. 
These \ac{CPU} cycles could instead be used by other threads or processes to perform more useful tasks.
Moreover, as the state of enclave threads are kept in protected memory, the amount of active threads is limited.
Besides, the optimal number of worker threads also depends on workloads, which are not always easy to estimate.
For these reasons, Intel \ac{SGX} \ac{SDK} leaves for the user to decide whether edge routines are switchless by marking them with the keyword \texttt{transition\_using\_threads} in \ac{EDL} files. 

Regarding memory usage, we performed an experiment to observe caching effects of employing enclaves~\cite{gottel:2018:sevsgx}.
It consists of reading through a fixed amount of memory in sequential and random access patterns within an enclave and natively.
Both runs were done in the same machine, an Intel Xeon E3-1275 v6, with \SI{16}{\gibi\byte} of \ac{RAM}.
\Cref{fig:sgxcache} shows the observed throughput averaged over 10 runs.
Within the cache, performance in the trusted environment is strictly equivalent to native, particularly up to the L2 cache limit.
When the amount of memory surpasses the \ac{LLC}, throughput is greatly affected, with random accesses incurring in bigger overheads than sequential reads.
	
\begin{figure}
\centering
\includegraphics{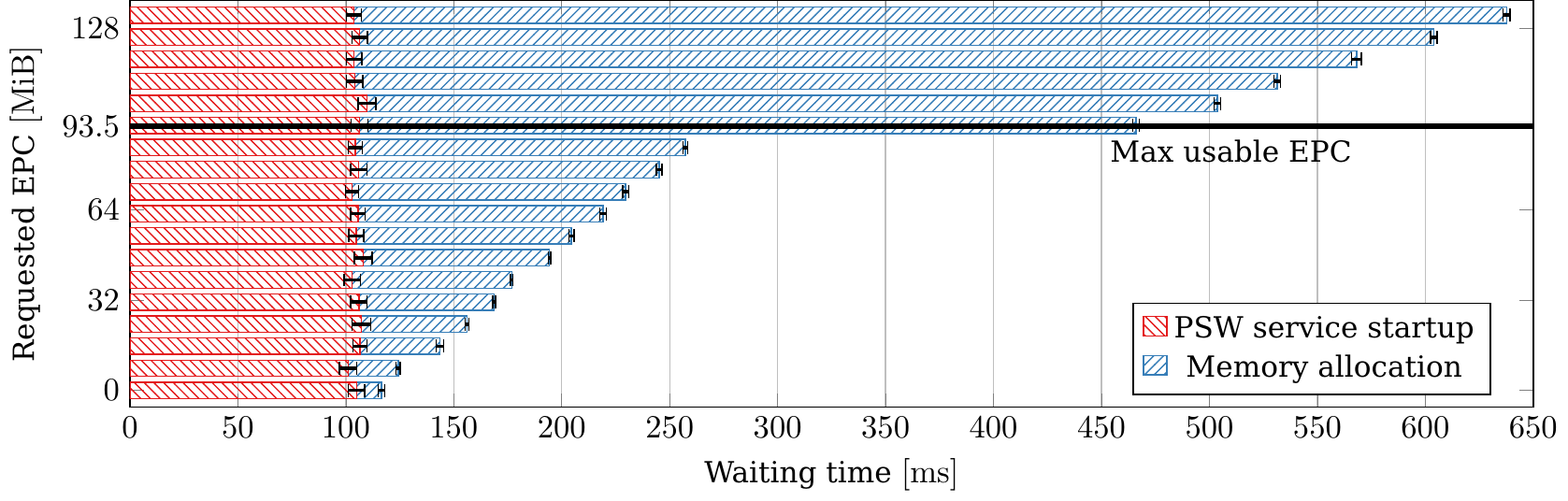}
\vskip 5mm
\caption{\label{fig:inittime}Overhead when starting SGX jobs.}
\end{figure}

In the first version of \ac{SGX}, the whole memory reserved to an enclave must be allocated at the initialisation of the trusted environment.
This constraint imposes some issues, as some services react to instant loads that vary over time.
Having to set memory allocation beforehand may lead to exhaustion in case of underestimation or to underused pages otherwise.
Besides, it also brings some start-up waiting time that would be diluted across the process lifetime in case it supported dynamic allocation.
This limitation, along with the impossibility of changing page permissions (read, write and execute), guarantees that the initial state that is certified by attestation (see \Cref{sec:sgx:attestation}) remains unchanged throughout the enclave's execution. 
SGX2 extensions~\cite{mckeen:2016:sgxdynmem}, on the other hand, support dynamic allocation. To this date, we had no access to SGX2 hardware.

To quantify the described initial overhead, we conducted an experiment~\cite{vaucher:2018:sched}.
Apart from memory allocation, we also consider the support service initialisation time.
Intel \ac{SGX} \ac{SDK} provides the \ac{PSW} that includes the \ac{AESM}, which needs to be bootstrapped before enclaves are deployed.
It assists \ac{SGX} processes in enclave initialisation, attestation and supporting the access to platform services (see \Cref{sec:sgx:platservices}).
\Cref{fig:inittime} shows the average results for \num{60} runs of docker containers that first start \ac{AESM} and launch enclaves that allocate different amounts of memory.
Error bars represent the \SI{95}{\percent} confidence interval.
As expected, the service startup time is virtually the same in all runs, accounting for about \SI{100}{\milli\second}.
Memory allocation time, on the other hand, shows two clear linear trends: before and after reaching the usable \ac{EPC} memory limit.
Until this limit, the time increase rate is \SI{1.6}{\milli\second\per\mebi\byte} after which it jumps to \SI{4.5}{\milli\second\per\mebi\byte}, plus a fixed delay of about \SI{200}{\milli\second}.
Note that these times are just for \emph{allocating} memory, before any real use.
Higher overheads are expected when memory pages need to be swapped (see \Cref{sec:sgx:memory}), and we evaluate this behaviour in a full-fledged \ac{pubsub} system in \Cref{sec:scbr}.

\vspace{-0.5em}
\subsection{Enclave signing and attestation}
\label{sec:sgx:attestation}

The development of applications targeted to run within \ac{SGX} enclaves, besides the usual iterations on coding and compiling, also includes a mandatory signing step before the executables are able to be deployed and used in production.
This essentially serves two purposes:
\emph{(i)}~the code is uniquely associated to an \ac{ISV}, making it recognisable by customers and accountable for any consequence originated from its product; and
\emph{(ii)}~whoever communicates with the application can have guarantees that the endpoint inside the enclave has loaded and is actually running the expected code within a genuine \ac{SGX} platform.

The signing material includes information about the vendor, the date, some attributes, a version number and the enclave measurement, which corresponds to a digest (\ac{SHA}-256) made upon the enclave's initial state, including data, code and metadata (security flags associated with memory pages)~\cite{anati:2013:attestation,costan:2016:sgxexplained}.
When the enclave is loaded, a hardware implementation of the same measurement operates on the actual content of the running enclave (\texttt{MRENCLAVE} register), which has to precisely match the one that was computed during the signing step.
Upon measurements matching, a hash of the signer's public key is computed and stored in the \texttt{MRSIGNER} register.
These two registers reflect the enclave code and its author.

\begin{figure}[b!]
	\centering
	\includegraphics{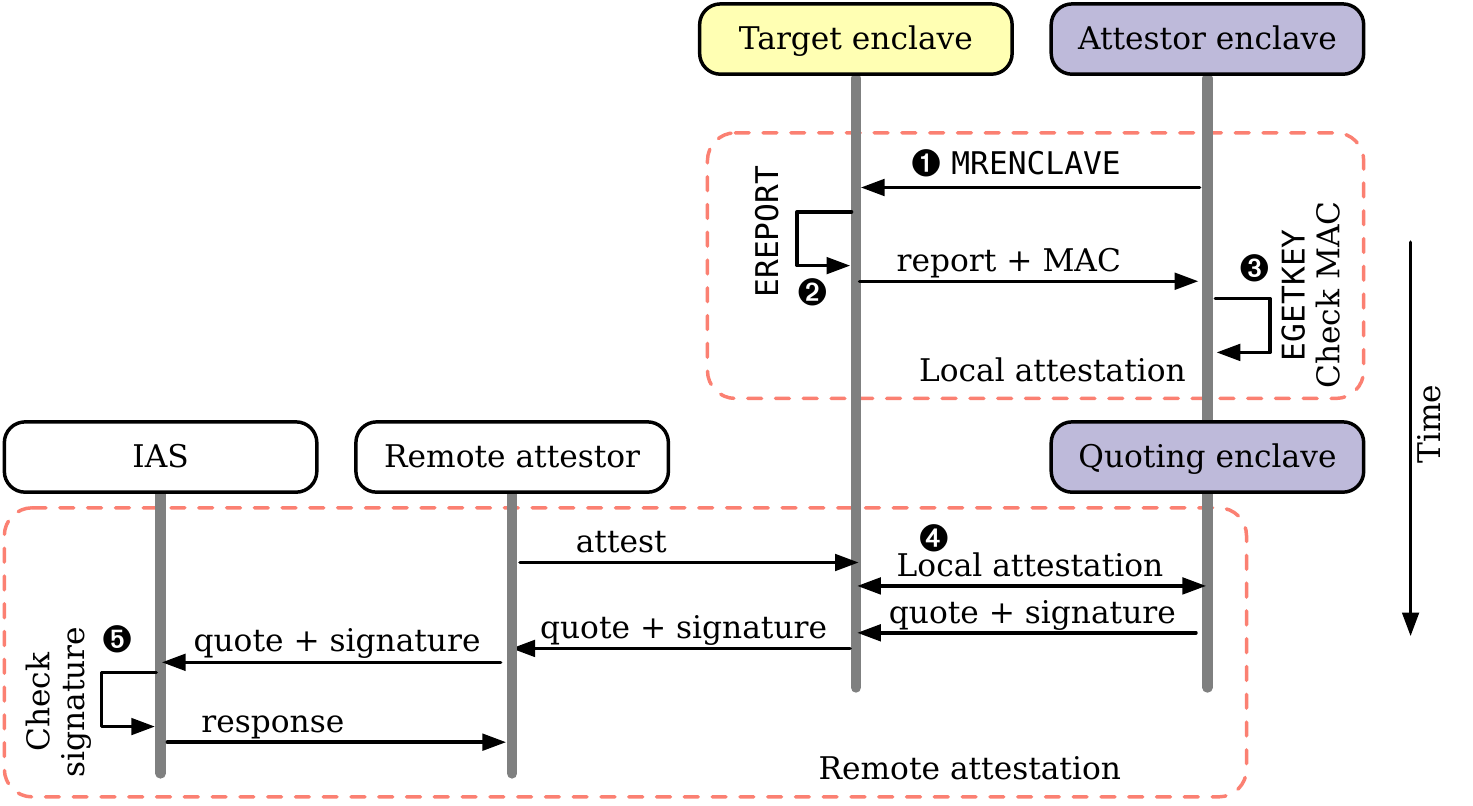}
	\caption{\label{fig:sgx:attestation}SGX local and remote attestations.}
\end{figure}

Later on, when interlocutors want to communicate with a running enclave, they should first attest it before sharing sensitive data with it.
Attestation is therefore an essential requirement for virtually any enclave deployment.
It can be performed locally or remotely, the latter being dependent on the former (\Cref{fig:sgx:attestation}).
The local procedure is based on a symmetric key and starts when the attestor sends its identity (\ding{202}) to the enclave being attested (target).
This, in turn, calls the instruction \texttt{EREPORT} (\ding{203}), that cryptographically binds its identity (\texttt{MRENCLAVE} and \texttt{MRSIGNER}) and other attributes in a \ac{MAC} tag that can only be locally checked by the attestor.
This guarantee comes from the fact that the key used for computing the \ac{MAC} can only be obtained by the \texttt{EGETKEY} (\ding{204}) instruction issued by the attestor on the same \ac{SGX} platform. 
Such instruction retrieves keys that derive from the processor specific key hierarchy and can only be executed inside an enclave.

In case of remote attestation, asymmetric keys are used.
A \emph{quoting enclave} (installed along with the \ac{PSW}) first performs a local attestation (\ding{205}) and creates another report called a \emph{quote}.
The quote, in turn, may be checked by the remote party with the aid of an online service called \ac{IAS}, which checks the signature affixed to the quote (\ding{206}).
The protocol also allows the enclave being attested to append arbitrary additional data to the quote.
As the establishment of a secure channel is often desirable once the trust is settled, a sensible approach would consist of the following steps:
\begin{enumerate*}[label={(\roman*)}]
\item target enclave generates a pair of asymmetric keys;
\item the public part of such key is sent along with the quote;	
\item once the attestation is successful, the key from the quote is used by the attestor to provide the enclave with any secrets, possibly a symmetric communication key. As the corresponding private key is securely shielded within the enclave, no third party would be able to inspect further interactions.
\end{enumerate*}
Alternatively, \ac{DH} could analogously be used for the same purpose.

\vspace{-0.5em}
\subsection{Sealing}

Confidentially and integrity guarantees are provided during execution. Once destroyed, enclave data are lost.
For this reason, \ac{SGX} provides mechanisms for secure data persistence.
Apart from storing application data, the sealing feature typically aids on certificates' storage, which obviates the need of new remote attestations every time an enclave application restarts.
Like the keys used for attestation, sealing keys also depend on the platform hardware key derivation and are obtainable through the \texttt{EGETKEY} instruction.

Sealing key derivation can either consider \texttt{MRENCLAVE} or \texttt{MRSIGNER}.
This choice determines who is able to decrypt the sealed data later on.
If the enclave measurement is used, only the same enclave deployed on the same platform will be able to perform the unsealing.
Any code modification would render inaccessible data sealed under this policy.
This is useful when old data needs to be invalidated once new versions of the enclave are released, \eg, due to a vulnerability mitigation.
Instead, if the signer identity is used, every other enclave in the platform which was also signed by the same \ac{ISV} is allowed to access the data.
In this case, offline transfer of sealed data is possible as long as the same \ac{CPU} performs the unsealing.

\vspace{-0.5em}
\subsection{Platform services}
\label{sec:sgx:platservices}

Sealing preserves confidentiality of persistent data.
It thus allows the storage of sensitive results and loading sealed secrets or certificates across enclaves re-initialisations.
However, an attacker could still try to serve an enclave with previous versions of sealed data that are properly encrypted and authenticated.
That can have severe consequences as the enclave could be misled to reflect some past state.
To illustrate, imagine your bank account balance getting back to that of your pay day's eve.
To prevent such replay attacks, an enclave can use monotonic counters provided by the platform.
Each time an enclave writes a new state on disk, it increments a monotonic counter and stores the new value inside the sealed state.
When the enclave restarts, it reads the monotonic counter and checks that it matches the value stored after unsealing the  persistent data.

Due to the \ac{SGX} restriction of limiting enclave executions to user-mode, trusted time and persistent monotonic counters are provided separately, as both require the usage of \ac{IO} either to access a clock or non-volatile memory.
Intel decided to implement these services in the \ac{ME}, a subsystem that allows remote system administration tasks like turning the machine on and off, regardless of having a running operating system. It consists of a proprietary firmware\footnote{Minix \ac{OS}, according to \cite{positive:2017:meiminix}} running in a coprocessor in the chipset that is always active as long as the machine is connected to a power source.
Because of that, these services may not be available in all platforms (particularly servers), as it depends on machine vendors.

To use such services, enclaves first need to open a session with a \ac{PSW} enclave called \ac{PSE} and have access to Internet~\cite{intel:2019:sgxsdk}.
Most likely, such connection provides a key that allows the \ac{PSE} to communicate with the \ac{ME} and then intermediate user enclave requests.
Once the session is established, trusted timer and monotonic counters may be used.
The trusted timer has a resolution of one second, which is substantial. 
It is mostly useful for determining the expiration of session keys or certificates rather than being used for profiling or performance measurements, since these often require finer grained timers.

Once an enclave requests the creation of a monotonic counter, it is given a \ac{UUID} associated to the newly created counter.
This id must be remembered, so that the counter value may be incremented or queried. 
Similar to sealing, access control to counters may be associated to the enclave measurement (\texttt{MRENCLAVE}) or its signer (\texttt{MRSIGNER}).
Such control is performed by the \ac{PSE}.

Intel platform counters are known to be very slow: between \SI{60}{\milli\second} and \SI{250}{\milli\second} per increment, which are equivalent to alternative solutions such as \acp{TPM}~\cite{strackx:2016:ariadne}.
Besides, since they are stored in flash memory they suffer from wear-out.
Due to this, they can possibly stop working after a few hundreds of thousands write cycles~\cite{parno:2011:memoir}.
Moreover, all counters are lost upon the re-installation of \ac{PSW}~\cite{matetic:2017:rote} which can be done by privileged users, who are not trusted in the \ac{SGX} threat model.
Altogether, these reasons make Intel platform counters unsafe depending on security requirements.
Distributed approaches are safer~\cite{matetic:2017:rote}, although they may render poor performances for demanding applications.
On the performance front, faster (and unsafe) layers backed by slower (and safer) ones may be used, although they are still vulnerable to rollbacks during the so-called \emph{stability time}~\cite{bailleu:2019:speicher}. 

\vspace{-0.5em}
\subsection{\label{sec:sgx:vulnerabilities}Vulnerabilities}

A number of attacks to \ac{SGX} enclaves were proposed.
One class of attacks exploits software vulnerabilities like memory safety violations~\cite{kuvaiskii:2017:sgxbounds} (stack overflows or dangling pointers) to hijack the software control-flow~\cite{weichbrodt:2016:asyncshock}.
Once that is accomplished, the attacker may find \emph{gadgets} consisting in benign pieces of code (like the libc~\cite{shacham:2007:roplibc}) and reuse them to perform \ac{ROP}~\cite{bletsch:2011:rop} in order to leak data or change the enclave behaviour.
Such approach succeeds even on top of encrypted code provisioned to enclaves~\cite{lee:2017:hackdarkness} or when countermeasure techniques like \ac{ASLR}~\cite{seo:2017:sgxalsr} are employed~\cite{biondo:2018:dilemma}.

This kind of attacks lay down strong arguments for having small \acp{TCB}, as software bugs are proportional to code size.
Nevertheless, several proposals provide complete runtimes in favour of usability, as they require little or no modification of legacy applications.
Notable examples are Haven~\cite{baumann:2014:haven}, \acs{SCONE}~\cite{arnautov:2016:scone}, Graphene-\ac{SGX}~\cite{tsai:2017:graphene}, Panoply~\cite{shinde:2017:panoply} and SGX-LKL~\cite{priebe:2019:sgxlkl}.
Indeed, software vulnerabilities fall out of the \ac{SGX} threat model, which assumes to shield bug-free applications.

Another class of attacks consists of using alternative sources of information like power, sound, electromagnetic or, most importantly, time analysis to infer behaviour of targets and finally reveal some sensitive data that they handle.
Processors nowadays are very complex machines and offer shared resources to several logical entities like \acp{VM}, \acp{OS} and processes.
Usage of such components may leave traces that can be exploited by malicious agents~\cite{xu:2015:controlledchannel}.
One of these shared resources is the cache hierarchy, notably the most targeted component by \ac{SGX} side-channel attacks~\cite{brasser:2017:sgxcacheattack}.
Since the \ac{MEE} sits at the edge of the on-chip memory hierarchy (see \Cref{sec:sgx:memory}) and therefore inside the trust boundary, \ac{SGX} cannot prevent software side-channel attacks that exploit cache leaks.
Like for software correctness, Intel claims that ``\textit{preventing side channel attacks is a matter for the enclave developer}''~\cite{intel:2017:sidechannels} and, in fact, some solutions for preventing them were proposed~\cite{shih:2017:tsgx,fu:2017:sgxlapd}.

Recently, micro-architectural implementation bugs were exploited for breaching privilege barriers and getting access to the whole memory address space, including kernel pages.
They are based on components that are supposed to improve performance like speculative execution~\cite{kocher:2019:spectre,chen:2019:sgxpectre,bhattacharyya:2019:smtspectre} and out-of-order execution~\cite{lipp:2018:meltdown}.
The biggest issue is that these attacks can leak sensitive data of perfectly secure bug-free applications.
Foreshadow~\cite{vanbulck:2018:foreshadow}, for instance, was ultimately able to retrieve the long-term attestation signing keys from the \ac{SGX} quoting enclave, making it possible to decipher all communication made by an attested enclave through a man-in-the-middle attack.
Countermeasures were released both in system software and microcode updates, although future microarchitecture versions are supposed to handle those issues in the silicon.

We also proposed an attack~\cite{mogage:2019:malware}.
Instead of targeting running enclaves, we exploit the separation between building and signing stages of the supply chain.
The attack is based on the assumption that a malicious agent is able to infect the machine where the enclave binary is signed.
Once there, it suspends the signer process and diverts its input to a component that injects malicious code within the binary.
The tampered version is then given back to the signer, which continues to execute as if nothing had happened, therefore giving authenticity to the infected binary.
Every subsequent integrity and signature checks would then succeed.
We also proposed mitigation measures by atomically shielding compilation and signing stages within enclaves.

Despite all the reported issues, throughout this work we explore Intel \ac{SGX} in a range of distributed systems. 
We consider these flaws to be orthogonal to our research, and hence do not consider them in our evaluations.
From an architectural vantage point, system designs that count on \acp{TEE} do not lose their relevance due to occasional breaches.
In principle, they could use different (even future) hardware implementations that retain equivalent or comparable features to \ac{SGX}.
 
\vspace{-0.5em}
\section{\label{sec:securecloud}SecureCloud project}

\begin{figure}[b!]
	\centering
	\includegraphics[width=\linewidth]{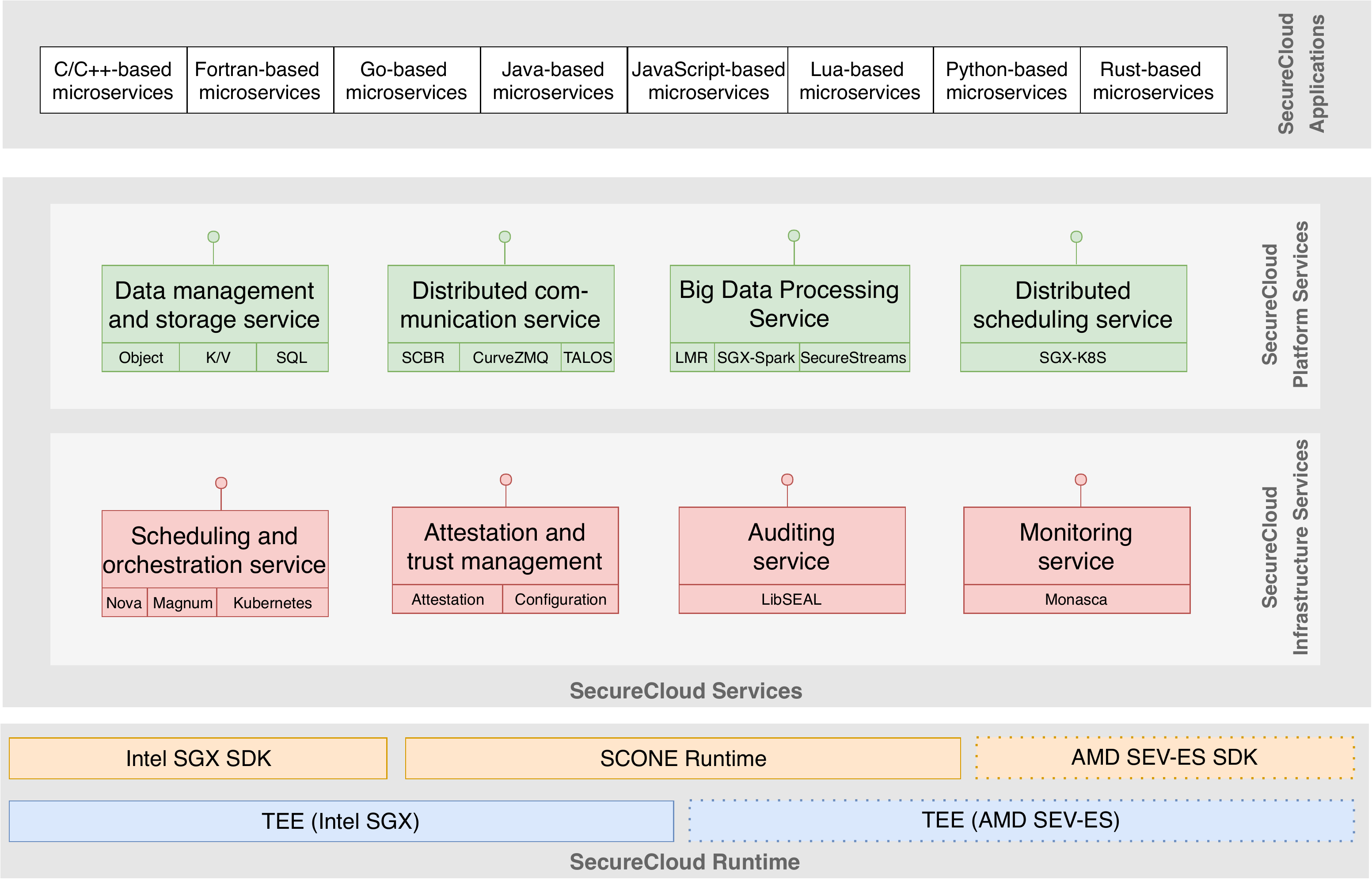}
	\caption{\label{fig:securecloud}SecureCloud platform architecture.}
\end{figure}

A substantial part of this work was done in the context of the SecureCloud project~\cite{kelbert:2017:securecloud} (2016-2018), whose goal was to enable the deployment of sensitive applications in the cloud while having strong security and privacy guarantees.
That was built upon a layered architecture on top of \ac{SGX}.
Each layer provides a set of micro-services which can be combined into full-fledged cloud systems.
The platform was validated with smart metering applications~\cite{brito:2018:securecloud}.
The consortium was composed by \num{14} partners in Europe, Israel and Brazil, including universities, research institutes, industry and government agencies.

\Cref{fig:securecloud} shows the architecture of SecureCloud's platform, including \acs{SCBR} (\Cref{sec:scbr}), LMR (\ie, lightweight \mapreduce---\Cref{sec:lwmp}) and SecureStreams (\Cref{sec:sstreams}).
Besides, we also contributed with SGX-K8S~\cite{vaucher:2018:sched}, a SGX-aware solution for  orchestrating containers in Kubernetes clusters.
That was achieved by modifying the \ac{SGX} driver to report the actual usage of \ac{EPC} memory by each process to help on scheduling placement decisions.
Moreover, the protected memory report was used for preventing containers from over-committing the \ac{EPC} thus avoiding performance losses due to memory swaps (see \Cref{sec:sgx:memory}) in the benefit of all other users who share the same platform.

Apart from our work, SecureCloud project produced a number of other relevant scientific contributions. The \ac{SCONE}~\cite{arnautov:2016:scone}, for instance, provides a complete runtime for deploying \ac{SGX} containers.
Glamdring~\cite{lind:2017:glamdring} aids on partitioning \ac{SGX} applications through code annotations.
LibSEAL~\cite{aublin:2018:libseal}, in turn, allows auditing enclave operations and checking invariants to find integrity violations.
More information about the project, along with deliverables, demonstrators and list of publications can be found in the official website~\cite{2019:securecloud}.

\section{\label{sec:back:cloud}Communication and data processing}

This section provides background information concerning the \Cref{chap:clouds}.
It describes system designs that fit in cloud deployment scenarios and can take advantage of \acp{TEE}.
Specifically, \Cref{sec:back:pubsub} covers secure \ac{pubsub} systems while \Cref{sec:back:mr} and \Cref{sec:back:stream} discuss about distributed processing frameworks.

\subsection{\label{sec:back:pubsub}Secure publish/subscribe}

Content-based routing (\acs{CBR}) is a powerful communication model that supports scalable asynchronous communication among large sets of geographically distributed nodes~\cite{eugster:2003:manyfaces}.
Yet, preserving privacy represents a major limitation for the wide adoption of \ac{CBR}, notably when the routers are located in public clouds.
This represents a major deterrent for companies for which data is a key asset, as for instance in the case of financial markets.

In the \ac{pubsub} model, publisher nodes submit data to a routing service as publications formed of a header describing the data and a payload containing the effective data.
The routing service matches the publications header with subscriptions previously registered by subscriber nodes and further routes the matching publications towards their destinations.
Content-based \ac{pubsub} appears to be incompatible with privacy preservation, as messages must be filtered based on their content, \ie, the routing engine should be able to see both the payload of publications and subscriptions.
Additionally, by structuring the containment relations between subscriptions, one can build efficient data structures to store subscriptions and match publications.
Containment allows for a significant reduction of the number of subscriptions stored as well as the number of matching evaluations executed per publication.
As a consequence, containment is used in most \ac{CBR} systems in use today \cite{carzaniga:2001:siena, chand:2004:xnet, li:2005:pubsub}, which makes them largely incompatible with classical cryptographic techniques for privacy preservation.

There is vast amount of literature about secure content-based \ac{pubsub}, including major surveys \cite{onica:2016:CPPSsurvey, uzunov:2016:surveysecps}, but no solution provides at the same time powerful filtering capabilities and high performance.
While there exist some techniques for privacy-preserving computation, they are either prohibitively slow or too limited to be usable in real systems.

Specialised solutions, like \ac{ASPE}~\cite{choi:2010:aspe}, allow for a direct match of encrypted publications with encrypted subscriptions.
Publication attributes and subscriptions constraints are represented as coordinates of multidimensional points.
\ac{ASPE} is based on an exact relation preserving isomorphism and supports subscription containment, although it is vulnerable to known-plaintext attacks.
Given that \ac{ASPE}'s matching complexity is prohibitively high when using a large number of attributes, an alternative~\cite{barazzutti:2012:thrifty} was proposed to enhance it with a pre-filtering approach that expresses equality constraints using Bloom filters~\cite{bloom:1970:filters}.
This allows for quickly identifying subscriptions that do not match the publication as their equality constraints cannot be satisfied.

Solutions based on traditional techniques encrypt sensitive information that traverse untrusted domains, hence basically preventing routers from filtering publications based on their content.
They propose architectures designed to respond to specific security threats by combining access control and key management mechanisms.
Most of these solutions integrate elaborated access control models into existing event-based middleware \cite{bacon:2008:pubsubac, zhao:2006:dynacpubsub, wun:2007:policycbps}, organising routing brokers in sets that share keys to encrypt or decrypt data.
Different encryption granularities are used, ranging from individual attributes to entire messages.

In \Cref{sec:scbr}, we follow a different strategy by taking advantage of \acp{TEE}.
We implement a \ac{CBR} engine in a secure enclave, so that its compute-intensive operations can work on decrypted data and leverage efficient matching algorithms.
Our experimental evaluation shows that \ac{SGX} adds only limited overhead to insecure plaintext matching outside secure enclaves while providing much better performance and more powerful filtering capabilities than alternative software-only solutions.
Since its publication~\cite{pires:2016:scbr}, further research explored our scheme~\cite{sampaio:2018:dissemination} or proposed alternatives~\cite{arnautov:2018:pubsub}.

\subsection{\label{sec:back:mr}Batch processing}

\mapreduce is a programming model used extensively for parallel data processing in distributed environments.
A wide range of algorithms were implemented using \mapreduce, from simple tasks like sorting and searching up to complex clustering and machine learning operations.
Many of these implementations are part of services externalized to cloud infrastructures, where concerns regarding security guarantees are common.
In \Cref{sec:lwmp}, we explore the use of Intel \ac{SGX} for providing privacy guarantees for \mapreduce operations, and based on our evaluation we conclude that it represents a viable alternative to cryptographic mechanisms.
We present results based on the widely used k-means clustering algorithm, but our implementation can be generalized to other applications that can be expressed using the \mapreduce model.

A number of security schemes for \mapreduce were proposed.
SecureMR~\cite{wei:2009:securemr} focuses on data integrity, verification of results and prevention of replay attacks, but it does not handle privacy of data and code.
Airavat~\cite{roy:2010:airavat} and \textsc{Gupt}~\cite{mohan:2012:gupt} combine differential privacy~\cite{dwork:2006:calibrating} with access control policies and therefore incur on the classical trade-off between utility and privacy, \ie, too much noise produces bad results while no obfuscation renders no privacy.
Through static code analysis, MrCrypt~\cite{tetali:2013:mrcrypt} selects distinct homomorphic encryption schemes for every data column. As expected, it is very slow: some benchmarks show performance penalties of one order of magnitude with respect to unencrypted execution.
Sedic~\cite{zhang:2011:sedic} and Tagged \mapreduce~\cite{zhang:2014:taggedmr,xu:2015:framework} split tasks to be deployed on private and public clouds based on how sensitive are the data they compute.
Besides the requirement of partitioning applications and failing to handle the case in which all computation happens in shared infrastructures, they impose higher latencies due to data transfers between different clouds.

Closer to our approach, verifiable confidential cloud computing (VC3)~\cite{schuster:2015:vc3} is a distributed, secure execution environment extending the Apache Hadoop \mapreduce framework, originally implemented and evaluated in an \ac{SGX} emulator.
Each worker node hosts a trusted loader that runs a key exchange protocol in the enclave before decrypting and executing map/reduce functions.
VC3 guarantees global integrity by generating verifiable work summaries within trusted workers.
The user code is written in C++, which can make the implementations prone to potential faults like illegal memory accesses.
In this regard, authors provide an optional compiler through which the programmers can enforce self-integrity invariants for memory regions.

An extension of VC3 appears in \cite{ohrimenko:2015:mrleakage}, which focuses on security issues generated by traffic analysis attacks on the exchanges between mapper and reducer nodes.
We do not consider such attacks in our evaluation, but we believe that the analysis and the solutions proposed are also applicable to our system.
$M^{2}R$ \cite{dinh:2015:m2r} also presents a secure framework for \mapreduce, which takes a more general approach with a design that can be implemented on any \ac{TEE}.
The authors refer to \ac{SGX}-enabled processors as one potential target, but like VC3, the evaluation is conducted on simulated \acp{TEE}.
As in \cite{ohrimenko:2015:mrleakage}, $M^{2}R$ specifically focuses on attacks exploiting the leakage between mappers and reducers.

The general idea behind the above frameworks is close to our proposal in the sense that users can write their own \emph{map} and \emph{reduce} functions that execute in \acp{TEE}.
However, we take a different approach with respect to the processing mechanism of interpreted code and to the data dissemination on top of \ac{SCBR}.
Besides the fact that we evaluate on real \ac{SGX} hardware, our approach provides additional flexibility and ease-of-use through the high-level Lua-based programming environment (see \Cref{sec:lwmp}).

\subsection{\label{sec:back:stream}Stream processing}

Stream processing represents another class of computations.
It consists in the data treatment of continuous events, such as sensor measurements, user activity in social networks or financial transactions.
Like batch processing, it is also subject to security and privacy concerns when deployed in hostile environments such as multi-tenant clouds.

Several industrial players introduced their own stream processing solutions, \eg, Twitter's Heron~\cite{kulkarni:2015:twitterheron} and Google's Cloud DataFlow~\cite{akidau:2015:dataflow}.
These systems are mainly used to ingest massive amounts of data and efficiently perform real-time analytics.
They are typically deployed on the provider's premises and are not offered as a service to end-users.

Some open-source middleware frameworks like Apache \textsc{Spark}~\cite{apachesparkstreaming}, Apache \textsc{Storm}~\cite{apachestorm} and \textsc{Infinispan}~\cite{infinispan} introduced \acp{API} to allow developers to quickly set up and deploy stream processing infrastructures.
These systems rely on the \ac{JVM}~\cite{lindholm:2014:jvm}, whose memory requirements impose considerable challenges to achieve good performance in the face of \ac{SGX} constraints.
Spark~\cite{zaharia:2013:discrstreams} is the most prominent solution.
It leverages resilient distributed datasets (RDDs) to provide a uniform view on processing data.
Despite its popularity, vanilla Spark only handles unencrypted data and hence does not offer security guarantees.
Some proposals extend it to provide security for data at rest~\cite{shah:2016:secspark}.
More recently, researchers ported it to run in \ac{SGX} enclaves: Sgx-Spark~\cite{lsds:2019:sgxspark} which is based on SGX-LKL~\cite{priebe:2019:sgxlkl} and SGX-PySpark~\cite{lequoc:2019:sgxpyspark} based on \ac{SCONE}~\cite{arnautov:2016:scone}.

Some proposals rely on a hybrid model with trusted and untrusted infrastructures, where critical processing is done in private clouds.
Under this model, \textsc{Styx}~\cite{stephen:2016:styx} uses partial homomorphic encryption and show overheads of \SI{25}{\%} in comparison to Apache Storm. However, they cannot guarantee data integrity.
A few dedicated solutions exist today for distributed stream processing using reactive programming.
For instance, \textsc{Reactive Kafka}~\cite{reactivekafka} allows stream processing atop of Apache \textsc{Kafka}~\cite{apachekafka,kreps:2011:kafka}.
Such solutions do not support secure execution in \acp{TEE}.

\Cref{sec:sstreams} introduces \securestreams, a reactive middleware framework to deploy and process secure streams at scale by decrypting data inside enclaves and performing plaintext processing.
Its design combines the high-level reactive data-flow programming paradigm with \acp{TEE} in order to ensure privacy and integrity of the processed data.
The experimental results of \securestreams are promising: while offering a fluent scripting language based on \textsc{Lua}, our middleware delivers high processing throughput, thus enabling developers to implement secure processing pipelines in just few lines of code.
To the best of our knowledge, \securestreams was the first lightweight and low-memory footprint stream processing framework that can fully execute within \ac{SGX} enclaves.
Later on, StreamBox-TZ~\cite{park:2019:streambox} was proposed for stream processing at the edge with \ac{ARM} TrustZone, which counts on weaker security guarantees (see \Cref{sec:back:trustzone}).
 
\section{\label{sec:related:datasharing}Data sharing}

Using public cloud services for storing and sharing confidential data requires mechanisms to cryptographically protect the data and limit the access to them. We use \acp{TEE} to tackle this subject in \Cref{chap:sharing}.
In \Cref{sec:ibbe}, we introduce \ibbesgx, a new cryptographic access control extension that is efficient both in terms of computation and storage even when processing large and dynamic workloads of membership operations.
\ibbesgx addresses the impracticality of \ac{IBBE} by exploiting Intel \ac{SGX} to derive cuts in the computational complexity of encryption operations.
We also propose a group partitioning mechanism that attenuates the computational cost of decryption, so that it becomes bound to a constant partition size rather than the entire group's size.
As a result, \ibbesgx performs membership changes much faster than the traditional approach of \ac{HE} while producing less metadata.

In some cases, the identity of end users needs to remain confidential against the cloud provider and fellow users accessing the data.
As such, the underlying cryptographic access control mechanism needs to ensure the anonymity of both data producers and consumers.
We introduce \asky in \Cref{sec:asky}, a cryptographic access control extension capable of providing confidentiality and anonymity guarantees.
Thanks to \acp{TEE}, we are able to handle the impracticality of \ac{ANOBE} schemes with simple cryptographic constructs.
We achieve faster execution times and shorter ciphertexts, being hence able to efficiently scale to large organisations.

We split the related work on 
\begin{enumerate*}[label=\emph{(\roman*)}]
	\item cryptographic schemes for access control, 
	\item cryptographic protection for storage, and
	\item confidential messaging systems.
\end{enumerate*}
We describe next what are the main features in each class of solutions and how they relate to our proposals.

\subsubsection{Cryptography and access control}

Hybrid encryption (\acsu{HE}) consists of using a public key to cipher a symmetric key.
Once the resulting ciphertext is shared with an interlocutor who holds the corresponding private key, further communication is performed using symmetric encryption, which is faster and produces smaller ciphertexts.
Such approach combined with a \ac{PKI} and \ac{IBE} was used in  a role-based access control scheme. Its high overhead makes it unsuitable for reasonably dynamic cloud storage scenarios~\cite{garrison:2016:daccloud}.
Attribute-based encryption (\acsu{ABE})~\cite{sahai:2005:fuzzy} is a cryptographic construction that allows a fine-grained access control by matching labelled attributes to users and content.
Depending on the location of labels, one can distinguish between key-policy \ac{ABE}~\cite{goyal:2006:abe} and ciphertext-policy \ac{ABE}~\cite{bethencourt:2007:abe}.
Even when employed for simple access control policies, \ac{ABE} costs are substantially greater than \ac{IBE}.

Hierarchical identity-based encryption~(\acsu{HIBE})~\cite{boneh:2005:hierarchicalibe} and \ac{FE}~\cite{boneh:2011:functional} are two cryptographic schemes offering functionalities for access control that rely on pairing-based cryptography, like \ac{IBE} and \ac{ABE}.
\ac{HIBE} is designed to handle hierarchical setups where each node may issue private keys to its descendants.
\ac{FE} is a powerful construct that can arbitrarily encapsulate programs as access control, but is unsuitable for practical use.
Iron~\cite{fisch:2017:iron}, which handles \ac{FE}, is the closest to our proposals in the sense that it takes advantage of \ac{SGX} to build a practical encryption scheme for an unpractical strategy thus far.
Like \ibbesgx, they use an enclave that holds a master secret as root for later key derivations.
The enclave generates a key that is associated to a function, so that the computation can be performed without revealing the data on top of which it is applied.
However, results of applying such function are presented in clear.

Proxy re-encryption~\cite{ateniese:2006:reencryption} allows a data owner to delegate re-encryption to a proxy by sharing a transformational key with it, with the intent of sharing these data with another user.
It can possibly be combined with \ac{IBE}~\cite{green:2007:identity} or \ac{ABE} ~\cite{green:2011:outsourcing, sahai:2012:dynamic} and is suitable for cloud environments, as the re-encryption and data storage can happen in the same premises.
\asky (\Cref{sec:asky}) uses a similar proxying approach for writing operations, although we do not require users to share keys.

Multicast communication security~\cite{stinson:2005:cryptography, canetti:1999:multicast} defines efficient schemes that focus on revocation aspects.
Logical Key Hierarchy~\cite{wallner:1999:rfc2627} is a re-keying approach in which communications costs for revocation operations are logarithmic.
Other schemes~\cite{fiat:1994:be, naor:2010:revoke} exploit secret sharing mechanisms that tolerate up to a maximum numbers of colluding revoked users.

Like \ac{HE}, \ibbesgx (\Cref{sec:ibbe}) encrypts a shared key with a more complex cryptogrpahic scheme, \ie, \ac{IBBE}, so that simpler and faster approaches like \ac{AES} can further leverage that key.
\asky (\Cref{sec:asky}), on the other hand, employs symmetric encryption to encapsulate the shared key. That is done, however, once per group member.
Composing the groups in both systems is responsibility of an administrator who has no access to keys.
We adopt a \emph{lazy} revocation policy, \ie, revoked users lose access to future group keys but keep being able to read old data.
In our systems, collusion between any number of revoked users cannot grant access to keys encapsulated after their revocations.

\subsubsection{Cryptographic cloud storages}

A number of storage and sharing system designs have been proposed for mitigating the lack of trust in cloud providers.
DepSKY~\cite{bessani:2013:depsky} proposes a client-side object store  that encrypts and redundantly stores ciphertexts on multiple untrusted storages.
Encryption keys are split by using a secret sharing scheme~\cite{shamir:1979:howsharesecret} and dispersed over multiple storage premises that are assumed to not collude with each other.
SCFS~\cite{bessani:2014:scfs} extends DepSKY by using a trusted metadata coordination service that also encapsulates access control, which, in turn, is not cryptographically protected. 
This coordinator service is trusted and can be therefore compromised if an attacker breaches into it.

Other systems cryptographically enforce access control using \emph{key envelopes}.
CloudProof~\cite{popa:2011:cloudproof} is a cloud storage system with client-side encryption that solves access control by using broadcast encryption~\cite{boneh:2005:collusion} to envelope two keys: one for encryption and another for signing.
It offers confidentiality, integrity, freshness and write-serializability. 
CloudProof does not discuss how the authenticity of users' identity is established and checked.
Hypothetically, \ac{PKI} could be used, thus requiring a trusted entity in the system.
To avoid that, \ibbesgx relies on the identity-based version of broadcast encryption.

Sieve~\cite{wang:2016:sieve}, instead, uses \ac{ABE} for access control and key homomorphism for providing a zero knowledge guarantee against the storage provider.
It allows users to store encrypted data in the cloud and to delegate access to consuming web services.
REED~\cite{li:2016:rekeying} also uses \ac{ABE}~\cite{bethencourt:2007:abe} to envelope symmetric keys that cipher deduplicated content.
Performance overheads of rekeying operations drastically increase to several seconds when the number of users is as low as 500.
This reinforces the inability of \ac{ABE}'s practicality for large scale access control.

Our protocols are agnostic to the storage system.
Both \ibbesgx and \asky cryptographically protect a key that can only be deciphered by rightful group members.
To do that, these members need a ciphertext and additional metadata that could possibly be disseminated through any channel.
We leverage the envelope idea in \asky, so that each encrypted file has corresponding metadata that are sufficient for retrieving its plaintext content.
\ibbesgx, on the other hand, has both a global piece of information (the public key) and individual pieces of metatada per group.
In both systems, infrastructure providers learn nothing about keys.

\subsubsection{Confidential messaging systems}

Encrypted messaging systems share a common initial phase with our file sharing model by requiring the construction of a group key that shields group communication.
Popular messaging systems (\eg, WhatsApp, Threema, Signal) use a \ac{DH} group key agreement and derivation~\cite{rosler:2018:more}.
Such protocols require all active participants to contribute to the creation of the group key, albeit with no anonymity guarantees.
Pung~\cite{angel:2016:unobservable} uses \ac{PIR} in conjunction to a group \ac{DH} key derivation, thus achieving anonymity. 
Such mechanisms are different from our target model, in which active users do not need to participate in the group key's creation.

Pretty good privacy~(\acsu{PGP})~\cite{zimmermann:1995:pgp} is used for cryptographic protection of files or emails.
It addresses anonymity (hidden recipient mode, in its nomenclature) by performing symmetric encryption of the content and then several public key encryptions of the shared key, one per user in the sharing group.
An outside adversary only sees ciphertext and cannot infer who are the recipients.
At decryption time, rightful recipients must try to decrypt each fragment until they succeed ($\frac{n}{2}$ trials on average, where $n$ is the group size).
We took a similar approach with \asky (\Cref{sec:asky}), but using symmetric cryptography inside enclaves.
\Cref{tab:gpg} shows the results of a simple benchmark of \ac{GPG} v.\,1.4.2 in hidden recipient mode.
Encryption and decryption have enormous latencies of \SI{12}{\second} and \SI{60}{\second} for groups of \num{1000} members, respectively.
Moreover, the inner implementation of \ac{PGP}'s hidden recipient mode is reputed as insecure against chosen ciphertext attacks~\cite{barth:2006:privacy}.

\begin{table}
	\centering
	\caption[GPG latency.]{\label{tab:gpg}GPG latency in hidden recipient mode.}
	\begin{tabular}{lSSS}
		\toprule
		Group size & {Avg. encrypt [\si{\second}]} & {Avg. decrypt [\si{\second}]} & {Envelope size [\si{\kilo\byte}]} \\
		\midrule
		10 & 0.1 & 0.6 & 5.3\\
		$10^2$ & 0.7 & 5.8 & 16.5\\
		$10^3$ & 12 & 60 & 129\\
		\bottomrule
	\end{tabular}
\end{table}

The problem of devising a cryptographic scheme that can guarantee both confidentiality and anonymity is referred to as \emph{anonymous} (or \emph{private}) \emph{broadcast encryption} (\acs{ANOBE}).
A few of such schemes were proposed, however without assessing their practicality within real systems.
Barth \etal~\cite{barth:2006:privacy} (\bbw, per authors' initials) achieves inner and outer anonymity, in addition to guaranteeing \ac{IND-CCA}.
It extends the public key enveloping model of \ac{PGP} by incorporating signatures~\cite{boneh:2006:signatures} such that an attacker who is a group member cannot reuse envelopes to broadcast arbitrary messages to the group.
\bbw can handle key enveloping throughputs of only few hundreds of users per second.

Additionally, they propose the construction of publicly-known labels to decrease the number of decryption trials.
The ciphertext fragments in the envelope are sorted by labels and can therefore be located in logarithmic time at decryption.
This way, only one asymmetric deciphering operation is performed.
The scheme was further extended by Libert \etal~\cite{libert:2012:abe} by suggesting the use of \emph{tag-based encryption}~\cite{mackenzie:2004:alternatives} to hint users about their ciphertext fragment.

\asky provides indexing mechanisms for achieving better performance at decryption time, so that the specific ciphertext of a given user can be located inside the envelope in logarithmic time.
Like in \ac{PGP}, \asky also encrypts the same data several times, once per group member.
We use instead more efficient symmetric cryptographic algorithms.
Unlike the messaging systems mentioned above, key generation in both \ibbesgx and \asky is made inside enclaves and is simply based on random number generators.
Therefore, such procedure does not require that individual users be online to contribute in the constitution of shared keys.
 
\section{\label{sec:back:privacy}Privacy assurance}

Not every system can be designed from scratch to properly ensure security and privacy.
Sometimes users must use established services that are unlikely to change, specially if monitoring user activity is at the heart of the service provider's economic model.
That is the case of Web search engines whose business depend on targeted advertisements.
\Cref{chap:privacy} discusses privacy-preserving systems that use \acp{TEE} as building blocks.

By regularly querying Web search engines, users disclose large amounts of their personal data as part of the queries, possibly unconsciously.
Among these data, they may reveal sensitive information like health issues, sexual, political or religious preferences.
Nowadays, there is no satisfactory approach to enable users to access search engines in a privacy-preserving way.
Existing solutions are either subject to attacks, too costly due to heavy use of cryptographic mechanisms, dependent on weak adversarial models or unable to provide accurate results.

Privacy-preserving proposals can be classified in three main categories.
One of them is composed of alternative search engines which implement specific privacy-preserving protocols generally based on \ac{PIR}, thus enforcing privacy by design. 
In such systems, users access data stored on the remote server without revealing what information they access.
In general, \ac{PIR} protocols consist of three algorithms: building protected queries with encrypted keywords, performing the information retrieval in such a way that the search engine has no access to the query and results, and finally reconstructing the result list on the client side. 
They generally rely on heavy cryptographic protocols that are still unpractical~\cite{aguilar:2016:xpir}, especially when data stores contain millions of documents, which is the case of search engines.

The other two categories of privacy-preserving systems enable clients to use existing search engines while offering them a set of privacy guarantees.
In the text that follows, we describe related work in private Web search and discuss the limitations of existing solutions. 

\subsubsection{Enforcing unlinkability}

\begin{figure}[t]
	\centering
	\includegraphics{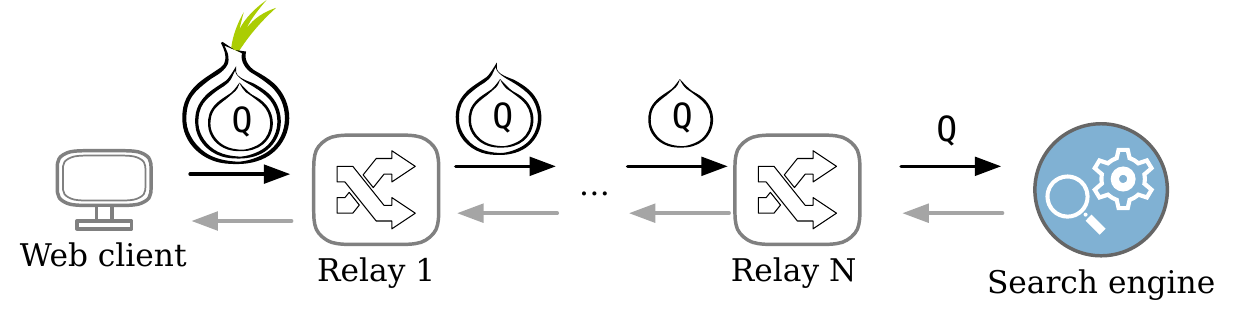}
	\caption{Unlinkability: Tor.}
	\label{fig:tor}
\end{figure}

This category of solutions includes a set of protocols enabling users to anonymously send queries to a search engine, thus enforcing unlinkability between queries and the identity of their issuing users (\eg, IP address).
The most popular protocol among these solutions is \tor~\cite{dingledine:2004:tor}, an implementation of the onion routing protocol~\cite{goldschlag:1999:onion}.
\tor (\Cref{fig:tor}) sends each query through randomly-selected nodes using a cryptographic protocol where queries are encrypted using a public key of each node in the chain.
The ciphertext can hence be associated to an \emph{onion}, \ie, with multiple layers.
Each relay node deciphers the outermost layer of the onion and further forwards the remaining blob until it reaches the exit node.
The exit node, in turn, can finally send the query to the search engine on behalf of the original user.
One of the limitations of this protocol is that participating relays are assumed to faithfully execute the forwarding of onions, which might not be true as some may behave selfishly, \eg, by dropping them, or even maliciously, \eg, by injecting fake traffic to slow down the system.

\begin{figure}[b]
	\centering
	\includegraphics{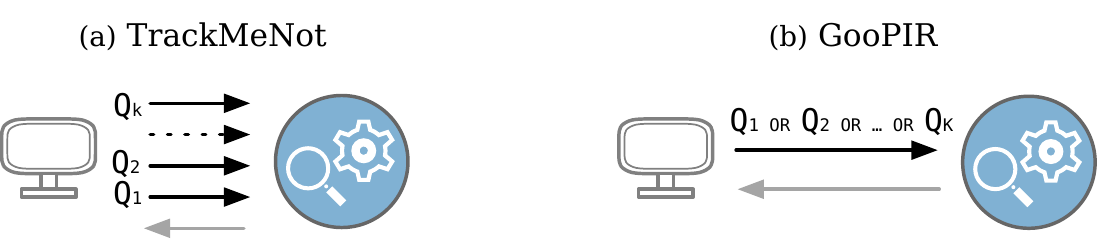}
	\caption{\label{fig:priv:indist}Indistinguishability: \textit{(a)} \tmn and \textit{(b)} \goopir.}
\end{figure}

To mitigate negative effects caused by selfish or malicious users, RAC~\cite{mokhtar:2013:rac} was proposed.
In this protocol, nodes are organised on several virtual rings in a way that for each ring a node has a predecessor and a successor node. 
A node might be part of several rings and thus have multiple predecessors and successors.
To ensure that no message is dropped by a freerider, nodes have to broadcast all messages they relay.
If a node does not receive a message from a given predecessor, it is considered as a freerider.
RAC suffers from severe performance limitations, presenting throughputs that are orders of magnitude lower than \tor.
The dissent protocol~\cite{corrigan:2010:dissent,wolinsky:2012:dissent}, in turn,  enforces accountability in presence of malicious and selfish participants.
However, its performance is even worse than the one of RAC as it is a combination of two heavy cryptographic protocols~\cite{chaum:1988:dining,brickell:2006:anonymity}.

In addition to the performance issue, protocols enforcing unlinkability have been shown not to resist to re-identification attacks~\cite{peddinti:2014:web,petit:2016:simattack}.
Assuming a set of user profiles built from user past queries, user re-identification attacks try to link anonymous queries to a profile corresponding to their originating user. 
The issue comes from the fact that search queries themselves disclose enough information for breaking the unlinkability property.
The reason is that users tend to look for similar things even when they have a different \ac{IP} address.
Besides, browser metadata included in the \ac{HTTP} headers can also help to identify users.

Unlikability is enforced in both systems we propose in \Cref{chap:privacy}.
While \xsearch (\Cref{sec:xsearch}) hides the identities of users from search engines by proxying their requests, \cyclosa (\Cref{sec:cyclosa}) creates a \ac{P2P} network of users who also act as relays.
Despite the flaws we listed above, service providers cannot tell with certainty where requests come from if the endpoint is not really the issuer, unless they have additional information.
This is why we further try not to provide them with this supplementary data, as we discuss in the next section.

\subsubsection{Enforcing indistinguishability}

Another class of solutions aim at making real user interests indistinguishable from fake ones. 
\queryscrambler~\cite{arampatzis:2013:scrambler} protects users by replacing their requests by semantically related queries.
For each request, it generates a set of related queries by generalizing the concepts used in the initial query.
By merging and filtering all the results obtained with these related queries, it retrieves the most plausible results for the initial query.
Unfortunately, results' accuracy might be impaired.

TrackMeNot~\cite{howe:2009:trackmenot} (\Cref{fig:priv:indist}a) is a browser extension which periodically sends fake queries to the search engine on behalf of users, independently of their real queries.
The intent is that the user profile stored at the search engine will eventually get obfuscated by mixing the user's real interests with fake ones.
Instead, GooPIR~\cite{domingo:2009:goopir} (\Cref{fig:priv:indist}b) obfuscates each user query by aggregating $k-1$ fake queries with the real one using the logical \texttt{OR} operator.
As such, the search engine cannot distinguish the real query from fake ones.
However, these solutions suffer from two limitations:
\begin{enumerate}[label={(\roman*)}]
	\item the user's identity is still known to the search engine; and
	\item they are subject to attacks as the fake queries they generate (based on \acsu{RSS} feeds in TrackMeNot and dictionaries in GooPIR) are easily distinguishable from real ones.
\end{enumerate}

\begin{figure}[tb]
\centering
\includegraphics{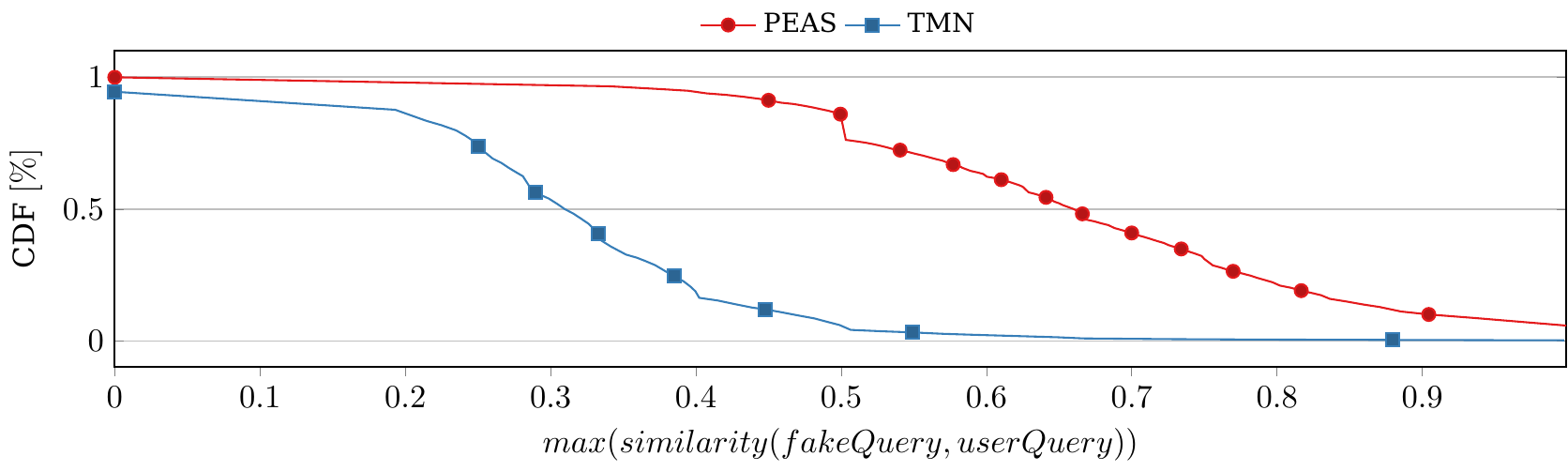}
\vskip 4mm
\caption[Quality of fake queries.]{\label{fig:xsearch:similarity}Similarity between fake queries generated by \tmn and \peas with the \acs{AOL} dataset.}
\end{figure}

\begin{figure}
	\centering
	\includegraphics{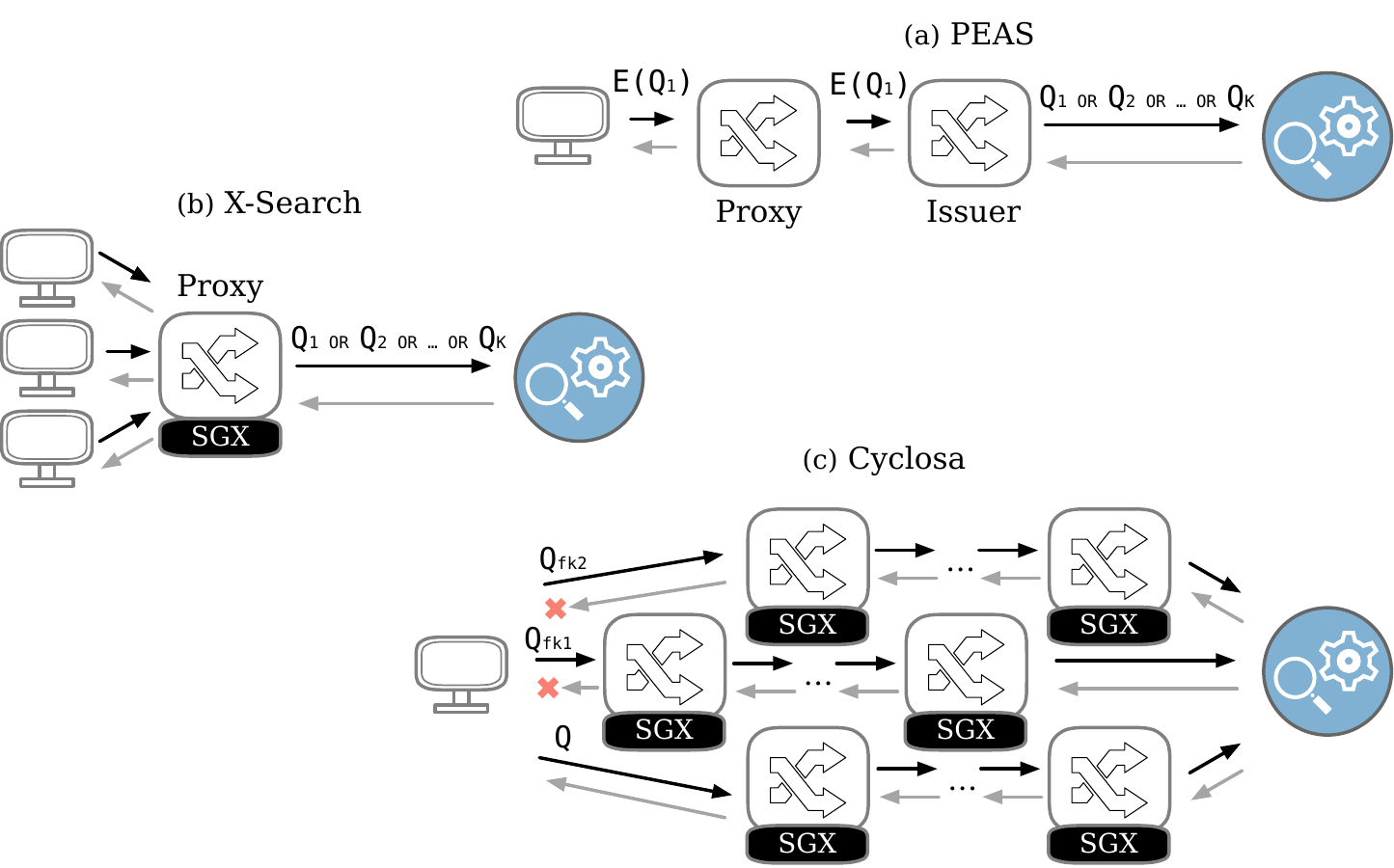}
	\caption[Systems combining indistinguishability and unlinkability.]{\label{fig:priv:unlink:indist}Indistinguishability and unlinkability: \textit{(a)} \peas, \textit{(b)} \xsearch and \textit{(c)} \cyclosa.}
\end{figure}

To overcome these limitations a solution named \peas~\cite{petit:2015:peas} (\Cref{fig:priv:unlink:indist}a), combining both unlinkability and indistinguishability was proposed, based on two non-colluding servers. 
The first, called \emph{proxy}, has access to the requester's identity but not to the query's content since it is encrypted with the public key of the second server.
This server, in turn, is called \emph{issuer} and has access to the query but does not know the originating user.
In addition to forwarding the query on behalf of the user, the issuer generates $k-1$ fake queries and aggregates them with the original query to enforce indistinguishability. 
Differently from \goopir and \tmn, \peas's fake queries are generated using a co-occurrence matrix of terms built by the issuer from other users past queries.
Hence, \peas better resists re-identification attacks as its fake queries are syntactically closer to real ones.

To highlight how challenging it is to generate efficient fake queries, we show in \Cref{fig:xsearch:similarity} the \ac{CDF} of the maximum similarity between fake queries generated by \peas (\ie, based on the co-occurrence of terms in past queries) and \tmn (\ie, based on \ac{RSS} feeds) and past queries on the \ac{AOL} dataset~\cite{pass:2006:picture} (the similarity metric is further detailed in \Cref{sec:xsearch}).
Even though \peas produces better fake queries (median similarity of \num{0.65} against \num{0.3} of \tmn), most of the fake queries are significantly different from real ones, \ie, they have never been requested to \ac{AOL}.

\Cref{sec:xsearch} introduces \xsearch (\Cref{fig:priv:unlink:indist}b), a novel private Web search mechanism that builds upon \ac{SGX} for proxying user requests.
\xsearch enforces both unlinkability and indistinguishability.
It runs query obfuscation based on past queries on untrusted proxy nodes within enclaves.
Apart from improvements on performance, it operates under a stronger adversarial model than its alternatives and better resists to re-identification attacks.

One of the above solutions' limitations is that all user queries get obfuscated with the same intensity, \ie, by generating $k-1$ fake queries in \goopir, \peas and \xsearch regardless of their sensitivity.
As a consequence, a small value of $k$ may lead to under protecting sensitive queries, which increases the risk that they get linked back to the original user.
Conversely, a large value of $k$ may unnecessarily generate large amounts of traffic.
\cyclosa (\Cref{fig:priv:unlink:indist}c), presented in \Cref{sec:cyclosa}, handles this issue while being scalable and returning accurate results.
It is fully decentralised, spreading the load for distributing fake queries among other \ac{SGX} peers.
The number of fake queries used for obfuscation is dynamically adapted to the user query's sensitivity.
As it handles the real and fake queries separately, it achieves perfect results' accuracy.

\subsubsection{Wrapping-up private web search}

With regards to unlinkability, existing protocols are either efficient but assume honest but curious servers or robust to malicious adversaries but have unpractical performance (\eg, Dissent, RAC).
In terms of indistinguishability, the challenge is to generate realistic fake queries that are as close as possible to real ones.

Enforcing indistinguishability by aggregating the user query with fake queries (\eg, using the \texttt{OR} operator) generates noise in the responses sent by the search engine as the results corresponding to fake queries get merged with those related to the real one. 
This noise is generally filtered out at the client side (in \peas and \goopir) or by the proxy (for \xsearch) by removing the responses that do not contain words composing the original query. 
Despite this, relevant responses of the original query may be lost while noise may be returned to the user.
Furthermore, the logical \texttt{OR} operator for multiword-based queries is not natively supported by all search engines.

Practical private Web search mechanisms must scale.
This is not the case of centralised schemes such as \peas or \xsearch.
In addition to the ability of private Web search to sustain the load coming from Internet users, a more concrete problem comes from the rate limitations imposed by search engines to counter bots and D\acs{DoS} attacks.
Google's bot protection, for instance, triggers after about \num{1000} queries before either asking to fill a captcha or refusing the request.
Another problem of approaches like \peas and \xsearch is the monetary cost of deploying proxies.
In contrast, \cyclosa leverages client machines that require no deployment.
\Cref{tab:websearch:summary} summarizes the comparison between private web search solutions.

\begin{table}[t]
	\setlength{\tabcolsep}{2.3pt}
	\centering
	\caption{Comparison of private Web search mechanisms.}
\begin{tabular}{r c  c  c  c  c  c }
    \toprule
    &\tor   &   \tmns & \goopir  &  \peas & \xsearch    &   \cyclosa    \\
    \midrule
    Unlinkability & \y  &       \n      &       \n      &       \y      &    \y     &  \y\\
    Indistinguishability & \n   &       \y      &       \y      &   \y          &       \y  & \y\\
    Accuracy & \y   &       \y      &       \n      &       \n      &    \n     & \y\\
    Scalability & \y    &           \y  &       \y      &       \n      &   \n      & \y\\
    \bottomrule
\end{tabular}
 	\label{tab:websearch:summary}
\end{table}

\vspace{-1em}
\subsubsection{Web and SGX}

\textsc{TrustJs} is a framework for trustworthy execution of security-sensitive JavaScript code inside commodity browsers~\cite{goltzsche:2017:trustjs}.
It leverages enclaves to protect the client-side execution of JavaScript, enabling a flexible partitioning of web application code.
Being attested by the server, the trusted interpreter can be used to offload its computation, which results in lower latencies in the user experience and lower performance demand for the application servers.

Kim et al.~\cite{kim:2015:tee4network} explored the possibility of using enclaves to provide security and privacy in network applications.
They initially demonstrate how to use enclaves to prevent software-defined inter-domain routers to disclose their routing policies.
They also show how the \tor anonymity network~\cite{dingledine:2004:tor} can be strengthened by using enclaves to run its directory authorities to attest each other.
Attackers can still launch denial-of-service attacks but they cannot alter the directory behaviour.
Also, by putting onion routers within enclaves, they can attest their integrity and their admission can be done automatically so directory authorities can be eliminated, and the routers can simply keep track of their membership in a distributed hash table.
Finally, they present how enclaves can be used to securely introduce in-network functionality into \ac{TLS} sessions.

\cyclosa resembles \textsc{TrustJs} in its client-side nature.
Rather than processing scripts locally though, we forward user queries to the \ac{P2P} network of relays.
\Wrt \tor, \cyclosa is similar in the sense of establishing the network of collaborating users.
The difference, besides the usage of symmetric encryption instead of layered ciphertexts, is that the enclave execution ensures fair behaviour.
As the \ac{P2P} network establishment is outside the scope of \cyclosa, we do not discuss about peer directories.
Nevertheless, the aforementioned strategies in this regard could be employed.

\vspace{-1em}
\section{Summary}

In this chapter, we presented several commercially available \acp{TEE} and gave a general overview of how they operate.
Special emphasis was given to Intel \ac{SGX}.
Due to its ensemble of security guarantees currently unmatched by alternatives, \ac{SGX} is the \ac{TEE} technology we further explore throughout the rest of this dissertation.

To contextualise part of this work's achievements, we briefly described the SecureCould project in \Cref{sec:securecloud} before covering the related work.
In \Cref{sec:back:cloud} we gave a general overview of potential cloud services materialised by communication and data processing frameworks, as per our contributions in this domain presented in \Cref{chap:clouds}.
First,  in \Cref{sec:back:pubsub}, we covered security and privacy in \iac{pubsub} middleware and next, batch (\Cref{sec:back:mr}) and stream processing (\Cref{sec:back:stream}) frameworks.

\Cref{sec:related:datasharing} outlined research work in cryptographic schemes that can be used for group data sharing in untrusted channels.
We discussed access control, cloud storages and messaging systems related to our work detailed in \Cref{chap:sharing}, where we design innovative cryptographic protocols that count on \acp{TEE} for achieving better performance and security guarantees.

Finally, \Cref{sec:back:privacy} explained the state of the art in private Web search.
It described the concepts of unlikability and indistinguishability while presenting advantages and drawbacks of systems that enforce them.
Doing so, it gave grounds for the contributions we report in \Cref{chap:privacy}.

\chapter{Communication and processing}
\label{chap:clouds}
\acresetall

To reach our goals, we start with empirically exploring the usage of \acp{TEE} by designing, implementing and evaluating systems that naturally fit into cloud deployment scenarios, where many potential threats are possible (see~\Cref{chap:background}).
Particularly, we describe the implementation of a secure content-based routing engine in~\Cref{sec:scbr} and two distributed processing frameworks: one for batch executions, in~\Cref{sec:lwmp}, and another for handling event streams in~\Cref{sec:sstreams}.
The goal is to quantify the \ac{SGX} performance implications as discussed in~\Cref{sec:sgx:perfissues} in the context of practical systems, besides learning about the main design concerns involved in building secure applications with the aid of \acp{TEE}.
Results show that there are performance penalties whenever cache and \ac{EPC} limits are surpassed.
However, horizontal scalability helps to overcome the impact of memory constraints.
In addition, the ability to replace complex cryptographic primitives with simple ones allows considerable gains in processing time.

\section{Communication}
\label{sec:communication}

The choice of \ac{CBR}, a flexible paradigm for scalable communication among distributed processes, conforms well to our target scenario because it handles sensitive data and is well suited for off-site deployment.
\ac{CBR} decouples data producers from consumers by dynamically routing messages depending on their content.
For this purpose, routers must filter messages (publications) by matching their content against a collection of stored predicates (subscriptions).
Such scheme thus requires the router to see the content of both publications and subscriptions, which represents a considerable privacy threat.
In the canonical example of stock market, for instance, quotes published by exchange platforms have commercial value, while subscriptions may reveal the client's interests and portfolio. Both the publisher's assets and user's privacy are at stake, hence why their sensitive information must be protected from leakage.

Since interacting processes are geographically scattered and routers are supposed to be deployed somewhere in between data producers and consumers, it makes sense to deploy them in third-party infrastructure providers.
Moreover, they could serve several organisations (data providers), each of which having their own clients. 
All combined, we have the main elements for reinforcing the security of a system: third-party servers, multi-tenancy and sensitive data processing.

In this section, we describe an original \ac{CBR} architecture that exploits \ac{SGX} for executing a routing engine in a secure enclave.
We also propose a protocol for securely exchanging keys among data producers, consumers, and the routing engine in the enclave.
Publications and subscriptions are encrypted and signed, thus protecting the system from unauthorised parties willing to observe or tamper with the information.
Our system, called \ac{SCBR} \cite{pires:2016:scbr}, thus combines a key exchange protocol and a state-of-the-art routing engine to provide both security and performance while executing under the protection of the secure enclave.
We then evaluate our implementation with a few workloads to observe the sources of performance overheads and the various trade-offs of \ac{SGX}. We provide comparative results against plain-text (insecure) matching as well as with an encrypted filtering alternative.
To the best of our knowledge, \ac{SCBR} was the first system to experimentally evaluate and demonstrate the benefits of executing \ac{CBR} in a \ac{TEE}.

\subsection{Secure content-based routing}
\label{sec:scbr}

Even if one can rely on trusted environments, designing a secure, privacy-preserving \ac{CBR} system is not trivial.
Consumers who want to protect their confidential data have to trust the code of the \ac{CBR} engine.
If it handles multiple publishers, from different administrative domains, each of which with their own clients, consumers might be discouraged of using the system fearing that their private subscription data could be accessible by other organisations.
Besides, that would require the maintenance of multiple data structures within routers, one per group of data provider and their clients.
Although technically feasible, we consider a simpler model based on the following assumptions:
\begin{itemize}
\item The \ac{pubsub} system operates as a service under the control of a single ``data provider'' that publishes data.
\item Consumers are the clients of the service and typically pay recurring fees to have access to the data.
\end{itemize}

Producers must be able to control which subscribers can join and read data from the service, and to exclude clients who stop paying their fees or give any other cause for contract dissolution.
The publishers operate within the administrative domain of the service provider from which data originates, and they are trusted by the clients for the purpose of the considered service.
This model closely maps, for instance, to the aforementioned scenario of a stock exchange.

Given the trust relationships between different components of \ac{SCBR}, it is clear that publishers and clients must share cryptographic keys that are not known by the infrastructure provider.
Furthermore, there should be some mechanisms for the publishers to include and revoke clients.
Excluded clients should be prevented from receiving new data, independently of whether they previously were legitimate customers.

\subsubsection{System Model}

\begin{figure}
\centering
\includegraphics{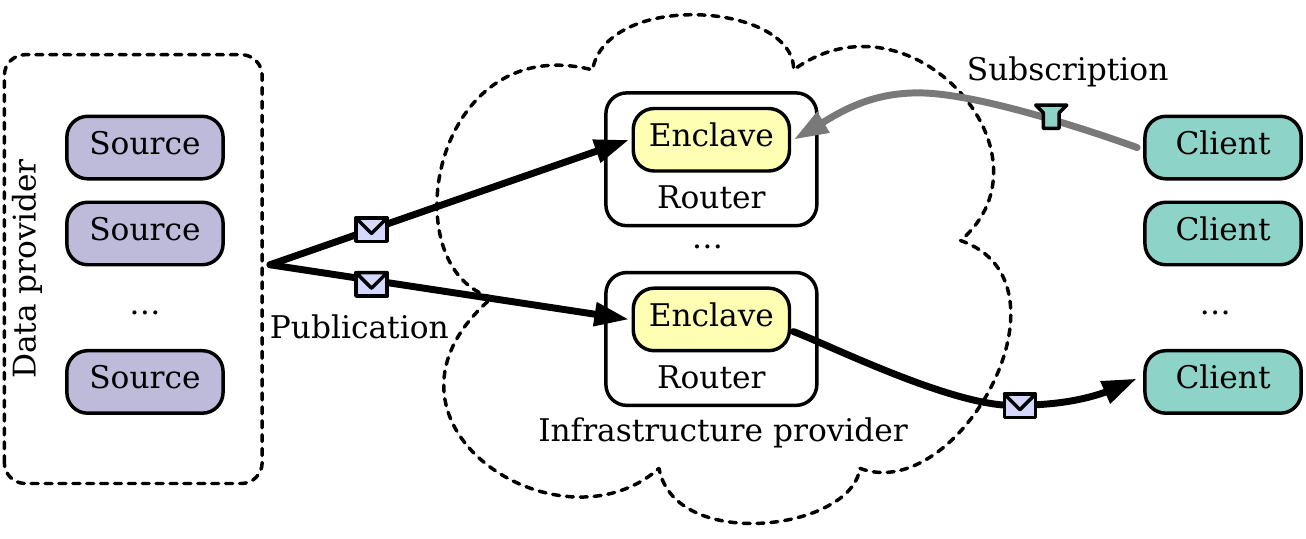}
\caption{\label{fig:roles}SCBR overview.}
\end{figure}

Following the considerations discussed above, we distinguish three main roles in our architecture, as illustrated in Figure~\ref{fig:roles}:
\begin{itemize}
\item
The \emph{data provider} produces information streams for the clients, typically ``as a service'' and for a fee.
The data may be produced by multiple sources (publishers) operating within the same administrative domain.
\item
The \emph{infrastructure provider} hosts the \ac{CBR} engines in the cloud.
It provides secure hardware and performs the actual data routing and transmission through its network.
As it operates under a different administrative domain and may share its resources among several customers in a multi-tenant configuration, the infrastructure provider is not trusted, although it accordingly performs its services.
\item The \emph{clients} of the service are the end users who are interested in the actual data and subscribe to information flows via the \ac{CBR} engines.
They trust the data providers but not the infrastructure.
\end{itemize}

Messages are composed of a payload, which is of interest to the end users but opaque to the router, and a header that contains several \emph{attributes} and associated values.
Since \ac{SCBR} is not concerned about payload contents, its encryption is out of scope of this chapter. We discuss alternatives for encrypted group communication in~\Cref{chap:sharing}.
The \ac{SCBR} router filters publication messages based on the attribute values in their header.

Subscriptions are composed of \emph{predicates} specifying constraints over the attributes.
Predicate expressions can include equality constraints or any kind of ranges over the attribute values, \ie, they can use the operators $=$, $>$ or $<$.
For instance, a subscriber interested in specific quotes for a company when they fall below a certain price can register a subscription such as ``$symbol = \texttt{"ABC"} \wedge price < 40$''.
A publication message \emph{matches} a subscription if its header satisfies the constraints expressed in the subscription predicate.

Subscriptions are typically stored by the \ac{CBR} engine in a dedicated data structure that operates as an \emph{inverted} database. Rather than actual data, the queries are stored instead, and data occasionally comes to be matched against them.
By exploiting relationships between the different predicates~\cite{carzaniga:2001:siena}, one can both reduce the memory footprint of the subscription index and improve the matching speed.
In particular, the property of \emph{containment} (or coverage) can be leveraged to avoid unnecessary tests.
Essentially, we say that a subscription $s$ contains or covers another subscription $s'$
if any event that matches $s'$ also matches $s$.
That is, $s$ is more general than $s'$.
For instance, predicate ``$x>0$'' covers both predicates ``$x=1$'' and ``$x>0 \wedge y=1$''.
Note that the containment relationships create a partial order on subscriptions that can be represented as a \ac{DAG}.
In \ac{SCBR}, we use a matching algorithm that exploits containment to minimise the footprint of stored subscriptions. This is particularly useful in enclaves, where only a limited amount of memory is available.

\ac{SCBR} makes use of both symmetric and asymmetric (public key) cryptography.
The former is more efficient and is used for communication between the publishers and the routers, while the latter is used between clients and the service provider when registering subscriptions, as will be detailed next.

\subsubsection{The subscription process}

\begin{figure}[tb]
\centering
\includegraphics{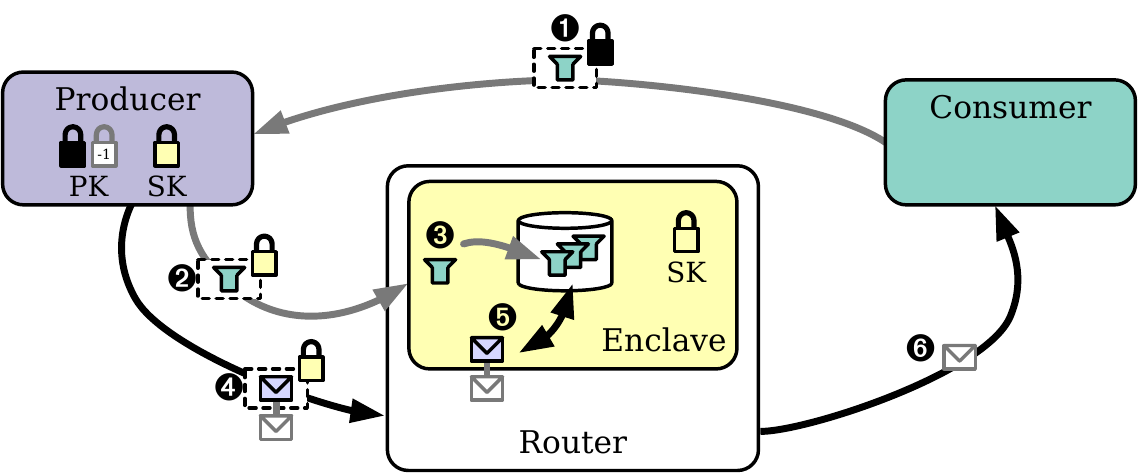}
\caption{\label{fig:interactions}\acs{SCBR} messages cycle.}
\end{figure}

We designed the system so that producers are the owners of the generated data.
They have therefore the ability to decide whether they accept a subscription from a client, as well as to subsequently remove it.
To control access to the service, we rely upon an additional admission phase when registering a new subscription.
The client cannot freely submit its subscriptions to the \ac{CBR} engines in the cloud, but has to go through a data producer.
The registration process works as follows (see Figure~\ref{fig:interactions}).

Consider a client $c$ that wants to register a subscription $s$ by the routing engine $r$ and subsequently receive message with header $h$ sent by the data producer $p$.
The publisher has a public/private key pair ($\mathit{PK}$/$\mathit{PK}^{-1}$), as well as symmetric key ($\mathit{SK}$) that is shared with the router running in the enclave, but unknown to clients and to the infrastructure provider.
This is made possible thanks to \ac{SGX}. Particularly, the symmetric key exchange happens during the attestation phase, as explained in~\Cref{sec:sgx:attestation}.

\begin{enumerate}
\item
The client first encrypts its subscription $s$ using the data provider's public key, hence preventing unauthorised parties to see it, and sends the resulting encrypted subscription $\{s\}_{\mathit{PK}}$ to $p$.
\item
Then, after decrypting and verifying that the subscription is valid, as well as verifying the client's status, the publisher re-encrypts $s$ using $\mathit{SK}$ and signs it.
It then sends the encrypted subscription $\{s\}_{\mathit{SK}}$ to the routing engine $r$.
\item
Finally, $r$ validates and decrypts the subscription inside the enclave and inserts it in its index.
\end{enumerate}

Subscriptions also embed location information about the clients that it visible to the code running outside the enclave.
This allows the router to correctly address the forwarding of publications payload to matching consumers.

\subsubsection{The publication process}

Once subscriptions have been registered by the routers, data can flow along the reverse path.
The publication process works as follows.
\begin{enumerate}
\setcounter{enumi}{3}
\item
The publisher encrypts the header $h$ of the message using $\mathit{SK}$, which is only known to the code running inside the enclave.
The encrypted message $\{h\}_{\mathit{SK}}$ is then sent to the routing engine $r$.
\item
Upon receiving the message, $r$ decrypts the header in the enclave, leaving the opaque payload outside, and matches it against its subscription index.
The result of this operation is a list of clients that have registered a matching subscription.
\item
Finally, $r$ forwards the encrypted message payload to all clients that have been identified as part of the matching operation.
\end{enumerate}

The payload of messages is encrypted separately. There are different approaches for doing so. A simple solution would be to use a symmetric group key shared between the publisher and all its consumers.
Depending on the number and turnover of clients, this can incur in excessive communication overhead since each revocation would require a new key to be propagated across the set of users. This would allow publishers to prevent clients that have cancelled their membership from accessing newly published messages.
This process is orthogonal to the privacy-preserving \ac{CBR}, \ie, the encryption performed for protecting the publications' header and subscriptions.
We further discuss cryptographic protocols for group communication in~\Cref{chap:sharing}.

Having multiple routers in the path would increase the complexity of the key management between publishers and matchers.
We believe that an overlay broker network is not the best architecture for a scalable privacy-preserving pub/sub engine, and we would rather advocate for a similar structure to StreamHub~\cite{barazzutti:2013:streamhub}, where system components are specialized in order to enhance performance.
In such an architecture, the current publisher-router key management scheme could be simply replicated.

\subsubsection{Evaluation}
\hyphenation{Rivest-Sha-mir-Adleman}

We evaluated the system as described in Figure~\ref{fig:interactions}, with both the producer and consumer running in one machine and the filtering engine in another.
Measurements were collected at the machine running the filter, which was equipped with an Intel Skylake \ac{CPU} model i7-6700 running at \SI{3.4}{\giga\hertz} with an \SI{8}{\mebi\byte} cache and \SI{8}{\gibi\byte} of main memory.
We allocated \SI{128}{\mebi\byte} of main memory to \ac{EPC} (maximum allowed).
In \ac{SCBR}, encryption outside the enclave was implemented using the Crypto++ library~\cite{cryptopp:2019} using respectively \ac{AES} in \ac{CTR} and Rivest-Shamir-Adleman (\acsu{RSA}) respectively for symmetric and asymmetric encryption.
We use the ZeroMQ library~\cite{zeromq:2019} for communication, and we encode messages in Base64 text format.
Information about page faults is obtained via the Linux system's \texttt{getrusage} function (attribute \texttt{minflt}).
Similarly, we rely on a Linux system call to configure and read the processor's performance counters for cache misses.

\begin{table}[b]
\caption{\label{tab:scbr:workloads}\ac{SCBR}: Workloads description.}
\begin{center}
\newcommand{\rbx}[1]{\raisebox{1.2ex}[0pt]{#1}}
\setlength\tabcolsep{2pt}
\renewcommand{\arraystretch}{1.1}
\begin{tabular}{r|rcl|l}
\hline
\multicolumn{1}{c|}{{Workload}} & \multicolumn{3}{c|}{{Proportion of}}  & \multicolumn{1}{c}{{Number of}}  \\[-1pt]
\multicolumn{1}{c|}{{name}}     & \multicolumn{3}{c|}{{equality predicates}}          & \multicolumn{1}{c}{{attributes}}  \\
\hline
\texttt{e100a1}        & 100\% & : & 1 eq. pred.                  &  \\
\cline{1-1}            \cline{2-4}
\texttt{e80a1}         & {20\%} & {:} & {0 eq. pred.}  & \rbx{8--11 (original)} \\
\cline{1-1}                                                       \cline{5-5}
\texttt{e80a4}      & {80\%} & {:} & {1 eq. pred.} & $4\times$ more            \\
\cline{1-1}            \cline{2-4}                                \cline{5-5}
\end{tabular}
 \end{center}
\end{table}

\begin{figure}[tb]
\center
\includegraphics{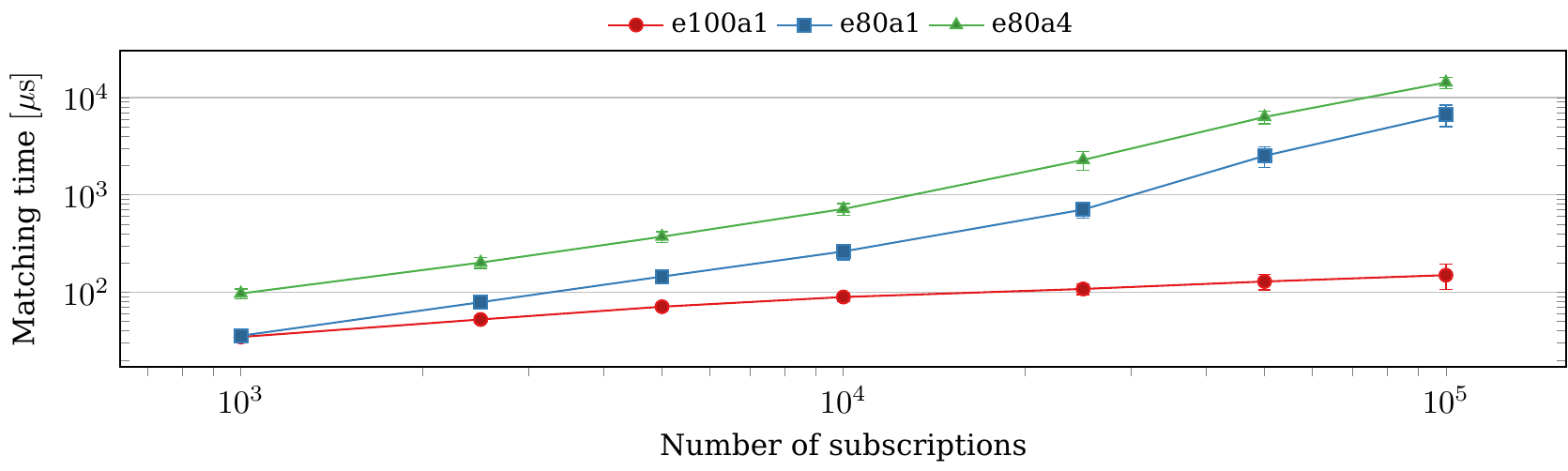}
\vskip 4mm
 \caption[Distinct workloads' characterisation.]{\label{fig:workloads}Performance of the containment-based algorithm applied to the different workloads in plaintext, outside enclaves.}
\end{figure}

To evaluate \ac{SCBR} and facilitate comparison, we reused the workloads from previous work~\cite{barazzutti:2012:thrifty} by picking 3 out of the 9 datasets used then.
They were chosen based on the diversity of performance output when applying our containment-aware matching algorithm to each dataset (\Cref{fig:workloads}).
They were built based on real data corresponding to randomly selected stock quotes from the Yahoo!\ finance website~\cite{yahoofinance:2019}.
Approximately 250,000 entries were collected in a period of 5 years, with publications composed by 8 to 11 attributes.
The entries collected were used to produce synthetic subscription datasets containing an assortment of equality and range predicates on the quotes' attributes according to a uniform random distribution.
In order to assess the algorithms' performance with a greater number of variables and different levels of containment, one workload was synthesised with four times the number of attributes of the original publications, by merging data from multiple quotes.
Table~\ref{tab:scbr:workloads} summarises the characteristics of the datasets used.

\begin{figure}[tb]
\center
\includegraphics{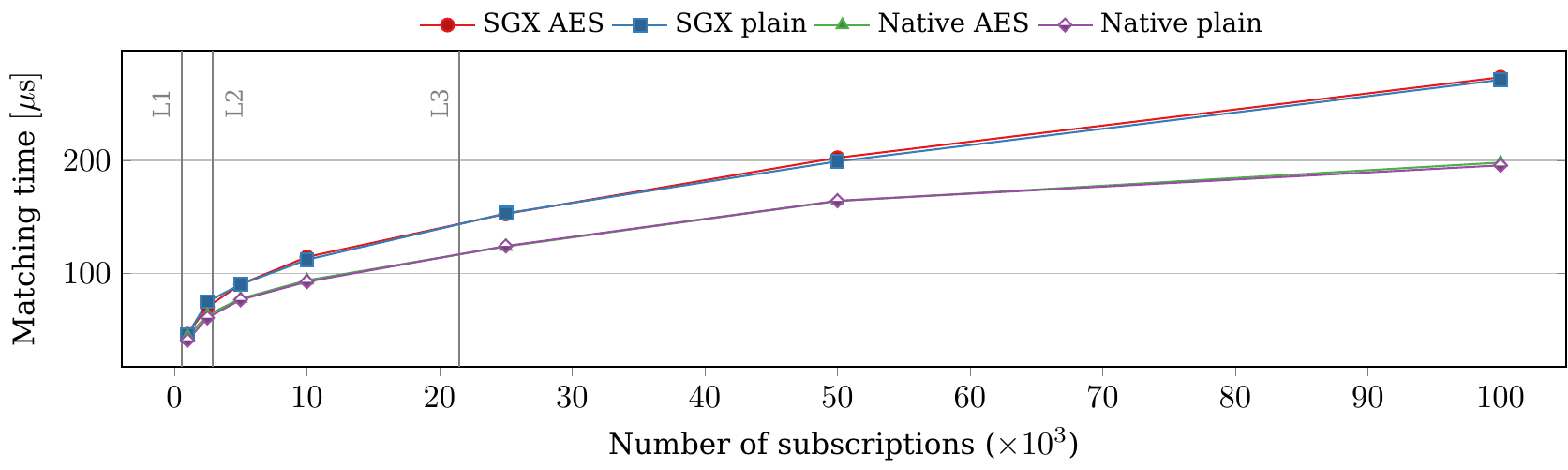}
\vskip 4mm
 \caption[\acs{SCBR} comparison: Native and \acs{SGX}.]{\label{fig:scbr:inoutcrypto}Comparison between native and \acs{SGX} executions, with and without encryption.}
\end{figure}

Our first experiment aimed at evaluating the performance overhead caused by executing our filter inside an enclave.
We filled the subscription database with datasets with up to 100,000 subscriptions, reaching a total memory size of approximately \SI{43}{\mebi\byte}.
Then we sent a batch of 1,000 publications to be matched against the subscriptions and measured the time it took to accomplish each filtering operation.
We ran an identical set-up with and without encryption, inside and outside an enclave, using the same filtering code.
When using encryption, publications and subscriptions were encrypted in the producer and decrypted in the filter using \ac{AES}-\ac{CTR}.
The average results for the first workload (\emph{e100a1}) are shown in \Cref{fig:scbr:inoutcrypto}.
By considering the proximity of the lines with and without encryption, we can see that encryption overhead is small and nearly constant.
Indeed, this overhead remains below \SI{5}{\micro\second} for both \ac{SGX} and native executions, which is negligible when compared to the matching time given a reasonably large database size.
The overhead resulting from the enclave is more significant, reaching nearly 40\% for the largest set of subscriptions considered in this experiment,
which is explained by enclave mode transitions and the occurrence of cache misses. Cumulative cache sizes are shown by the vertical lines.

We then focused on the influence of the workloads.
In order to understand the effect of different datasets on \ac{SCBR} performance, we first executed each of them without encryption outside secure enclaves.
Results are shown in \Cref{fig:workloads}.
The first (\emph{e100a1}) workload show the best performance, since all subscriptions contain equality predicates and the subscription set forms deeper containment trees.
In contrast, datasets with more attributes (\emph{e80a4}) perform worse because they yield indexes with more roots and shallow trees, therefore inducing more comparisons to traverse the whole subscription graph.

\begin{figure}
\begin{center}
\includegraphics{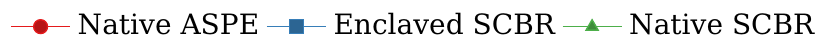}\hspace{5mm}\includegraphics{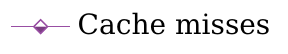}
\includegraphics{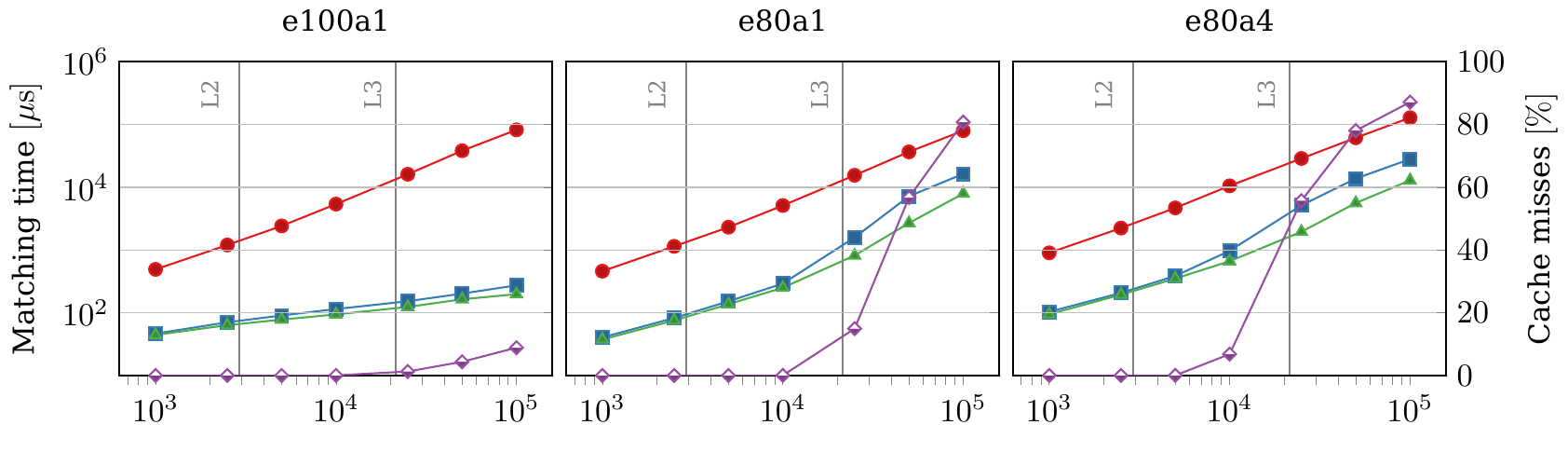}
\vskip -4mm
Number of registered subscriptions\\
\end{center}
\caption{\label{fig:scbr:wloads}SCBR response to different workloads.}
\end{figure}

Figure~\ref{fig:scbr:wloads} displays separate measurements for each workload running \ac{SCBR} inside and outside an enclave, both using \ac{AES} encryption.
We also measured, for each workload, the performance of our implementation of \ac{ASPE} \cite{choi:2010:aspe, barazzutti:2012:thrifty} as a baseline for a software-only alternative that does not use enclaves.
We measured only the matching step, and not the encryption or decryption of \ac{ASPE} messages.
The presented \ac{ASPE} performance cost was therefore inherent to its matching algorithm, which
grows faster than any other strategy when increasing the size of the subscription database.
The difference is more substantial for the first workload, although it remains close to at least one order of magnitude in all setups.
These observations indicate that the performance penalties of \ac{SGX} are largely tolerable when considering software-only alternatives for secure filtering, at least when the amount of memory used by the routing engine remains below the \ac{EPC} assigned size.

Another interesting aspect is the gap between the curves corresponding to native and \ac{SGX} executions.
After approximately 10,000 subscriptions, the versions inside and outside enclaves begin to drift apart due to the number of memory accesses necessary to accomplish every comparison.
At some point, the filtering data does not fit completely in the processor's cache memory and cache misses start to occur more frequently.
When this happens, data must be fetched from system memory and, in the case of enclave executions, it must be decrypted and checked for integrity and freshness.
Moreover, the evicted enclave's cache node must be encrypted before being sent to system memory.
That behaviour is consistent with cache miss rates (measured outside the enclaves), which are also reported in Figure~\ref{fig:scbr:wloads}.
We only measured cache misses outside the enclaves, because our Linux version failed to properly monitor the cache performance counters inside enclaves.
Since the code running inside and outside the enclaves is the same, it is reasonable to assume that cache miss rates would be similar.

We finally observe the performance penalties when exceeding the maximum protected memory size and memory swapping begins to happen.
Since \ac{EPC} memory is limited, whenever it is full and more space is required, pages must be evicted from the protected area to the main (untrusted) memory.
Accordingly, a page swap occurs every time a previously evicted page is accessed.
Besides the fact that system memory is slower than the processor's cache, which already imposes performance costs, memory page swaps are serviced by the \ac{OS} and hence incur an even higher overhead.

\begin{figure}
\center
\includegraphics{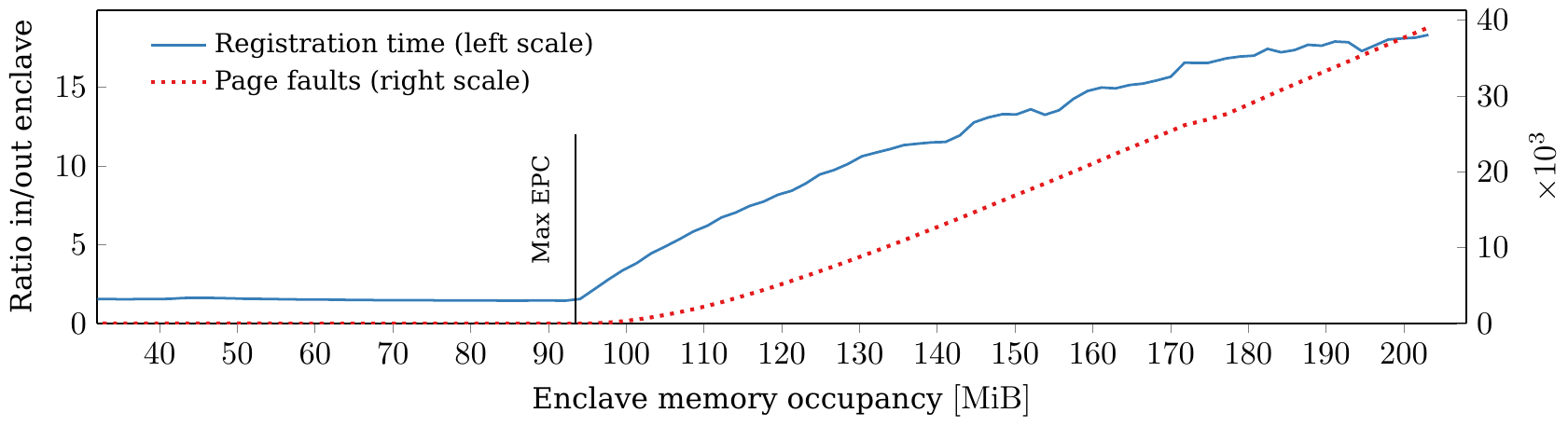}
\vskip 5mm
\caption{\label{fig:scbrepc}Performance loss when surpassing the EPC limit.}
\end{figure}

Figure~\ref{fig:scbrepc} shows the combined results of two executions when populating the in-memory subscription storage.
In one execution we registered subscriptions inside an enclave, and outside in the other.
We used the workload \emph{e80a1} in plain-text format, and we executed the same registration code in both experiments.
The graph accounts for a moving average of 5,000 points, registering up to 500,000 subscriptions.
We plotted the page fault rates by dividing the number of occurrences from the \ac{SGX} execution over native.
The values measured outside are very large for the largest database size, reaching up to 40,000 more page faults.

We also divided the time it took to register one subscription inside the enclave by the time required outside.
We can clearly see the point where paging kicks in, when memory consumption reaches just over \SI{93.5}{\mebi\byte}.
The vertical line shows the usable \ac{EPC} memory limit, excluding the reserved memory for \ac{SGX} internal data structures.
At the maximum size of our experiment (\SI{213}{\mebi\byte}), registering a subscription inside the enclave took 18 times more time than doing it outside.
These results show that the overhead grows outrageously when paging starts to happen, and they make a strong case for further studies on optimising the memory footprint of applications running inside secure \ac{SGX} enclaves.

This concludes our evaluation on \ac{SCBR}. We summarize the results as follows:
\begin{itemize}
\item The cost of symmetric encryption is negligible, less than \SI{5}{\micro\second} of increase in matching time, be it in enclaves or not. In fact, we did not observe any performance difference when using \ac{AES} in enclave mode in comparison to native execution.
\item Exceeding L3 cache level, on the other hand, brings some costs. We observed overheads of 38\%, 167\% and 164\% for the workloads e100a1, e80a1 and e80a4, respectively, with subscription databases of up to \SI{43}{\mebi\byte}.
\item The biggest performance issues come when overstepping \ac{EPC}'s limit. In our experiments, up to 18$\times$ for a memory usage of \SI{213}{\mebi\byte}.
\end{itemize}

\vskip 2mm
\section{Processing}
\label{sec:processing}

With the growth of cloud computing, distributed data processing has been extensively studied.
Offloading processing to public clouds is a natural choice when organisations do not possess their own data centres, or some specific hardware like \acp{GPU}.
Cost cutting can also be one important reason for supporting this decision.
Security, however, is still an important concern when it comes to data crunching in third-party infrastructure providers.
Even if it is possible to use effective encryption during data transmission and storage, there is no performant way of doing the same when processing it.
We have therefore good reasons for enhancing the security in distributed data processing systems.
We chose to port a \luavm to run within \ac{SGX} enclaves and use it as a building block in such systems. \Cref{sec:luavm} describes this port and presents some micro-benchmarks.
Next, we use it in two processing frameworks: one that follows the \mapreduce programming model for batch processing, described in \Cref{sec:lwmp}, and another event stream processing scheme that uses a library inherently asynchronous called RxLua that implements the paradigm of \emph{reactive programming}, described in \Cref{sec:sstreams}.

In order to provide processing services, there must be a way of receiving the operations that clients want to perform and the data supposed to be processed.
Then, these services need to apply the computation on the data and output the results to the next stage in the pipeline.
The trusted environment runtime must provide a way to load user instructions, integrate them into the engine and discard them to make room for the following rounds.
We discuss here the tradeoffs between different formats of these instructions: binary machine code or \emph{scripts}.

\ac{SGX} enclave binaries must be previously signed (see \Cref{sec:sgx:attestation}), so that when interlocutors want to attest it before sharing sensitive information, they are able to know the exact initial state of the protected code, its author and some flags.
Based on this, they assess whether to trust it and to continue the data exchange.
Although dynamic linking is possible after the enclave creation \cite{tsai:2017:graphene,silva:2017:dynsgx}, its use brings some security concerns.
In order to execute binary code received after initialisation, the enclave has to allocate memory pages with read, write and execution permissions, or else it would be prevented by the \ac{MMU} from running such code.
In a way, this weakens the assurance given by the attestation, since code that was not present during initialisation---and therefore not attested---can still be executed.
In this case, it is up to the enclave developer to care for the security of dynamically loaded code, which is not an easy task.
Moreover, if attackers are able to rewrite these pages by exploring some vulnerability (\eg, stack overflow or control-flow hijacking), they would be able to execute arbitrary code inside the protected environment.
Because of such threats, page access rights are signalized in the flags that compose the attestation data, so that an attestor can choose not to interact with enclaves that make use of writable and executable heap.

Interpreted programs are executed by software.
Although inherently slower because of that, they are usually more flexible in terms of abstractions and memory management.
The code, in this case, is actually input data to the interpreter.
Even if the execution of malicious code is still possible, its effects are easier to mitigate.
Since there is an intermediary between code and machine, forbidden instructions like system calls or out-of-bounds memory accesses can be previously checked and acted upon.
Besides, the machine binary code is entirely available during initialisation, allowing the enclave to have read-only executable pages and therefore inspiring more confidence to whichever process attesting it.
For these reasons, we decided to use an interpreted language to operate our distributed processing prototypes.

\begin{table}[b]
\centering
\caption{\label{tab:luabmarks}Parameters and memory usage for \textsc{Lua} benchmarks.}
\begin{tabular}{r|c|c|c}
\hline
           benchmark   &configuration &memory      &ratio \\
                name   &parameter     &peak        &SGX/Native \\
\hline
\textsf{dhrystone}     &\num{5e3}   &\SI{275}{\mebi\byte} & \num{1.14} \\
                       &\num{5e6}   &\SI{275}{\mebi\byte} & \num{1.04} \\
\hline
\textsf{fannkuchredux} &\num{10}    &\SI{28}{\mebi\byte}  & \num{0.99} \\
                       &\num{11}    &\SI{28}{\mebi\byte}  & \num{1.04} \\
\hline
\textsf{nbody}         &\num{2.5e6} &\SI{38}{\mebi\byte}  & \num{0.99} \\
                       &\num{25e6}  &\SI{38}{\mebi\byte}  & \num{1.00} \\
\hline
\textsf{richards}      &\num{10}    &\SI{106}{\mebi\byte} & \num{1.02} \\
                       &\num{100}   &\SI{191}{\mebi\byte} & \num{0.97} \\
\hline
\textsf{spectralnorm}  &\num{500}   &\SI{52}{\mebi\byte}  & \num{1.00} \\
                       &\num{5e3}   &\SI{404}{\mebi\byte} & \num{0.99} \\
\hline
\textsf{binarytrees}   &\num{14}    &\SI{25}{\mebi\byte}  & \num{1.18} \\
                       &\num{19}    &\SI{664}{\mebi\byte} & \num{4.76} \\
\hline
\end{tabular}
 \end{table}

\subsection{Lua within enclaves}
\label{sec:luavm}

We settled on a lightweight yet efficient embeddable runtime, based on the \textsc{Lua} \ac{VM}~\cite{ierusalimschy:1996:lua} and the corresponding multi-paradigm scripting language~\cite{ierusalimschy:2016:lua}.
Porting legacy code to \ac{SGX} means that every system call or input/output instruction have to be dealt with, since they are not allowed inside enclaves.
To achieve this, we traced all system calls made by the interpreter to the standard C library and replaced them by alternative implementations that either mimic the real behaviour or discard the call.
As a result, attempts made by malicious code to access the hard disk or make network connections are frustrated.

The \textsc{Lua} runtime requires only few kilobytes of memory and is designed to be embeddable. 
The language API provides the possibility to call Lua functions from C/C++ code.
As such, it represents an ideal candidate to execute in the limited space allowed by the \ac{EPC}.
Moreover, application-specific functions can be easily expressed in \textsc{Lua}, including complex algorithms~\cite{leonini:2009:splay}.
Our changes to the vanilla \textsc{Lua} source code consist of the addition of about \SI{600}{} \acp{LoC}, or \SI{2.5}{\percent} of its total size.

\begin{figure}
\centering
\includegraphics{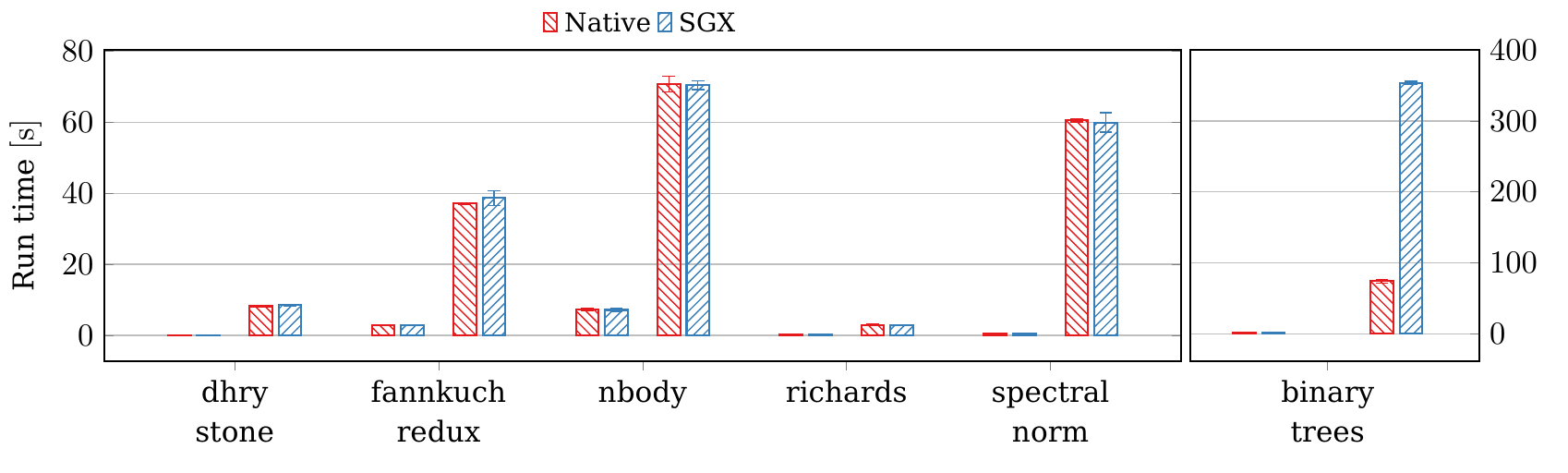}
\caption{Enclave versus native running times for \textsc{Lua} benchmarks.}
\label{fig:luabenchs}
\end{figure}

We evaluate the raw performances of our \ac{SGX} \luavm by selecting six available benchmarks from a standard test suite~\cite{bolz:2015:metatracing}.
We based this choice on their library dependencies (by selecting the most standalone ones) and the number of input/output instructions they execute (selecting those with the fewest \ac{IO}).
Each benchmark runs \SI{20}{} times with the same pair of parameters of the original paper, shown in~\Cref{tab:luabmarks}.
Figure~\ref{fig:luabenchs} depicts the total time (average and standard deviation) required to complete the execution of the \SI{6}{} benchmarks.
We use a bar chart plot, where we compare the results of the \emph{Native} and \emph{SGX} modes.
For each of the benchmarks, we present two bars next to each other (one per executing mode) to indicate the different configuration parameters used.
Finally, for the sake of readability, we use a different y-axis scale for the \textsf{binarytrees} case (from \SI{0}{} to \SI{400}{\second}), on the right-side of the figure.

In the first version of \ac{SGX}, it is required to pre-allocate all the memory area to be used by the enclave at initialisation.
The most memory-eager test --- \textsf{binarytrees} --- used more than \SI{600}{\mebi\byte} of memory.
If we compared the running time including the initial allocation, its duration would preponderate for shorter tests.
Because of that, we subtracted the allocation time from the measurements of \ac{SGX} executions, based on the average for the $20$ runs.
Fluctuations on this event produced slight variations in the execution times, sometimes producing the unexpected result of having \ac{SGX} executions faster than native ones (by at most $3\,\%$).
Table~\ref{tab:luabmarks} lists the parameters along with the maximum amount of memory used and the ratio between run times of \ac{SGX} and Native executions.
When the memory usage is low, the ratio between the Native and \ac{SGX} versions is small, less than \SI{20}{\percent} in our experiments.
However, when the amount of memory usage increases, performance drops to almost $5\times$ worse, as in the case of the \emph{binarytrees} experiment.
As we already observed in previous sections, the smaller the memory usage, the better performance we can obtain from \ac{SGX} enclaves when compared to native executions.

\vskip 6mm 
\subsection{Lightweight \mapreduce}
\label{sec:lwmp}
\vskip 2mm 

Since its adoption by Google~\cite{dean:2008:mapreduce}, the \mapreduce programming model consistently gained ground as a viable solution for assuring the necessary scalability of distributed data processing.
The generic model, composed of the \emph{map} and \emph{reduce} functions, was widely used to implement applications that can leverage parallel task processing.
The data to be processed are made available to a set of \emph{mapper} nodes, which apply in parallel a \emph{map} function responsible for converting individual data items to a finite set of key and value pairs.
The output is redistributed based on the keys, so that all values for a given key is grouped in one single \emph{reducer} node, in a step called \emph{shuffle}. 
Reducer nodes, in turn, execute in parallel a \emph{reduce} function for processing each dataset and outputting a final result.
This model can be used in processing tasks that range from simple compoundable operations like counting, sorting, and searching data; to more complex algorithms like cross-correlation or page rank.
\mapreduce was adapted in different ways to fit a wide diversity of scenarios and deployment platforms (see \Cref{sec:back:mr}).

We propose a self-contained framework for securing \mapreduce that leverages \ac{SGX}~\cite{pires:2017:lwmp}.
Our system combines \ac{SCBR} for communication (see \Cref{sec:scbr}), a \luavm as processing engine (see \Cref{sec:luavm}), and a \mapreduce library.
It is independent of the particular characteristics of the \emph{map} and \emph{reduce} functions and can hence be used for any problem that can be made parallel with \mapreduce.
The specific code to be executed in the \mapreduce service can be integrated in simple scripts, which run in isolation using \ac{SGX} over data decrypted only inside the enclave.
Our focus is on providing a flexible framework for securely running \mapreduce applications that can be easily implemented and deployed.
The basic \emph{word count} \mapreduce example, for counting the number of occurrences of different words in a given text, can be implemented in our framework with less than 30~\acp{LoC}.
We evaluate our approach using the widely used k-means clustering algorithm, showing that the overhead incurred is minimal and that our solution is applicable to other use cases.

\subsubsection{Solution architecture}

\begin{figure}[!b]
\centering
\includegraphics{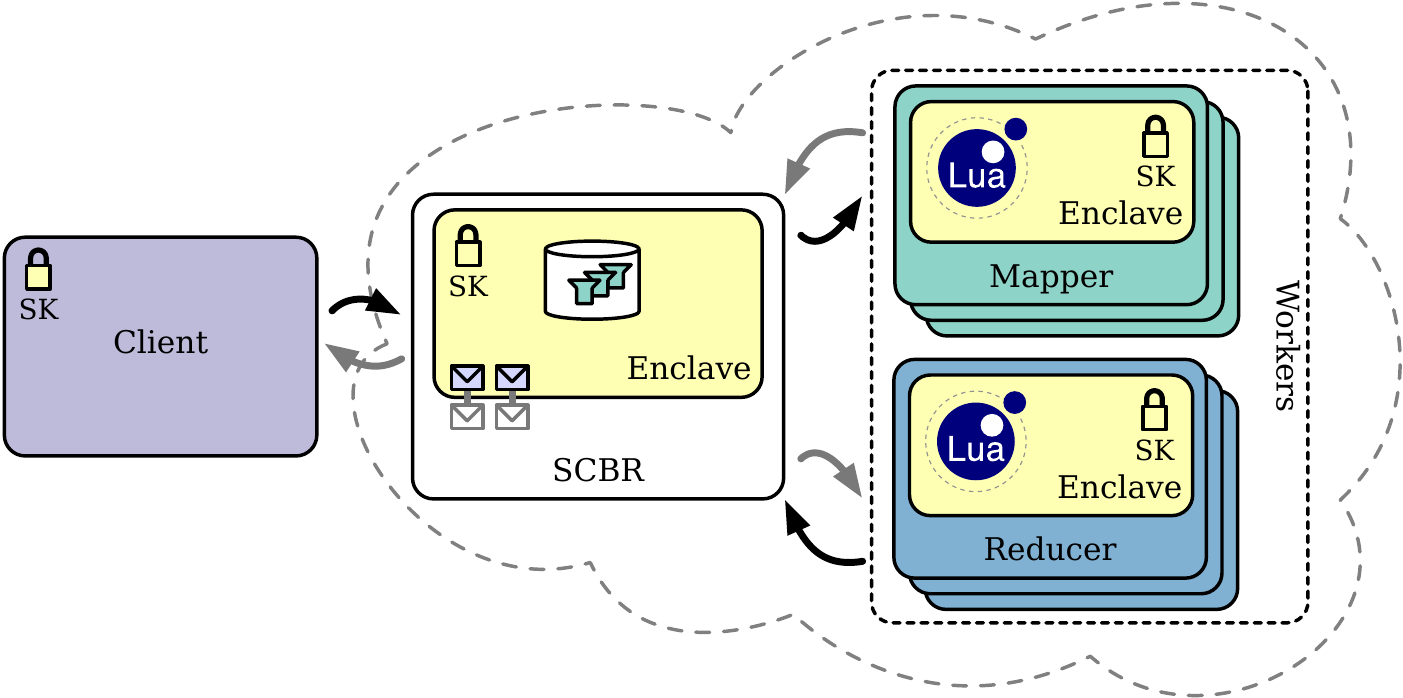}
\caption{\label{fig:entities}Lightweight \mapreduce components.}
\end{figure}

\Cref{fig:entities} displays the entities composing our solution: clients, \ac{SCBR} pub/sub engine and workers, which can assume the role either of a mapper or
a reducer.
Clients provide the code to be executed, the data to be processed and gather the results after completion.
In our work, we are concerned with assuring the privacy and integrity of code and data processed within the \emph{map} and \emph{reduce} functions, while offering to the programmer an accessible lightweight environment of implementing various use-cases.
Data or code is only seen in plain-text form at the client premises or inside enclaves. For simplicity, we assume a shared key $SK$ was previously established between clients and workers. 

All communication channels use the \zmq~\cite{hintjens:2013:zeromq} message passing library, having a central point in the \ac{SCBR} engine.
Although such a centralised approach is not suitable to large-scale data processing, it is arguably useful for modest quantities of highly sensitive data that could be, for instance, partitioned from higher amounts of non-sensitive data.
Nevertheless, it has been demonstrated~\cite{barazzutti:2013:streamhub,barazzutti:2014:elastic} that it is possible to elastically scale a pub/sub engine by specializing its functional steps into replicable operators.
That could dramatically improve the network performance of such a centralised approach.

Batch processing is the act of transforming one large and finite group of data that is entirely available upfront.
It contrasts with the processing of events, which are smaller pieces of data that are individually processed and timely unbounded.
In our system, a batch process is bootstrapped with an initial protocol among all parties and with the provisioning of code and data made by the client, when he sends chunks of the input data to mappers. 
We describe bellow what happens next.
\begin{enumerate}
\item Mapper nodes independently execute a \emph{map} function on their input data chunk and output a collection of $n$ key and value pairs $\{<k_0,v_0>, <k_1,v_1>, ..., <k_{n-1},v_{n-1}>\}$.
Optionally, pairs with common keys can be aggregated by a combiner function in order to optimize the redistribution, \eg, if $k_0=k_1$ then a \emph{combine} function may be called with $<k_0, \{v_0,v_1\}>$ as input.
Conceptually, \emph{combine} acts like a local \emph{reduce} function, in the context of a single mapper.
\item $<k,v>$ tuples are shuffled according to their keys and redistributed to reducer nodes. All tuples corresponding to the same key arrive on the same reducer.

\begin{table}[b]
\centering
\caption{\label{tab:mrio}\mapreduce: Payload used as inputs and outputs for each role.}
\begin{tabular}{c|c|c}
\toprule
Role                     & Input                          & Output                          \\ \midrule
\multirow{3}{*}{Client}  & \multirow{3}{*}{Final Results} & Map code                        \\
                         &                                & Reduce code                     \\
                         &                                & Mappable data                   \\ \cline{1-3}
\multirow{2}{*}{Mapper}  & Map code                       & \multirow{2}{*}{Reducible data} \\
                         & Mappable data                  &                                 \\ \cline{1-3}
\multirow{2}{*}{Reducer} & Reduce code                    & \multirow{2}{*}{Final results}  \\
                         & Reducible data                 &                                 \\ \cmidrule(lr){1-3}
\end{tabular}
 \end{table}

For instance, if the outputs of three mappers $m_i$ are:
\begin{itemize}
\item $m_0 \rightarrow \{<k_0,v_0>,<k_1,v_1>\}$
\item $m_1 \rightarrow \{<k_1,v_2>,<k_2,v_3>\}$
\item $m_2 \rightarrow \{<k_0,v_4>,<k_2,v_5>\}$
\end{itemize}
As a result of the shuffling phase, three reducers $r_i$ could receive as input:
\begin{itemize}
\item $r_0 \leftarrow <k_0,\{v_0,v_4\}>$
\item $r_1 \leftarrow <k_1,\{v_1,v_2\}>$
\item $r_2 \leftarrow <k_2,\{v_3,v_5\}>$
\end{itemize}
\item Reducer nodes independently execute the \emph{reduce} function on all data associated to a given key and output the result of their computation, which is forwarded back to the client.
\end{enumerate}

\subsubsection{\mapreduce as topic-based pub/sub}

The \ac{SCBR} engine is responsible for securely storing subscriptions that contain the conditions under which each message is forwarded to the corresponding interested party.
To design the session establishment protocol, we first list in \Cref{tab:mrio} the interests of each role, \ie, what data kind each entity is interested in.
This will later define their subscriptions.

\begin{table}[t]
\centering
\caption{\label{tab:mrtopics}\mapreduce: Topics for the \acs{pubsub} based protocol.}
\begin{tabular}{@{}l|l|l|l@{}}
\toprule
Description        & Symbol           & Subscribers          & Publishers           \\ \midrule
Map code           & \texttt{MAP\_CODETYPE}    & Mappers              & Clients              \\
Reduce code        & \texttt{REDUCE\_CODETYPE} & Reducers             & Mappers              \\
Mappable data      & \texttt{MAP\_DATATYPE}    & Mappers              & Clients              \\
Reducible data     & \texttt{REDUCE\_DATATYPE} & Reducers             & Mappers              \\
Final results      & \texttt{RESULT\_DATATYPE} & Clients              & Reducers             \\
\midrule 
Job advertisement  & \texttt{JOB\_ADVERTISE}   & Workers              & Clients              \\
Apply for position & \texttt{JOB\_APPLY}       & Clients              & Workers \\ \bottomrule
\end{tabular}
 \end{table}

\begin{figure}[b!]
\centering
\includegraphics{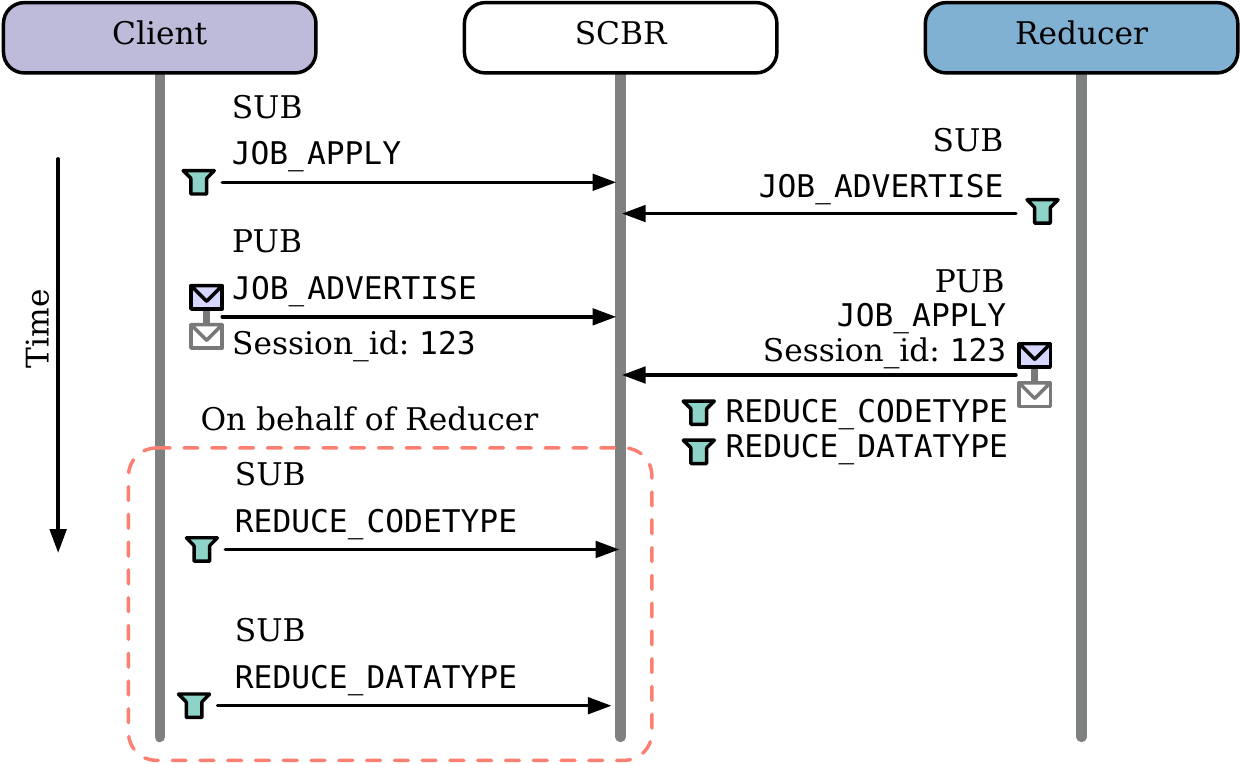}
\caption{\label{fig:initprotocol}\mapreduce: Session establishment protocol.}
\end{figure}

We decided to map payload data into topics~\cite{eugster:2003:manyfaces}. 
By tagging each message with the corresponding content, subscriptions and publications are respectively made according to the columns \emph{Input} and \emph{Output} in \Cref{tab:mrio}.
Besides these messages, we also identified the need of having two more topics, so that idle workers (mappers and reducers) are notified when there is a new client willing to process a batch of data (job advertisement).
Conversely, clients need to know when a worker is willing to serve them (apply for position).
\Cref{tab:mrtopics} lists all topics. \emph{Workers} stands for both mappers and reducers.

Worker nodes will act as subscribers registering queries to find out about new \mapreduce job openings, and also as publishers to signal their availability and type (mapper or reducer) in a job application.
The client of the \mapreduce service, which is also the data owner, will both act as subscriber and publisher, registering subscriptions for job applications and publishing advertisements on new \mapreduce jobs to be executed.
Moreover, the client will publish code and data to the registered workers and obtain the results after job completion.

The \mapreduce processing starts with an initial message exchange, as shown in \Cref{fig:initprotocol}. For simplicity, we use a reducer node as worker, but an analogous procedure applies for mappers.
Worker nodes register their intent of being notified for \mapreduce job openings through subscriptions on the topic \texttt{JOB\_ADVERTISE}.
The client registers its interest in knowing when workers are ready to perform a task by subscribing to the topic \texttt{JOB\_APPLY}.
When the client has a new job to execute, it advertises it through a \texttt{JOB\_ADVERTISE} publication, which is received by workers that previously registered for this topic.
Idle workers notify then their readiness to execute the advertised job through a \texttt{JOB\_APPLY} publication. They also include in the message payload their subscriptions for code and data (particular to the role they choose: mapper or reducer).
At the end of this negotiation, if the client decides to hire a worker, it registers on \ac{SCBR} the received subscriptions for code and data on the worker's behalf.
By doing so, the client establishes the \mapreduce chain and keeps track of how many workers it has hired and of which kind (mappers and reducers).

\begin{figure}[tb]
\centering
\includegraphics{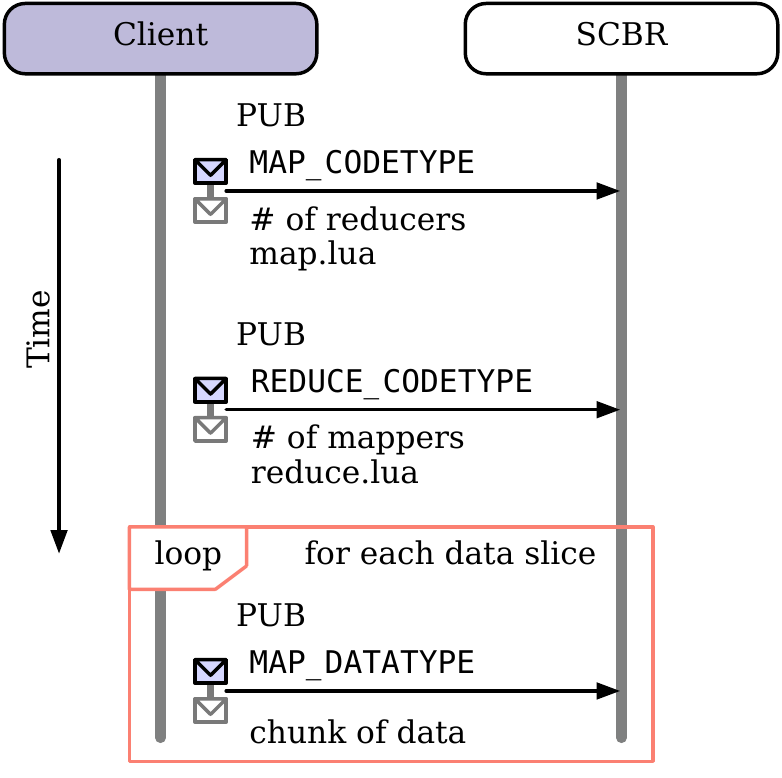}
\caption{\label{fig:provisioning}\mapreduce: Provisioning of code and data.}
\end{figure}

The provisioning of code and data is shown in \Cref{fig:provisioning}.
Besides the code itself, the client includes the number of reducers along with the Lua scripts in case of \texttt{MAP\_CODETYPE} publication topic, or the number of mappers in case of type \texttt{REDUCE\_CODETYPE}.
The purpose is to make the workers aware of how many messages indicating the stream's termination that they have to wait before considering the work done.
This is important because the reduce phase can only start once all data for a given key is routed to the intended worker.
Additionally, the amount of reducers (\texttt{rcount} in \Cref{lst:lua:map}) received by a mapper is used in a hash function that takes as argument the tuple key and returns the indication of which reducer it has to be forwarded to.
After sending the code, data is split by the client among the mappers.
The destination identifier is included in the header of the \texttt{MAP\_DATATYPE} publication.

Workers decrypt the received code and store it inside the enclaves.
When data arrive, mappers perform the processing and each one of the resulting key-value pairs is forwarded to the proper reducer.
The reducer's identifier is obtained after providing the key and number of reducers as arguments to the hash function that comes along with the code of the mapper.
The shuffling phase is hence conducted by the mappers.  
In order to forward data to the following step, all that the Lua script has to do is calling a special function called \emph{push(key,value)}, and the framework handles all the communication aspects of forwarding the data.

\Cref{lst:lua:map} shows the sample code of a mapper of a word count application.
The script can contain as many helper functions as desired.
The following special functions, however, are called by the framework:

\begin{figure}[b]
\begin{multicols}{2}
\begin{minipage}{0.45\textwidth}
\begin{lstlisting}[style=lua,caption={Map code in Lua for word count.},label={lst:lua:map}]
function hash(key,rcount)
    return string.byte(key,1) % rcount
end

function combine(key,values)
    local sum = 0
    for k,v in pairs(values) do
        sum = sum + v
    end
    push(key,sum)
end

function map(key,value)
    for word in value:gmatch("%w+") do
        push(word,1)
    end
end
\end{lstlisting}
\end{minipage}
\columnbreak

\null \vfill
\begin{minipage}{0.45\textwidth}
\begin{lstlisting}[style=lua,caption={Reduce code in Lua for word count.},label={lst:lua:reduce}]
function reduce(key, values)
    local sum = 0
    for k,v in pairs(values) do
        sum = sum + v
    end
    push(key,sum)
end
\end{lstlisting}
\end{minipage}

\vfill \null
\end{multicols}
 \end{figure}

\begin{tabular}{p{0.25\linewidth}p{0.6\linewidth}}
   \texttt{\textbf{map}(key, value)} &
    Contains the functional implementation of mapper. \\[3pt]
   \texttt{\textbf{combine}(key, values)} &
    [Optional] Post-processing on values grouped by keys. \\[3pt]
   \texttt{\textbf{hash}(key, rcount)} &
    Returns the reducer id that is supposed to receive a given \texttt{key}
    considering that there are \texttt{rcount} reducers in total. \\
\end{tabular}

Likewise, listing \ref{lst:lua:reduce} shows a sample code for the reduce step that contains a single special function:

\begin{tabular}{p{0.25\linewidth}p{0.6\linewidth}}
   \texttt{\textbf{reduce}(key, values)} &
    Contains the functional implementation of reducer. \\[3pt]
\end{tabular}

\subsubsection{Evaluation}

\begin{figure}[tb]
\centering
\includegraphics{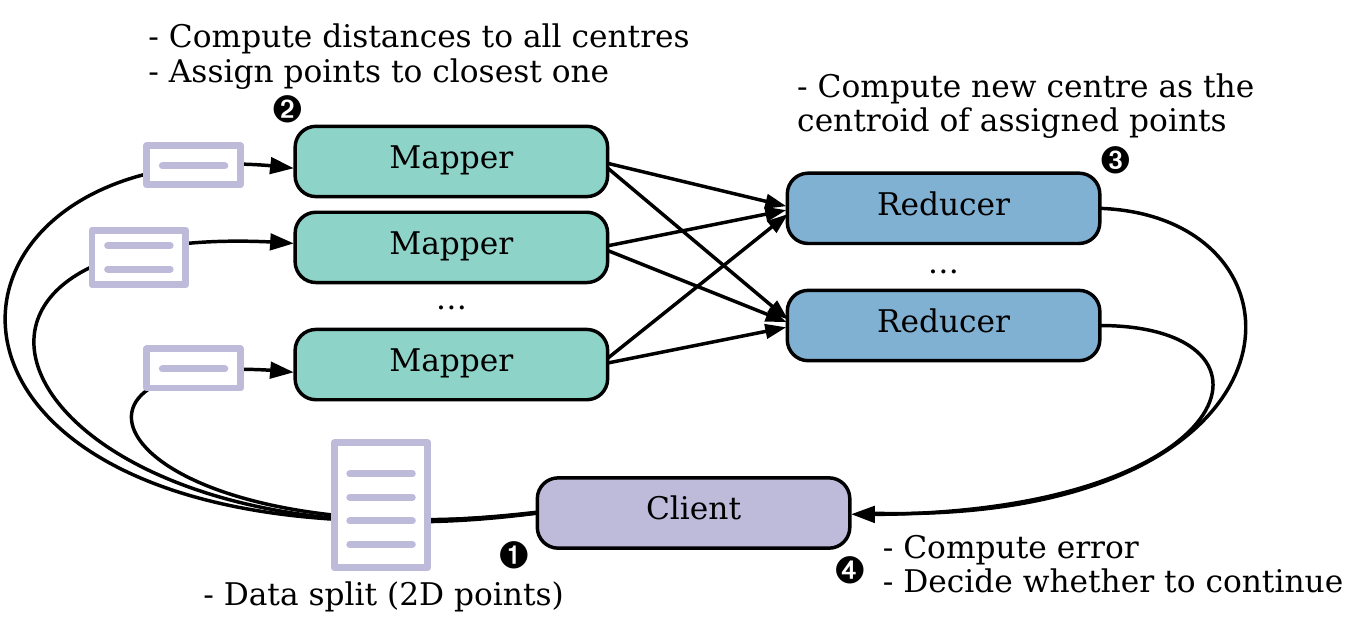}
\caption{\label{fig:mapreduce}K-means with \mapreduce.}
\end{figure}

To test our solution, we use the k-means clustering method for data classification.
K-means~\cite{macqueen:1967:classification} is an unsupervised learning algorithm (\ie, it does not depend on training data) widely used for data classification. It operates as following:
\begin{enumerate}[label={(\roman*)},ref={(\roman*)}]
\item \label{enum:kmeans1} A certain $k$ number of clusters is fixed \emph{a priori} and $k$ corresponding centres are defined for each one of them;
\item \label{enum:kmeans2} Data items are iterated and assigned to the nearest cluster centre (typically through Euclidean distance); and
\item \label{enum:kmeans3} Every cluster centre is recomputed as the centroid of the assigned data items (the mean of those points).
\end{enumerate}
Steps~\ref{enum:kmeans2} and~\ref{enum:kmeans3} repeat until a termination criterion is reached
(\eg, the sum of distances between the old and the new cluster centres is below a given threshold).
For implementing k-means in the \mapreduce model we put~\ref{enum:kmeans2} in the \emph{map} function and~\ref{enum:kmeans3} in the \emph{reduce} function as displayed in~\Cref{fig:mapreduce}.
The termination criterion is checked by the client, who decides whether to iterate again.
If it does, the centres input for mappers are replaced by the most recently calculated ones.

We conducted all the experiments using two \ac{SGX}-capable machines, both with processor Intel i7-6700 64bits, clock of 3.4GHz, 8MB cache, 4 cores, 8 threads, with 8GB of installed memory and \ac{SSD} of 256GB.
In terms of software, we used the Intel \ac{SGX} \ac{SDK} 1.7.100 over Ubuntu 14.04.1, kernel 4.2.0-42.
Unless mentioned otherwise, messages were all encrypted with \ac{AES}-\ac{CTR}~\cite{daemen:2002:rijndael} with key and input vector both of 128 bits and were decrypted only inside the enclaves.
The process placement was made as follows:
Machine 1: the client, 8 mappers and 5 reducers.
Machine 2: 8 mappers and 5 reducers.
The number of mappers was chosen to be twice as much as the number of cores in each machine to take advantage of parallelism, while the number of reducers was set to be a divisor of input data size to stimulate an even distribution of work among them.
To illustrate how small our final code-base is, \Cref{tab:codesize} shows the memory section sizes of executables and shared libraries that are loaded into enclaves.

\begin{table}[tb]
\centering
\caption{\label{tab:codesize}Lightweight \mapreduce: binaries' size.}
\begin{tabular}{c|c|c|c|c}
\hline
            &text       &data       &bss    &sum     \\
\hline
client          &\SI{371}{\kibi\byte} &\SI{26}{\kibi\byte} &\SI{376}{\byte}     &\SI{397}{\kibi\byte} \\
worker          &\SI{282}{\kibi\byte} &\SI{26}{\kibi\byte} &\SI{768}{\byte}     &\SI{308}{\kibi\byte} \\
worker enclave  &\SI{663}{\kibi\byte} &\SI{59}{\kibi\byte} &\SI{82}{\kibi\byte} &\SI{803}{\kibi\byte} \\
scbr            &\SI{271}{\kibi\byte} &\SI{14}{\kibi\byte} &\SI{72}{\byte}      &\SI{285}{\kibi\byte} \\
scbr enclave    &\SI{272}{\kibi\byte} &\SI{7}{\kibi\byte}  &\SI{79}{\kibi\byte} &\SI{358}{\kibi\byte} \\
\hline
\end{tabular}
\vspace{-10pt}
\end{table}

\begin{figure}
\centering
\includegraphics{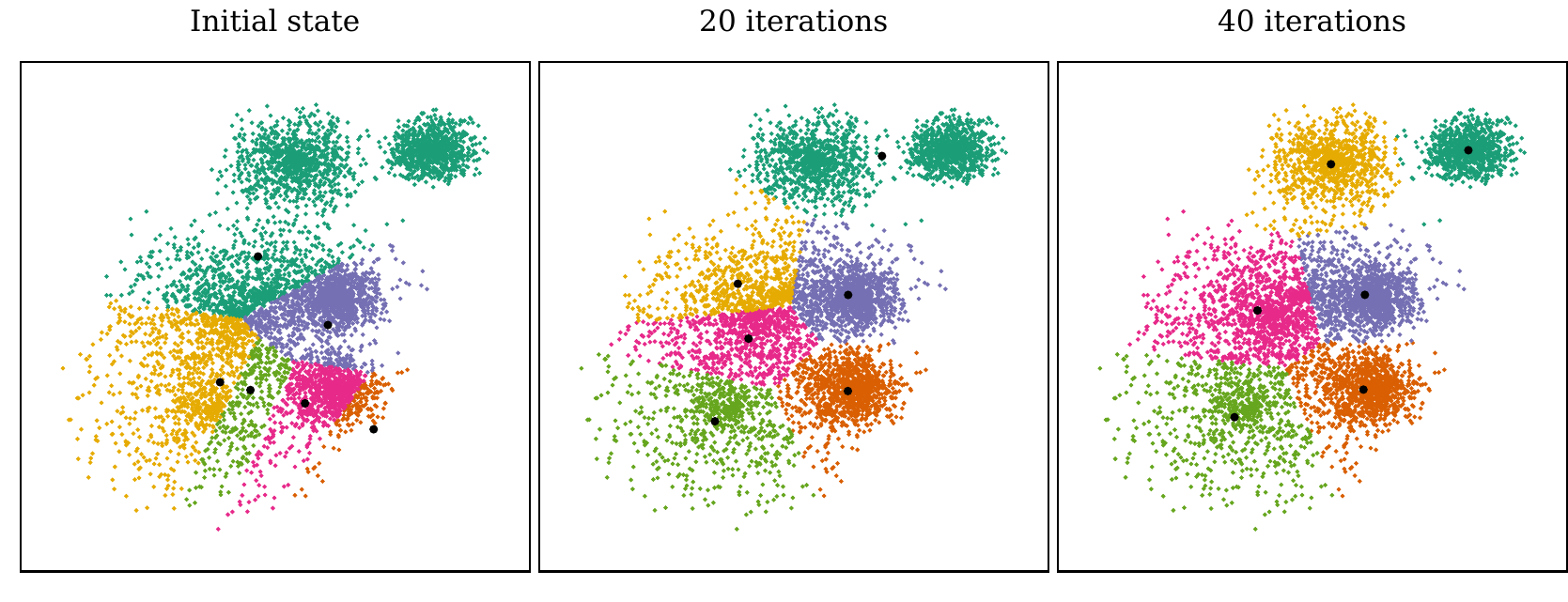}
\caption[K-means example.]{\label{fig:kmeans}K-means example with 6 centres and 7000 data points.}
\end{figure}

\Cref{fig:kmeans} illustrates 3 out of 40 iterations that k-means took to converge.
In this example, we synthetically generated 7,000 observation points and 6 centres.
As it can be seen in the first frame, the initial centroids are far from an even distribution across the grid.
After $20$ iterations, the centroids assume closer positions to those of the final result, which is achieved at the $40^{th}$ iteration.
We arbitrarily set the threshold to be one thousandth of the diagonal of the rectangle that contains all the observed points.
That means that the iterative process finishes when the average distance of centroids between two subsequent iterations is less than that fraction.
In such algorithm, the final result and convergence speed depend on the initial points.
Besides, there is no guarantee that the solution is the global optimum.

\begin{figure}[tb]
\center
\includegraphics{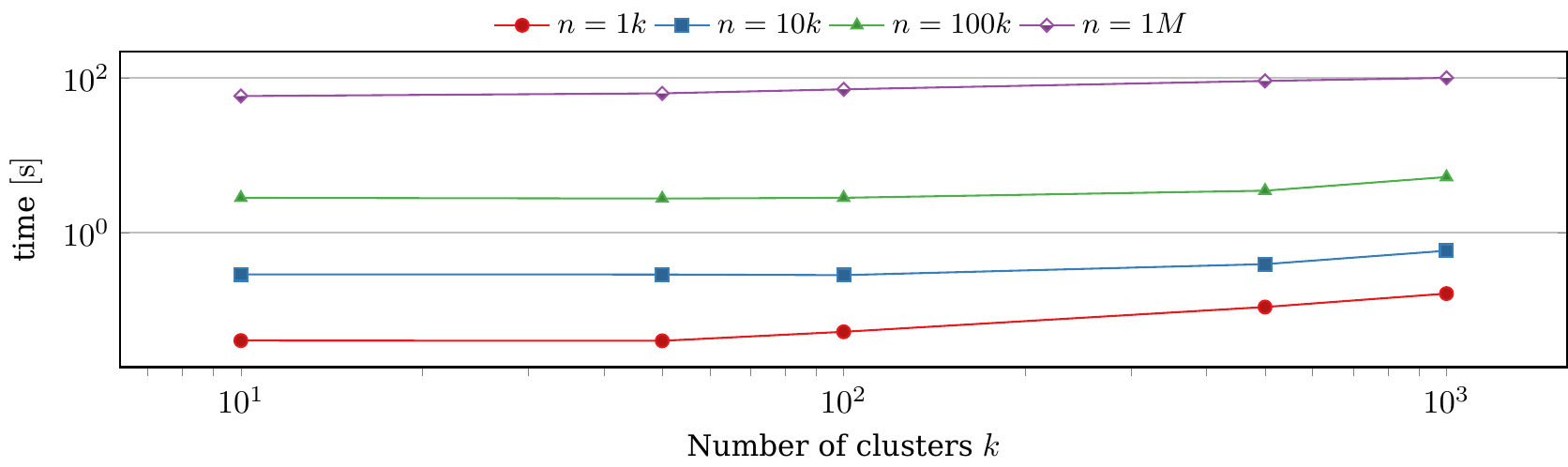}
\vskip 4.5mm
 \caption[\mapreduce: Iteration duration.]{\label{fig:varinput}\mapreduce: Average time to run one iteration with varying input sizes.}
\end{figure}

Next, we conducted experiments to assess the influence of input data sizes, \ie, the number of observation points $n$ and centroids $k$, on the memory usage and processing time.
\Cref{fig:varinput} shows the average time it took to complete one iteration of Kmeans with varying input sizes.
It can be noticed that, although the variation on the number of clusters can cause some inflection in the curves, the completion time is mostly affected by the number of observed data points.
Moreover, while the two first increments on the number of data points ($n=10k$ and $n=100k$) caused a proportional increase on consumed time regarding the data growth (ten times), the last one ($n=1M$) induced a twenty-fold rise.
That can be explained by the growth in the occurrence of cache misses within each worker.
When that happens, data must be fetched from main memory.
When using \ac{SGX} protected executions, this means one page has to be evicted from cache (and hence, encrypted), while the one that is fetched must be decrypted and checked for integrity and freshness.

\begin{figure}[tb]
\center
\includegraphics{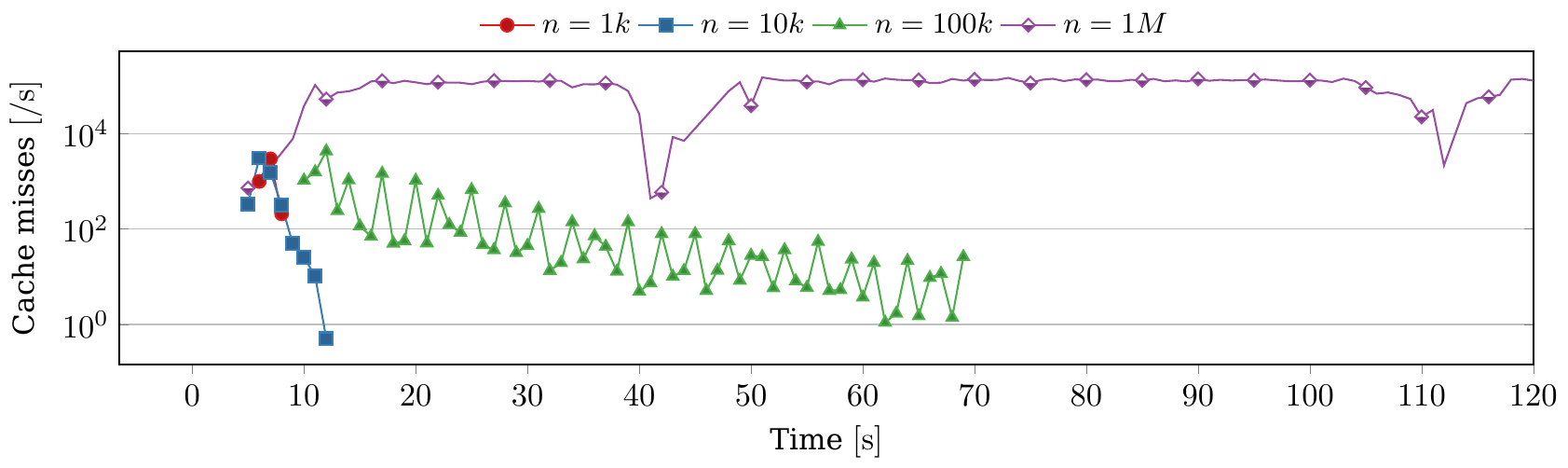}
\vskip 3.5mm
 \caption[\mapreduce: Cache misses.]{\label{fig:cachemisses}\mapreduce: Cache miss rates for different input sizes.}
\end{figure}

To better analyse these cache effects, we decided to measure cache miss rates.
Since the \emph{reduce} phase is more memory intensive, we chose to make the average on the cache miss rates per second of all 10 reducers in each execution, \ie, for the same second, cache miss rates of reducers were summed and divided by 10.
Wall clocks were synchronized with a common NTP server, so that the resulting skew was at the range of tens of milliseconds and should not affect the sampling resolution of \SI{1}{\second}.
\Cref{fig:cachemisses} shows that measurement as reported by the tool \texttt{pidstat} when the number of centroids is $k=50$.
Note that \emph{y} scale is logarithmic, so that the cache miss rates for $n=1M$ is at least two orders of magnitude higher than $n=100k$. 
Valleys in the curves represent the interval between iterations, which is more apparent in the larger experiments.

\begin{figure}[t]
\center
\includegraphics{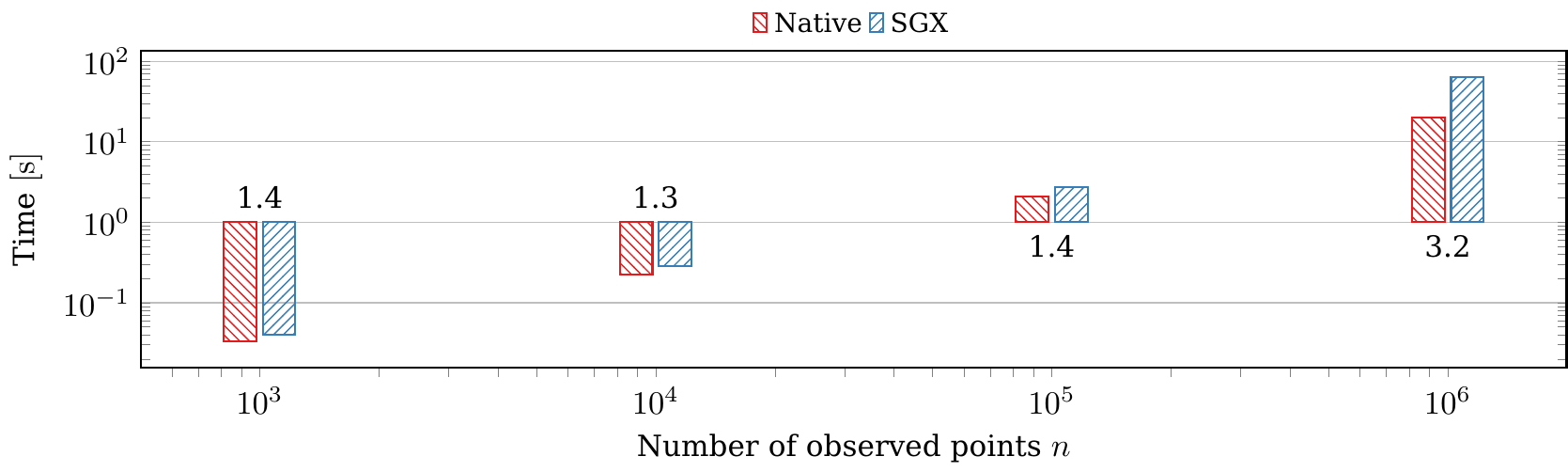}
\vskip 5.5mm
 \caption{\label{fig:sgxandenc}\mapreduce: \ac{SGX} overhead.}
\end{figure}

Finally, we measured the influence of \ac{SGX} when compared to native executions, \ie, with no hardware protection.
We ran the same datasets with a fixed number of clusters $k=50$ and varying the number of observed points from $n=1000$ until $n=1M$.
Results are plotted in \Cref{fig:sgxandenc}. \Cref{tab:datasize} shows the data volume exchanged in each \mapreduce step for these experiments.
We also include the ratio between \ac{SGX} and native run times.
The time corresponds to the average of all iterations in a single run of k-means (until the threshold was reached). Coefficient of variation across multiple runs was negligible.
Enclave execution overhead is kept around 35\% until we start to get high occurrence of cache misses, as discussed before, when it reaches more than 200\%.

\begin{table}[b]
\centering
\caption[\mapreduce: Data volume per iteration.]{\label{tab:datasize}\mapreduce: Data volume exchanged per iteration.}
\begin{tabular}{c|c|c|c}
\hline
            &Split       &Shuffle     &Output \\
\hline
$n=1k$      &\SI{58.7}{\kibi\byte}     &\SI{112.1}{\kibi\byte}    &\SI{4.3}{\kibi\byte} \\
$n=10k$     &\SI{257.2}{\kibi\byte}    &\SI{1.1}{\mebi\byte}      &\SI{4.5}{\kibi\byte} \\
$n=100k$    &\SI{2.2}{\mebi\byte}      &\SI{11}{\mebi\byte}       &\SI{4.6}{\kibi\byte} \\
$n=1M$      &\SI{19.1}{\mebi\byte}     &\SI{96.5}{\mebi\byte}     &\SI{4.6}{\kibi\byte} \\
\hline
\end{tabular}
\end{table}

This concludes our experiments with lightweight \mapreduce.
As in \Cref{sec:scbr}, we notice again some expressive overhead after using more than the processor's cache limit.
Based on its size, we can establish the maximum amount of data that a single \ac{SGX}-capable machine would be able to handle before incurring in too much overhead.
In our experiments, we perceived that behaviour when processing amounts somewhere in between \SI{11}{\mebi\byte} and \SI{96.5}{\mebi\byte} shared between two machines (average of around \SI{54}{\mebi\byte}, or \SI{27}{\mebi\byte} per machine).
A rough estimation based on our empirical evaluations would be to limit those amounts to three times the cache size, or \SI{24}{\mebi\byte} in our case.
Scalability could be achieved horizontally, with the addition of more machines.
This reinforces the lightweight aspect of our approach, both in terms of the framework code size (\SI{2}{\mebi\byte} summing up all components) and its capacity before incurring into important overheads (\SI{24}{\mebi\byte} per machine in the shuffling phase).
 
\vskip 4mm
\subsection{SecureStreams}
\label{sec:sstreams}
\newcommand{\rxl}{\mbox{\textsc{RxLua}}\xspace}
\definecolor{darkgreen}{rgb}{0.3,0.5,0.3}
\definecolor{darkblue}{rgb}{0.3,0.3,0.5}
\definecolor{darkred}{rgb}{0.5,0.3,0.3}
\lstdefinelanguage{YAML}{
  sensitive=true,
  keywordstyle=[1]{\color{darkblue}\bfseries},
  keywordstyle=[2]{\color{darkgreen}\bfseries},
  morekeywords=[1]{image, entrypoint, environment, devices, hostname},%
  morekeywords=[2]{},%
  otherkeywords={.,=,~,*,>,:},
  morestring=[b]",
  breaklines=true,
  breakatwhitespace=true,
  linewidth=\columnwidth,
  comment=[l]{--},
  escapeinside={(*@}{@*)}
}

\lstdefinelanguage{LUA}{
  sensitive=true,
  keywordstyle=[1]{\color{darkblue}\bfseries},
  keywordstyle=[2]{\color{darkgreen}\bfseries},
  morekeywords=[1]{and,break,do,else,elseif,end,for,function,if,in,local,
    nil,not,or,repeat,return,then,until,while,require,alias},%
  morekeywords=[2]{},%
  otherkeywords={.,=,~,*,>,:},
  morestring=[b]",
  stringstyle={\color{darkred}\itshape},
  breaklines=true,
  breakatwhitespace=true,
  linewidth=\columnwidth,
  comment=[l]{--},
  escapeinside={(*@}{@*)}
}

The \ac{IoT} has fostered the emergence of novel data analytics and processing technologies to support the continuous flow of information gathered by a large amount of sensing devices.
Since these data streams may convey sensitive information, stream processing requires support for end-to-end security guarantees in order to prevent third parties from accessing restricted data.
We present \securestreams~\cite{havet:2017:securestreams}, a middleware framework for developing and deploying secure stream processing on untrusted distributed environments.

\securestreams supports the implementation, deployment, and execution of stream processing tasks in distributed settings.
It employs a message-oriented middleware~\cite{curry:2005:mom}, the \ac{TLS} protocol~\cite{dierks:2015:rfc5246} for communication and \ac{SGX} to deliver end-to-end security guarantees along data stream processing pipelines.
\securestreams can scale vertically and horizontally by adding or removing processing nodes at any stage of the pipeline.
Its design is inspired by the dataflow programming paradigm~\cite{uustalu:2006:dataflow}, where the developer combines independent processing components (\eg, mappers, reducers, sinks, shufflers, joiners) to compose specific processing pipes.
Regarding packaging and deployment, \securestreams smoothly integrates with a lightweight virtualisation technology, namely Docker~\cite{merkel:2014:docker}.

\vskip 3mm
\subsubsection{Architecture}

\begin{figure*}[!t]
  \centering
  \includegraphics{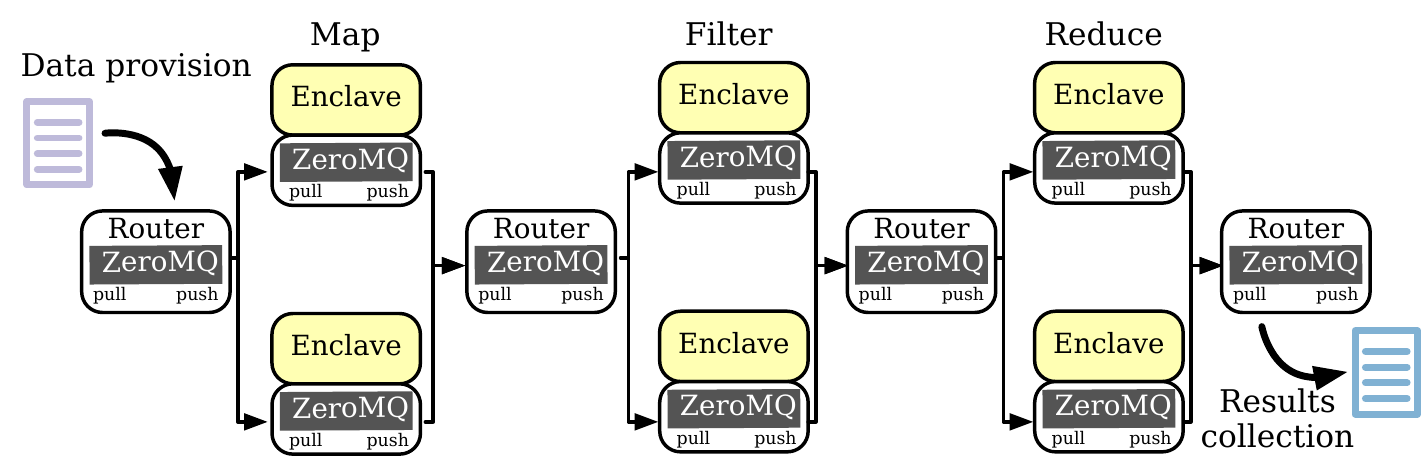}
  \caption{Example of \securestreams pipeline.}
  \label{fig:architecture_pipeline}
\end{figure*}

\securestreams combines two base components: \emph{worker} and \emph{router}.
A worker continuously listens for incoming data by means of non-blocking \ac{IO}.
As soon as data flows in, an application-dependent business logic is applied.
Router components act as message brokers between workers according to a given \emph{dispatching policy}.
A typical use-case is the filter/map/reduce pattern from the functional programming paradigm~\cite{bird:1988:functional}, in which each worker executes one of these functions.
Figure~\ref{fig:architecture_pipeline} depicts a possible implementation of this dataflow using the \securestreams middleware.

\securestreams is designed to support the processing of sensitive data inside \ac{SGX} enclaves.
Each component is wrapped inside a lightweight Linux container (in our case, the industrial standard Docker~\cite{merkel:2014:docker}).
Each container embeds all the required dependencies, while guaranteeing the correctness of their configuration, within an isolated and reproducible execution environment.
By doing so, a \securestreams processing pipeline can be easily deployed without changing the source code on distinct public or private infrastructures.
Containers' deployment can be transparently executed on a single machine or a cluster, using a Docker network and the Docker Swarm scheduler~\cite{docker:swarm_2016}.

Like in \ac{SCBR} (\Cref{sec:scbr}), communication is done with \zmq{}, a high-performance asynchronous messaging library~\cite{zeromq:2019}.
Each router component hosts inbound and outbound queues, according to \zmq's pipeline pattern~\cite{zeromq:2019:pipeline}.
Messages are streamed from a set of \emph{push} peers, \ie, the upstream workers in the pipeline.
They use \emph{push} sockets that send messages to downstream workers in a round-robin fashion.
These, in turn, have an inbound \textsc{pull} socket, which uses fair-queuing scheduling to deliver messages to upper layers.

\begin{figure}[t!]
  \centering
  \includegraphics{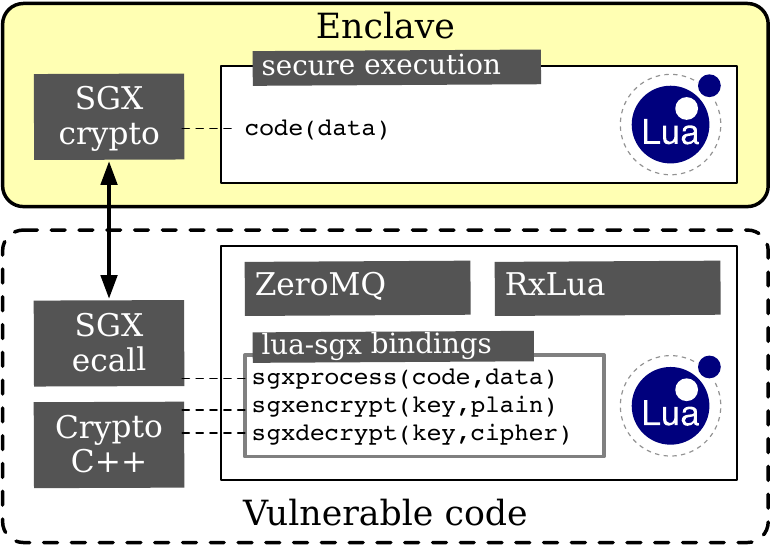}
  \caption{Integration between \textsc{Lua} and Intel \ac{SGX}.}
  \label{fig:arch-luasgx}
\end{figure}

We define the processing pipeline components and their chaining by means of Docker's Compose~\cite{docker:compose} description language.
Once the processing pipeline is defined, the containers can be deployed on the computing infrastructure.
We use the \texttt{constraint} placement mechanism to make the Docker Swarm's scheduler deploy workers requiring \ac{SGX} on appropriate hosts.

\subsubsection{Implementation}

\securestreams is implemented in \textsc{Lua} version 5.3.
The implementation of the middleware itself requires careful engineering, especially with respect to the integration with \ac{SGX} enclaves.
A \securestreams application, on the other hand, can be implemented in remarkably few lines of code.
For instance, the implementation of the map/filter/reduce pipeline accounts for only $120$ \acp{LoC} (not including dependencies).
The framework partially extends \rxl~\cite{github:rxlua}, a library for reactive programming in \textsc{Lua}.
\rxl provides the required API to design a data stream processing pipeline following the dataflow programming pattern~\cite{uustalu:2006:dataflow}.

\Cref{lst:lua:sstreams} provides an example of a \rxl program (and consequently a \securestreams program) to compute the average age of a population by chaining \texttt{:map}, \texttt{:filter}, and \texttt{:reduce} functions.
The \texttt{:subscribe} function performs the subscription of 3 functions to the data stream.
Following the \emph{observer} design pattern~\cite{szallies_using_1997}, these functions are observers, while the data stream is an observable.

\begin{figure}[tb]
\centering
\begin{minipage}{0.6\textwidth}
\begin{lstlisting}[style=lua,caption={Example of process pipeline with RxLua.},label={lst:lua:sstreams}]
Rx.Observable.fromTable(people)
 :map(
   function(person)
     return person.age
   end
 )
 :filter(
   function(age)
     return age > 18
   end
 )
 :reduce(
   function(accumulator, age)
     accumulator[count] = (accumulator.count or 0) + 1
     accumulator[sum] = (accumulator.sum or 0) + age
     return accumulator
   end
 )
 :subscribe(
   function(data)
     print("Adult people average:", data.sum / data.count)
   end,
   function(err)
     print(err)
   end,
   function()
     print("Process complete!")
   end
 )
\end{lstlisting}
\end{minipage}
\end{figure}

\securestreams ships the business logic for each component into a dedicated Docker container and executes it.
Communication between routers and workers happens through \zmq (version 4.1.2) and the corresponding \textsc{Lua} bindings~\cite{github:lzmq}.
The framework safely forwards data and code to enclaves, so that they do not have to operate on files or any other \ac{IO}. In case such attempts occur, they are frustrated by sanitized versions of \emph{libc} procedures. 
By adopting this set of measures, \securestreams safely abstracts the underlying network and computing infrastructure from the developer perspective.

We reuse here the \ac{SGX} \luavm described in \Cref{sec:luavm}.
Additionally, we include in the enclave both \emph{json}~\cite{bray:2014:rfc8259} and \emph{csv}~\cite{shafranovich:2005:rfc4180} parsers to ease the development of \securestreams applications.
With these libraries, the enclave size containing the complete runtime remains small, approximately \SI{220}{\kibi\byte} ($19\,\mathit{\%}$ larger than the original).
Besides the \ac{SGX} \luavm, we still had to provide support for communication and the reactive streams framework itself.
For this, we use an external vanilla \textsc{Lua} interpreter, with a couple adaptations to allow the interaction with the enclaves and the \luavm therein.
Figure~\ref{fig:arch-luasgx} shows the resulting scheme.
We extend the \textsc{Lua} interface with 3 functions: \texttt{sgxprocess}, \texttt{sgxencrypt}, and \texttt{sgxdecrypt}.
The first one forwards the encrypted code and data to be processed in the enclave, while the remaining two provide cryptographic functionalities.
We assume that attestation and key establishment was previously performed.

\subsubsection{Evaluation}

We conducted our experiments on 2 machines using the Intel Core i7-6700 processor~\cite{intel:i7_6700} and \SI{8}{\gibi\byte} \ac{RAM} running \textsc{Ubuntu} 14.04.1 \ac{LTS} (kernel 4.2.0-42-generic).
Each node runs Docker (v1.13.0) and joins a Docker Swarm~\cite{docker:swarm_2016} (v1.2.5) using the Consul~\cite{consul} (v0.5.2) discovery service.
The Swarm manager and the discovery service are deployed in a third machine.
Containers that compose the pipeline leverage the Docker overlay network to communicate with each other. Machines are physically interconnected using a switched \SI{1}{Gbps} network.

To evaluate our system, we chose a dataset released by the \emph{American Bureau of Transportation Statistics}~\cite{rita:bts}.
It reports on flight departures and arrivals of $20$ air carriers~\cite{statistical_computing:data}.
We implemented a \securestreams application to compute average delays and the amount of delayed flights of each air carrier (inspired by~\cite{webber}). \Cref{tab:appsize} specifies the code size of each application component.
We implemented a processing pipeline that
\begin{enumerate}[label={(\roman*)}]
\item \emph{maps} the input dataset, in \ac{CSV} format, into a data structure;
\item \emph{filters} out irrelevant data, \ie only let data concerning delayed flights go through; and
\item \emph{reduces} the filtered data by computing the output information.
\end{enumerate}

\begin{table}[t!]
\centering
\caption{Code size of a \securestreams application.}
\begin{tabular}{l|r}
\hline
\textbf{System layer}          & \textbf{Size (LoC)} \\
\hline
\textsc{DelayedFlights} app    & $86$ \\
\textsc{SecureStreams} library & $350$ \\
\textsc{RxLua} runtime         & $1,481$ \\
\hline
Total                          & $1,917$ \\
\hline
\end{tabular}
\label{tab:appsize}
\end{table}

\begin{figure}[tb]
\center
\includegraphics{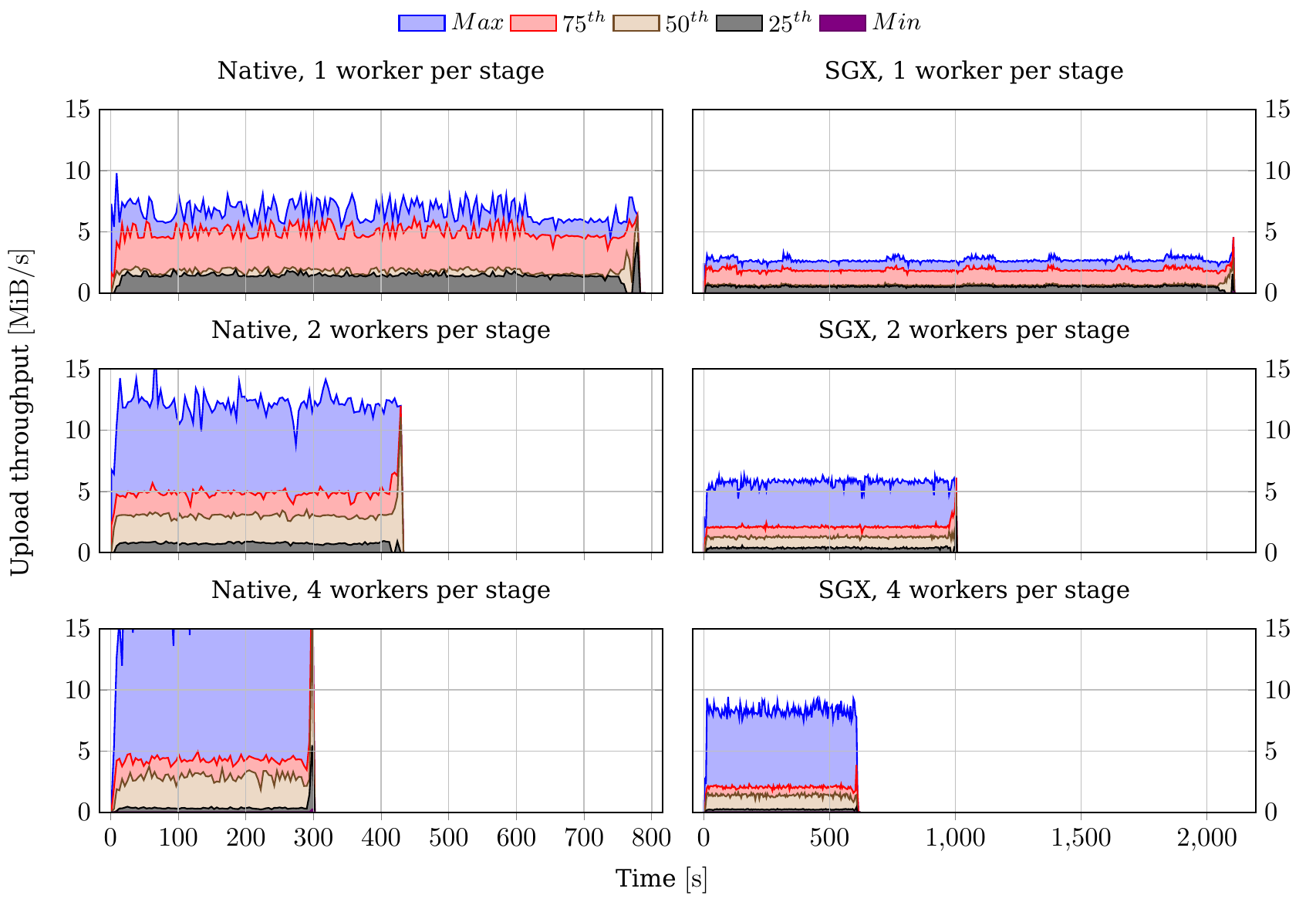}
 \caption[\securestreams: Throughput comparison.]{\label{fig:sstreams:throughput}\securestreams: Throughput comparison between native and \ac{SGX} versions with 1 and 4 workers per stage.}
\end{figure}

We use the $4$ last years of the available dataset (from 2005 to 2008), for a total of $28$ million entries and \SI{2.73}{\gibi\byte} of data.
To measure the achievable throughput and the network overhead of our system, we deploy the \securestreams pipeline in 2 configurations:
\begin{enumerate}
\item The baseline does not use enclaves. The input dataset is encrypted before its injection in the pipeline, so that data transmission is secure. This approach is however unsafe for deployment in untrusted infrastructures since data is decrypted before being processed.
\item In the \ac{SGX} deployment, the input dataset is encrypted and the data processing is operated inside enclaves.
\end{enumerate}

Data nodes inject the input dataset as fast as possible while bandwidth measurements are collected from Docker's internal monitoring and statistical module.
For each configuration, we vary the number of workers per stage, from one to four.
\Cref{fig:sstreams:throughput} shows the results in the form of stacked percentiles.
To exemplify, the median ($50^{th}$ percentile) throughput at \SI{200}{\second} when operating with four nodes per stage corresponds to \SI{3.1}{\mebi\byte\per\second} and  \SI{1.5}{\mebi\byte\per\second} for native and \ac{SGX} executions, respectively.
This means that $50\%$ of the nodes output data at no more than these rates at that moment in time.

The baseline configuration, \ie, native execution with $1$ worker per stage, completes in \SI{794}{\second} with median of \SI{1.8}{\mebi\byte\per\second} and peak of \SI{9.8}{\mebi\byte\per\second}.
Doubling the number of workers reduces the processing time down to \SI{442}{\second}, a speed-up of \SI{1.8}{}$\times$ (median: \SI{3}{\mebi\byte\per\second}, peak: \SI{16.5}{\mebi\byte\per\second}).
Scaling up the workers to 4 per stage in the native execution makes the dataset be consumed in \SI{302}{\second}  (median: \SI{2.9}{\mebi\byte\per\second}, peak: \SI{27}{\mebi\byte\per\second}\footnote{not visible in the figure due to visualisation purposes, so that y-axis scale is kept consistent for comparison while allowing some noticeable detail in lower throughput experiments}), or \SI{1.5}{}$\times$ speed-up considering the previous experiment.

When using \ac{SGX} enclaves, data processing slows down by \SI{2.7}{}$\times$ (\SI{2120}{\second}) for the setup with one worker per stage.
We also pay a penalty in terms of overall throughput---\ie, the median rarely exceeds \SI{650}{\kibi\byte\per\second}, peak of \SI{4.6}{\mebi\byte\per\second}.
Increasing the workers per stage renders speed-ups of \SI{2.1}{}$\times$ (median: \SI{1.3}{\mebi\byte\per\second}, peak: \SI{5.9}{\mebi\byte\per\second}) and \SI{1.6}{}$\times$ (median: \SI{1.4}{\mebi\byte\per\second}, peak: \SI{9.9}{\mebi\byte\per\second}), with $2$ and $4$ workers, respectively.
In comparison to native execution though, they present slow-downs of \SI{2.3}{}$\times$ and \SI{2.1}{}$\times$.

\begin{figure}[t!]
\center
\includegraphics{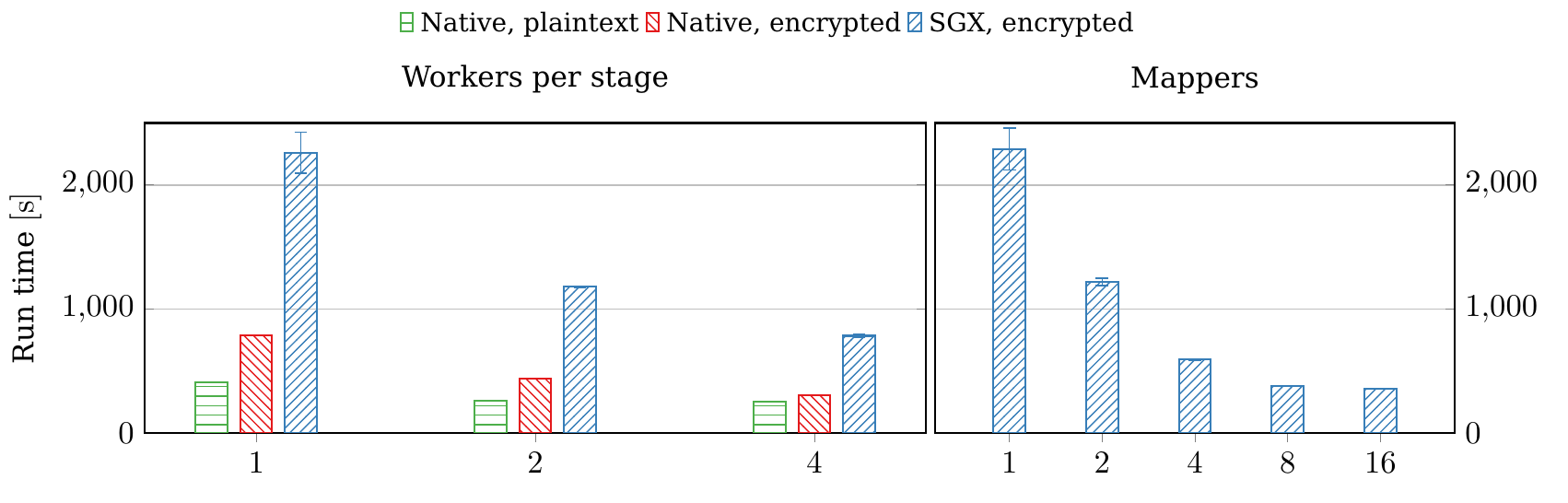}
 \caption[\securestreams: processing time.]{\label{fig:sstreams:scalability}Average processing time when varying number of workers.}
\end{figure}

We further evaluate  \securestreams in terms of scalability by establishing correlations among running time, number of worker nodes and available hardware resources. Specifically, the number of processing units. 
We do so by changing the number of workers, like in the previous experiment, and also by only varying the number of mappers, the most significant stage in terms of overall performance.
Additionally, as baseline, we experiment with plain-text data across the whole pipeline.
We repeat each configuration for $5$ times.
\Cref{fig:sstreams:scalability} shows the mean runtime, with error bars corresponding to the standard deviation.

First, we increase the number of workers in each stage of the pipeline.
We observe an ideal acceleration when going from the configuration using $1$ worker per stage to $2$.
In the setup using $4$ workers, on the other hand, we do not reach the same speed-up.
This mainly happens due to the number of deployed containers, which becomes greater than the amount of available cores in each processor.
Containers account for the sum  $s$ of input data streams $i$, routers $r=4$ and workers.
The amount of the latter is equal to the number of stages in the pipeline ($3$ in our setup, see \Cref{fig:architecture_pipeline}) multiplied by the number of workers per stage $w$. Besides, we define the number of input data streams as $i=w$.
The total number of containers is therefore $s=4w+4$, or $s=20$ when $w=4$, which is greater than the number of physical processing units in our setup ($2$ machines with $8$ cores each, \ie, $16$ cores).

Finally, we observe the running time when varying from $1$ to $16$ the number of mappers in the first stage of the pipeline, as it is the most computational intensive one number of filters and reducers is kept constant.
Again, we observe an ideal speed-up ($2\times$) until the number of deployed containers reaches the amount of physical cores.
Beyond that, the improvement is only marginal ($1.07\times$ from $8$ to $16$ mappers). These experiments show that \securestreams' scalability is primarily limited by the total number of physical cores available, for a map/filter/reduce pipeline.

Apart from that, other factors contribute for limited streaming throughput.
We observed, for instance, that the system does not saturate the available network bandwidth.
We believe this behaviour can be explained by the lack of optimisations in the application logic and tuning options of the inner \zmq{} queues.
For this reason, we did not reach memory occupancy beyond the \ac{EPC} limit, which would dramatically deteriorate performance (see \Cref{sec:scbr}).
Nevertheless, we reached our primary goal of finding out the overhead when using \ac{SGX} enclaves for processing real data in a streaming pipeline, along with a couple insights about scalability. 
As expected, one should anticipate longer processing times (roughly $2-3\times$ in our experiments) and lower throughputs when executing stream processing within \ac{SGX} enclaves.
Disadvantages of this trade-off between security and runtime can be mitigated by parallelisation, as long as the workload is adapted to hardware resources, as our experiments indicate.
 
\section{Summary}

In this chapter, we proposed and empirically evaluated a set of distributed systems that use \acp{TEE} and naturally fit in cloud scenarios. They are therefore supposed to be deployed in third-party administrative domains, which are commonly reputed as hostile and unsafe environments.

First, we presented the architecture and evaluation of \ac{SCBR} (\Cref{sec:scbr}), a secure content-based routing engine that takes advantage of \ac{SGX} enclaves.
In doing so, we were able to leverage state-of-the-art techniques for efficient filtering in plaintext, since the trusted perimeter is limited to the \ac{CPU} die.
Outside that boundary, private data in main memory is always encrypted and protected from tampering and replay attacks, even from \acp{OS}, hypervisors, and administrators with physical access to machines.
As a result, we do not suffer from the prohibitive performance and space overheads of software-based secure approaches, such as homomorphic encryption or dedicated algorithms like \ac{ASPE}.

As part of our experiments, we tested the system with different workloads and analysed the influence of cache misses and page faults on the code running within secure enclaves.
While both events introduce some overhead (as compared to insecure matching outside the enclave), performance degrades much more heavily with the latter, which occurs when exceeding the available amount of protected memory.
We also compared the performance of \ac{SGX} against the software-based \ac{ASPE} alternative and observed that \ac{SGX} performs systematically better as long as memory usage is kept below \SI{93.5}{\mebi\byte}.

Next, we proposed a lightweight framework for implementing secure \mapreduce applications in untrusted environments (\Cref{sec:lwmp}).
From the user's perspective, our approach does not require any particular programming knowledge of cryptographic mechanisms or communication aspects of data distribution.
Also, preserving privacy and integrity is not dependent in any way on the specific characteristics of \emph{map} and \emph{reduce} functions, which are defined and easily maintainable by relying on a standard Lua interpreter for code execution (\Cref{sec:luavm}).
In the security aspect, we simply took advantage of isolation guarantees provided by \ac{SGX} enclaves.

Our objective was to show the viability of such batch processing framework and to assess its behaviour under different conditions.
We observed a correlation between the number of cache miss occurrences and the slow down of \ac{SGX} executions in comparison to native ones.
Besides, based on our results, we established an upper bound limit of memory usage after which it would be advisable to horizontally scale out in order to minimise the referred slow down.

We moved on to secure stream processing by  introducing the design and evaluation of \securestreams, a concise middleware framework to implement, deploy and evaluate processing pipelines for continuous data streams (\Cref{sec:sstreams}).
We reused our Lua port that operates in \ac{SGX} enclaves with a couple adaptations aiming at interacting with a vanilla interpreter that operates outside enclaves.
This, in turn, runs libraries for communication message queuing and reactive programming.

Empirical results based on real-world traces showed performance penalties of $2-3\times$ when using enclaves.
We also assess \securestreams scalability capabilities.
In our setup, it reaches theoretical speed-ups as long as the number of worker nodes is kept below the total amount of processing units.

All in all, we clearly noticed performance deterioration when exhausting different memory levels: first, the L3 cache and later, the \ac{EPC} limit.
On the bright side, this limitation can be overcome through horizontal scalability or future hardware evolutions of \ac{SGX}.
Security-wise, these systems' safety relies entirely on \ac{SGX} and it could be compromised by trapdoors, design flaws or hardware bugs (see \Cref{sec:sgx:vulnerabilities}).
Our results open the way for further research on \ac{TEE}- and cloud-based distributed systems.
We keep analysing different design strategies targeted at distinct scenarios in the upcoming chapters.

\chapter{Group communication and data sharing}
\label{chap:sharing}
\acresetall

Looking into the adjustments that one should make to leverage secure enclaves in distributed communication and processing systems, we observed considerable performance implications under memory-intensive scenarios.
Notwithstanding, the benefits of \acp{TEE} outreach such niche.
Confidentiality and isolation can be used in less memory-eager applications.

In this direction, we turn our attention to designing or adapting cryptographic schemes in order to profit from \acp{TEE}.
The principle is basically the same and in fact derived from the construction of \ac{SGX} itself: it conceals a master key somewhere very hard to find (the processor die) and uses it to derive other secrets.
We do the same in \Cref{sec:ibbe}, where a master secret is generated within an enclave and never leaves it, thus allowing for its usage instead of more complex asymmetric encryption derivations in the context of group access control. As a consequence, we are able to lower the computational cost of an~\ac{IBBE} scheme.
Likewise, in \Cref{sec:asky}, the enclave is used to harbour keys and group membership data, so that users can share information yet guaranteeing anonymity among them.
Through empirical analysis, we show the practicality of both systems and discuss their outcomes.

\vspace{-1em}
\section{Cryptographic group access control}

Cloud storage services have largely grown in the last decade.
Many approaches rely on cryptographic solutions where data is secured on the client side before reaching the storage premises~\cite{bessani:2014:scfs}, therefore extenuating concerns caused by the lack of trust in the cloud provider.
In the case where there is only one single reader and writer, keys can be simply shared beforehand or established over public channels according to protocols like \ac{DH}.
To enable collaborative operations, however, one needs to enforce access control policies, so that only rightful users get access to keys.

A number of cryptographic constructions have been proposed for achieving access control.
The simplest, known as \ac{HE}, uses symmetric and public-key cryptography by employing the former on the actual data and the latter on the symmetric key~\cite{goh:2003:sirius}. The drawback is the huge amount of metadata composed by the cumulative ciphertext produced when using the public keys of each and every group member to protect the group shared key.
Other approaches rely on \ac{PBC} as a substitute for public-key cryptography.
Even though pairing-based approaches produce small metadata volume irrespective of group sizes, they suffer from important performance issues, one order of magnitude slower than public-key cryptography~\cite{garrison:2016:daccloud}.

We present here a cryptographic access control scheme that is both computationally- and storage-efficient considering large sets of users and dynamic membership operations~\cite{contiu:2018:ibbe}.
This is achieved by cutting the computational complexity of an \ac{IBBE} construction~\cite{delerablee:2007:ibbeconstantsize}, since we profit from the isolation guarantees provided by \acp{TEE} and use a \emph{master key} for performing membership updates.
When users need to perceive such changes though, we still would obtain high overheads considering that we assume they may not have access to hardware protection (\eg, mobile devices, \ac{IoT}, distinct chip manufacturers).
To mitigate this, we propose a group partitioning mechanism that imposes a complexity upper bound on the user side, limited by the partition size.

We implemented our access control scheme using Intel \ac{SGX} as \ac{TEE}.
To do so, we adapted a \ac{PBC} library~\cite{lynn:2006:pbc} and its underlying dependency, \ac{GMP}~\cite{granlund:1991:gmp}, to run within \ac{SGX} enclaves.
Our evaluation shows that our scheme performs better than \ac{HE} \wrt metadata expansion, group creation and user removal from a group; and worse for user addition to a group and key decryption time.
The benefits of having small metadata, though, goes beyond access control latencies, since it leads to less storage and network usage.
Additionally, the fact that we run the access control inside secure enclaves prevents curious administrators to snoop on group keys, an extra security feature not present in traditional \ac{HE}.

Even though the main motivation for this work is to securely share data in cloud environments, the proposed solution can be applied for encrypting arbitrary information that is securely broadcast to a group of users in any shared media, \eg, peer-to-peer networks and pay-per-view TV.
In the remaining of this section, we 
\begin{enumerate*}[label={(\roman*)}]
\item propose a new approach to \ac{IBBE} encryption by relying on Intel \ac{SGX};
\item propose an original partitioning scheme that lowers the time required by users to absorb access control changes;
\item implement and evaluate our system in a realistic setup; and 
\item compare it with state-of-the-art solutions.
\end{enumerate*}

\vspace{-1em}
\subsubsection{Model}
\begin{figure}[t]
    \centering
    \includegraphics{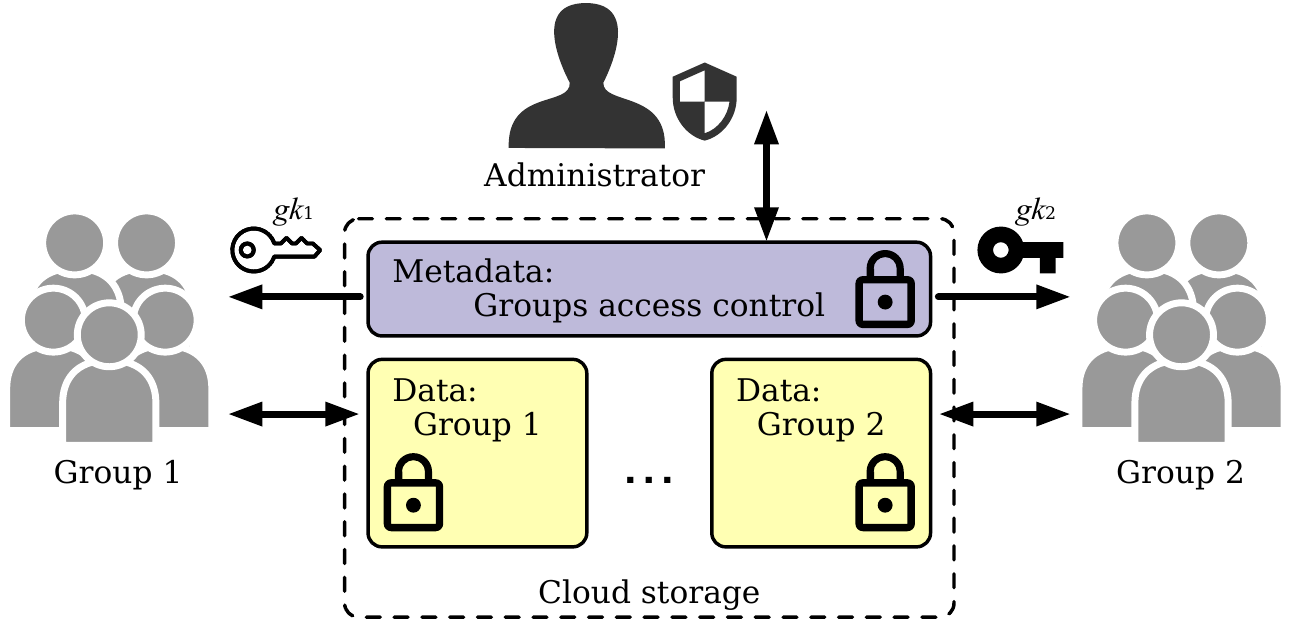}
    \caption{\label{fig:model_diagram}\ibbesgx model diagram.}
\end{figure}

\ibbesgx targets at managing groups of users who perform collaborative editing on cryptographically protected data stored on untrusted cloud storages.
Data are protected using a block cipher encryption algorithm such as \ac{AES}, using a symmetric group key $gk$ that is derivable by every legit group member based on metadata accessible to him.

As illustrated in Figure~\ref{fig:model_diagram}, we distinguish three actors: 
\begin{itemize}
\item \textit{administrators}, who perform membership operations: group creation, group members' addition and revocation. They may express an honest-but-curious behaviour by correctly performing their duties and maliciously trying to spy on group keys;
\item \textit{clients}, who derive keys for groups to which they belong and use them to add or modify data in the group's shared content. They are trustworthy, meaning that they do not disclose group keys; and
\item \textit{cloud storage}, which stores metadata that hold group access control information along with encrypted shared files. Besides the role of storage medium, we also use the cloud storage as communication channel between administrators and users. When the formers update membership information, this action is propagated through the respective changes in the storage. Users, in turn, may listen to modifications in files that are relevant to them.
The cloud storage may try to eavesdrop on users' data and collude with administrators or revoked users.
\end{itemize}

We enforce authenticity only with respect to administrator identities on membership operations.
Managing encryption and authenticating data created or altered by users is out of scope.
Identities of group members are not secret, nor the type of membership operations, as they can be inferred by the cloud storage from traffic access patterns.
Privacy constructions offering such guarantees~\cite{mayberry:2014:efficient,devadas:2015:onionoram,apon:2014:vobliviousstorage} are orthogonal to our work. We propose a system that considers members anonymity in \Cref{sec:asky}.

\begin{figure}[t]
	\centering
	\includegraphics{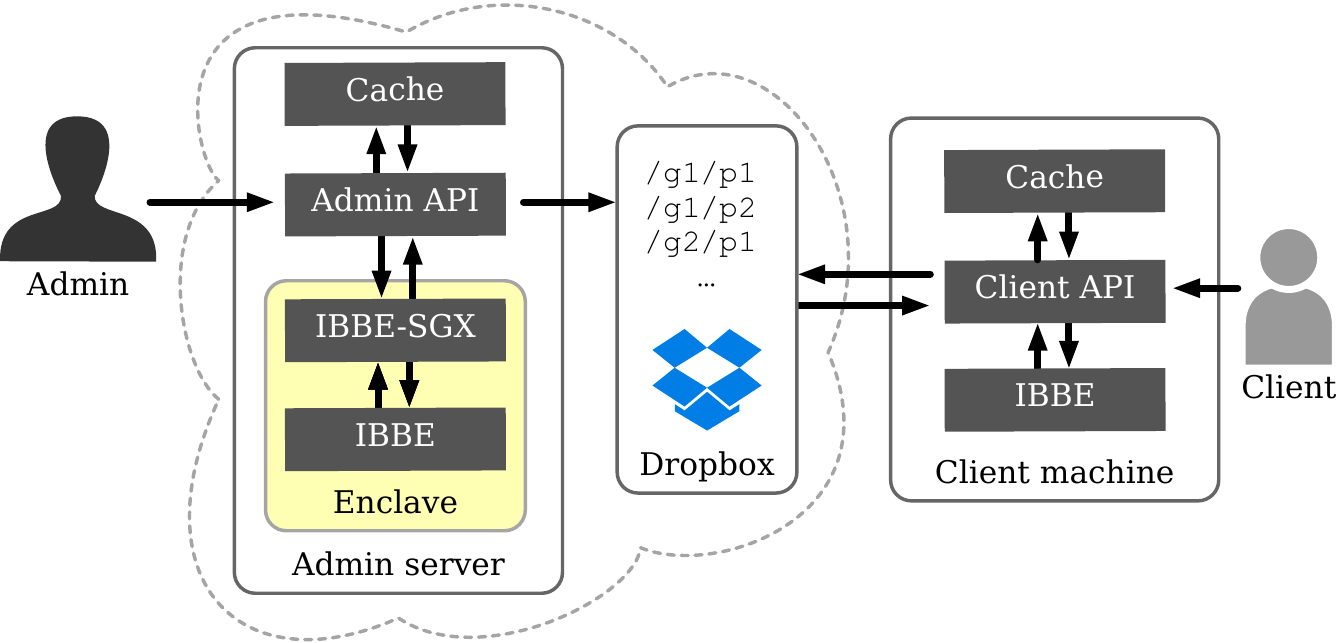}
	\caption{\label{fig:ibbemodel}\ibbesgx: System components.}
\end{figure}

\Cref{fig:ibbemodel} illustrates the system components, including a client and an administrator which use Dropbox as public cloud storage provider.
The administrator's \ac{API}  makes calls to the underlying \ac{SGX} enclave that contains \ibbesgx. This, in turn, uses the \ac{IBBE} component.
Clients are not required to have a \ac{TEE}, although that could enhance the protection of their private key. \ac{IBBE} decrypting functionalities are directly called by the client's \ac{API}.
Both administrators and clients may use local caches in order to save round-trips to the storage provider.
Administrators use the \ac{HTTP} verb \texttt{PUT} to send data to the cloud, whereas clients listen to updates on relevant files through \ac{HTTP} \textit{long polling}.
In Dropbox, long polling works at the directory level.
As a consequence, we index group metadata in a two-level hierarchy: parent folders represent one group, and each child a partition.

\vspace{-1em}
\subsection{\ibbesgx}
\label{sec:ibbe}

\ibbesgx access control scheme can be broadly described in 3 steps:
\begin{enumerate*}[label={(\roman*)}]
\item trust establishment and private key provisioning;
\item membership definitions and group key provisioning; and
\item membership changes and key updates.
\end{enumerate*}
\ac{IBBE} schemes generate a single public key that can be paired with several
private keys, one per user.
Users, in turn, need to be sure that the private key they receive is indeed
generated by someone they trust, otherwise they would be vulnerable to malicious
entities trying to impersonate the key issuer.
To achieve that, we rely upon \iac{PKI} to provide
verifiable private keys to users.

Another security requirement of \ibbesgx is that the key management must be
kept in \iac{TEE}, so that the master key is never accessible by attackers (not even administrators).
Therefore, there must be a way of checking whether that is the case.
On that front, Intel \ac{SGX} makes it possible to attest enclaves (\Cref{sec:sgx:attestation}).
Running this procedure gives the assurance that a given piece of binary code
is truly the one running within an enclave, on a genuine Intel \ac{SGX} processor.

\begin{figure}
	\centering
	\includegraphics{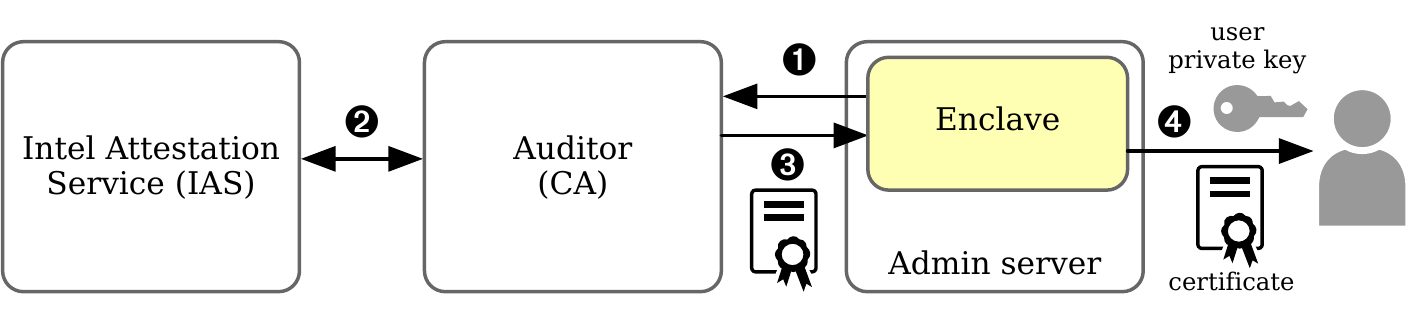}
	\caption{\label{fig:ibbe:startup}\ibbesgx initial setup.}
\end{figure}

\Cref{fig:ibbe:startup} illustrates the initial setup of trust that must be
executed at least once before any key leaves the enclave.
Initially, the enclave generates a pair of asymmetric keys.
While the private one never leaves the trusted domain, the public key is sent
along with the enclave measurement to the Auditor \ding{202}, who is both responsible
for attesting the enclave and signing its certificate, thus also acting as a
\ac{CA}.
Next, the Auditor checks with \ac{IAS} \ding{203} if the
enclave is genuine.
Being the case, it compares the enclave measurement with the expected one, so
that it can be sure that the code inside the shielded execution environment is trustworthy.
Once that is achieved, the \ac{CA} issues the enclave's certificate \ding{204}, which also
contains its public key.
Finally, users are able to receive their private keys and the enclave's
certificate \ding{205}.
Users' keys will be encrypted by the enclave's private key generated in the
beginning.
To be sure they are not communicating with rogue key issuers, users check
the \ac{CA}'s signature in the certificate and then use the enclave's public key contained therein.
All communication channels described in this scheme must be encrypted by
cryptographic protocols such as \ac{TLS}.

Once user private keys are established, they can be used to retrieve group keys from metadata held in a shared storage.
This can only happen, though, if the administrator has first included a given user as a member of some group and updated the respective group metadata in the storage.
Likewise, if metadata were updated after a user revocation, such user will not be able to derive the group key any longer.

\vspace{-1em}
\subsubsection{Cryptographic operations}

Suppose that we want to come up with a simple, yet secure, cryptographic scheme to protect a group key $gk$ by using an asymmetric encryption primitive~\cite{ferguson:2003:cryptography}, based on \ac{RSA} or \ac{ECC}.
As each user in the system has a public-private key pair, the scheme consists in encrypting $gk$ using the public key of each member in the group.
Group members can access $gk$ by decrypting the resulting ciphertext using their private key.
This construction is referred to as trivial broadcast encryption~\cite{stinson:2005:cryptography}, or \ac{HE}~\cite{garrison:2016:daccloud}.

In such scheme, the amount of group metadata grows linearly with the number of members in the group, making it impractical in the context of very large groups.
Besides, when revoking group members, a new key $gk$ needs to be created. As a consequence, the entire group metadata needs to be regenerated and updated.
This causes the propagation of the linear increase both to the cipher generation and to data transmission latencies.

Additionally, when performing group membership operations, administrators must check the authenticity of public keys that are linked to members' identity.
A \ac{PKI}~\cite{ferguson:2003:cryptography} can be used to solve this issue.
Apart from risks that \ac{PKI} brings~\cite{ellison:2000:ten}, one needs to account for the practical costs of setting up, running and accessing \iac{PKI}.

Alternatively, one could replace public-key primitives with identity-based ones.
Using \ac{IBE}~\cite{boneh:2001:ibe, waters:2005:efficient} allows for the adoption of arbitrary strings (\eg, user name or e-mail) as public keys, alongside public known parameters.
It is even possible to encrypt messages which are addressed to users who have not yet interacted with the system. 
The user secret key is generated at setup phase or later by \iac{TA}, which guards a private master key for that purpose.

In conclusion, \ac{HE} combines symmetric (\ac{AES}) and asymmetric (\ac{RSA} or \ac{ECC}) encryptions for group communication. It shows a linear growth on metadata with respect to the number of group members. \ac{RSA} requires \ac{PKI} for checking authenticity of each user's public key, whereas \ac{IBE} replaces the \ac{PKI} for publicly known parameters plus a user identifier string, therefore obviating the need of checking the public key authenticity for every user.

\vspace{-1em}
\subsubsection{Identity-based broadcast encryption (\acs{IBBE})}

Broadcast encryption (\acsu{BE})~\cite{fiat:1994:be} allows a node to encrypt a message that is only accessible to a subset of all users who were provided with a private key when they initially joined the system.
In such scenario, a system-wide public key (as opposed to one per user as in \ac{HE}) is used to encrypt messages, which typically contain a group key $gk$. The cipher is then decrypted by private keys and used by rightful group members in order to get access to protected content.

\begin{figure}[b]
	\center
\includegraphics{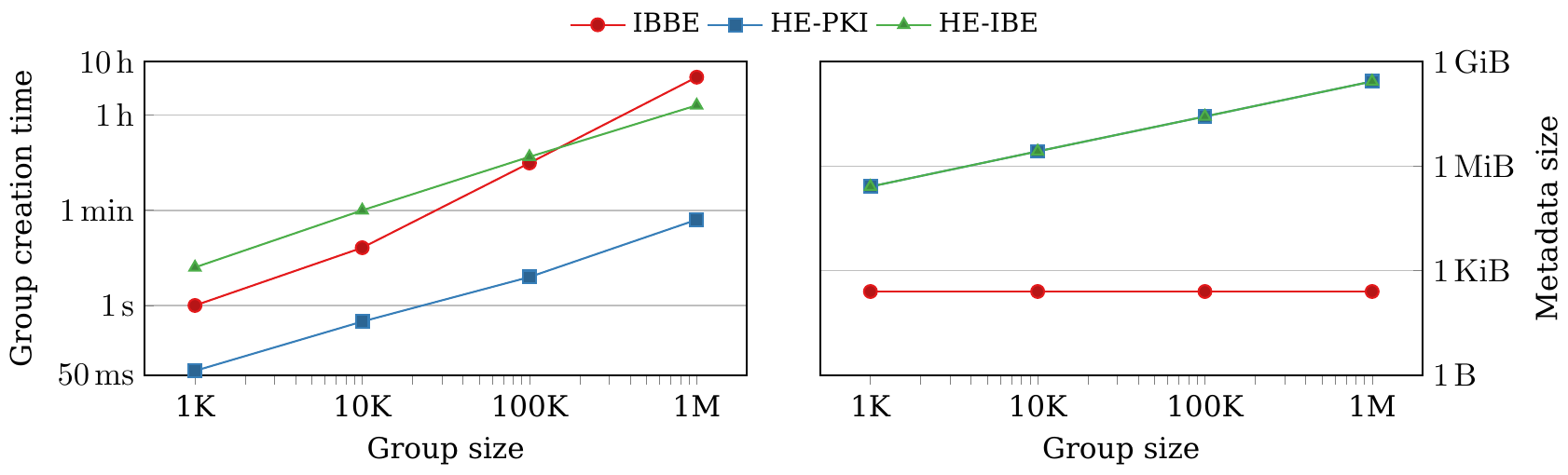}
 	\caption[Comparison between distinct cryptographic protocols.]{\label{fig:bb_test}Comparison between HE-PKI, HE-IBE and IBBE regarding group creation time and metadata produced.}
\end{figure}

There are \ac{BE} solutions that tolerate any coalition of illegitimate members~\cite{boneh:2005:collusion} or allow dynamic changes of broadcast groups~\cite{delerablee:2007:collusionsecurebe}.
We chose, however, \ac{IBBE}~\cite{delerablee:2007:ibbeconstantsize} that besides contemplating both features, also integrates with \ac{IBE}, therefore taking advantage of small sized metadata: the system-wide public key is linear to the maximum group size, while ciphertexts and private keys have constant sizes.
The drawback, however, is the quadratic complexity of crypto operations in the number of group members.
In short, even though the scheme brings a tremendous gain in the size of group metadata, the computational cost of \ac{IBBE} might render it unpractical.

Figure \ref{fig:bb_test} compares \ac{HE-PKI} (\ac{RSA}-2048, not considering signature check), \ac{HE-IBE} (as in \cite{boneh:2001:ibe}) and \ac{IBBE} schemes when encapsulating a \SI{32}{\byte} key.
We see, on the left, the total time taken for group creation when varying the number of members and, on the right, the group metadata expansion.
\ac{IBBE} always produces \SI{256}{\byte} of metadata, regardless of the group members amount.
That is preferable when compared to \ac{HE-PKI} and \ac{HE-IBE}, which produce
increasingly larger values: \SI{27}{\mebi\byte} for groups of 100,000 users, and \SI{274}{\mebi\byte} for the largest group size.
On the other hand, \ac{IBBE} performs much worse than \ac{HE-PKI} when considering the execution time. It is $150\times$ and $144\times$ slower for groups of 10,000 and 100,000 users, respectively.

It is clear that \ac{IBBE} would be a better choice if it were not so slow.
Since we can take advantage of the shielding provided by \acp{TEE}, we are able to improve its performance during encryption.
Moreover, we propose a mitigation technique for lower decryption times, when we assume there is no \ac{TEE} available.
We succinctly explain how traditional \ac{IBBE} works before introducing the proposed modifications. Let

\begin{itemize}
	\item $ e: \mathbb{G}_1 \times \mathbb{G}_2 \rightarrow \mathbb{G}_T$ be a bilinear map defined by cyclic groups of prime order $p$.
	\item $g \in \mathbb{G}_1$, $h \in \mathbb{G}_2$ and $\gamma, k \in \mathbb{Z}_p^*$ be random values. Besides, $w = g^\gamma$ and $v = e\left(g, h\right)$.
	\item $n$ be the largest possible group size.
	\item $\mathcal{H}:\mathbb{Z}^* \rightarrow \mathbb{Z}_p^*$ be a cryptographic hash function
\end{itemize}

The \ac{IBBE} scheme~\cite{sakai:2007:ibbe,delerablee:2007:ibbeconstantsize} consists of the following operations.

\begin{enumerate}
\item \textbf{System setup}: \ac{TA} runs it once by generating a master secret key $M_{K}$ and a system-wide public key $P_{K}$.
\begin{align}
M_{K}&=\lbrace g, \gamma\rbrace\\
P_{K}&= \lbrace w, v, h, h^\gamma, h^{\gamma^2}, ... , h^{\gamma^n}\rbrace \label{eq:ibbe:pubkey}
\end{align}
\item \textbf{Extract user secret}: \ac{TA} uses the $M_{K}$ to extract the secret key $U_{K}$ for each user.
\begin{align*}
U_{K}&=g^{\left(\gamma + \mathcal{H}\left(u\right)\right)^{-1}}
\end{align*}
\item \textbf{Encrypt  broadcast key}: The broadcaster generates a randomized broadcast key $bk$ for a given set of receivers $\mathcal{S}$.
Along with $P_{K}$, this operation turns $bk$ into a public broadcast ciphertext $\mathcal{C}$.
\begin{align*}
bk&=v^k\\
\mathcal{C}&=\lbrace C_1,C_2\rbrace\mathtt,~where:\\
C_1 &= w^{-k}\\
C_2 &= \left(\left(h^{\gamma^N}\right)\cdot\left(h^{\gamma^{N-1}}\right)^{\mathcal{E}_1}\cdot\left(h^{\gamma^{N-2}}\right)^{\mathcal{E}_2}\cdot ... \cdot\left(h^{\gamma}\right)^{\mathcal{E}_{N-1}}
\right)^k,~where:\\
N&=|\mathcal{S}|\\
\mathcal{E}_1 &= \sum_{u \in \mathcal{S}} \mathcal{H}\left(u\right) \\
\mathcal{E}_2 &= \sum_{u_1, u_2 \in \mathcal{S}, u1 \neq u2} \mathcal{H}\left(u_1\right) \cdot \mathcal{H}\left(u_2\right) \\
&... \\[0in]
\mathcal{E}_{N-1} &= \prod_{u \in \mathcal{S}}\mathcal{H}(u)
\end{align*}
\item \textbf{Decrypt broadcast key}:  any member of $\mathcal{S}$ can derive $bk$ from $\left(\mathcal{S}, \mathcal{C}\right)$ using $U_{K}$.
\end{enumerate}

In contrast to traditional \ac{IBBE} that requires \iac{TA} to perform the \emph{system setup} and \emph{extract user secret} operations, we use \ac{SGX} enclaves instead.
By doing so, the master secret key $M_{K}$ can be used in plaintext form inside the enclave, and securely sealed when stored outside for persistence.

The operation to \emph{encrypt} the broadcast key rely on the system-wide public key $P_{K}$, hence it can be performed by any user of the system in traditional \ac{IBBE}.
We, instead, require that all group membership changes and group key encryption are performed by an \emph{administrator}. 
Since administrators operate on the same \ac{SGX} machines where system setup and user key generation take place, encryption in \ibbesgx may use $M_{K}$ instead of $P_{K}$, which dramatically reduces the computational complexity of such operation.
The decryption operation, however, remains identical to the traditional \ac{IBBE} approach, executable by any user.

By using $M_{K}$ inside the enclave, we are able to bypass the polynomial expansion of quadratic cost shown in step $3$ of \ac{IBBE}.
As a consequence, the encryption operation's complexity drops from $O(N^2)$ in \ac{IBBE} to $O(N)$ in \ibbesgx. The value $C_2$ is simply computed by:

\vspace{-1em}
\begin{align}
C_2 &= h^{k \cdot \prod\limits_{u \in \mathcal{S}}\left(\gamma + \mathcal{H}\left(u\right)\right)} \label{eq:ibbesgxcut}
\end{align}

This complexity cut is sufficient to tackle the \ac{IBBE}'s impracticality highlighted in \Cref{fig:bb_test}. Moreover, operations for re-keying, adding or removing a user from a broadcast group are done in $O(1)$. 
In order to remove a user and update the broadcast key $bk$ in constant time, we had to add a third component $C_3=(C_2)^{k^{-1}}$ to the ciphertext, which is completely safe since it may be entirely derived from public key's components (\Cref{eq:ibbe:pubkey}). 
This is due to the fact that the administrator has no access to the random component $k$ which is required for updating the cipher upon key update, which happens in both user removal and re-key operations.
\Cref{tab:constantmath} shows how to recompute the ciphertext for each of them.

\begin{table}
\centering
\caption{\label{tab:constantmath}\ibbesgx: Constant time operations.}
\begin{tabular}{c|c|c}
    \toprule
    Add user $u_a$  & Remove user  $u_r$ & Re-keying \\
    & and update key to $k_n$ &with a new $k_n$\\
    \midrule
    {$\!\begin{aligned}
        C_2 &\gets \left(C_2\right)^{\gamma + \mathcal{H}\left(u_{a}\right)} \\
        C_3 &\gets \left(C_3\right)^{\gamma + \mathcal{H}\left(u_{a}\right)} \end{aligned}$}
    & {$\!\begin{aligned}
        C_1 &\gets w^{-k_n} \\
        C_3 &\gets \left(C_3\right)^{\left(\gamma + \mathcal{H}\left(u_{r}\right)\right)^{-1}} \\
        C_2 &\gets \left(C_3\right)^{k_n} \end{aligned}$}
    & {$\!\begin{aligned}
        C_1 &\gets w^{-k_n} \\
        C_2 &\gets \left(C_3\right)^{k_n} \end{aligned}$}\\
    \bottomrule
\end{tabular}

 \end{table}

Unfortunately, \ibbesgx maintains the quadratic complexity when a group member needs to decrypt the ciphertext by performing a polynomial expansion similar to the encrypt operation described above.
We address this issue by introducing a partitioning mechanism.

\subsubsection{Partitioning}

\begin{table}[b]
\caption{\label{tab:complexity_drop} \ibbesgx: Operations complexities comparison.}
\centering
\begin{tabular}{ lccc }
    \toprule
    Operation & \ac{IBBE}~\cite{delerablee:2007:ibbeconstantsize} & \ibbesgx & \ibbesgx with partitioning \\
    \midrule
    System setup  & $O(N)$ & $O(N)$ &$O(n)$ \\
    Extract user key & $O(1)$& $O(1)$ &$O(1)$  \\
    Create group key & $O(N^2)$& $O(N)$ &$O(mn)$ \\
    Add user to group & & $O(1)$ & $O(1)$ \\
    Remove user from group & & $O(1)$ & $O(m)$ \\
    Decrypt group key & $O(N^2)$ & $O(N^2)$ & $O(n^2)$ \\
    \bottomrule
\end{tabular}
{\\
\footnotesize Cardinalities: $N$: global number of users. $n$: members in one partition. $m$: number of partitions.}
 \end{table}

\begin{figure}
\centering
\includegraphics{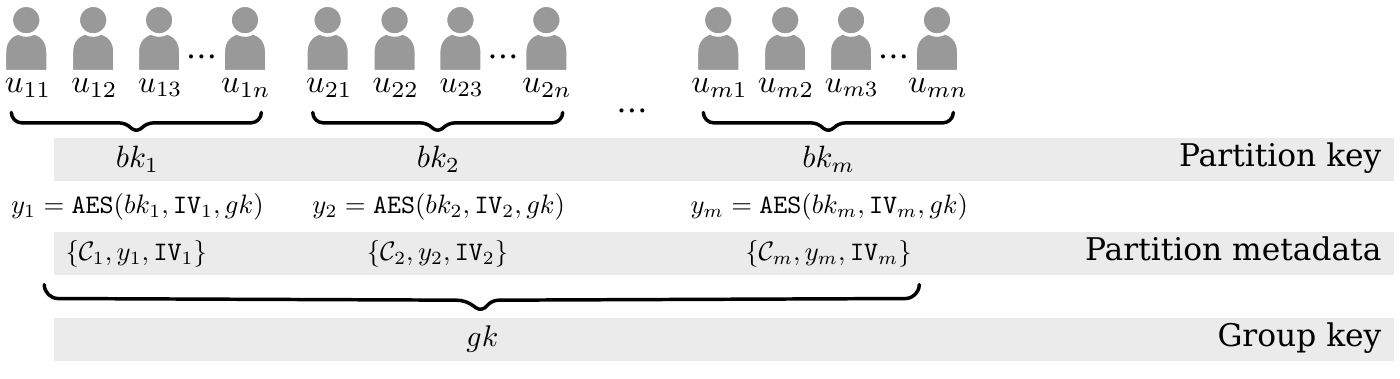}
\caption{\label{fig:ibbe:partitioning}\ibbesgx: Partitioning mechanism.}
\end{figure}

As the decryption time is bound to the number of users in the receiving set, we split the group into $m$ partitions (sub-groups) of at most $n$ users each, thus limiting the user decryption time to the number of members in a single partition.
Each partition $p \in \mathcal{P}$ corresponds to a broadcast group such that $m=|\mathcal{P}|$  with its respective broadcast key $bk_i$ encapsulated in a ciphertext $\mathcal{C}_i$, where $i \in \lbrace 1, \ldots, m \rbrace$.
Such key $bk_i$ is used to encrypt the group key $gk$ shared across all partitions.
To encrypt $gk$, we use symmetric encryption, such as \ac{AES}, and produce a new component $y_i=\mathtt{AES}(bk_i,\mathtt{IV}_i,gk)$ where $\mathtt{IV}_i$ is a random input vector, which are both appended to the partition metadata, \ie, $\lbrace\mathcal{C}_i,y_i,\mathtt{IV}_i\rbrace$. 
A partition $p$ groups a set of users $u_{ij} \in p$ such that $n=|p|$ and $j \in \lbrace 1, \ldots, n \rbrace$.
\Cref{fig:ibbe:partitioning} illustrates the scheme.

\definecolor{sgx}{HTML}{DAA520}
\tcbset{
	title=Enclaved,
	arc=2mm,
	boxsep=0.6mm,
	top=0.6mm,
	bottom=0.2mm,
	left=5mm,
	text width=\linewidth-8mm,
	enlarge left by=-5.8mm,
	colframe=sgx,
	coltitle=black,
	boxrule=0.2mm,
	halign title=right,
}

\newcommand{\nextnr}{\stepcounter{AlgoLine}\ShowLn}
\begin{figure}
\fontsize{8}{9.6}
\setlength\columnsep{0mm}
\begin{multicols}{2}
\null \vfill
\begin{minipage}{0.43\textwidth}
\begin{algorithm}[H]
    \caption{\label{alg:ibbe:creategroup}\ibbesgx: Create group.}
    \Input{ $g$: group \\
            $\mathcal{S} = \{u_1, ... , u_N\}$: members \\
            $n$: partition size
    }
    $\mathcal{C} \gets \emptyset$\;
    $\mathcal{P} \gets \{\{u_1,...,u_n\}, \{u_{n+1},...,u_{2n}\} ,... \}$ \label{alg:ibbe:defpartit} \;
    \begin{tcolorbox}
    \nextnr $gk \gets RandomKey()$  \label{alg:ibbe:encl1st}\;
    \nextnr \ForEach{$i \in \mathcal{P}$}{
         \nextnr $\lbrace bk_i, \mathcal{C}_i \rbrace \gets create\_partition(M_k, i)$\;
         \nextnr $\mathtt{IV}_i \gets RandomIV()$\;
         \nextnr $y_i \gets \mathtt{AES}(\mathtt{SHA}(bk_i), \mathtt{IV}_i, gk)$\;
         \nextnr $\mathcal{C} \gets \mathcal{C} \cup \lbrace i, \mathcal{C}_i, y_i, \mathtt{IV}_i \rbrace$ \;
    }
    \nextnr $sealed\_gk \gets sgx\_seal(gk)$ \label{alg:ibbe:encllst}\;
    \end{tcolorbox}
    \addtocounter{AlgoLine}{-1}
    $store\_key(g,sealed\_gk)$ \label{alg:ibbe:upload1}\;
    $\forall c \in \mathcal{C} : store\_meta(g,c)$ \label{alg:ibbe:upload2}\;
\end{algorithm}
\end{minipage}
\vfill \null
\columnbreak

\tcbset{
    text width=\linewidth-7mm,
    left=10mm,
    enlarge left by=-10.8mm,
}
\begin{minipage}{0.45\textwidth}
    \begin{algorithm}[H]
        \caption{\label{alg:ibbe:add}\ibbesgx: Add user to group.}
        \Input{$g$: group\\
            $\mathcal{P}$: partitions of $g$\\
            $n$: partition size\\
            $u_{a}$: user to add\\
            $sealed\_gk$: sealed group key\\}
        $p_a \gets \exists p \in \mathcal{P},\textnormal{ $such$ $that$ } \vert p \vert < n$ \label{alg:ibbe:findpart}\;
        \If{$p_a = \emptyset$}{
            \ShowLn $p_{a} \gets \{u_{a}\} $ \label{alg:ibbe:newpart} \;
            \begin{tcolorbox}
                \nextnr $\lbrace bk_{a}, \mathcal{C}_{a} \rbrace \gets create\_partition(M_k, p_{a})$ \label{alg:ibbe:envkey1st}\;
                \nextnr $gk\gets sgx\_unseal(sealed\_gk)$ \;
                \nextnr $\mathtt{IV}_a \gets RandomIV()$\;
                \nextnr $y_{a} \gets \mathtt{AES}(\mathtt{SHA}(bk_{a}), \mathtt{IV}_a, gk)$ \label{alg:ibbe:envkeylst}\;
            \end{tcolorbox}
            \addtocounter{AlgoLine}{-1}
            $store(g,\{p_{a}, \mathcal{C}_a,y_{a}, \mathtt{IV}_{a}\})$ \label{alg:ibbe:storeciph}\;
        }\Else{
            $p_{a} \gets p_a \cup \{u_{a}\} $ \label{alg:ibbe:addexist1st} \;
            \begin{tcolorbox}[notitle]
                \nextnr $\mathcal{C}_{a}\gets add\_user\_to\_partition(M_k, p_{a}, u_{a})$ \label{alg:ibbe:addexist2nd}\;
            \end{tcolorbox}
            \addtocounter{AlgoLine}{-1}
            $update\_meta(g, \lbrace p_{a}, \mathcal{C}_{a}, -, - \rbrace)$ \label{alg:ibbe:updatemeta}\;
        }
    \end{algorithm}
\end{minipage}
\end{multicols}
 \end{figure}

From the administrator perspective, there is an impact when removing a user from the group, since that would provoke an update on the group key $gk$ and therefore the recalculation of $y_i$ for all partitions, along with $bk_i$ of the implicated one. This renders the complexity $O(m)$ to the remove user operation instead of $O(1)$ with no partitioning.
Diversely, all other operation complexities are kept or reduced.
\Cref{tab:complexity_drop} compares them. In order to distinguish cardinalities, we use $n=|p|$ for partitions and $N=|\mathcal{S}|$ for a single broadcast group, \ie, without the partitioning scheme.

Extracting a user key and adding a user to a group remain $O(1)$. The addition of a user can cause the creation of a new partition if all existing ones are full. In such case, the constant complexity is kept since it corresponds to that of encrypting the broadcast key $bk_i$ for the new partition with a single member (\Cref{eq:ibbesgxcut}).
Creating a group has the cost of creating the first partition, or $O(n)$ (\Cref{eq:ibbe:pubkey}).
The biggest gain comes, however, in key decryption. 
Instead of being quadratic in the total number of users $O(N^2)$, it becomes quadratic in the number of users per partition $O(n^2)$.

Partitioning also impacts storage footprint.
The public key $P_K$ is linear in the maximal number of users per partition $O(n)$ instead of the total user amount $O(N)$.
Concerning group metadata, the footprint is augmented by the symmetrically encrypted group key, \ie, $y_i$, and the respective \ac{IV}.
For an entire group, metadata storage corresponds to the number of partitions times the size of $\lbrace\mathcal{C}_i,y_i,\mathtt{IV}_i\rbrace$, in addition to a data structure that keeps the mapping between users and partitions.
Although this induces a slight overhead, the number of partitions in a group is relatively small compared to the group size.
Besides, administrators alone manipulate partition metadata.
They can therefore use a local cache to bypass the cost of accessing remote storage.

Determining the optimal value for the partition size mostly depends on the dynamics of the group.
There is a trade-off between the number and frequency of group membership operations performed by the administrator and those performed by regular users for decrypting the broadcast key.
A small partition size reduces the decryption time on the user side while a larger one reduces the number of operations performed by the administrator to run \ibbesgx and to maintain the metadata.
We further evaluate this trade-off in the upcoming sections.

\begin{figure}
\centering
\small
\begin{minipage}{0.53\textwidth}
\begin{algorithm}[H]
\caption{\label{alg:remove}\ibbesgx: Remove user from group.}
\Input{ $g$: group\\
        $\mathcal{P}$: partitions of $g$\\
        $u_{r}$: user to remove\\}

    $p_{r} \gets p \in \mathcal{P}, \ such \ that \ u_{r} \in p$ \label{alg:ibbe:finddelpart} \;
    $p_{r} \gets p_{r} \setminus \{u_{r}\} $ \label{alg:ibbe:extract} \;
    \begin{tcolorbox}
    \nextnr $gk \gets RandomKey()$ \label{alg:ibbe:newkey} \;
    \nextnr $(bk_{r}, \mathcal{C}_{r}) \gets remove\_user(M_k, p_{r}, u_{r})$ \label{alg:ibbe:upmeta}\;
    \nextnr $\mathtt{IV}_r \gets RandomIV()$\;
    \nextnr $y_{r} \gets \mathtt{AES}(\mathtt{SHA}(bk_{r}), \mathtt{IV}_r, gk)$ \label{alg:ibbe:reenvelope} \;
    \nextnr \For{$p \in \mathcal{P} \setminus p_{r}$}{
        \nextnr $(bk_{p}, \mathcal{C}_{p}) \gets rekey\_partition(p)$ \label{alg:ibbe:rekey1st}\;
        \nextnr $\mathtt{IV}_p \gets RandomIV()$\;
        \nextnr $y_{p} \gets \mathtt{AES}(\mathtt{SHA}(bk_{p}), \mathtt{IV}_p, gk)$ \label{alg:ibbe:rekeylst}\;
    }
    \nextnr $sealed\_gk \gets sgx\_seal(gk)$ \label{alg:ibbe:reseal} \;
    \end{tcolorbox}
     \addtocounter{AlgoLine}{-1}
     $update\_key(g,sealed\_gk)$ \label{alg:ibbe:storereseal}\;
     $update\_meta(g,\lbrace p_r, \mathcal{C}_r, y_r, \mathtt{IV}_r \rbrace)$ \label{alg:ibbe:upmeta1}\;
     $\forall p \in \mathcal{P} \setminus p_{r} : update\_meta(g,\lbrace p , \mathcal{C}_p, y_{p}, \mathtt{IV}_{p} \rbrace)$ \label{alg:ibbe:upmeta2}\;
\end{algorithm}
\end{minipage}
 \end{figure}

\subsubsection{Membership operations}

In order to create a group, we execute the instructions in \Cref{alg:ibbe:creategroup}.
Once the partitions are determined (line~\ref{alg:ibbe:defpartit}), the execution enters the \ac{SGX} enclave (lines~\ref{alg:ibbe:encl1st} to~\ref{alg:ibbe:encllst}), when the group key is encrypted with the partition broadcast key.
The ciphertext and the sealed group key leave the enclave to be later pushed to the shared storage (lines~\ref{alg:ibbe:upload1} and~\ref{alg:ibbe:upload2}).

The operation of adding a user to a group, shown in \Cref{alg:ibbe:add}, starts by finding one partition which is not yet full (line~\ref{alg:ibbe:findpart}).
If none is found, a new partition is created with a single user (line~\ref{alg:ibbe:newpart}) and the group key is enveloped by the broadcast key of this new partition (lines~\ref{alg:ibbe:envkey1st} to~\ref{alg:ibbe:envkeylst}), before storing the corresponding ciphertexts (line~\ref{alg:ibbe:storeciph}).
Otherwise, the user is added to the not-yet-full partition found (lines~\ref{alg:ibbe:addexist1st} and~\ref{alg:ibbe:addexist2nd}).
Since both the partition broadcast and the group key remain unchanged, only the ciphertext needs to be updated (line~\ref{alg:ibbe:updatemeta}).

When a user needs to be removed from a group (\Cref{alg:remove}), his identifier is extracted from the partition to which he belongs (lines~\ref{alg:ibbe:finddelpart} and~\ref{alg:ibbe:extract}) and a new group key is randomly generated (line~\ref{alg:ibbe:newkey}).
Next, metadata from the partition that used to contain the removed user is updated (line~\ref{alg:ibbe:upmeta}) and the new partition key $bk_r$ is used for enveloping the new group key $gk$ (line~\ref{alg:ibbe:reenvelope}).
For all the remaining partitions, a constant time re-keying operation (see \Cref{tab:constantmath}) regenerates each partition broadcast key, which is used to encrypt the new group key (lines~\ref{alg:ibbe:rekey1st} to~\ref{alg:ibbe:rekeylst}).
After sealing the new group key (line~\ref{alg:ibbe:reseal}), the resulting value is persisted (line~\ref{alg:ibbe:storereseal}) along with all updated metadata (lines~\ref{alg:ibbe:upmeta1} and~\ref{alg:ibbe:upmeta2}).

\tcbset{
	left=5mm,
	enlarge left by=-5.8mm,
}

As many removal operations may cause sparsely occupied partitions, we propose a re-partitioning scheme whenever partition occupancies are low.
We implement a heuristic to detect low occupancy when half of the partitions in one group are two thirds full or less. In such case, re-partitioning is triggered by simply re-creating the implicated group according to \Cref{alg:ibbe:creategroup}.
The client decrypt operation works by first using the standard \ac{IBBE} to discover the broadcast key, whose hash is then used to obtain the group key.

\subsubsection{Implementation and evaluation}

We used the \ac{PBC}~\cite{lynn:2006:pbc} library which depends on \ac{GMP}~\cite{granlund:1991:gmp}.
They both have to be used inside \ac{SGX} enclaves to implement the \ac{IBBE} component (see \Cref{fig:ibbemodel}).
Luckily, since both \ac{PBC} and \ac{GMP} mostly perform computations rather than input
and output operations, the challenges on adapting them were chiefly restrained
to tracking and adapting calls to \ac{glibc}.
The adaptations were done either by relaying operations to the operating
system through \acp{ocall}, or performing them with equivalent operations inside enclaves.
Outside calls, in turn, do not perform any sensitive action that could compromise security.
Aside from source code, we also adapted the compilation toolchain, since enclaves must use curated versions of standard libraries (we used Intel \ac{SGX} \ac{SDK}). 
Moreover, specific code generation flags must be used, along with the prevention of using the compiler's built-in functions.
The modifications account for 32 \ac{LoC} for \ac{PBC} and 299 for \ac{GMP}.

\begin{figure}[tb]
	\center
\includegraphics{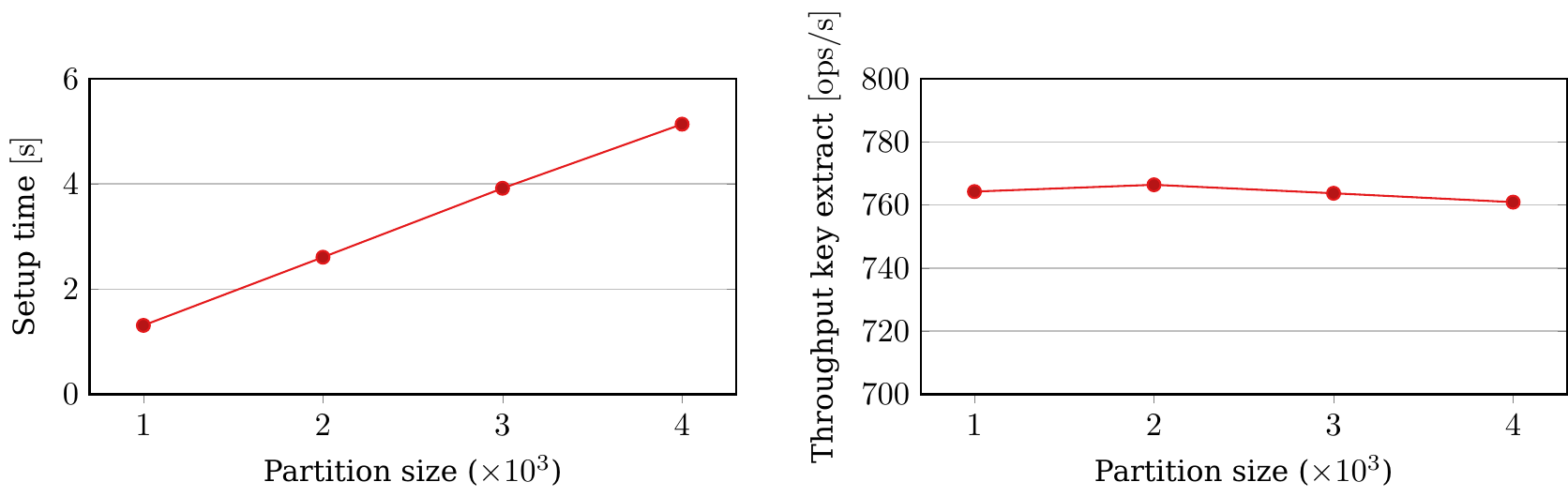}
 	\caption{\label{fig:micro_bootstrap}Performance of \ibbesgx's bootstrap phase.}
\end{figure}

Apart from these changes, we also needed to use common cryptographic libraries.
Due to limitations in Intel \ac{SGX} \ac{SDK} v.1.9, we used an OpenSSL \ac{SGX} port~\cite{intel-sgx-ssl}.
The end-to-end system including both \ibbesgx and \ac{HE} schemes consists of 3,152 lines of C/C++ code and 170 lines of Python.

\subsubsection{Micro-benchmarks}
\vskip 1mm

We benchmark the performance of \ibbesgx in isolation, by comparing it to \ac{HE}, and by using access control traces.
Experiments were performed on a quad-core Intel i7-6600U machine with a processor at 3.4 GHz, 16 GB of \ac{RAM} and Ubuntu 16.04 \ac{LTS}.

First, we evaluate the bootstrap's performance. It consists of setting up the system and generating user private keys. Results are shown in \Cref{fig:micro_bootstrap}.
The setup latency increases linearly according to partition size, with a growth of \SI{1.2}{\second} per \SI{1,000}{} users.
Differently, extracting secret user keys gives an average throughput of \SI{764}{\operation\per\second}, irrespective of the partition size.
This is unsurprising, since we have seen that such operations are $O(n)$ and $O(1)$, respectively (see \Cref{tab:complexity_drop}).

\begin{figure}[tb]
\centering
\includegraphics{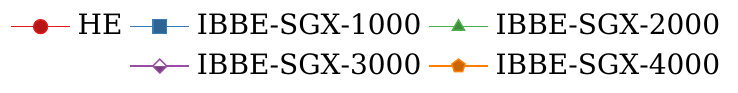}
\includegraphics{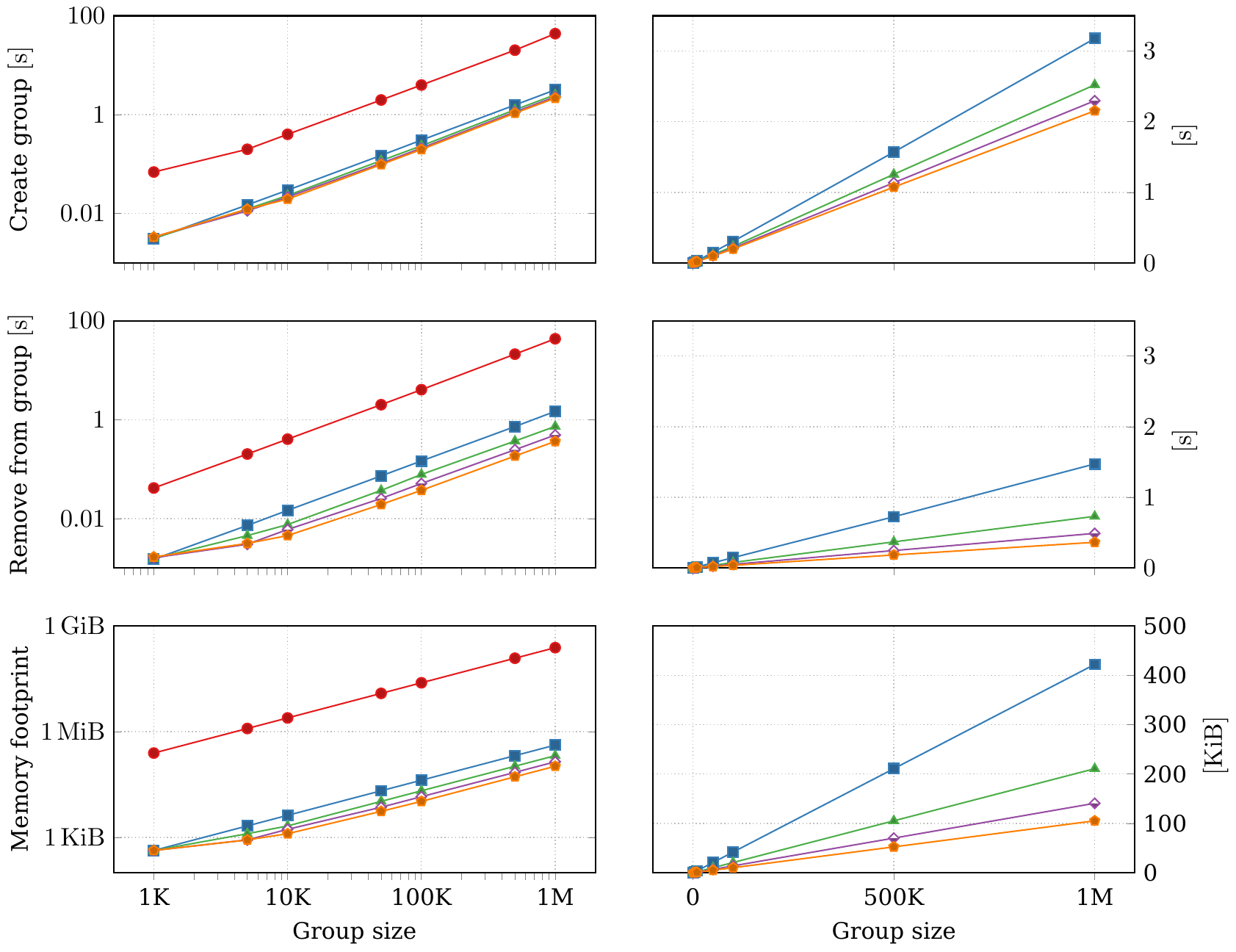}
\vskip 6mm
\caption[Create and remove operations in \ibbesgx.]{\label{fig:partitioning}Evaluation of create and remove operations and storage footprint, by varying the partition size for \ibbesgx (1000, 2000, 3000 and 4000).}
\end{figure}

Next, we compare \ibbesgx with \ac{HE}.
\Cref{fig:partitioning} displays performance measurements of operations for creating a group and removing a user from a group, along with the storage footprint of metadata.
\ibbesgx is better than \ac{HE} in all three by approximately a constant factor.
Create and remove operations with \ibbesgx is \SI{1}{} order of magnitude faster than \ac{HE}.
If we compare to the original \ac{IBBE} scheme, \ibbesgx is better by \SI{2}{} orders of magnitude for groups of 1,000 users and \SI{3}{} for one million users (see Figure~\ref{fig:bb_test}).
Concerning memory usage, \ibbesgx is up to 6 orders of magnitude better than \ac{HE}.
The plots on the right-hand side in \Cref{fig:partitioning} show the same data in linear axis but disregarding \ac{HE}, so that we are able to observe the effect of different partition sizes.
One can notice that the remove operation takes half the time of creating a group.
Considering the storage footprint, the increase in memory use brought by using smaller partition sizes is fairly small. For instance, \SI{422}{\kibi\byte} with \SI{1000}{} users per partition vs. \SI{105}{\kibi\byte} with \SI{4000}{} users per partition, both for groups with 1 million members.

\begin{figure}[tb]
\centering
\includegraphics{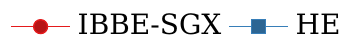}
\includegraphics{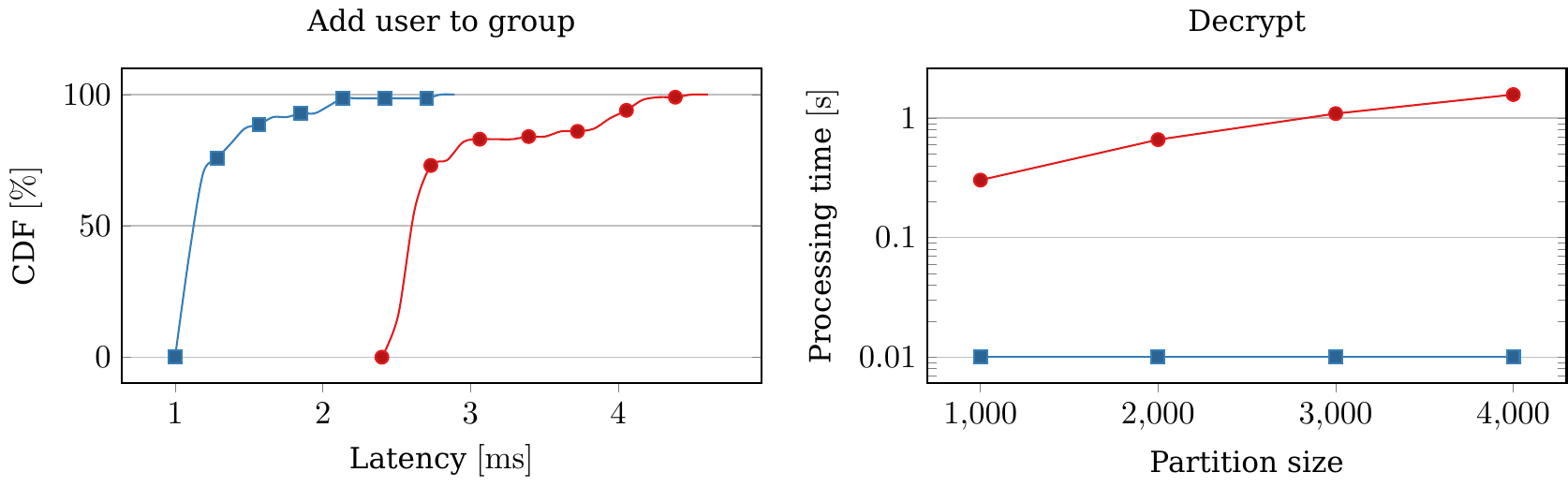}
 \caption[\ibbesgx: Adding user to group and decryption.]{\label{fig:perf_final}Performance of adding a user to a group and decrypt operations.}
\end{figure}

The \ac{CDF} of latencies for adding a user to a group is shown in \Cref{fig:perf_final}.
The operation has a constant time complexity for both \ibbesgx and \ac{HE}.
As noted in \Cref{alg:ibbe:add}, the user addition operation in \ibbesgx can take two paths: to an existing partition or to a new one if all the others are full.
Such effect can be perceived in the plot from $80^{th}$ percentile on.
Besides, the \ac{HE} add operation is generally twice as fast as \ibbesgx.

The client decryption performance is shown on the right-hand side in \Cref{fig:perf_final}.
This operation is also faster with \ac{HE}.
The difference of 2 orders of magnitude is caused by the quadratic cost of \ibbesgx decryption operation.
We argue that a slower decryption time for \ibbesgx can be acceptable in practice, since it is preceded by metadata updates which involve cloud communication and is therefore minimised when put into perspective.
Additionally, the decryption cost is bounded to the partition size, no matter the number of users in the group.

\subsubsection{Macro-benchmarks}

\begin{figure}[tb]
\center
\includegraphics{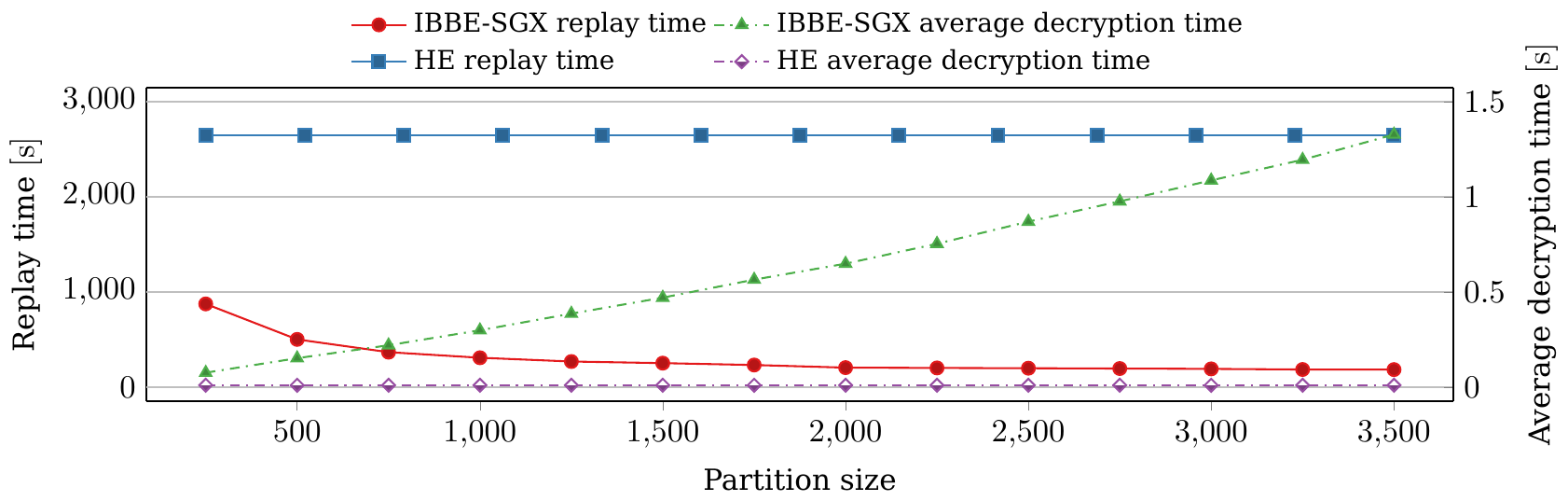}
 \caption[\ibbesgx replay time for a given dataset.]{\label{fig:macro_linux}Measuring total administrator replay time and average  user decryption time per different partition sizes using the Linux Kernel access control list data set.}
\end{figure}

To capture the performance of the \ibbesgx scheme within a realistic scenario, we replay an access control trace based on the membership changes in the version control repository of the Linux Kernel~\cite{_linux_dataset}.
We derive the membership trace by considering one single group where the first commit of a user represents his addition. Conversely, the last commit is interpreted as his removal from the group.
This crafted trace contains 43,468 membership operations that spawn across a period of 10 years, during which the group size never exceeds 2803 users.
We replay the generated trace sequentially for both \ac{HE} and \ibbesgx by varying partition sizes. We measure the total time spent by the administrator to replay the whole trace and the average user decryption time.
\Cref{fig:macro_linux} shows the results.

Considering the administrator replay time, \ibbesgx performs better when the partition size converges to the number of users in the group because, in this case, the overhead of maintaining partitions is minimised.
Using a small partition size, \eg 250, is $2.8\times$ more inefficient when compared to partitions of 1000 users.
Compared to \ac{HE} replay time, \ibbesgx is \SI{1}{} order of magnitude faster for partitions of 1250 users and beyond.
Contrarily, decryption time for \ibbesgx grows quadratically according to the partition size while in HE it remains constant.
This highlights the \ibbesgx's trade-off caused by different partition sizes.
Satisfactory outcomes both in terms of administrator performance and user decryption times may be hence achieved by finding a partition size (such as 750 users) that balances both metrics given the application requirements (\eg, decryption times of \SI{250}{\milli\second}).

In order to observe the impact of different workloads of group membership access control, we generated a set of synthetic traces with incremental revocation rates.
Namely, eleven traces consisting of 10,000 membership operations.
Each trace is composed by randomly generated operations according to different revocation rates.
We replay the traces and measure the total time required by the administrator to perform all membership changes and repeat the experiment for distinct partition sizes.

\begin{figure}[tb]
\centering
\includegraphics{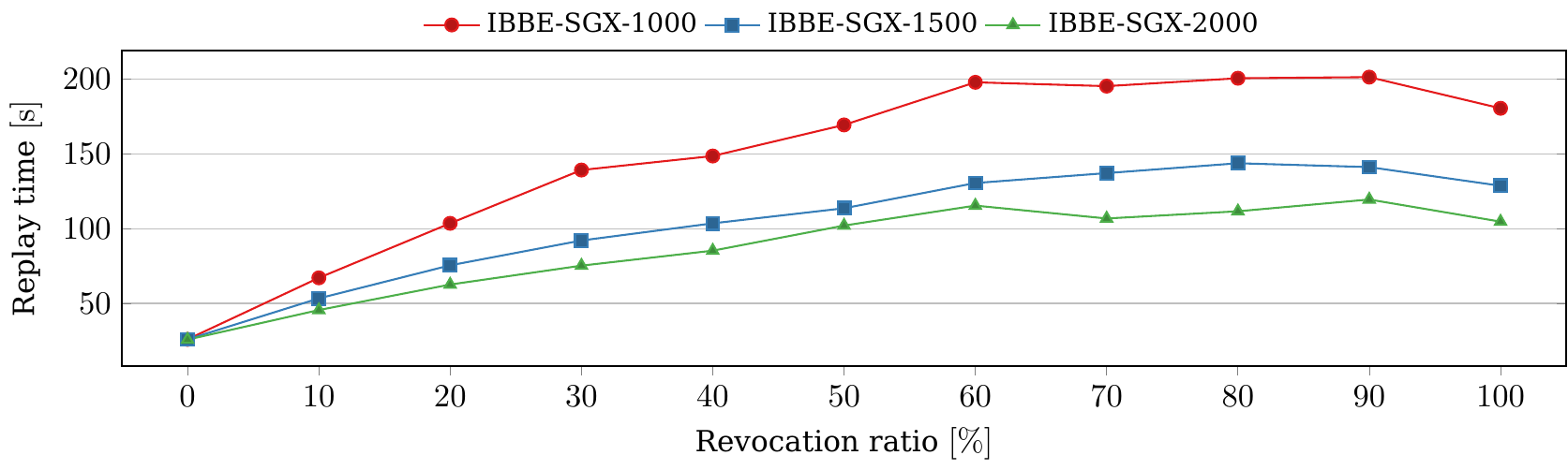}
 \caption[\ibbesgx: Replay time of synthetic datasets.]{\label{fig:macro_syn}\ibbesgx: Replay time of synthetic datasets varying partition sizes and revocation rates.}
\end{figure}

Results are shown in \Cref{fig:macro_syn}.
We observe a linear increase in the total time when incrementally increasing the revocation ratio roughly up to 50\%, \ie, in workloads dominated by constant-time addition operations.
After this point, the total time stabilizes and finally decreases when the revocation ratio is more than 90\%.
This behaviour is caused by the merging of sparse partitions, which happens more frequently with the increase of revocations.
With fewer partitions, \ibbesgx's remove operation becomes faster, therefore decreasing the total time.

In conclusion, although \ibbesgx loses in performance for some operations (\ie, decryption and user addition) when compared to \ac{HE}, its gains arguably outweigh the drawbacks.
Due to \ac{TEE} usage, administrators are prevented from accessing group keys, and are therefore restricted to only perform membership operations, \ie, they could not possibly inspect into groups' shared content.
Apart from the operations of creating groups and removing users from groups, which are more performant in \ibbesgx, the biggest advantage comes with metadata reduction --- orders of magnitude lower than \ac{HE}, which saves resources both in bandwidth and storage space.
Although shared content is protected by this scheme, group membership information is not. This raises privacy concerns when the very fact of belonging to a group is sensitive information. In order to address this issue, we propose in the following section a scheme where this is not leaked.
 
\vspace{-1em}
\section{Anonymous file sharing}

Besides shared data, identity of users may also be sensitive information.
For instance, \acp{VDR}~\cite{iDeals} are used for exchanging confidential documents during business acquisitions. Apart from having to enforce access control, stakeholders' identities must be protected, or else they could have business strategies disclosed to competitors.

Confidential systems that focus on group communication offer anonymity guarantees by group key exchange methods~\cite{angel:2016:unobservable}, requiring all active members to be present and take part in a multi-phase protocol (\eg, Diffie-Hellman) each time a key is derived.
Such an approach is indeed suitable for instant group communication, but impractical for file sharing that generally does not require online users.
Theoretical anonymous file sharing extensions have been proposed~\cite{barth:2006:privacy,libert:2012:abe} without ever turning into functional systems.

Anonymous sharing of confidential content was practically addressed in an unsophisticated manner by \ac{GPG}.
It drops public keys from the resulting ciphertext, therefore keeping no references to the identity of recipients.
The drawback comes at decryption time, when recipients must perform asymmetric decryption trials until the portion of the ciphertext matching their private key is found, if any.
Due to this, \ac{GPG} works well for groups of few users but quickly becomes impractical for larger ones (\Cref{sec:related:datasharing}, Table~\ref{tab:gpg}).

To tackle that, we propose an anonymous access control scheme that leverages \ac{SGX} for a narrow scope: enforcing anonymity during the publishing operation (\ie, upon \emph{writing}).
Our scheme does not require a \ac{TEE} on the user side for performing the \emph{read} operation, nor does it require that users connect to a designated proxy.
To demonstrate the feasibility of our solution, we propose a scalable system design leveraging micro-services that can possibly scale depending on the workload.

By doing so, we achieve an improvement of three orders of magnitude when compared to state-of-the-art \ac{ANOBE}~\cite{barth:2006:privacy}.
Besides, our end-to-end implementation, \asky~\cite{contiu:2019:asky}, can scale to cope with a realistic amount of administrative and user operations.

In the remaining of the section, we
\begin{enumerate*}[label=\emph{(\roman*)}]
	\item define a theoretical anonymous cryptographic access control extension that relies on \acp{TEE} for a minimal subset of operations (\ie, \emph{writes} but not \emph{reads});
	
	\item propose an end-to-end design that leverages micro-services which can scale according to the undergoing workloads; and
	
	\item implement and evaluate the system, first in isolation showing its benefits against state-of-the-art cryptographic schemes, and then by benchmarking its scaling capabilities and practical feasibility.
\end{enumerate*}

\vspace{-1em}
\subsubsection{Model and use case}

We consider users (humans or software agents) that are uniquely identified within the premises of an organisation and share files.
They are clustered into uniquely identifiable \emph{groups} by organisation-specific policies.
We consider two functional categories: \emph{access control} and \emph{content management}.
The first represents group membership operations (adding and removing users from groups) and is performed by \emph{administrators}, while the latter comprises file creation and consuming by group members (writes and reads).  
A group member can perform one or both roles of \emph{writer} or \emph{reader} within one or multiple groups.
The cloud object storage holds uniquely-identified binary objects (\eg, Amazon S3).

We define four security properties for our confidential and anonymous file sharing system: 

\begin{enumerate}%
	\item \textbf{Confidentiality and authenticity}: content of shared files is exclusively accessible by group members.
	\item\textbf{Forward secrecy}: compromising a group secret does not compromise past sharing sessions.
	\item\textbf{Recipients privacy}: no recipient except the group administrator is able to know other recipients' identities (\ie \emph{readers}).
	\item\textbf{Sender privacy}: no recipient except the group administrator is able to know the sender's identity (\ie \emph{writer}).
\end{enumerate}

Revoked and external users behave arbitrarily.
They may try to read shared content and identify group members by intercepting, replaying, deciphering and altering exchanged messages (\ie Dolev-Yao~\cite{dolev:1981:pubkeyprotocols} adversarial model).
User anonymity is not only threatened by external adversaries, but also internally by peer group members.
Being so, active users that can rightfully decrypt group content may maliciously try to infer their peers' identities.

The storage provider is \emph{honest-but-curious}, \ie, it accordingly provides its services while possibly trying to inspect file contents and user identities (\eg, by traffic analysis).
Revoked users may collude with the cloud storage to get access to content shared after their revocation.
Lastly, our privacy model enforces anonymity guarantees solely with respect to user identities, and not to group sizes, content lengths and traffic patterns.

Virtual data rooms (\acsp{VDR})~\cite{iDeals}, for instance, strictly control repositories of electronic documents for company \acp{MA}.
Thanks to \acp{VDR}, sellers, supporting parties who assist them and bidders can confidentially exchange documents (\eg, terms and valuations) atop of untrusted remote storage mediums.
The seller acts as \emph{administrator} and enforces access control.
User roles are composed by \emph{writers} (sellers and supporting parties) and \emph{readers} (bidders).
As enforced by confidentiality agreements, supporting parties operate in the seller's interest, and remain unidentifiable from one another.
Similarly, bidders identities are concealed among themselves.
As such, \emph{inner} anonymity guarantees need to be enforced within \emph{writers} and \emph{readers}, while \emph{outer} anonymity has to withstand against any external entity to the \ac{MA} process.
Additionally, withdrawing bidders or misbehaving supporting parties can be \emph{revoked} by the seller, therefore becoming unable to access the document corpus any further.

\subsection{\asky}
\label{sec:asky}

\begin{figure}[t]
	\centering
	\includegraphics{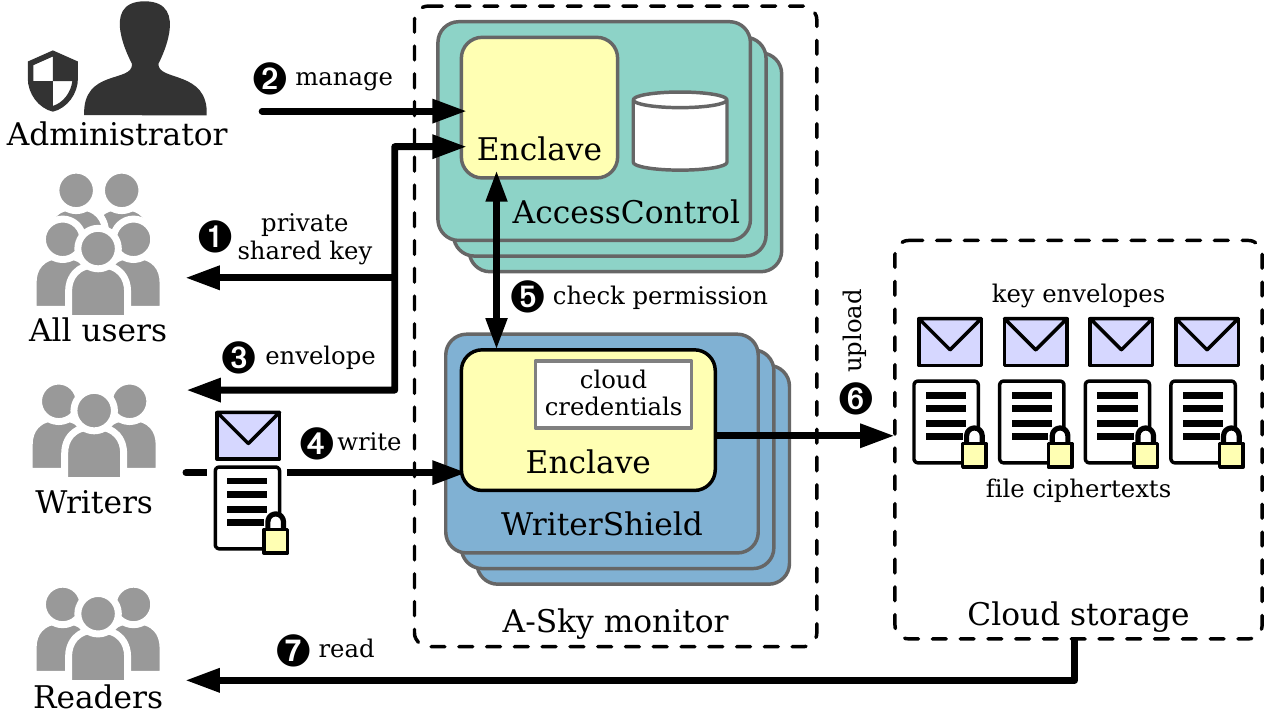}
	\caption{\asky architecture.}
	\label{fig:sys_arch}
\end{figure}

\asky is composed of two components: a cryptographic key management and a data delivery protocol, both designed to combine performance, data confidentiality and user anonymity. 
In order to avoid passing all the system operations through a \ac{SGX} \emph{monitor}, we propose a design in which only \emph{writers} are constrained to pass through such a proxy (\Cref{fig:sys_arch}, step \ding{205}).
\emph{Readers} consume confidential content without communicating with the \ac{TEE}-enabled monitor (\ding{208}), therefore reducing service time penalties.

The monitor acts as an outbound \ac{TA} that authenticates all the content passing by.
Being a proxy, it also masks the identities of data writers. 
Moreover, since it executes in a \ac{TEE}, traditional anonymous key management schemes~\cite{barth:2006:privacy,libert:2012:abe} can count on a trusted entity for the key enveloping operation, therefore allowing the usage of simpler and more efficient encryption schemes. 

The \emph{monitor} sits in between end-users and the cloud storage. It is split into two micro-services, whose amount of instances can be independently adapted according to the undergoing load:
\begin{enumerate}[label={(\roman*)}]
\item the \accessmonitor provides a cryptographic mechanism for storing and enforcing access control to data.

Our cryptographic key management solution requires the connection to a \ac{TEE} monitor.
It generates keys that are securely shared with users at their first interaction with the monitor (\ding{202}).
This allows the usage of symmetric encryption during the enveloping step, therefore taking advantage of hardware acceleration and smaller ciphertexts while achieving equivalent security in comparison to asymmetric approaches.
The \accessmonitor ensure that only authorized entities can write and decrypt data, according to memberships established by an administrator (\ding{203}).
Based on this, envelopes are generated with metadata that allow rightful users to decipher shared content  (\ding{204}).

\item the \writerproxy acts as an outgoing proxy for write operations.

\asky requires that users write the encrypted shared content through the \writerproxy service (\ding{205}), which checks with the \accessmonitor whether a user has the permission to write in a given group (\ding{206}).
Being the case, it authenticates the outgoing content and does the writing itself (\ding{207}).
Since it runs in an enclave, the cloud storage credentials are safely stored by the \writerproxy service.
\end{enumerate}

In traditional anonymous key sharing solutions~\cite{barth:2006:privacy}, a \ac{TA} sets up the key management system and extracts user private keys, while end users perform key enveloping and content encryption by employing asymmetric cryptographic primitives, \ie, using the public keys of group members.
Differently, we leverage \acp{TEE} and are therefore able to use more efficient symmetric constructs.
This change also brings advantages with respect to creating indexes that make de-enveloping more efficient, which will be described later.

Before relying on any service of the \asky \emph{monitor}, it is necessary to check whether the service is running on a trustworthy Intel \ac{SGX} platform and that the instances of the \accessmonitor and \writerproxy are genuine.
This validation phase is performed by administrators and users using the typical \ac{SGX} attestation procedure and establishment of a secure channel (see \Cref{sec:sgx:attestation}).

\vskip 2mm
\subsubsection{System design and implementation}

\vskip 2mm
\subsubsection{\accessmonitor}

The \accessmonitor is responsible for generating and storing user keys, maintaining group membership information and generating envelopes.
Since it deals with sensitive information, its core runs entirely within enclaves.
All exchanges are encrypted with keys only known in the trusted environment: persisted state is ciphered and stored in a database while external communication is done through \ac{TLS} connections terminated inside the enclave.

Its methods are invoked by regular users and administrators after secure channels are established upon successful attestation processes.
All these interactions happen through a \acs{TLS}-encrypted \acs{REST}-like protocol.
Exchanges are represented in \ac{JSON}, for which we slightly modified a C++ library~\cite{json-cpp}.
In order to terminate \ac{TLS} connections in the enclave, we use a port of OpenSSL for \ac{SGX}~\cite{openssl-sgx}.

Users interact with the \accessmonitor either to obtain 
\begin{enumerate*}[label=\emph{(\roman*)}]
	\item their credentials, which are randomly generated \SI{256}{\bit} secret shared keys that are stored by the monitor; or
	\item envelopes, to be attached to shared encrypted files.
\end{enumerate*}
Given a unique user identifier~$u_{id}$, the service generates a key~$u_{k}$ and stores it in a container $keys$ of pairs $\langle u_{id},u_{k} \rangle$, \ie, $keys[u_{id}] \gets u_{k}$.
Administrators, on the other hand, may create and populate groups.
Depending on granted access capabilities, users can be content readers, writers or both.
Such information is kept within persistent dictionaries, $group^r$ and $group^w$, which store sets of users belonging to each group identifier $g_{id}$, \eg, $group^w[g_{id}] = \{u_a, ... , u_z\}$.
Only administrators can modify these dictionaries.

Once groups are defined, metadata that grants access to files may be generated upon user requests.
We call this operation \emph{key enveloping}, and it consists of encapsulating an \emph{access key} such that only group members with read permission are able to retrieve it.
An envelope contains a file access key encrypted several times, once per group member.
Just like user keys, access keys are \SI{32}{\byte} long.
We use \ac{AES} \ac{GCM}, which generates a \emph{tag} of \SI{16}{\byte} for integrity.
Considering the addition of \SI{12}{\byte} for the \ac{IV}, each group member adds \SI{60}{\byte} to the envelope.

\setlength{\columnsep}{-2mm}
\begin{figure}
\begin{multicols}{2}
\null \vfill
\begin{minipage}{0.45\textwidth}
\begin{algorithm}[H]
	\caption{\label{alg:key_envelop}\asky: \accessmonitor key enveloping.}
	\SetKwFunction{KeyEnveloping}{KeyEnveloping}
	\DontPrintSemicolon
	\Input{ $u_{id}$: writing user identity \\
		$g_{id}$: group identifier \\
		$k$: file access key \\
	}
	\Output{$envelope$: ciphertext of the access key}
	\Procedure{\KeyEnveloping{$u_{id}$, $g_{id}$, $k$}}{
		$envelope \gets \emptyset$ \; 
		\If {$u_{id} \in group^w[g_{id}]$}{ \label{alg:env:checkwpermis}
			\ForAll {users $u \in group^r[g_{id}]$}{ \label{alg:env:first}
				$u_{sk} \gets keys[u]$ \;
				$(c_k, t) \gets AE({u_{sk}},k)$ \;
				$envelope \gets envelope \cup \{(c_k, t)\}$\label{alg:env:last}
			}
		}
		\Return $envelope$
	}
\end{algorithm}
\end{minipage}
\vfill \null
\columnbreak
\begin{minipage}{0.5\textwidth}
\begin{algorithm}[H]
	\caption{\asky: Key enveloping with efficient decryption.}
	\label{alg:key_envelop_eff}
	\DontPrintSemicolon
	\Input{	$u_{id}$: writing user identity \\
		$g_{id}$: group identifier \\
		$k$: file access key}
	\Output{ $envelope$: ciphertext of the access key}
	\SetKwFunction{KeyEnveloping}{KeyEnveloping}
	\Procedure{\KeyEnveloping{$u_{id}$, $g_{id}$, $k$}}{
		$envelope \gets \emptyset$ \;
		\If {$u_{id} \in group^w[g_{id}]$}{
			$nonce \gets Random$ \; \label{alg:asky:nonce}
			\ForAll {users $u \in group^r[g_{id}]$} {
				$u_{sk} \gets keys[u]$ \;
				$l_u \gets \mathcal{H}(u_{sk}\ ||\ nonce)$ \; \label{alg:asky:label}
				$(c_k, t) \gets AE({u_{sk}},k)$ \;
				$envelope \gets envelope \cup \{(l_u, c_k, t)\}$\;
			}
			$\mathtt{Sort}(envelope)$\tcp*{by label $l_u$}\label{alg:asky:sortlabels}
		}
		$envelope \leftarrow nonce~||~envelope$\;
		\Return {$envelope$} \; \label{alg:asky:nonceenv}
	}
\end{algorithm}
\end{minipage}
\end{multicols}
\end{figure}

As shown in \Cref{alg:key_envelop}, given the identity of the writing user $u_{id}$, the group unique identifier $g_{id}$, and the file access key $k$, the algorithm produces a ciphertext called \emph{envelope}.
The \accessmonitor first checks if the user $u_{id}$ has writing permission for the group $g_{id}$ (line \ref{alg:env:checkwpermis}).
If it is the case, the \emph{envelope} is built by the union of ciphertexts and authentication tags resulted from encrypting the access key $k$ using the secret key $u_{sk}$ of each group member $u$ (lines \ref{alg:env:first}-\ref{alg:env:last}).

We establish the following operations and notations for the Algorithms:
\begin{itemize}
	\item $E(k,p) \to c $ and $D(k,c) \to p$ are symmetric encryption and decryption algorithms (\eg, \ac{AES}-\ac{CTR}) using the key $k$, where $p$ is plaintext and $c$ is ciphertext;
	
	\item $AE(k,p) \to (c, t)$ and $AD(k,c,t)\to \{p, \bot\}$  are authenticated encryption and decryption algorithms (\eg, \ac{AES}-\ac{GCM}). Besides encrypting, it also produces an authentication tag $t$ that ensures the integrity of the ciphertext $c$ under the key $k$. During decryption, in case the integrity check fails, no plaintext is generated ($\bot$). 
	
	\item $S(PK^{-1},p) \to \sigma$ and $V(PK,p,\sigma) \to \{true, false\}$ are digital signature and verification algorithms (\eg, \acs{RSA}-based or \acs{ECDSA}) employing an asymmetric public/private key pair ($PK$ and $PK^{-1}$).
	
	\item $\mathcal{H}$ is a unidirectional cryptographic hash function, \eg, \ac{SHA}; and
	
	\item $||$  is the literal concatenation operation.
\end{itemize}

As in traditional \ac{ANOBE} schemes~\cite{barth:2006:privacy,libert:2012:abe}, we propose a method that reduces user decryption times by circumventing the need to perform $O(n)$ key decryption trials (line \ref{alg:asky:read:trials} of \Cref{alg:user_read}) by trading it off for an increase in key enveloping times and in the envelope size (these trade-offs are evaluated later).
To this end, publicly known labels are appended to each user fragment in the envelope and used as its index.
Such fragments are lexicographically sorted by labels and can be hence located in logarithmic time with respect to the group size (\eg, binary search).
Consequently, users retrieve their keys performing one single decryption of the located fragment.

Labels are computed as the \acs{SHA}-224 hash of the user's secret shared key (also known by the \accessmonitor service) salted with a \emph{nonce}. This adds \SI{28}{\byte} to the envelope for each group member.
Our approach is more efficient in comparison to traditional ones that perform modular exponentiations~\cite{barth:2006:privacy} or tag-based encryption~\cite{libert:2012:abe}. 
\Cref{alg:key_envelop_eff} details the efficient variant of key enveloping with the creation of labels  as the salted hash of the user secret key (line \ref{alg:asky:label}) and the fragments sorting (line \ref{alg:asky:sortlabels}). 
The random \emph{nonce} to be used as salt is generated for each envelope (line \ref{alg:asky:nonce}) and publicly attached to it (line \ref{alg:asky:nonceenv}).

For implementation purposes, we split the \accessmonitor service into two components.
The first, developed in C++ for a total of \num{3000} \acp{LoC}, constitutes the entry-point for service requests.
The other holds users and groups metadata within a replicated database.
For this purpose, we use a cluster of MongoDB~\cite{mongodb} servers with three replicas, since it offers out-of-the-box scale-out support and is well suited to store denormalized documents.
In order to perform queries against it from the first sub-component, we ported the official MongoDB client library~\cite{mongoc} to run inside an enclave.
To secure the data stored in the MongoDB back-end, each replica of the entry-point sub-component is provisioned with a master key $M_k$ at attestation time.

\begin{figure}
	\centering
	\includegraphics{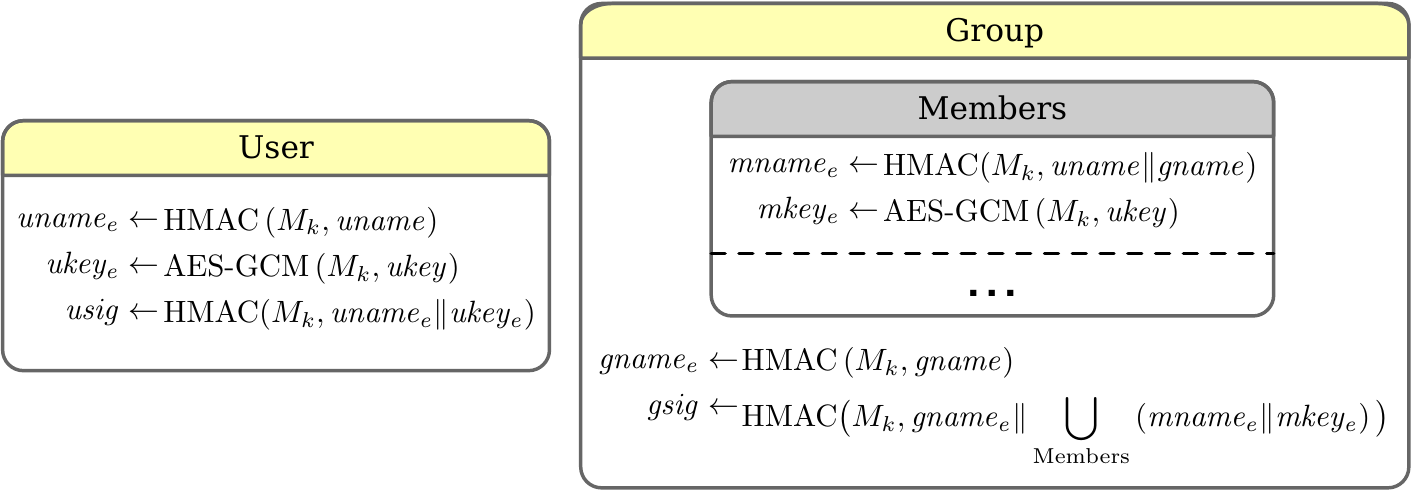}
	\caption[\asky data model.]{\label{fig:mongo-storage}Data model of user and group \emph{documents} stored in MongoDB.}
\end{figure}

Since the storage back-end runs outside of enclaves, we make sure that every piece of stored data is either hashed using the \acl{HMAC} \acs{HMAC}-\acs{SHA}256 or encrypted using \ac{AES}-\ac{GCM}.
We thus guarantee that providers that host MongoDB instances cannot infer any information about users or groups (apart from the size of each group, which is already leaked in the \emph{envelopes}).
\Cref{fig:mongo-storage} shows how we organize data in MongoDB: one collection of users and other of groups.
Users are stored once in the users collection and again in each group of which they are member.
This de-normalisation prevents the service provider from inferring to which groups a user belongs, since the attributes of a given user are hashed or encrypted differently for each representation (\ie, we use the name of the group as salt when hashing, and different \acp{IV} when encrypting).
To ensure integrity, each document is signed using \ac{HMAC}.

\subsubsection{\writerproxy}

The \writerproxy serves the purposes of
\begin{enumerate*}[label=\emph{(\roman*)}]
	\item protecting cloud storage credentials;
	\item signing the ciphertext content;
	\item hiding user identities by proxying their requests to the cloud storage; and
	\item generating access tokens to the cloud storage to improve performance.
\end{enumerate*}
When forwarding user requests to write files, the \writerproxy checks with the \accessmonitor whether the query comes from a user who has the correct permissions.

User requests, including file contents, cross over the enclave boundary.
This obviously slows down transmission rates because of content re-encryptions and trusted/untrusted edge transitions.
To tackle that, we have also implemented a variant where temporary access tokens are given to users.
Such tokens allow them to upload content without crossing through the \ac{TEE}.
The ciphertext digest, however, still needs to be authenticated by the signing key available in the \writerproxy.
In such case, users are responsible for using appropriate proxies that can conceal the request's origin.
Also, communication with the cloud storage must happen through encrypted connections.
Even if the data files are encrypted, metadata can leak group information along the network path.

We modelled the cloud storage component using Minio~\cite{minio}, a distributed object store that is fully compatible with the \acp{API} of Amazon \ac{S3}.
As we need to perform operations in the cloud storage from enclaves, we ported the Java version of the Minio client library to C++ so that it can run together with the \writerproxy.
These modifications account for \num{4000}~\acp{LoC} of C++.
Without accounting for external libraries, the \writerproxy consists of \num{800} \acp{LoC}.

As the \writerproxy is the sole service possessing the write credentials for the cloud provider, it constitutes a necessary hop for uploading the file: either for relaying the upload operation to the cloud storage (\Cref{alg:write}) or for obtaining the token and signature of the file content.
The proxying routine first verifies that the invoking user has write capabilities for the desired group (line \ref{alg:proxy:permit}).
If positive, the content is authenticated (line \ref{alg:proxy:sign}) by using long term credentials ${PK_{TA}^{-1}}$ provisioned during attestation.
The file ciphertext and the corresponding signature are then uploaded to the cloud (line \ref{alg:proxy:up}).

\setlength{\columnsep}{-3mm}
\begin{figure}
\begin{multicols}{2}
\null \vfill
\begin{minipage}{0.44\textwidth}
\begin{algorithm}[H]
	\caption{\label{alg:write}\asky: \writerproxy proxy file.}
	\Input{ $f$: file reference\\
		$u_{id}$: writing user identity\\
		$g_{id}$: group identifier\\
		$\mathcal{C}$: file ciphertext\\
		$\mathcal{A}$: \accessmonitor instance\\
	}
	\DontPrintSemicolon
	\SetKwFunction{ProxyFile}{ProxyFile}
	\Procedure{\ProxyFile{$f$, $u_{id}$, $g_{id}$, $\mathcal{C}$, $\mathcal{A}$}}{
	\If {$u_{id} \in \mathcal{A}.group^w[g_{id}]$}{ \label{alg:proxy:permit}
		$\sigma \gets S({PK_{TA}^{-1}},\mathcal{C})$ \; \label{alg:proxy:sign}
		$\mathtt{UploadToCloud}(f, \mathcal{C}, \sigma)$ \; \label{alg:proxy:up}
	}
	}
\end{algorithm}
\end{minipage}
\vfill \null
\columnbreak
\begin{minipage}{0.5\textwidth}
\begin{algorithm}[H]
	\caption{\asky: User write file to group.}
	\label{alg:user_write}
	\Input{ $f$: file reference\\
		$u_{id}$: writing user identity \\
		$g_{id}$: group identifier \\
		$P$: file plaintext \\
		$\mathcal{A}$: \accessmonitor instance \\
		$\mathcal{W}$: \writerproxy instance \\
	}
	\SetKwFunction{WriteToGroup}{WriteToGroup}
	\DontPrintSemicolon
	\Procedure{\WriteToGroup{$f$, $u_{id}$, $g_{id}$, $P$, $\mathcal{A}$, $\mathcal{W}$}}{
		$fk \gets Random$\tcp*[r]{file access key}\label{alg:asky:writeakey}
		$envelope \gets \mathcal{A}.KeyEnveloping(u_{id}, g_{id}, fk)$\\\tcp*[r]{\ie, Alg.~\ref{alg:key_envelop} or Alg.~\ref{alg:key_envelop_eff}}\label{alg:asky:wenv}
		$cipher \gets E(fk,P)$\;\label{alg:asky:wenc}
		$\mathcal{C} \gets envelope\ ||\ cipher$\;\label{alg:asky:wcat}
		$\mathcal{W}.ProxyFile(f, u_{id},g_{id},\mathcal{C},\mathcal{A})$\\\tcp*[r]{\ie, Alg.~\ref{alg:write}}\label{alg:asky:wup}
	}
\end{algorithm}
\end{minipage}
\end{multicols}
\end{figure}

\vspace{-1em}
\subsubsection{Users}

Users 
\begin{enumerate*}[label=\emph{(\roman*)}]
	\item get their secret key from the \accessmonitor;
	\item write shared content; and
	\item read content shared with them.
\end{enumerate*}
The write operation is shown in \Cref{alg:user_write}.
The writing user first randomly creates a file access key $fk$ (line \ref{alg:asky:writeakey}) and asks the \accessmonitor service $\mathcal{A}$ to envelope this key  (line \ref{alg:asky:wenv}) for the group $g_{id}$, so that it can be anonymously deciphered by any member of it.
He then encrypts the file using $fk$ (line \ref{alg:asky:wenc}) and concatenates the two ciphertexts, key envelope and encrypted file (line \ref{alg:asky:wcat}), before transmitting the result to the \writerproxy, so that it is uploaded to the cloud storage (line \ref{alg:asky:wup}).

\asky satisfies the \emph{lazy} revocation model~\cite{backes:2006:lazy}, where a revoked user can continue accessing data created prior to revocation but not beyond that.
Additionally, past data becomes inaccessible upon the first succeeding write to the same resource.
The revocation is triggered by an administrator removing the user's id from the $group^r$ and $group^w$ access lists.
Later, when new content is published in that group, a new random key is derived for encrypting the content and a new envelope is attached to it (\Cref{alg:user_write}, lines \ref{alg:asky:writeakey}-\ref{alg:asky:wenv}).
The revoked user's key will not be included in the envelope, and therefore he will be unable to access the new group key along with the newly published content.

Users read files according to \Cref{alg:user_read}.
First, they download the ciphertext package from the cloud storage (line \ref{alg:asky:rdown}), that can be validated by the signature check (line \ref{alg:asky:rscheck}). 
In case the signature is valid, the user then splits the package between the key envelope and the file ciphertext (line \ref{alg:asky:rsplit}).
Next, the user iterates over user fragments in the envelope and tries to decrypt each of them with his secret key $u_{sk}$ (lines \ref{alg:asky:read:trials}-\ref{alg:asky:rdkey}).
If successful, the obtained file access key can be used to symmetrically decrypt the file ciphertext (lines \ref{alg:asky:rtest}-\ref{alg:asky:rdec}).

In case the indexed envelopes are used, the user read operation (\Cref{alg:user_read_eff}) requires the label reconstruction (line \ref{alg:asky:rlab}) followed by a binary search for the corresponding envelope fragment (line \ref{alg:asky:rbsearch}).
Once located, the file access key is retrieved and the shared file is deciphered (lines \ref{alg:asky:redeenv}-\ref{alg:asky:refdec}).

As part of our prototype implementation, we developed a full-featured client in \num{1200} \acp{LoC} of Kotlin.
The client can be set up to operate in all possible configurations of \asky: keys in linear or indexed envelopes, \emph{writes} through the \writerproxy, or through a standard proxy onto a Minio or Amazon \ac{S3} storage back-end with short-lived token-based authentication.
Kotlin's full interoperability with the Java ecosystem allows us to easily integrate with the \ac{JMH}~\cite{jmh} and \ac{YCSB}~\cite{cooper:2010:ycsb} frameworks that we use to perform the evaluation of \asky.

\setlength{\columnsep}{1mm}
\begin{figure}
\begin{multicols}{2}
\null \vfill
\begin{minipage}{0.43\textwidth}
\begin{algorithm}[H]
	\caption{\label{alg:user_read}\asky: User read file.}
	\DontPrintSemicolon
	\Input{$f$: file reference\\
		$u_{sk}$: user secret key}
	\Output{$P$: plaintext file content}
	\SetKwFunction{ReadFile}{ReadFile}
	\Procedure{\ReadFile{$f$, $u_{sk}$}}{
	$(\mathcal{C}, \sigma) \leftarrow \mathtt{DownloadFromCloud}(f)$ \; \label{alg:asky:rdown}
	\If{$V({PK_{TA}},\mathcal{C}, \sigma) = \mathit{true}$} { \label{alg:asky:rscheck}
		$envelope$, $cipher \gets split(\mathcal{C})$ \; \label{alg:asky:rsplit}
		\ForAll {pairs $(k_c, t)$ in $envelope$}{ \label{alg:asky:read:trials}
			$fk \gets AD({u_{sk}},k_c, t)$ \; \label{alg:asky:rdkey}
			\If {$fk \neq \bot$} { \label{alg:asky:rtest}
				$P \gets D({fk},cipher)$ \; \label{alg:asky:rdec}
				\Return $P$ \;
			}
		}
	}
	\Return $\bot$\;
	}
\end{algorithm}
\end{minipage}
\vfill \null
\columnbreak
\begin{minipage}{0.5\textwidth}
\begin{algorithm}[H]
	\caption{\label{alg:user_read_eff}\asky: User read file with efficient decryption.}
	\DontPrintSemicolon
	\Input{$f$: file reference\\
		   $u_{sk}$: user secret key}
	\Output{$P$: plaintext file content}
	\SetKwFunction{ReadFile}{ReadFile}
	\Procedure{\ReadFile{$f$, $u_{sk}$}}{
	$(\mathcal{C}, \sigma) \leftarrow \mathtt{DownloadFromCloud}(f)$ \;
	\If{$V({PK_{TA}},\mathcal{C}, \sigma) = \mathit{true}$} {
		$nonce$, $envelope$, $cipher \gets split(\mathcal{C})$ \;
		$l_u \gets \mathcal{H} (u_{sk}\ ||\ nonce)$ \; \label{alg:asky:rlab}
		$(k_c, t) \gets \mathtt{BinarySearch}(l_u, envelope)$\; \label{alg:asky:rbsearch}
		\If{$(k_c, t) \neq \bot$} {
			$fk \gets AD({u_{sk}},k_c, t)$ \; \label{alg:asky:redeenv}
			$P \gets D({fk},cipher)$ \; \label{alg:asky:refdec}
			\Return $P$ \;
		}
	}
	\Return $\bot$\;
	}
\end{algorithm}
\end{minipage}
\end{multicols}
\end{figure}

\subsubsection{Evaluation}

We evaluate the performance and scalability of our solution by first conducting micro-benchmarks.
Then, we use the well-known \ac{YCSB}~\cite{cooper:2010:ycsb} test suite to evaluate the overall system performance.
All our experiments run on a cluster of 5 \acs{SGX}-enabled Dell PowerEdge R330 servers, each having an Intel Xeon E3-1270\,v6 processor and \SI{64}{\gibi\byte} of memory.
Additionally, we use 3 Dell PowerEdge R630 dual-socket servers, each equipped with 2 Intel Xeon E5-2683\,v4 \acs{CPU}s and \SI{128}{\gibi\byte} of \acs{RAM}.
One of the latter machines is split in 3 \acp{VM} to handle the roles of Kubernetes master, Minio server and benchmarking client (when a second client is needed).
\ac{SGX} machines use the latest available microcode revision \texttt{0x8e}, and have the Hyper-threading feature disabled to mitigate the Foreshadow security flaw~\cite{vanbulck:2018:foreshadow}.

Communication between machines is handled by a Gigabit Ethernet network.
All our components can be independently replicated to provide availability, fault tolerance or cope with the load.
Therefore, we have packaged our micro-services as individual containers, which we then orchestrate using an \ac{SGX}-aware adaptation of Kubernetes~\cite{vaucher:2018:sched}.
When error bars are shown, they represent the \SI{95}{\percent} confidence interval.

\vspace{-1em}
\subsubsection{Cryptographic scheme}

We first isolate and measure the performance of the underlying cryptographic primitive of \asky and compare it to the \ac{ANOBE} scheme defined by Barth \etal~\cite{barth:2006:privacy} (\bbw).
Our implementation of \bbw uses an elliptic curve integrated encryption scheme as the \ac{IND-CCA}2 public key cryptosystem.
Both key materials (\ie, keys, curve) are chosen to meet \SI{256}{\bit} of \emph{equivalent security strength}~\cite{barker2007nist}.
We also implement the efficient decryption of \bbw as suggested in the paper by relying on the \ac{DH} problem hardness, however in the context of \ac{ECDH}.

\begin{table}
\centering
\caption[Throughput of asky and \bbw.]{\label{tab:compare} Throughput comparison between \asky cryptographic scheme and \bbw.}
\sisetup{table-format = 1.1e1, table-align-exponent = true}
{\setlength{\tabcolsep}{1.6mm}
	\begin{tabular}{lSSSS[table-format = > 1]}
		\toprule
		& {Enveloping} & {De-enveloping} & {Enveloping} & {De-enveloping} \\
		& {[\si{\groupsize\per\second}]} & {[\si{\groupsize\per\second}]} & {Efficient [\si{\groupsize\per\second}]} & {Efficient [\si{\micro\second}]} \\
		\midrule
		\bbw & 3.3e2 & 5e3 & 3e2 & <4 \\
		\asky & 1.9e6 & 2.5e6 & 1.2e6 & <4 \\
		\midrule
		Ratio & 5.8e3 & 5e2 & 4e3 & {\emph{n/a}} \\
		\bottomrule
\end{tabular}}\\
\si{\groupsize\per\second}: group members per second
\end{table}

Table~\ref{tab:compare} shows throughputs for cryptographic key enveloping and de-enveloping considering that user keys are available. Apart from the experiment with efficient de-enveloping (where only one decryption is performed), units are in number of group members handled per second.
While \bbw can create envelopes at the rate of \num{330} members per second, \asky sustain rates 3 orders of magnitude bigger.
Likewise, performance is improved by 2 orders of magnitude during de-enveloping operation.
Such considerable difference is due to the performance gap between asymmetric and symmetric encryption primitives.

\bbw provides an \emph{efficient decryption} mode that achieves less than \SI{4}{\micro\second} for the largest group size. However, this comes at the cost of a lower throughput during enveloping of 300 members per second, or \num{90}\% of the standard version.
\asky is able to support the same decryption speed with indexed envelopes. In contrast, we achieve the rate of 1.2 million members per second when building envelopes (\num{63}\% of standard version), three orders of magnitude faster than \bbw.
Furthermore, \asky produces ciphertexts of \SI{60}{\byte} and \SI{88}{\byte} per member for standard and efficient modes, respectively, whereas \bbw produces \SI{126}{\byte} and \SI{154}{\byte} for the equivalent methods.

\begin{figure}[tb]
\center
\includegraphics{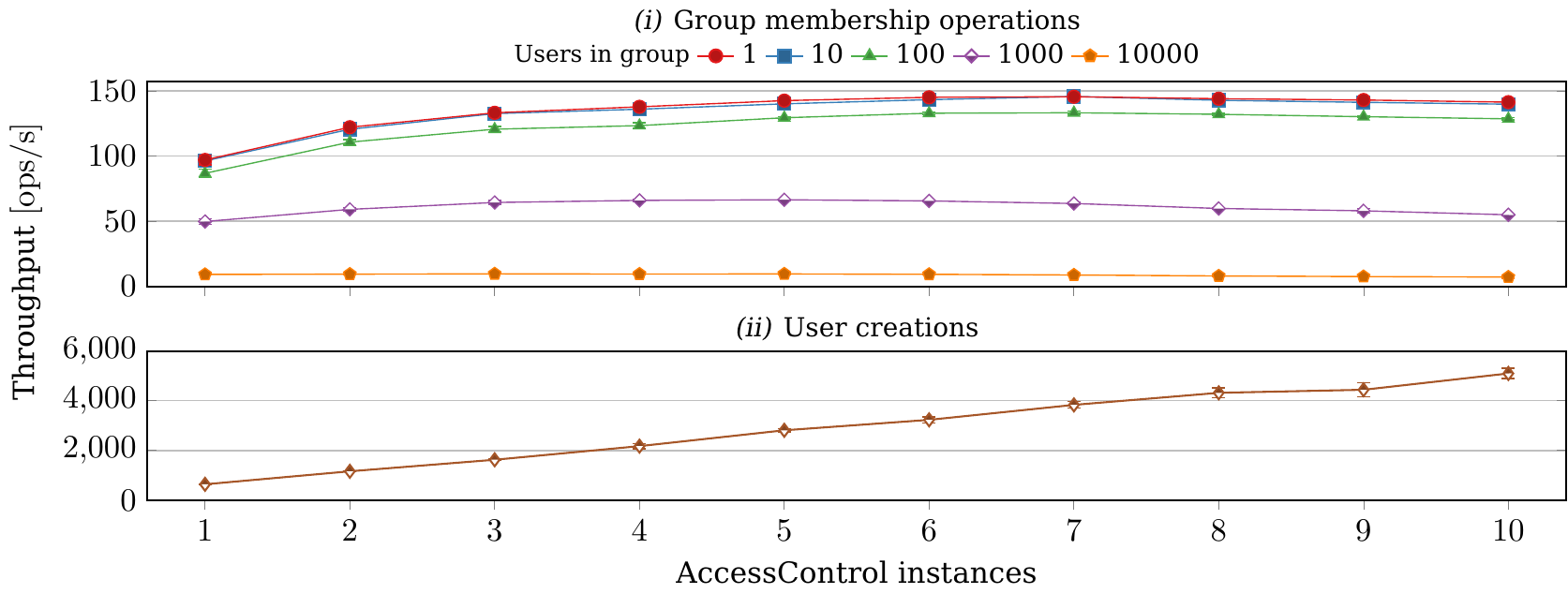}
\vskip 2mm
 \caption[Throughput for administrative actions.]{Throughput for administrative actions in the \accessmonitor: \emph{(i)}~adding or revoking users to/from groups of various sizes, and \emph{(ii)}~creating users.}
\label{fig:asky:admin}
\end{figure}

\subsubsection{Scalability}

In order to evaluate the throughput of operations performed by administrators when varying the number of \accessmonitor instances, we delegate to Kubernetes the distribution of requests among such instances, where a \emph{service} is exposed.
\Cref{fig:asky:admin} shows the results.
The scalability of adding users to a group or revoking their access rights is limited, as these operations require one \iac{RMW} cycle to check and update the signature of the group \emph{document}.
The larger the group, the more time the operation takes as each signature encompasses every user within the group, \eg, we reach an average of only \SI{8.8}{ops/s} for groups with \num{10000} users and \SI{135}{ops/s} for groups containing a single user.
This effect could be mitigated by, \eg, batching together multiple operations on a given group.
Diversely, the operation that creates users scales linearly with the number of \accessmonitor instances, allowing more than \num{5000} user creations per second with \num{10} replicas.

\begin{figure}[tb]
\centering
\includegraphics{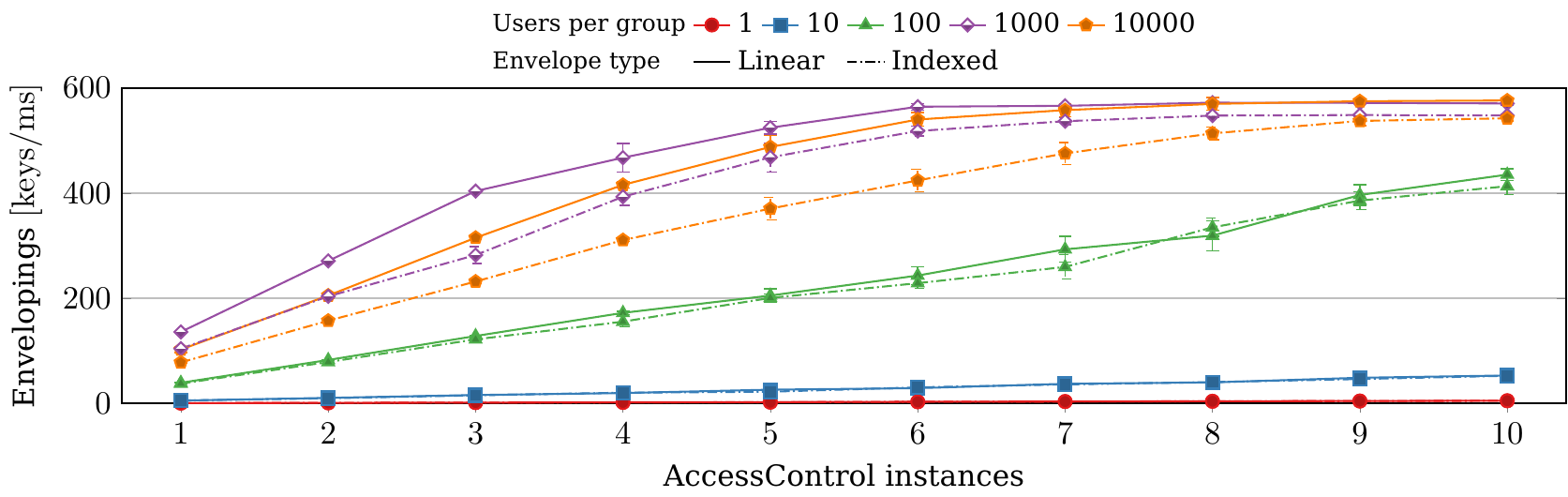}
 \caption[\asky enveloping throughput.]{\asky: Throughput of enveloping a message for groups of various sizes with varying instances of the \accessmonitor micro-service.}
\label{fig:envelope-tput}
\end{figure}

Next, we look at the number of keys that can be included in an envelope per unit of time, also when varying the number of instances of the \accessmonitor service.
A close-to-linear trend can be observed according to number of instances in \Cref{fig:envelope-tput}, showing that this operation also benefits from horizontal scalability.
For groups of \numrange{1000}{10000} members, the throughput ceases to increase with more than \num{7} instances as the MongoDB backend becomes a bottleneck. 
In smaller groups, the performance is diminished due to the overhead associated with each request (\eg, network connection, REST formatting, enclave transitions), although it benefits from more \accessmonitor instances.
Additionally, we ran the same experiment with the indexing feature turned on.
For groups of \num{10000} users, the throughput is reduced by \SIrange{6}{26}{\percent}, having a marginal impact on smaller groups where the performance mostly depends on fixed costs.

\begin{figure}[tb]
\centering
\includegraphics{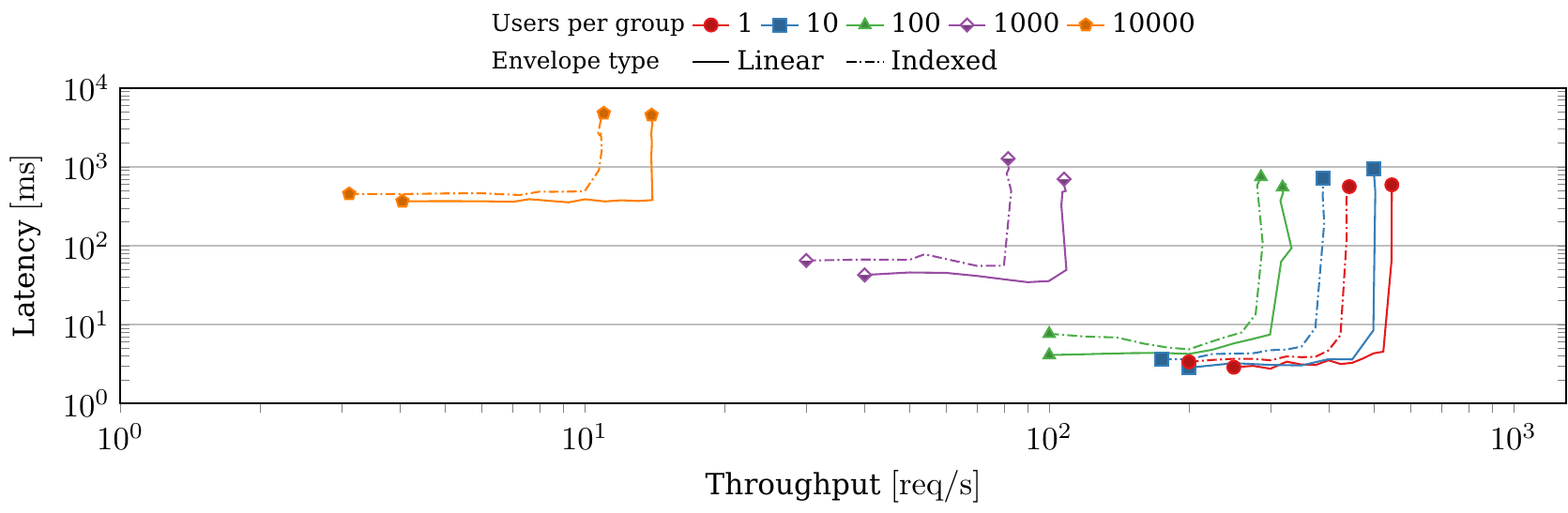}
 \caption[\asky: Throughput \emph{vs.} latency.]{\asky: Throughput \emph{vs.} latency of enveloping a message for groups of various sizes.}
\label{fig:envelope-tput-lat}
\end{figure}

We also evaluated the latency of the enveloping operation by increasing the throughput until saturation, again with indexing turned on and off.
Looking at \Cref{fig:envelope-tput-lat}, we notice that for groups which are larger than \num{100} users, latency increases linearly according to the group size, while the saturation throughput decreases linearly.

\begin{figure}[tb]
\centering
\includegraphics{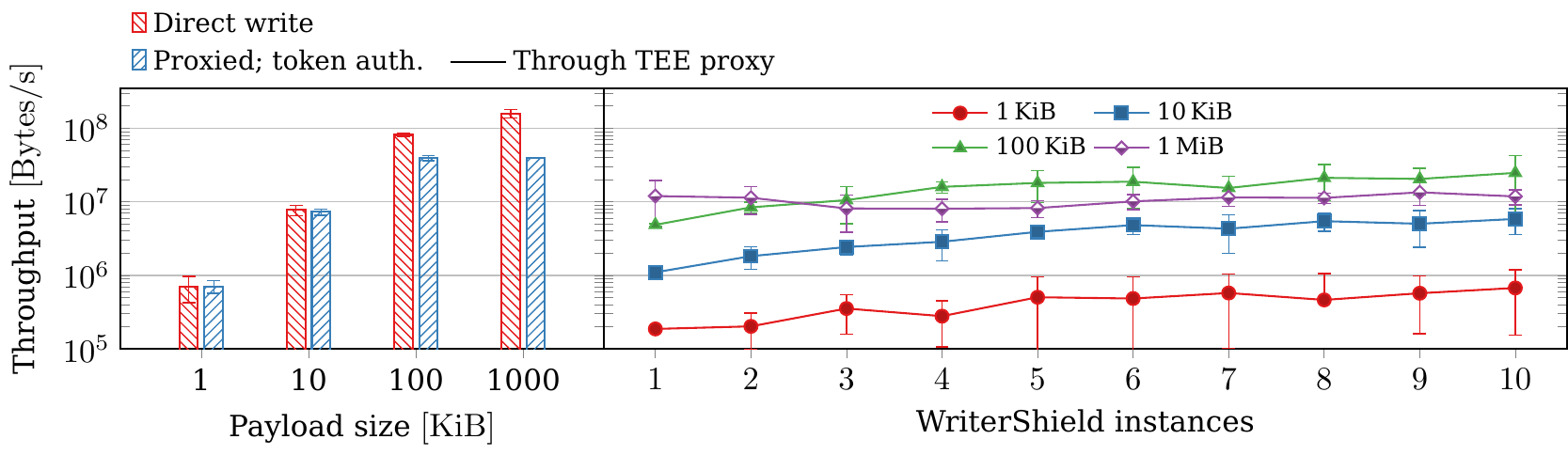}
\vskip 5mm
 \caption[\asky: Writing throughput to cloud storage.]{\asky: Throughput of writing data to the cloud storage in different ways: directly (baseline), through a TCP proxy using a temporary token for authentication, and through varying number of in-enclave \writerproxy instances.}
\label{fig:writer-proxy}
\end{figure}

To evaluate the performance of the \writerproxy, we conduct two experiments.
In the first, data written to the cloud is proxied through the \ac{TEE}, while in the other the \writerproxy is solely used for generating temporary cloud storage access tokens, with write operations being proxied through a NGINX server in \acs{TCP} reverse-proxy mode.
In order to establish a baseline, we also wrote directly to the cloud storage service, without intermediaries.
Results are shown in \Cref{fig:writer-proxy}.
The bar plot on the left-hand side shows that for files of \SI{1}{\kilo\byte} and \SI{10}{\kilo\byte} the difference in performance is negligible, whereas bigger files cause more performance degradation when using the token feature.
When the \writerproxy is used to forward data instead (right-hand side), we see that the throughput increases with the number of service instances until it plateaus at about the same values as with the tokenized variant.
For files of \SI{1}{\mega\byte}, adding \writerproxy instances renders no benefit.
This effect happens due to the saturation of enclave resources acting as a TLS bridge between clients and the cloud storage server.
Overall, using tokens would be the most efficient approach, although in this case the client would be responsible for using adequate proxies in order to hide its identity from the cloud storage.

\subsubsection{Macro-benchmarks}
\label{sec:evaluation:macro}

To observe the behaviour of \asky under specific conditions of data serving systems, we use \ac{YCSB}~\cite{cooper:2010:ycsb} workloads \emph{A} (update heavy\footnote{nomenclature used by \ac{YCSB}: \url{https://github.com/brianfrankcooper/YCSB/wiki/Core-Workloads}} 50/50), \emph{B} (read heavy 95/5) and \emph{C} (read only). Additionally, we include a write-only workload in our tests.
As our system is not capable of direct-access writes, updates are replaced by \ac{RMW} operations.
We implemented an interface layer to link the benchmarking tool to \asky and ran each workload with 3 different file sizes: \SI{1}{\kibi\byte}, \SI{100}{\kibi\byte} and \SI{1}{\mebi\byte}.
We simulate \num{100000} operations across 64 concurrent users and report the achieved throughput of user operations.
Additionally, we activate a concurrent instance of \ac{YCSB} that simulates 8 administrators doing group membership operations at three different rates: \num{0}, \num{50} and \SI{100}{\operation\per\second}.
The administrative operations are equally distributed between user additions and revocations, so that the user database size stays more-or-less constant.

\begin{figure}
	\centering
\includegraphics{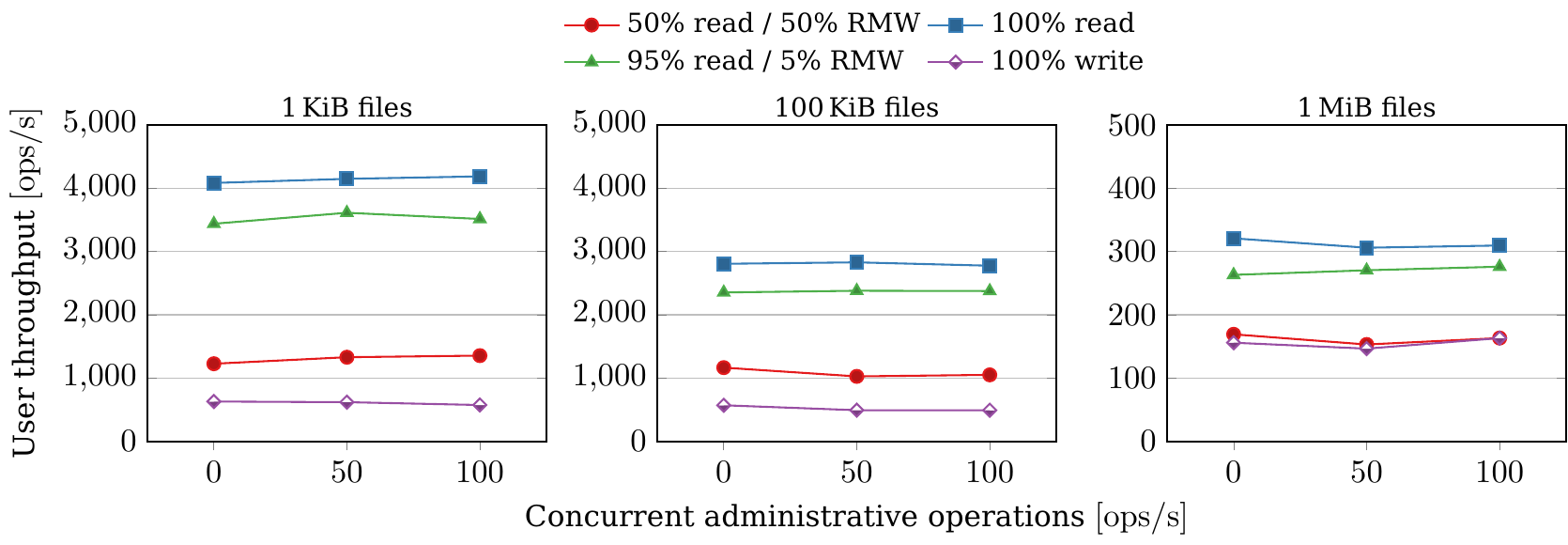}
\vskip 5mm
 	\caption[\asky: User throughput.]{\label{fig:macrobenchmark}\asky: User throughput observed by our YCSB-based macro-benchmark, with various file access patterns, varying file sizes, and additioning simultaneous administrative operations.}
\end{figure}

Results are shown in \Cref{fig:macrobenchmark}.
Overall, we notice that user throughputs are not influenced by concurrent administrative operations, as each type of operation involves separate components of our architecture.
For files of \SI{1}{\kibi\byte}, an increasing proportion of writes causes a performance degradation from \SI{4100}{\operation\per\second} (read-only) down to \SI{628}{\operation\per\second} (write-only).
For larger files (\SI{1}{\mebi\byte}), the difference is less dramatic, with throughputs going from \SI{320}{\operation\per\second} to \SI{155}{\operation\per\second}.
Therefore, fixed costs are dominant when writing small files (\eg, enveloping the file key), but are increasingly amortized for larger ones.
We can also observe that the throughput in \si{\byte\per\second} (\ie, multiplying the result in \si{\operation\per\second} to the file size) is largely superior for larger files, as we have already noticed in \Cref{fig:writer-proxy}.
In a nutshell, we retain that the end-user experience offered by \asky is not influenced by concurrent administrative operations, and that the overhead of the additional operations required for writing become smaller for larger files.
 
\section{Summary}

Diverging from intensive data processing protection within \acp{TEE}, this chapter rather focused on the generation and management of shared keys, which are obviously very sensitive information.
More specifically, we were interested in schemes that allow users to share content in a secure manner. Besides guaranteeing data confidentiality and integrity, we also enforced access control, \ie, being sure that only rightful users are able to read or modify shared content.
The trusted environment brings new design perspectives regarding cryptographic protocols since shielded computation may operate on computationally simpler yet secure encryption constructs. 
We took advantage of this property by proposing and evaluating two original systems.

\ibbesgx is a cryptographic access control extension that derives cuts in the computational complexity of \ac{IBBE} thanks to secure enclaves.
Such simplification came from the fact that the \ac{IBBE} encryption operation took place in a \ac{TEE} and could therefore use a master key that otherwise could not be available in untrusted environment. In such case, a much more complex operation involving a large public key would take place.
Turning \ac{IBBE} into a viable solution brought along its advantages regarding little metadata generation and the ability to dismiss the usage of \iac{PKI}.

Since we assumed that a variety of client devices should be able to access \ibbesgx, possibly with no \ac{TEE} available, the same encryption simplification could not be done for decryption. 
To mitigate the high performance cost, we proposed a group partitioning mechanism such that the complexity was bound to a fixed constant partition size rather than the size of the whole group.
We conducted experiments with real and synthetic benchmarks, demonstrating that \ibbesgx is efficient both in terms of computation and storage, even when processing large and dynamic workloads of membership operations.
Apart from producing group metadata six orders of magnitude shorter than the traditional \ac{HE}, which makes it more efficient both in terms of storage and bandwidth, our construction also performed membership changes at rates one order of magnitude faster than \ac{HE}. At the same time, administrators cannot infer user keys and snoop into sensitive shared content.

Next, we introduced \asky, an end-to-end system that brings the additional guarantee of anonymity among group members.
\asky leverages a \ac{TEE} intermediary exclusively for the content sharing operation, while users can consume the shared content by reading it directly from cloud storage providers.
We incorporated the cryptographic construction into a scalable system design that leverages micro-services that can possibly scale according to the undergoing access control and data sharing workloads.
Results indicated that our cryptographic scheme is faster than state-of-the-art \ac{ANOBE} schemes by \num{3} orders of magnitude and can serve groups of \num{10000} users with a throughput of \num{100000} key derivations per second per service instance.

\chapter{\label{chap:privacy}Privacy enforcement}
\acresetall

Until now, we described the design of different kinds of services in order to protect data owned by their clients, whether they be at rest or being processed.
These techniques are useful when one has access to the development stages of a system.
However, sometimes users must rely on a service provided by third parties, who in turn may not be able or willing to change it.
In such cases, privacy concerns may arise.
Despite possible legal agreements, there is no fail-proof way of being sure that service providers do not stealthily track user habits or preferences while accordingly servicing their requests.
The reasons for doing so can range from commercial (\eg, targeted advertisements) to espionage (\eg, by governments). 
To prevent that, we identify in this chapter ways of using \acp{TEE} for privacy assurance.

We chose the domain of web search as it is the most prominent instance of such scenario. In~\Cref{sec:xsearch}, we describe a centralised proxy that obfuscates queries in order to make it hard for search engines to keep accurate user profiles.
As any centralised approach, it suffers from the fact of simultaneously being a bottleneck and a single point of failure.
In response to such disadvantage, we propose in~\Cref{sec:cyclosa} a decentralised \ac{P2P} solution that is better in terms of performance and fault tolerance and optimal in terms of accuracy.

\section{Private web search}

Web search is the most widely used online service, with billions of queries sent on a daily basis to Google alone.
Indeed, search engines have become an essential service for finding content on the Internet.
By regularly querying these services, though, users disclose large amounts of personal data (\eg, \cite{aol-leak-scary-1,aol-leak-scary-2}).
Queries are generally stored by search engines to analyse user behaviour and to personalize responses according to profiles inferred from past queries~\cite{langville:2011:google,hannak:2013:perswebsearch}.
Additionally, the economic model of many online services heavily relies on personalized advertising~\cite{yang:2010:ads}.
Numerous studies point, however, that the collection of search queries opens a number of privacy threats as they possibly disclose sensitive information about individuals (\eg, age, gender, religion, political preferences, sexual orientation)~\cite{castelluccia:2010:privateweb}.

To limit personal information disclosure, solutions enabling users to query search engines in a privacy-preserving manner have been proposed.
They can be classified in categories, which enforce:
\begin{enumerate}
\item \textit{unlinkability} between users and their queries.

This is done by hiding identities of users through anonymous communication (\eg, \ac{Tor}~\cite{dingledine:2004:tor}, Dissent~\cite{corrigan:2010:dissent,wolinsky:2012:dissent} and RAC~\cite{mokhtar:2013:rac}).
Systems in such category are limited for two reasons:
\begin{itemize}
	\item they typically suffer from poor performance because of the heavy cryptographic mechanisms on which they rely;
	\item despite ensuring the requester's anonymity, it has been shown~\cite{peddinti:2014:web} that the actual content of search queries may be sufficient to link them back to the originating users' profiles by running \emph{re-identification attacks}~\cite{petit:2016:simattack}.
\end{itemize}

\item \emph{indistinguishability} between user profiles and queries.

To that end, they obfuscate user profiles in such a way that the search engine cannot distinguish between users' real interests and fake ones (\eg, Track me not~\cite{howe:2009:trackmenot}, GooPIR~\cite{domingo:2009:goopir}).
These approaches generally operate by sending fake queries on behalf of the user.

It has been shown~\cite{petit:2016:simattack}, however, that search engines may easily distinguish fake from real traffic due to external resources used for generating fake queries (\eg, \ac{RSS} feeds, dictionaries).

\item a combination between unlinkability and indistinguishability.

The only existing solution of which we are aware, PEAS~\cite{petit:2015:peas}, assumes a weak adversarial model of non-colluding proxy servers.

\item \emph{\acl{PIR}} (\acs{PIR}) (\eg, \cite{pang:2010:similarity,lindell:2010:private}).

These approaches build specialized search engines based on cryptographic techniques enabling to answer a user request without having access to its content.
These techniques are, however, still unpractical due to their limited performance. Response times can reach seconds for very large data stores~\cite{aguilar:2016:xpir}, which is the case of search engines.

\end{enumerate}

Based on these considerations, it appears that in order to fully support privacy-preserving Web search one must provide a protocol that
\begin{enumerate*}[label=\emph{(\roman*)}]
\item enables the protection of requesters' identities in a realistic adversarial model; and
\item provides effective indistinguishability with realistic fake queries, \ie, difficult to distinguish from real ones.
\end{enumerate*}

\vskip 4mm
\subsection{\xsearch}
\label{sec:xsearch}

\xsearch \cite{mokhtar:2017:xsearch} consists of a proxy that enables Internet users to access Web search engines in a privacy-preserving manner.
Instead of submitting queries directly to the search engine, a user sends them through an encrypted channel to the \xsearch proxy, where the connection endpoint lies inside \iac{SGX} enclave.
Once in there, queries are decrypted and manipulated in plain-text form before being forwarded to the search provider.
Such manipulation involves the generation of an obfuscated query by aggregating $k$ random past queries and the original one using the logical OR operator. As a consequence, the search engine is not able to distinguish which one is the original query.
Due to the obfuscation, results returned by the search engine are mixed.
The \xsearch proxy therefore filters them and forwards only the results related to the initial query back to the user.

The \xsearch protocol involves three entities:
\begin{enumerate}
\item the client, whose code and platform are trusted;
\item the \xsearch proxy, which runs on cloud platforms and may behave in a Byzantine manner~\cite{lamport:1982:byzantine}, \ie, they are subject to failures, bugs or malicious behaviour; and
\item the search engine, which is honest-but-curious~\cite{goldreich:2003:cryptoprotocols}, \ie, it correctly behaves when fetching answers to requests, while possibly collecting and exploiting the information they receive from clients. We also assume that search engines may run re-identification attacks (\eg, \cite{gervais:2014:quantifyingprivacy}) in order to associate the received request to a known user profile. 
Besides, they may collude with proxy nodes (\eg, \ac{Tor} relays or \xsearch proxies) in order to learn more about users.
\end{enumerate}

\vskip 4mm
\subsubsection{Protocol overview}

\begin{figure}[tb]
\centering
\includegraphics{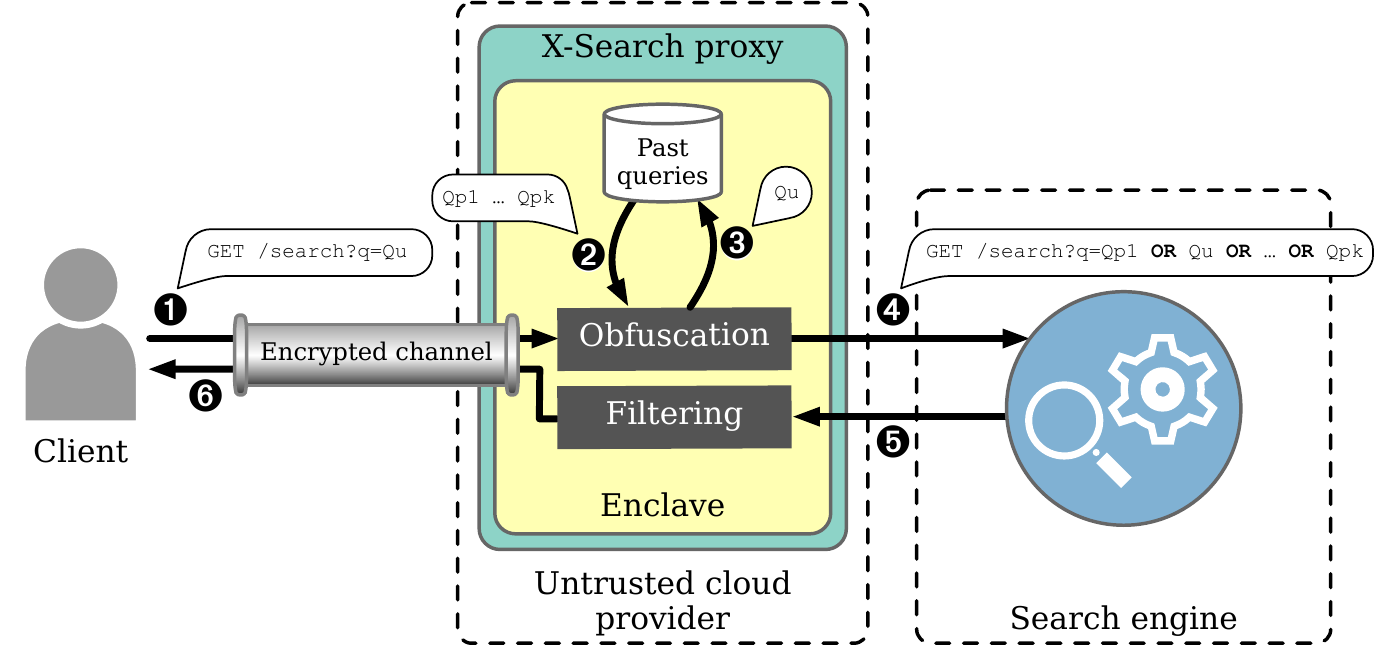}
\caption{\label{fig:xsearch:arch}\xsearch architecture and execution flow.}
\end{figure}

\xsearch combines two complementary schemes: unlinkability and indistinguishability. The former hides the identity of requesting users; the latter their queries.
\Cref{fig:xsearch:arch} depicts the architecture and execution flow of \xsearch.
The user interacts with the search engine through a \xsearch proxy hosted on untrusted public cloud services and deployed on physical \ac{SGX}-enabled nodes.
As the proxy intermediates the contact between search engine and users, it hides their identities (\ie, IP addresses). It is also in charge of obfuscating user queries and filtering the results returned from the search engine before forwarding them back to requesting users.

Initially, the user sends the query $Q_u$ to the \xsearch proxy (\ding{202}).
The proxy then generates a new obfuscated query by retrieving $k$ random past queries $Q_{p1},...,Q_{pk}$ (\ding{203}), which are aggregated to the original $Q_u$ in a random order with the logical OR operator.
Next, the proxy stores the initial query in the table of past queries (\ding{204}) and sends the  obfuscated one to the search engine (\ding{205}).
Unlike other indistinguishability protocols, \xsearch reuses past queries as fake queries.
Since they were sent by real users, they are effectively indistinguishable from the current user's real request.
This is possible because past queries are securely stored inside the \ac{TEE} with no correlation to the identity of their originating issuers. This database is therefore shielded against exploitation by malicious agents.

As the obfuscated query can alter the information returned by the search engine (\ding{206}) by mixing results for the original and fake queries, the proxy node performs a filtering step.
It consists of removing all the results which are not associated to the original query.
Finally, the remaining results are returned to the user (\ding{207}).

The \xsearch proxy does not maintain individual profile structures associated to each user.
Instead, it only updates a table containing the past queries.
Moreover, the user sends queries to the proxy through an encrypted tunnel terminated inside the \ac{SGX} enclave.
Consequently, the protection of the original query is ensured from the client until inside the \ac{TEE} of the proxy node.
From the proxy to the search engine, the original query is protected thanks to the obfuscation mechanism.

\subsubsection{Implementation details}

The search unlinkability provided by \xsearch relies on a query broker that runs within the client's domain.
It consists of a local daemon process executing alongside the client's Web browser and is in charge of the \ac{SGX} attestation step (see \Cref{sec:sgx:attestation}).
When users issue a Web search query, their Web clients first connect to this local broker, which encrypts the request and forwards the cipher to an \xsearch node hosted in an untrusted cloud provider.
The \xsearch proxy decrypts the cipher, generates the obfuscated query, and securely stores the original one in the \ac{SGX} reserved memory.
When the search engine sends back the response to the \xsearch node, relevant results are extracted and delivered backwards to the broker.
Finally, the broker decrypts the results and delivers it to the Web client.

To enforce indistinguishability, \xsearch relies on an obfuscation mechanism that hides user queries among multiple fake ones (\Cref{algo:generationObuscatedQuery}), so that an adversary cannot distinguish them.
More precisely, the original query is aggregated with $k$ randomly chosen fake queries connected by logical OR operators (lines \ref{alg:xsearch:addfakebeg}--\ref{alg:xsearch:addfakeend}).
Once the obfuscated query is generated, the initial query is stored in the history (line \ref{alg:xsearch:updateH}).
These fake queries come from the table of past requests maintained in the private memory of the \xsearch proxy ($H$).
Due to the \ac{EPC} limitation of \ac{SGX} enclaves (see \Cref{sec:sgx:perfissues}), we limit the size of $H$, so that a sliding window of the most recent queries are exploited.
Since they are real, each sub-query of the obfuscated request can potentially be mapped to an existing user profile by an adversary conducting re-identification attacks.
The attacker would thereby introduce some noise in their source, which would make it less accurate and consequently hinder re-identification attempts.

\setlength{\columnsep}{-1mm}
\begin{figure}
\fontsize{8}{9.6}
\begin{multicols}{2}
\begin{minipage}{0.46\textwidth}
\begin{algorithm}[H]
    \Input{$Q_u$: initial query \\
        $H=\lbrace Q_0,...,Q_m\rbrace$: query history \\
        $k$: number of fake queries\\}
    \Output{$Q_{obf}$: obfuscated query}
    $Q_{obf} \leftarrow \emptyset$\;\label{alg:xsearch:addfakebeg}
    $\mathit{index} \leftarrow$ $\mathtt{random}(k+1)$ \; \label{algo:fqg:random}
    \While{$\mathtt{sizeof}(Q_{obf})$ $<= k$\label{algo:fqg:loop}}{
        \If{$\mathit{index}=0$}{
            $Q_{obf} \leftarrow \mathtt{OR}(Q_{obf},Q_u)$ \; \label{algo:fqg:pushRealOne}
        }
        \Else{
            $Q_{obf} \leftarrow \mathtt{OR}(Q_{obf},H[\mathtt{random}(m)])$\; \label{algo:fqg:pushPastOne}
        }
        $\mathit{index} \leftarrow \mathit{index}-1$\;\label{alg:xsearch:addfakeend}
    }
    $H \leftarrow H \cup Q_u$ \;\label{alg:xsearch:updateH}
    \Return $Q_{obf}$\;
    \caption{\xsearch: Query obfuscation.}
    \label{algo:generationObuscatedQuery}
\end{algorithm}
\end{minipage}
\columnbreak
\null \vfill
\begin{minipage}{0.5\textwidth}
\begin{algorithm}[H]
	\SetKwInOut{Input}{input}
	\Input{$Q_u$: initial query, \\
		$\mathit{fakes} = \{Q_{p1}, \dots, Q_{pk}\}$: aggregated queries, \\
		$R$: set of results for $Q_u \lor Q_{p1} \lor \dots \lor Q_{pk}$.
	}
	\SetKwInOut{Output}{output}
	\Output{$\bar{R}$: filtered results}
	\DontPrintSemicolon
	$\bar{R} \leftarrow \emptyset$ \;
	$q^+ \leftarrow Q_u \cup fakes$ \;
	\For{$r \in R$}{\label{alg:xsearch:iterr}
		$s_{max} \leftarrow -\infty$ \;
		\For{$q_i \in q^+$}{ \label{alg:xsearch:qiter}
			$\mathit{score}[q_i] \leftarrow \mathtt{nbCommonWords}(q_i,r)$\; \label{alg:xsearch:score}
			$s_{max} \leftarrow \max(s_{max},\mathit{score}[q_i])$ \;
		}
		\If{$\mathit{score}[Q_u] =  s_{max}$}{ \label{alg:xsearch:maxscore}
			$\bar{R} \leftarrow \bar{R} \cup \{r\}$ \; \label{alg:xsearch:match}
		}
	}
	\Return $\bar{R}$ \;
	\caption{\xsearch: Results filtering.}
	\label{algo:filtering}
\end{algorithm}
\end{minipage}
\vfill \null
\end{multicols}
\end{figure}

In spite of that, the obfuscation mechanism has an impact on the results returned by the search engine.
This is due to the fact that they contain a mix of answers corresponding to~$(k+1)$ queries (\ie, $k$ fake queries and the real one).
The \xsearch proxy takes this into account by filtering the returned results, so that those which are not related to the initial query are removed (\Cref{algo:filtering}).
For each response $r$ from the result set (line \ref{alg:xsearch:iterr}), the algorithm determines whether it corresponds to the initial query based on a \emph{similarity score}
assigned to each query $q_i$ that composed the obfuscated request $q^+$ (line \ref{alg:xsearch:qiter}).
Such score is calculated as the amount of common words ($\mathtt{nbCommonWords}$) between the query $q_i$ and the result $r$ under evaluation (title and description---line \ref{alg:xsearch:score}).
A result $r$ is considered related to the initial query, and hence forwarded to the user, if the original query $Q_u$ has the largest score $s_{max}$ (lines \ref{alg:xsearch:maxscore}--\ref{alg:xsearch:match}).

To evaluate our system, we implemented in C++ a fully-functioning prototype based on Intel \ac{SGX} \ac{SDK} (v1.8) libraries and tools~\cite{intel:2019:sgxsdk}.
In order to avoid costly trusted/untrusted mode transitions, we limit the enclave interface to essential operations that deal with sensitive information (2 \acp{ecall} and 4 \acp{ocall}).
Concerning \ac{EPC} memory consumption, an excessive amount could potentially be caused by the management of the past queries inside the enclave's protected memory. We evaluate this aspect later on.

\vskip 6mm
\subsubsection{Experimental setup and metrics}
\vskip 3mm

To assess \xsearch, we use a real world Web search dataset from \ac{AOL} query logs~\cite{pass:2006:picture}.
It contains approximately 21 million queries, issued by 650,000 unique users between March and May 2006.
Our evaluation takes into account the 100 most active users (as in~\cite{petit:2015:peas}), since they are the most exposed to an adversary that aims at discovering their identities.
This is simply because they disclose more information to search engines, and therefore allow for meaningful profile histories.
In order to build these user profiles, we split the dataset in training (two thirds of queries) and testing sets (one third).

We compare the robustness and quality of \xsearch against two systems: \tor~\cite{dingledine:2004:tor} and \peas~\cite{petit:2015:peas}; and a baseline solution, where users directly send their queries to the search engine without any protection.
\tor provides unlinkability by leveraging several layers of encryption to hide user identities.
\peas, in turn, combines unlinkability---by relying on a trusted proxy--- and indistinguishability--- by obfuscating the original request with fake queries in random order.
These fake queries are generated from the graph of co-occurrence between terms in the user history.

We assess \xsearch along three dimensions: privacy (\ie, the protection of users’ data), accuracy (\ie, the results' quality), and performance (\ie, \xsearch's efficiency in terms of throughput, latency and memory usage).
To evaluate privacy, we leverage SimAttack~\cite{petit:2016:simattack} a re-identification attack that outperforms previous approaches.
To run it, we assume the attacker holds a set of user profiles built at the learning stage.
Next, \xsearch intermediates the requests (testing dataset) between users and search engine. 
For each obfuscated query the attacker receives, it tries to identify both the requesting user and the original query among fake ones.

SimAttack is based on a similarity metric $s(q, P_u)$ that corresponds to the proximity between a query $q$ and a user profile $P_u$.
Such profile is assumed to have been built by the adversary before users started using any privacy protection.
In our evaluation, $P_u$ contains queries that belong to the training set and were issued by a user $u$.
The metric accounts for the cosine similarity of $q$ to all queries belonging to the user profile $P_u$ and returns the exponential smoothing of these similarities.
We empirically set the smoothing factor to $0.5$ as this provided better re-identification rates.
Considering one obfuscated request $q^+$, the function $s(q, P_u)$ is applied for every pair composed of the sub-query $q \in q^+$ and user profiles. 
A successful match happens when a single pair $\langle q, P_u \rangle$ achieves a score beyond a given threshold.
Otherwise, the attack is considered as failed for the obfuscated query $q^+$ under scrutiny.

Once we obtain all successful matches, we assess privacy by establishing a metric called re-identification rate, which is defined as follow:

\begin{align}
\label{eq:reidentification}
\operatorname{\mathit{re-identification~rate}} = \frac{|Q_{id}|}{|Q|}
\end{align}

where $Q_{id}$ is the set of correctly re-identified queries: both the initial query and the associated user; while $Q$ is the set of original queries sent by users.
This metric is defined between $[0,1]$ where $0$ represents the best solution (\ie, no re-identification) and $1$ represents the worse solution (\ie, all queries are re-identified).

Since the obfuscation mechanism impacts the results returned by search engines, we evaluate the capacity of \xsearch to filter responses not related to the initial query, \ie, the results' accuracy.
This is done by comparing the search engine's reply (first 20 results) for a non-obfuscated query and the \xsearch's filtered results.
For each value of $k$ (\ie, the number of fake queries), we pick a random subset of the testing workload composed of 100 queries.
Our experiments use the Bing search engine\footnote{queries are directed to \texttt{http://www.bing.com/search=q?}}.
At the time of our work, Bing supported only single words with the $\mathtt{OR}$ operator, albeit our dataset comprises multi-word real queries. 
To circumvent that, we simulated the answer to an obfuscated query  by merging the result sets of each individual request.

The accuracy evaluation consists in comparing the lists of results associated to the original query $R_{or}$ and the ones returned by \xsearch $R_{xs}$, after obfuscation and filtering.
We consider precision (\ie, correctness) and recall (\ie, completeness) as defined below. Both metrics are within the interval $[0,1]$. The best accuracy is provided with precision and recall at $1$.

\begin{align} 
\label{eq:xsearch:accuracy}
precision = \frac{|R_{or} \cap R_{xs} |}{|R_{xs}|}  & &
recall = \frac{|R_{or} \cap R_{xs} |}{|R_{or}|} 
\end{align}

To evaluate \xsearch performance, we consider throughput, memory usage and round-trip time.
The throughput (requests/second) allows the assessment of \xsearch's scalability, since it indicates its capability to operate under adequate response times even with a growing number of requesting users.
Memory usage measurement, in turn, allows for verifying when the \ac{EPC} would become saturated, and therefore incur in larger overheads.
Finally, we measured response times considering the complete chain, including the search engine delays, so that we could compare our proposal to alternative approaches.

\subsubsection{Evaluation}
\label{sec:evaluation}

\begin{figure}[tb]
\centering
\includegraphics{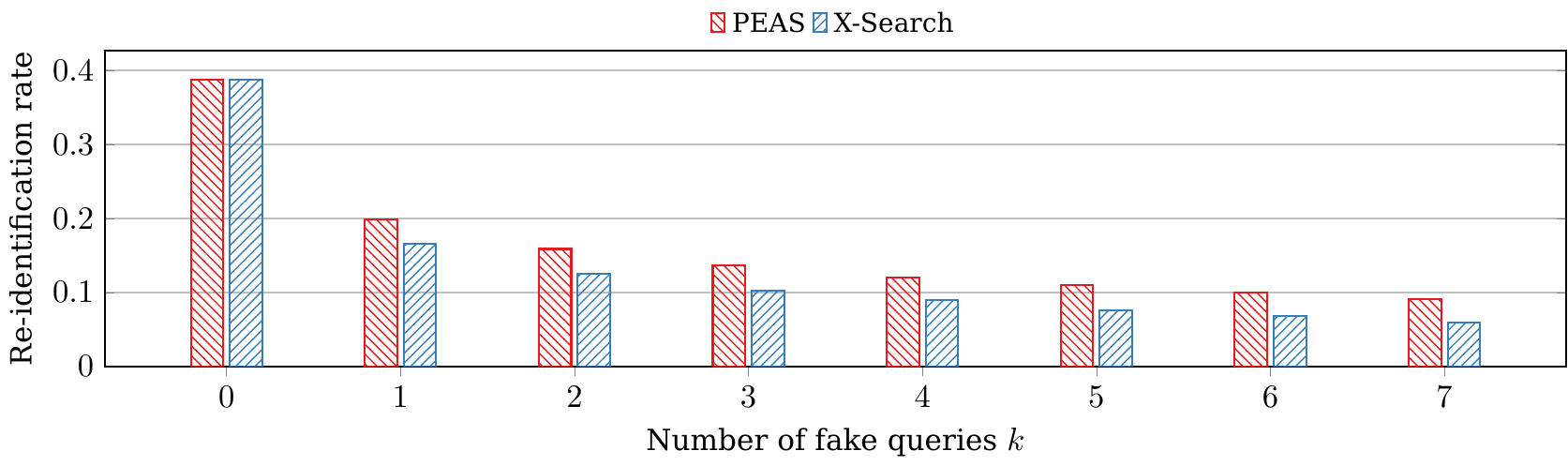}
\vskip 5mm
 \caption[Re-identification rates for \xsearch and PEAS.]{\label{fig:precision-privacy}\xsearch reduces the number of de-anonymized queries compared to PEAS.}
\end{figure}

\xsearch was deployed on a machine with Intel{\textregistered} Core\texttrademark~i7-6700 processor~\cite{intel:i7_6700} and \SI{8}{\gibi\byte} \ac{RAM} running on \textsc{Ubuntu} 14.04.1 \ac{LTS} (kernel 4.2.0-42-generic).
First, we evaluate the capacity of \xsearch to preserve user privacy and to improve user protection when compared to \peas.
To this end, we measure re-identification rates when leveraging SimAttack, as previously described.
\Cref{fig:precision-privacy} shows the results for \peas and \xsearch when varying the number of fake requests $k$.
When $k=0$, the evaluated systems enforce only unlinkability (\eg, \tor).
In such case, \ie, without query obfuscation, the adversary is able to re-associate almost $40\%$ of the queries to their originating user by making use of the profiles established with training data.
This confirms that unlikability alone is not enough to effectively protect users against re-identification attacks.

Adding only one fake query drops this re-identification rate to \SI{16}{\%} for \xsearch and almost \SI{20}{\%} for \peas.
This difference comes from the generation process of fake queries.
Using real past queries makes \xsearch more robust against re-identification attacks since all sub-queries can be mapped to past queries of other users, which creates confusion from the attacker's perspective.
Contrarily, generating fake queries on the basis of co-occurrence of terms does not ensure \peas to create the same kind of disorientation.
The re-identification rate decreases according to the augmentation of $k$ and \xsearch provides better protection than \peas in all scenarios. Such improvement varies from \SI{23}{\%} for $k=1$ to \SI{35}{\%} for $k=7$.

\begin{figure}[tb]
\centering
\includegraphics{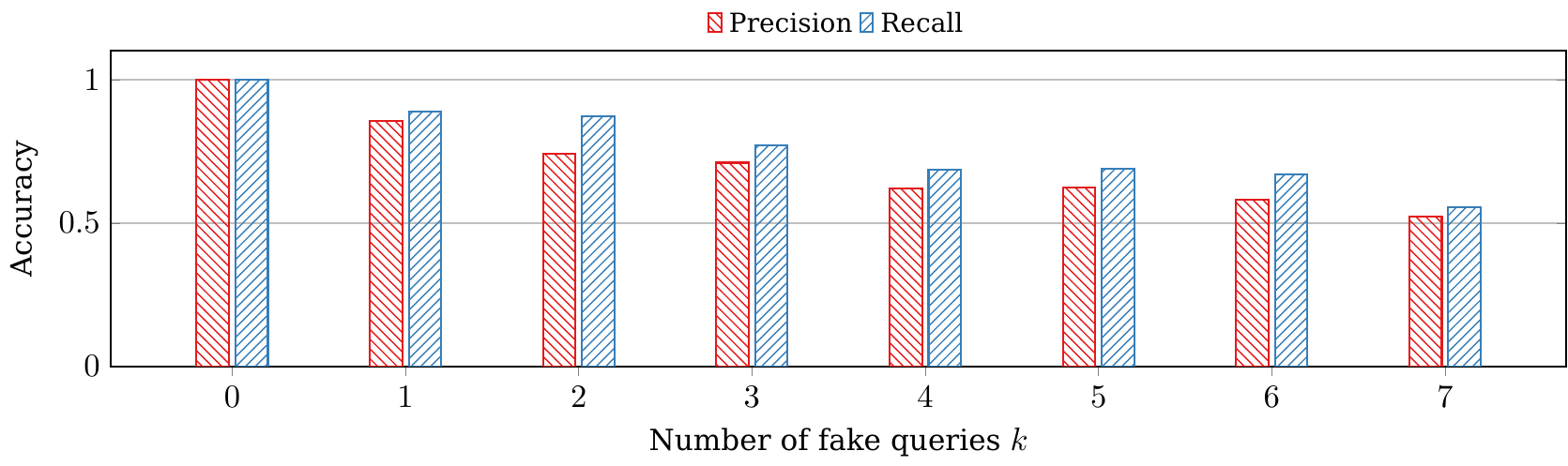}
\vskip 5mm
 \caption[\xsearch's accuracy.]{\label{fig:accuracy}Results returned by \xsearch are close to results associated to the original query.}
\end{figure}

\xsearch's accuracy is measured by quantifying its ability to remove results related to fake queries and keeping those resulted from the original query.
\Cref{fig:accuracy} depicts precision and recall according to raising values for $k$.
As expected, both decrease as $k$ increases, \ie, more noise in the input causes more incorrect and incomplete outputs.
Nevertheless, results returned to users are fairly accurate.
For instance, the recall value for $k=2$ is \SI{87}{\%}, \ie, only \SI{13}{\%} of the results delivered by \xsearch are not part of the set returned by the search engine if the original query was sent directly.
For the same value of $k$, precision is equal to \SI{74}{\%}, meaning that \SI{26}{\%} of the results returned to users are noise introduced by fake queries.
These numbers confirm that \xsearch preserves the quality of results.

\Wrt system performance, we begin by observing the throughput$~\times~$latency relation of the \xsearch proxy.
By iteratively increasing the rate at which requests are performed, we measure the latency to handle each request until it becomes too high.
To do so, we use the wrk2~\cite{wrk2} workload generator without actually hitting the web search engine, so that we understand the saturation point of the \ac{SGX}-based proxy in isolation.
We compare \xsearch against \tor and \peas in similar conditions.
From the usability perspective, unlike \peas and \tor which require custom clients to forge messages that follow their protocol, \xsearch can be used with third-party clients issuing regular \ac{HTTP} requests, such as \texttt{wget} or \texttt{curl}.

\begin{figure}[tb]
\centering
\includegraphics{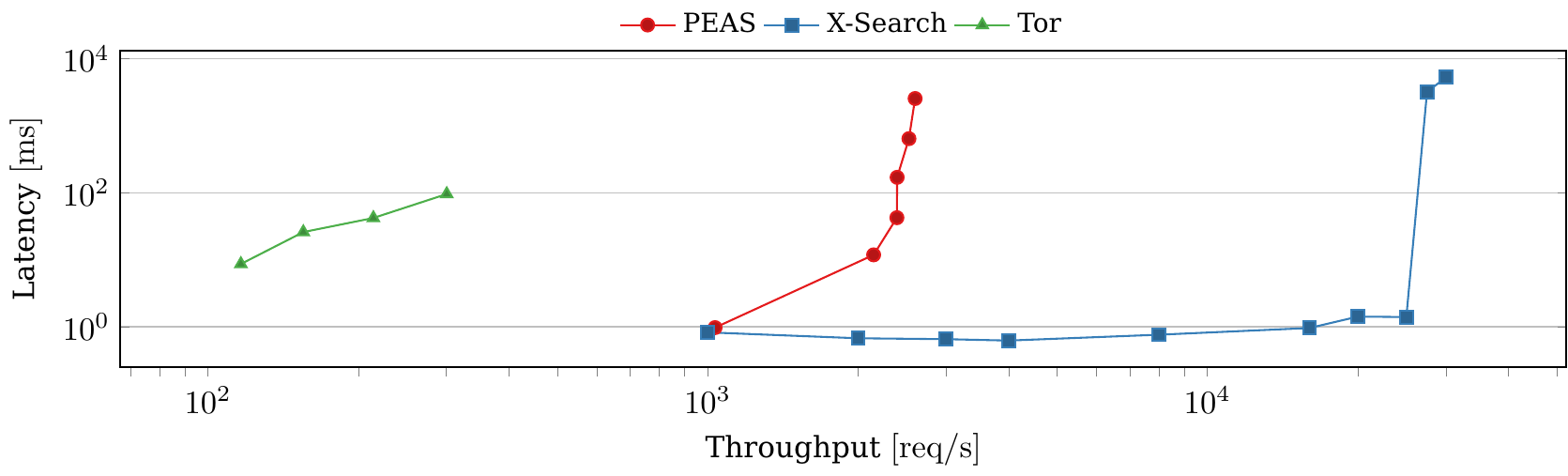}
\vskip 5mm
 \caption[Latency vs. throughput for \xsearch and others.]{\label{fig:xsearch:tputlat}Latency vs. throughput comparison for \xsearch proxy, \peas and Tor.}
\end{figure}

\Cref{fig:xsearch:tputlat} presents these results.
We observe that \xsearch scales well, being capable of serving up to \SI{25000}{\requests\per\second} with sub-second latencies.
Differently, \peas deteriorates much faster, with only \SI{1000}{\requests\per\second} being served with sub-second latency.
In our experiments, \tor performs very poorly: handling as few as \SI{100}{\requests\per\second} at an average reply latency of \SI{8.86}{\second}.
This result confirms our implementation to be fast and scalable.

\begin{figure}[tb]
\centering
\includegraphics{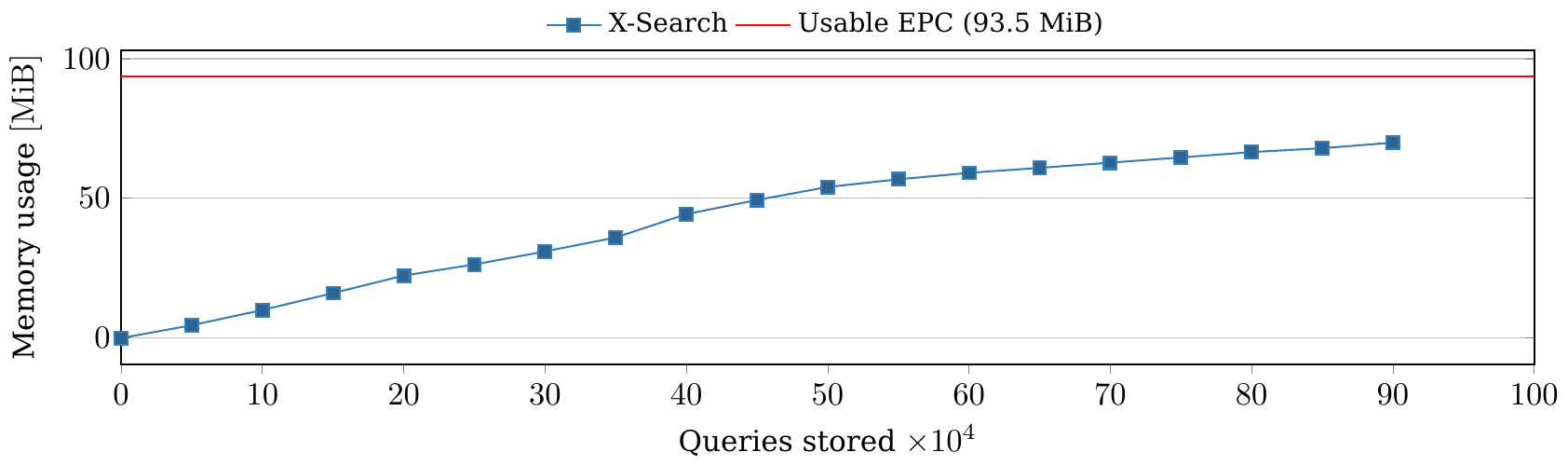}
\vskip 5mm
 \caption[\xsearch memory usage.]{\label{fig:xsearch:mem}\xsearch memory usage: The memory allowed to a single enclave can fit more than 1M queries before hitting the \ac{SGX} \ac{EPC} usable memory limit.}
\end{figure}

Next, we investigate how much memory is required by the obfuscation scheme.
For this experiment, we used a larger dataset.
Instead of considering only the \num{100} most active users, we use all the \num{6} millions unique queries available in the \ac{AOL} dataset.
To trace and profile the heap memory, we leverage Valgrind's Massif~\cite{seward:2008:valgrind}.
\Cref{fig:xsearch:mem} presents the results.
Observing the trend of the \xsearch curve, it is clear that the \ac{EPC} size is largely sufficient to store at least \num{1} million queries, a number that can support with ease the obfuscation mechanism.

\begin{figure}[tb]
\centering
\includegraphics{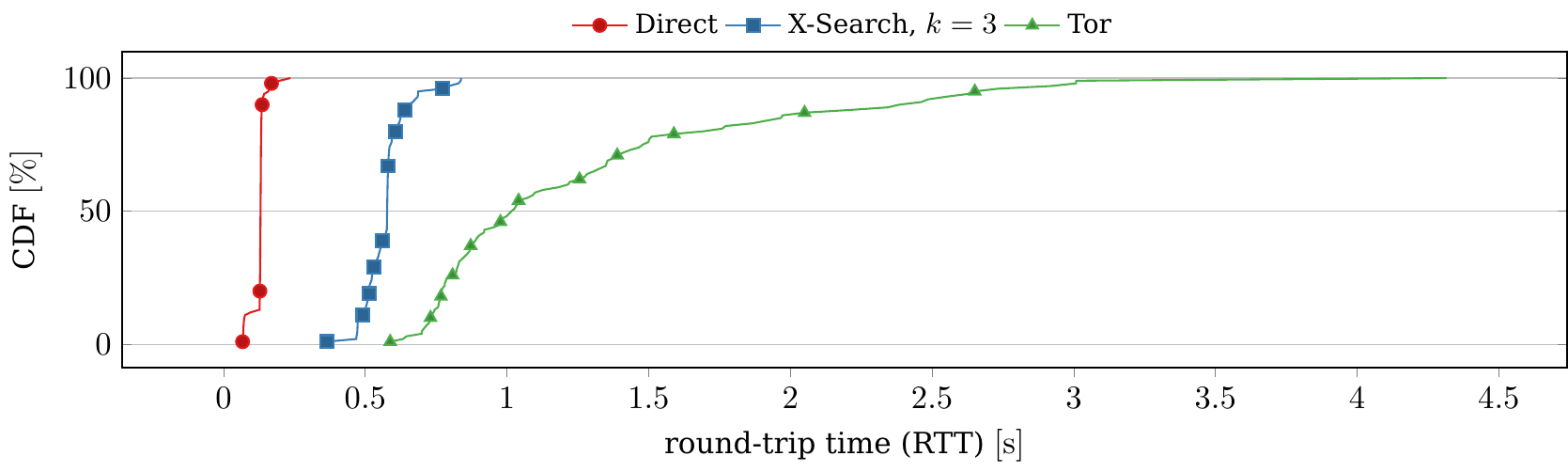}
\vskip 5mm
 \caption[Round-trip time for \xsearch and others.]{\label{fig:searchrtt}User-perceived web search round-trip time for 100 queries with \xsearch, over the \ac{Tor} network and directly contacting the web search engine.}
\end{figure}

Finally, we evaluate the end-to-end latency of a Web search query from submission to results' reception.
Due to limiting policies adopted by Bing's search engine, we only issue 100 queries,  picked at random in the \ac{AOL} dataset. Our experiments were done in May 2017, and repeated for weeks in order to mitigate possible occasional disturbances in the service provider.
We compare the observed round-trip time for three scenarios: 
\begin{enumerate*}[label=\emph{(\roman*)}]
\item client directly contacts the search engine with no privacy guarantees;
\item queries are routed through \tor network; and
\item using \xsearch.
\end{enumerate*}
\Cref{fig:searchrtt} presents the results as a \ac{CDF} of measured end-to-end latencies.
We observe that \xsearch allows for much faster replies in comparison to \tor.
\xsearch's median response time is \SI{0.58}{\second}, and the $99^{th}$ percentile is \SI{0.87}{\second}.
\tor, on the other hand, renders way worse results from the user perspective: the round-trip median time using onion routers was \SI{1.06}{\second}, while the $99^{th}$ percentile reached \SI{3}{\second}.
The \tor network largely exceeds well-known usability margins~\cite{palmer:2002:websitemetrics}, while \xsearch offers a usable yet secure browsing experience.

To summarize, we presented \xsearch, a novel architecture to allow privacy-preserving Web searches that exploits Intel \ac{SGX} to operate under stronger adversarial models than existing systems in literature.
We contribute with a novel query obfuscation mechanism and the implementation choices of our full prototype.
The assessment is made from three perspectives: privacy, accuracy, and performance.
We analytically show that \xsearch offers more vigorous privacy guarantees than its competitors as it operates under a stronger adversarial model.
Furthermore, we experimentally demonstrate using a real dataset that \xsearch is more resilient to a state-of-the-art re-identification attack than \peas by \SI{30}{\%} in average.
From the accuracy perspective, we show that the negative impact of \xsearch's obfuscation scheme remains limited.
Lastly, from the performance perspective, we show that \xsearch outperforms previous privacy-preserving systems both in terms of latency and throughput.

\section{TEE in the client-side}

Although \xsearch (see \Cref{sec:xsearch}) advances private Web search in some aspects, centralised approaches like itself and \peas~\cite{petit:2015:peas} are not easily scalable and constitute a single point of attack and failure.
For instance, they may be easily blocked by search engines that have anti-bot policies.
Besides, results filtering due to query obfuscation are not perfect and cause lower quality responses.
Moreover, in previous approaches, all queries are treated equally, \ie, with the same protection level. Taking query obfuscation as an example, all requests are appended with the same amount of fake queries.
This is not always effective as user searches may have different levels of sensitivity, \ie, non-sensitive queries may be overprotected, while the opposite may happen to sensitive ones.
We propose, in this section, a solution that tackles these issues: privacy, accuracy, scalability and employs adaptive protection.

The scalability aspect, in particular, is of special interest.
Up to now, all proposed solutions could be possibly deployed in a third-party service provider.
Contrarily, we now shift this design aspect to the client-side.
We assume that each client node supports \ac{SGX}, as it has been shipped in Intel processors targeting desktops and portable computers since 2015, therefore being by now fairly widespread.
By doing so, general user machines---which are more vulnerable to generic malwares and simpler infection vectors like social engineering---may be regarded as trusted.
This brings new possibilities like server processing offloading~\cite{goltzsche:2017:trustjs}, network monitoring~\cite{goltzsche:2018:endbox} or trusted decentralised \ac{P2P} networks, as we propose here.

\subsection{Cyclosa}
\label{sec:cyclosa}

We present \cyclosa \cite{pires:2018:cyclosa}, a decentralised private Web search solution that leverages \acp{TEE} and:

\begin{enumerate}[label={(\roman*)}]
\item enforces \textit{privacy} by protecting users against re-identification attacks through unlinkability and indistinguishability;
\item enforces \textit{accuracy} by providing users with similar responses to those they would get by directly querying the search engine; and
\item is \textit{scalable} to millions of users while enforcing service availability even if search engines limit bulk requests.
\end{enumerate}

To enforce unlinkability between a query and its sender, each node participating in \cyclosa acts both as a client when sending its own requests and as a proxy by forwarding queries on behalf of other nodes.
Each node has a \ac{TEE} such as \ac{SGX} wherein \cyclosa keeps secure inter-enclave communication endpoints and from where interactions with search engines are triggered.
These nodes are therefore regarded as \emph{untrusted} from the other peers' perspective and therefore no query information is leaked to them. The enclave they host, on the other hand, is trustworthy.

Re-identification attacks are mitigated by sending both fake queries and the real user query through multiple paths to the search engine.
Differently from \xsearch though, instead of blindly sending the same amount of fake queries regardless of the real query, \cyclosa assess its sensitivity and adapts the amount of noise to be sent along.
This query sensitivity evaluation is made by quantifying two attributes:
\begin{enumerate}[label={(\roman*)}]
\item \textit{linkability}, which relates to the current request's similarity with the user local profile: the more similar the bigger the chances of user re-identification; and
\item \textit{semantic sensitivity}, based on the query's topic. Out of a predefined set of topics, users pick a subset that they consider to be sensitive. 
\end{enumerate}

Whenever a query is sent to the search engine, \cyclosa checks whether it is linkable to its issuer profile and if the query's topic belongs to user-defined sensitive topics. 
Based on this, it accordingly adjusts the amount of fake queries, \ie, \cyclosa strongly protects sensitive queries and avoids overloading the system with fake ones when they are not sensitive.

In order to provide accurate results for clients, \cyclosa sends the real and fake queries through distinct relays.
In this way, filtering becomes straightforward, since results that correspond to fake requests can be entirely discarded.
Therefore, \cyclosa returns the same responses as when users directly query the search engine, hence reaching perfect accuracy.

\cyclosa's decentralised architecture consists of nodes with equal roles.
This allows the system to scale well, even for a large number of clients as the load gets evenly distributed between participating nodes.
Moreover, it brings advantages regarding fault tolerance: as there is no central point of failure, the system remains operative even in presence of localised attacks or blacklisting by search engines.

We implemented and evaluated \cyclosa using the same workload we used previously: the query logs extracted from the \ac{AOL} dataset~\cite{pass:2006:picture}.
Results show that \cyclosa meets the expectations concerning load balancing, economy of resources and privacy protection.

\subsubsection{Adversary model}

The computations performed by \cyclosa for each user query go through three premises, namely:
\begin{enumerate}[label={(\roman*)}]
\item the \textbf{client machine}, which is trusted by its owner but untrusted by peers that compose \cyclosa \ac{P2P} network.
Such trust includes all computations that involve client information (\eg, assessing the sensitivity degree of a query), which are performed locally outside of enclaves.
Furthermore, we assume that the interaction between a human user and \cyclosa is trusted, \ie, an adversary cannot modify a typed query, nor it can tamper with the local configuration of \cyclosa that a human user has set up, that is, the set of topics the user considers as semantically sensitive.

Clients cannot bypass the \ac{SGX} enclave.
Session keys are generated between pairs of \cyclosa enclaves after successful remote attestation.
Therefore, untrusted code is not able to forge requests correctly encrypted and signed.
If the client is breached though, the sensitivity analysis could be subverted.
As third-party sensitive data (\eg, table of past queries) is only handled inside enclaves, such attack would only compromise the infected peer;

\item a set of \textbf{proxies}, \ie, forwarding peers who mutually distrust each other. They can act in a Byzantine manner \cite{pease:1980:agreement,lamport:1982:byzantine}, \ie, they can arbitrarily  crash, be subject to bugs or under control of malicious adversaries.

Inter-enclave traffic as well as between enclaves and the search engine are protected by encrypted channels.
In addition, query forwarding performed by the proxy is done inside the \ac{SGX} enclave.
Consequently, a malicious or compromised proxy cannot hamper \cyclosa (side-channel attacks are not considered - see \Cref{sec:sgx:vulnerabilities}).
However, malicious hosts may replay past queries.
This threat can be mitigated by including a random identifier in each message to detect replays of a limited set of recent messages.
Also, a malicious proxy can deny initialisation or calls into enclaves.
\cyclosa tackles this issue by letting clients blacklist non-responding or slow peers; and

\item the \textbf{search engine}, which is honest but curious, \ie, it faithfully replies to search queries while possibly gathering information to build user profiles and run re-iden\-ti\-fi\-ca\-tion attacks~\cite{petit:2016:simattack}.

The search engine's capability of re-identifying users is very limited (more in the evaluation section).
However, the search engine could identify a real query when it receives this query for the first time. 
Later on, the request tends to become indistinguishable from others as it will be re-issued as fake.
Even in this case, the identity of the requesting user is still hidden by the proxy (unlinkability).
\end{enumerate}

\subsubsection{Protocol overview}

\begin{figure}[t]
	\centering
	\includegraphics{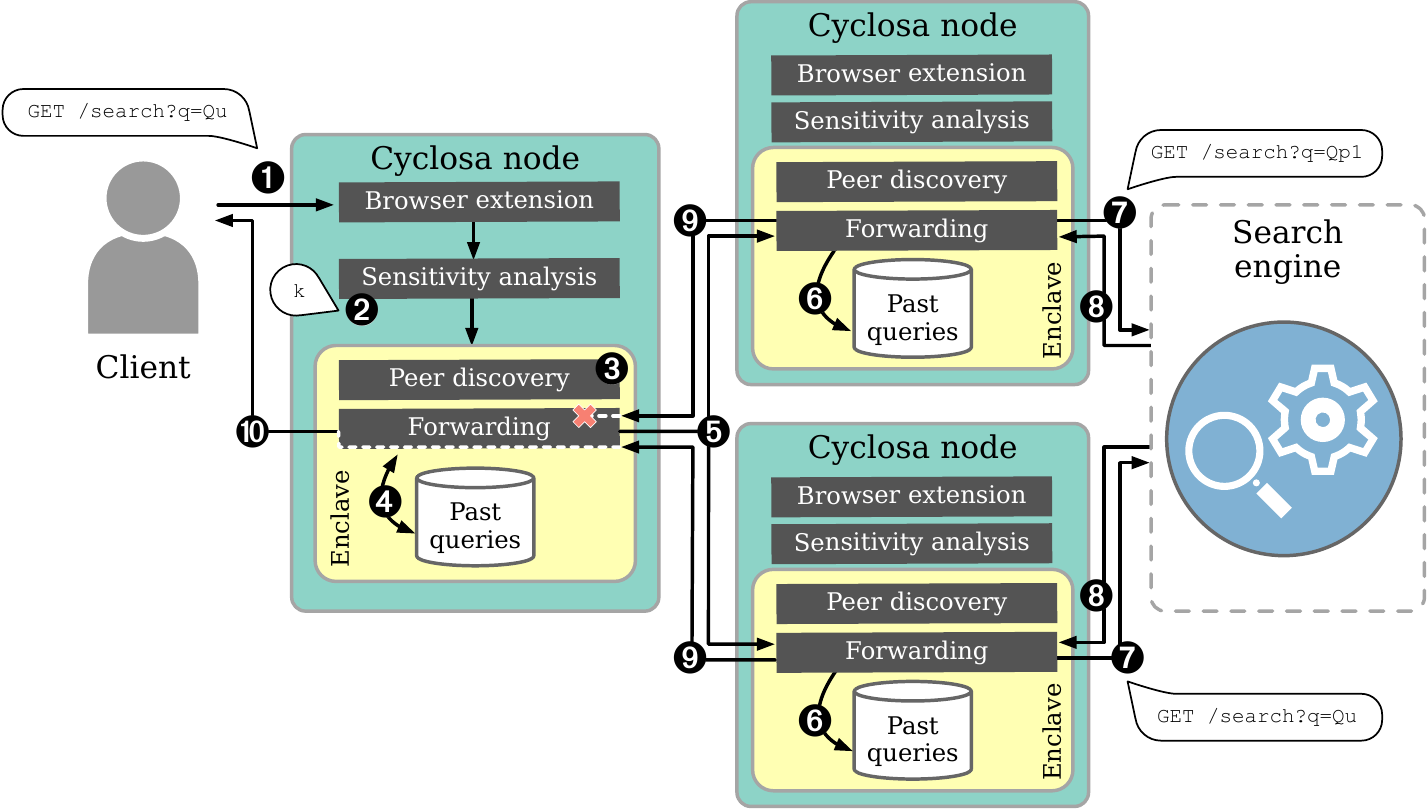}
	\caption{\cyclosa architecture and operating flow.}
	\label{fig:cyclosa:arch}
\end{figure}

To use our system, clients install the \cyclosa browser extension, so that they seamlessly get protection without changing their browsing habits.
\Cref{fig:cyclosa:arch} illustrates the operating flow.
When the user formulates a query $Q_u$ (\ding{202}), \cyclosa evaluates its sensitivity (\ding{203}) by checking whether the query is linkable to the user profile and if there is a match between the semantic analysis of $Q_u$ with the user-defined set of sensitive topics previously established.
As a result, a score $k$ is produced and used as the number of fake queries needed to make the real user query indistinguishable from others.

A peer discovery component (\ding{204}) then selects $k+1$ random peers $P_{p0},P_{p1},...,P_{pk}$ to which it sends (\ding{206}) the real query $Q_u$ and $k$ fake queries $Q_{p1},...,Q_{pk}$ obtained from the database of past queries (\ding{205}), issued by other users in the system. They are stored in a local table whenever a node acts as a relay. As indicated in \xsearch evaluation, real past queries used as fake make them seem real from the search engine's perspective. In contrast, systems such as \tmn or \goopir, where fake queries are generated using \ac{RSS} feeds or dictionaries, are more easily recognized as artificial.

When a request is received by peers acting as relays, it is stored in the local table of past queries (\ding{207}) before it is forwarded to the search engine (\ding{208}). 
Relays are not aware of whether the queries they handle are real or fake.
An external observer therefore learns nothing by analysing the encrypted network traffic.
In contrast, systems where fake queries are appended in the relays (\eg, \xsearch or \peas) allow an adversary to infer whether an outgoing message corresponds to a real query or an obfuscated one based on the request size, since obfuscated queries built with the $\mathtt{OR}$ operator are larger in length.

Upon receiving a request from the node acting as a relay, the search engine provides it with the query results (\ding{209}). Such responses are then routed back to the original client (\ding{210}).
Finally, \cyclosa drops the responses corresponding to fake queries before the user gets access to the initial query's results (\ding{211}).

To avoid information leakage, all \cyclosa components that process sensitive data, \ie, queries issued by other users, are located within enclaves. 
In order to minimize the amount of trusted code, however, components that process data related to the user who owns the machine are placed outside the enclave.
Specifically, the unit in charge of query sensitivity assessment issued by the local user is performed outside the enclave.
This allows to drastically minimise the amount of trusted code, which reduces the risk of having critical bugs and unnoticed vulnerabilities.
Since remote peers are untrusted, components that handle other users' queries in plain text are placed within enclaves, along with the communication endpoints where they are encrypted and decrypted before or after transmissions.
Components placed inside trusted environments include: peer discovery, query forwarding, encrypted communication endpoints and the past user queries database.

\subsubsection{Query sensitivity analysis}

Aiming at avoiding the network overload potentially caused by excessive noise injected by our proposal yet achieving indistinguishability of sensitive queries, \cyclosa performs a sensitivity analysis.
Specifically, it dynamically protects user queries according to their susceptibility by adjusting the amount of fake queries sent along with them.
To quantify query sensitivity, \cyclosa relies on two measurements computed outside the enclave:
\begin{enumerate}[label={(\roman*)}]
\item semantic assessment, \ie, indicates whether a query is related to sensitive topics previously declared by the user; and
\item linkability assessment, \ie, the risk that the query gets linked back to its issuer by means of comparison with past queries issued by such user.
\end{enumerate}

Since semantic sensitivity is subjective, \ie, one query might be considered as sensitive by one user and non-sensitive by another, \cyclosa proposes a user-centric approach.
Users select what they consider semantically-sensitive topics among health, politics, sex, and religion\footnote{Based on Google's privacy policy~\cite{google:2019:sensitive} which defines sensitive personal information as \textit{"a particular category of personal information relating to topics such as confidential medical facts, racial or ethnic origins, political or religious beliefs, or sexuality."}}.
Nevertheless, a user can import dictionaries to create other sensitive topics.
The \textbf{semantic-based} assessment's result is binary and indicates whether the query belongs to at least one topic marked as sensitive.
To achieve that, we use information retrieval approaches and build dictionaries of terms associated to each sensitive topic.

These dictionaries are built with keywords based on two datasets:
\begin{enumerate}[label={(\roman*)}]
\item WordNet~\cite{fellbaum:1998:wordnet}, a machine-readable lexical database that is organised by meanings, where words are grouped into sets of synonyms called \emph{synsets}; and
\item eXtended WordNet Domains~\cite{gonzalez:2012:xwnd}, which maps WordNet synsets to \num{170} domain labels. By using this, we identify keywords related to our privacy-sensitive topics.
\end{enumerate}

Additionally, we leverage a statistical approach to enrich our keyword dictionaries of sensitive topics.
It consists of latent topic models, or \ac{LDA}~\cite{blei:2003:dirichlet}, a generative probabilistic model which captures correlations among words and topics that is well adapted for modelling text corpora.
In the \ac{LDA} model, a topic is described through thematic vectors that indicate the topic's latent dimensions.
Once this model is trained with a text corpora associated to the user-defined sensitive topics, we build the dictionary by gathering the terms of all thematic vectors.

In the experiments, we consider sexuality as an example of sensitive subject in user queries.
We trained a \ac{LDA} statistical model using the Mallet toolkit~\cite{McCallumMALLET}, with 200 topics on 2 million titles and descriptions of videos related to the sensitive subject~\cite{mazieres:hal-00937745}.
Finally, a query is identified as semantically sensitive either if it contains a term in at least one \ac{LDA} topic or if it is linked to a WordNet domain tagged as sensitive.

The \textbf{linkability} assessment's goal is to determine if the query is vulnerable to a re-identification attack. 
In such attacks, an adversary tries to link an anonymous query to a specific user by measuring the distance between the query and a set of user profiles built from past queries that were collected, for instance, before users started using private Web search mechanisms.
Concretely, the linkability assessment performed on the client side provides a score $l$ within the interval $[0,1]$ that corresponds to the current query's proximity to the user profile.
To do that, we first represent the user query in a vector where each element is a term in the query. 
Then, the cosine similarity between this vector and the one corresponding to each past query made by the user is computed.
Finally, we use exponential smoothing as a window function\footnote{Window functions consider only a limited interval of the samples to compute a final score, usually based on time, \ie, the most recent samples. Exponential smoothing, unlike the moving average where samples are equally weighted, assigns exponentially decreasing weights to ``\emph{old}'' samples.} to produce an aggregated score. Instead of ordering the samples by time though, we order them by cosine similarities to compute this score, so that the similarity has more importance than how long ago a given query was submitted.

The number of fake queries $k$ may be defined based on one or both assessments.
As a result from the semantic evaluation, we simply obtain a boolean that indicates whether the current query is related to sensitive topics.
If it is the case, $k$ is defined to a maximum value $k_{max}$ and the linkability assessment can be skipped.
Otherwise, \ie, when the query is not semantically sensitive, $k$ is defined as the linear projection of the linkability score $l$ \wrt $k_{max}$, \ie, $k=l \cdot k_{max}$, with $l \in [0,1]$.

\subsubsection{Enclave operation and implementation details}

Once the number of fake queries $k$ is decided according to the user query sensitivity, the process continues in the \ac{SGX} enclave as it involves forwarding sensitive data to remote peers.
Each node discovers these peers and dynamically maintains a view of other alive nodes in the system.
When a query is issued, \cyclosa randomly picks $k+1$ out of these nodes to act as proxies for the original query and $k$ fake ones---randomly selected from the table of past queries.
Inter-enclave communication is always encrypted and the identity of the proxy handling the original query is kept in a table within the enclave.

Once a proxy receives a query forwarding request, it adds this query in its local table of past queries and routes it to the search engine.
Answers from the search engine are returned to the original user who, in turn, receives the result that came from the proxy which carried the real query, silently dropping all others.
Routing real and fake queries through different paths makes them indistinguishable from the search engine's perspective.

At bootstrap phase, besides the user declaration of sensitive topics, there are other key elements that need to be initialised when \cyclosa is first launched by a given user:
\begin{enumerate}
\item Initially, there are no past requests stored in the enclave.
Hence, \cyclosa fills the table with popular Google queries~\cite{google-trends}, as they are issued by real users regarding trendy topics.
\item Then, peer discovery is bootstrapped. We assume it is done as in classical \ac{P2P} systems using public repositories of IP addresses (\eg, \tor) from which a \cyclosa instance can select a first sample of random peers.
Peer discovery is done using classical algorithms and contributions to this field fall outside the scope of this work. For implementation purposes of our prototype, we use Zyre~\cite{zyre:2019}, a library based on ZeroMQ~\cite{zeromq:2019} for peer discovery, groups organisation and multicast of events.
\item Finally, \ac{SGX} remote attestation of connecting peers must be performed (see \Cref{sec:sgx:attestation}).
\end{enumerate}

To allow seamless integration to end-users' workflow, \cyclosa was designed as an extension to the Firefox browser.
It is JavaScript-based and integrates the \ac{SGX} enclave using \emph{js-ctypes}, which allows asynchronous calls between the enclave and the untrusted extension code.
Connections to search engines, in turn, are established with \ac{TLS} tunnels. 
They must originate from enclaves in order to prevent sensitive data disclosure.
To do so, we use a \ac{SGX}-compatible version of mbedTLS~\cite{mbed-tls}.
All in all, the \cyclosa enclave object accounts for only \SI{1.7}{\mebi\byte}.
This and a circular buffer for the past queries table prevents \cyclosa from suffering of \ac{EPC} paging.

\subsubsection{Experimental setup and metrics}

We compare \cyclosa against five approaches for protecting privacy in Web search: \tor~\cite{dingledine:2004:tor}, \tmn~\cite{howe:2009:trackmenot}, \goopir~\cite{domingo:2009:goopir}, \peas~\cite{petit:2015:peas} and \xsearch~\cite{mokhtar:2017:xsearch};
and a protection-free Web search scenario (see \Cref{sec:back:privacy}).
As in \Cref{sec:xsearch}, we use the \ac{AOL} query log dataset~\cite{pass:2006:picture} and the same methodology described in~\cite{gervais:2014:quantifyingprivacy} to evaluate privacy by considering a subset of the most active users.
Here, however, we manually extracted \num{198} users who sent at least one semantically sensitive query.
Again, to perform a re-identification attack, we split the queries into training set (two thirds) and testing set (one third).
In the training set, this represents an average of \num{487.6} queries per user out of \num{96,547} queries.

To determine user-perceived sensitivity regarding Web queries, we conducted a crowd-sourcing campaign using Crowdflower~\cite{crowdflower}.
We selected the first $10,000$ queries over all user queries in the testing set, and asked the workers to determine if these queries are related to sensitive topics (\eg,~health, politics, religion and sexuality). 
Each query was annotated by $5$ distinct workers.
It resulted that only \SI{15.74}{\%} of the queries are related to sensitive topics.
Since the great majority are non-sensitive, lots of resources (\eg, processing, storage, bandwidth) would be wasted should all queries be equally protected as in \xsearch.
This motivates \cyclosa's adaptive approach, which applies a dynamic protection scheme to sensitive queries.

To measure the accuracy of \cyclosa's ability to automatically determine if a query belongs to a sensitive topic, we consider the precision and recall metrics.
The \textit{precision} indicates the proportion of truly sensitive queries among those detected as such by \cyclosa (correctness), while the \textit{recall} express the ratio between the queries detected as sensitive \wrt all sensitive queries in the testing set (completeness). 
We consider the crowd-sourcing annotations as ground truth.
Let $Q_{s}$ be the set of actually sensitive queries (\ie, related to sensitive topics), and $Q_{m}$ the set of queries that are identified as sensitive by \cyclosa.
Precision and recall are defined as follows:

\begin{align*} 
precision = \frac{|Q_{m} \cap Q_{s} |}{|Q_{m}|}  & &
recall = \frac{|Q_{m} \cap Q_{s} |}{|Q_{s}|} 
\end{align*}

To evaluate privacy, we use SimAttack~\cite{petit:2016:simattack} in order to match queries with user profiles (see \Cref{sec:xsearch}).
Then, we calculate the overall re-identification success rate according to \Cref{eq:reidentification}, \ie, the proportion of queries for which the user profile is successfully re-identified with respect to all queries in the testing set.
Naturally, the lower the re-identification success rate is, the better is the privacy level.

By design, \cyclosa returns to users the same results associated to their search query as if no privacy-protection mechanism were employed.
However, other proposals such as \xsearch and \peas include an obfuscation scheme that impacts these results by adding some noise to what is returned by search engines.
Because of this, we evaluate the capacity of such approaches of returning results related to the initial query.
To achieve that, we measure accuracy in terms of precision and recall according to \Cref{eq:xsearch:accuracy}, \ie, by comparing results returned by direct requests to search engines and those resulted from using privacy-preserving strategies.

To evaluate the behaviour of \cyclosa from a systems perspective, we measure the end-to-end delays and throughput versus latency. 
Besides, we observe the impact of the amount of fake queries on latency and \cyclosa's ability to avoid being blocked by search engine's anti-bot prevention mechanisms in comparison to \xsearch.

\subsubsection{Evaluation}

\begin{figure}[tb]
	\centering
\includegraphics{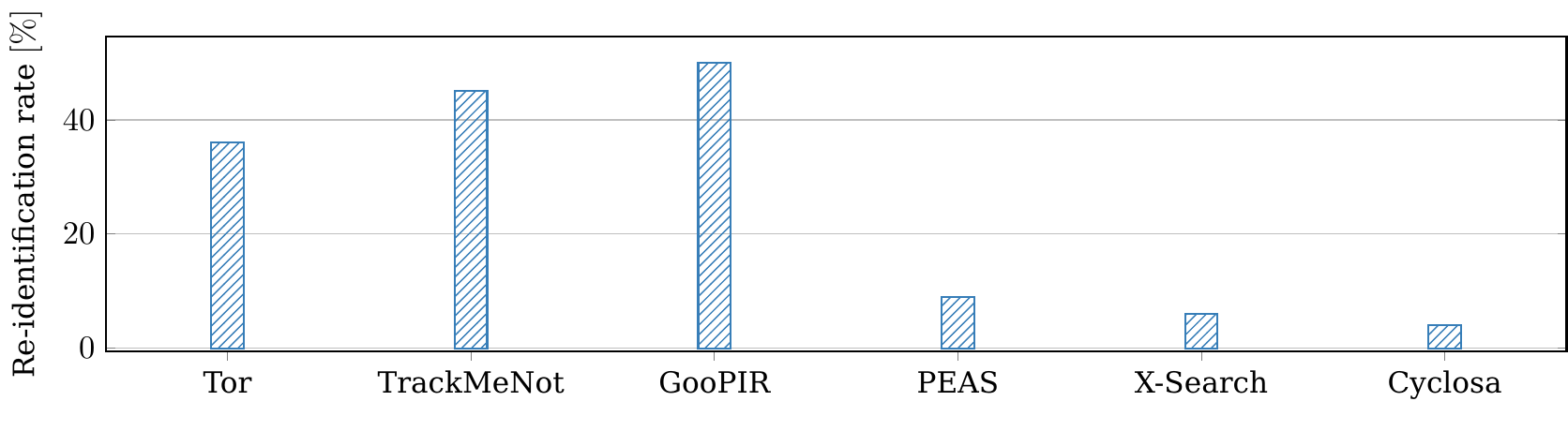}
 	\caption[Privacy comparison between \cyclosa and others.]{\label{fig:PrivacyEval-ReidentifAttack}Comparison of \cyclosa's privacy level with other approaches ($k=7$ when they employ obfuscation).}
\end{figure}

We first evaluate the capacity of \cyclosa to protect user privacy by measuring its robustness against an adversary conducting a re-identification attack.
\Cref{fig:PrivacyEval-ReidentifAttack} shows re-identification rates for \cyclosa, \tor, \tmn, \goopir, \peas, and \xsearch with $k=7$.

\begin{figure}[tb]
	\centering
\includegraphics{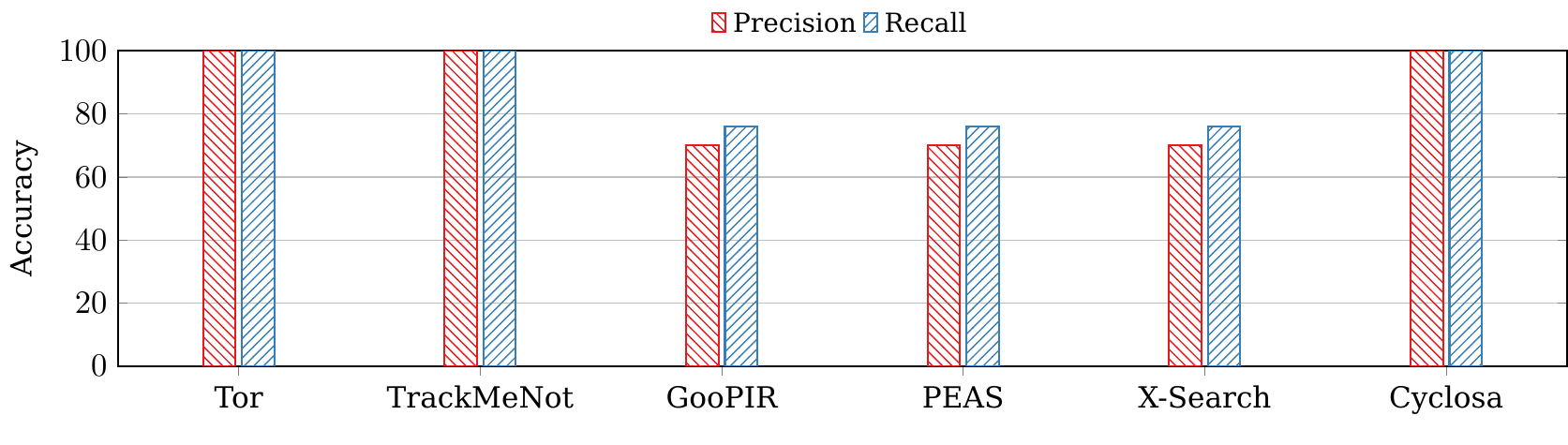}
 	\caption[Accuracy comparison between \cyclosa and others.]{\label{fig:AccuracyEval-WebSearch}Accuracy of results returned by \cyclosa and alternatives ($k=3$ when obfuscation is employed).}
\end{figure}

Without query obfuscation (\ie, \tor), an adversary is sure that every query has been issued by a user.
Consequently, the challenge in this case consists of mapping each query to user profiles built from previously collected user information.
Our results show that an adversary is able to re-affiliate around \SI{36}{\%} of the new queries to their correct original users.
Interesting enough, the re-identification rate for \tor is the same of \peas, \xsearch and \cyclosa with $k=0$ (not shown in the plot).

Without unlinkability (\ie, \tmn and \goopir), the adversary needs to distinguish real queries from the fake ones.
Results show that a large proportion of real queries are identified: 45\% and 50\% for \tmn and \goopir, respectively.
This high re-identification rate is mainly caused by the fake query generation process, which uses \ac{RSS} feeds.
When sources for fakes are far from the users' interests, the adversary can easily dissociate them.

Combining query obfuscation and unlinkability drastically drops re-identification rates.
Indeed, the adversary's challenge becomes harder.
It consists of retrieving both the user's identity and identifying real queries among fake ones.
By using real past queries as fake ones, \xsearch and \cyclosa provide lower re-identification rates than \peas, which uses a graph of co-occurrence of terms built from past queries. 
That is, terms that happen to be associated more often in real queries will compose the fake query along with the one issued by the user.

Moreover, \cyclosa slightly reduces this re-identification rate in comparison to \xsearch from \SI{6}{\%} to \SI{4}{\%}.
This difference comes from the obfuscation scheme adopted by them.
For \xsearch, the adversary receives a group of $k+1$ queries and has to identify the real one in this group.
In \cyclosa, however, the adversary receives individual queries from different proxies and has to decide whether it is real or fake.
Due to the increased amount of confusion it creates, the re-identification becomes harder.

Next, we evaluate \cyclosa's accuracy, \ie, its ability to return the same answers from the search engine as the ones resulted from unprotected direct queries.
\Cref{fig:AccuracyEval-WebSearch} shows the obtained accuracy in terms of correctness and completeness of answers returned by \cyclosa and alternatives, with $k=3$ when query obfuscation is used.
There are two sets of solutions.
\cyclosa and \tmn provide perfect accuracy as they differentially handle real and fake queries' responses, while \tor achieves the same because it does not employ obfuscation.
The other solutions provide lower accuracy because, to some extent, they are not able to distinguish between answers of fake and real queries.
Precision for \goopir, \peas and \xsearch reaches $65\%$ for a recall of $70\%$, when $k=3$ and decrease for larger values of $k$.

\begin{figure}[tb]
	\centering
\includegraphics{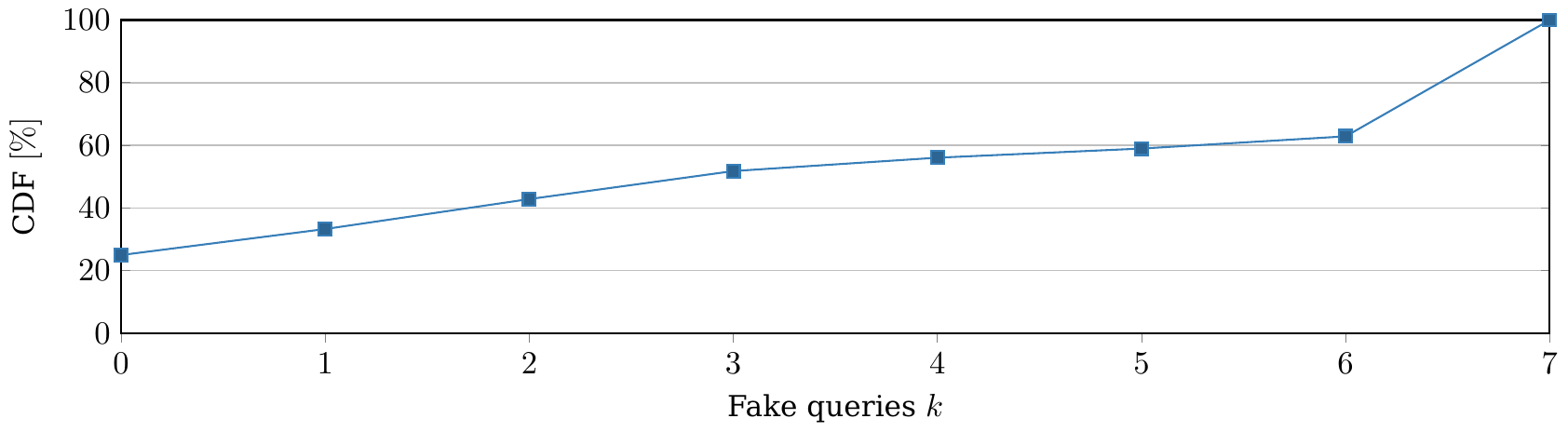}
 	\caption{\label{fig:kvalues}Actual number of fake queries in \cyclosa.}
\end{figure}

\begin{table}[!t]
	\setlength{\tabcolsep}{15pt}
	\centering
	\caption{\label{tab:detection}Detection of semantically sensitive queries.}
	\begin{tabular}{l c c}
		\toprule
		\textbf{Semantic tool} & \textbf{Precision} & \textbf{Recall} \\
		\midrule
		WordNet & 0.53 & 0.83 \\
		LDA & 0.84 & 0.89 \\
		WordNet + LDA & 0.86 & 0.85 \\
		\bottomrule
	\end{tabular}	
\end{table}

\cyclosa dynamically and adaptively protects queries according to their sensitivity by adjusting the amount of noise.
Figure~\ref{fig:kvalues} shows the \ac{CDF} of the number of fake queries induced by \cyclosa in our testing set when the maximum value for $k$ is \num{7}.
Results show that \SI{25}{\%} of queries do not need obfuscation, and \SI{50}{\%} of them use less than \num{3} fake queries.
The sharp increase for $k=7$ corresponds to requests identified as highly sensitive in the semantic-based assessment, and consequently scaled to the maximum protection level.
They amount to \SI{35}{\%} of our testing set. 
In contrast, \xsearch would have generated this maximum number of fake queries for all queries.

\begin{figure}[tb]
	\centering
\includegraphics{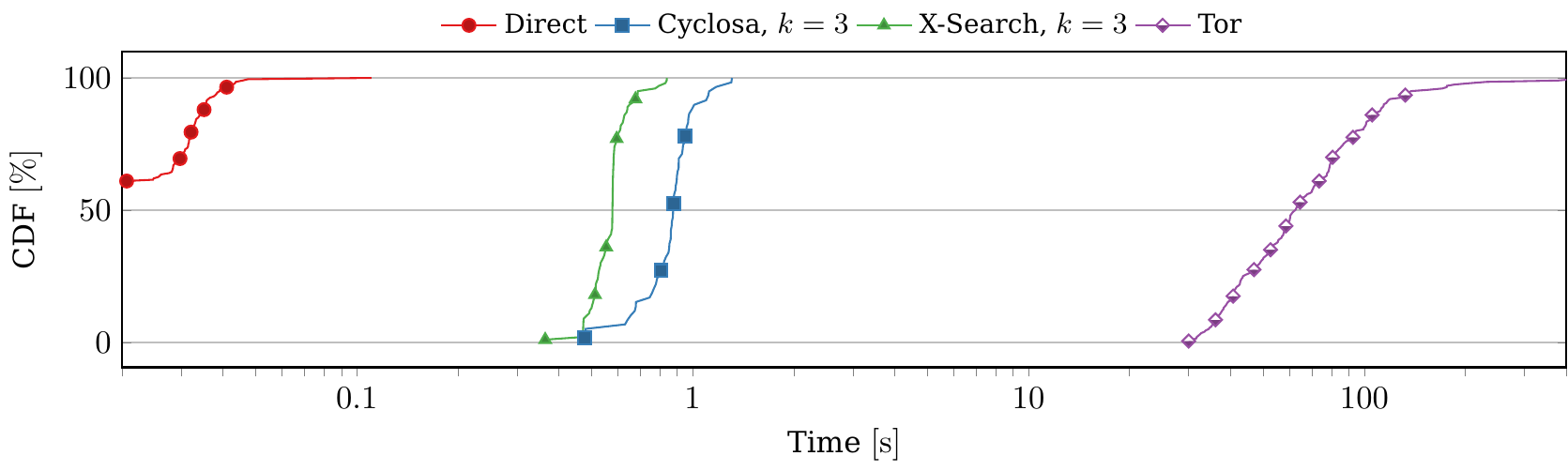}
\vskip 5mm
 	\caption[Distribution of end-to-end delays for privacy-preserving systems.]{\label{fig:endtoend-latency}Distribution of end-to-end delays for \num{200} queries, $k=3$.}
\end{figure}

As previously defined, the sensitivity of a query is evaluated in two dimensions: linkability to its issuer's profile and as to whether it belongs to user-defined sensitive topics.
This detection is based on both WordNet libraries and a trained LDA statistical model.
To evaluate their influence on the results' quality, we ran them individually and combined while obtaining precision and recall when detecting whether queries belonged to the sensitive topic related to sexuality.
\Cref{tab:detection} shows the results.
Overall, most of queries are detected, with a recall between \SI{83}{\%} for WordNet and \SI{89}{\%} for LDA.
Precision varies from \SI{53}{\%} (WordNet) to \SI{86}{\%} (WordNet + LDA).
In our experiments, combining WordNet and LDA provides the best trade-off between correctness and completeness.

To analyse \cyclosa's performance, we start by showing in \Cref{fig:endtoend-latency} the observed end-to-end latency in comparison to \xsearch and \tor, including the time it takes to the search engine for processing the requests.
We further include the measurements achieved without any protection by directly contacting the search engine.
As expected, \tor is the slowest, with a median latency of \SI{62}{\second}.
Both \cyclosa and \xsearch allow for sub-second delays for the large majority of the queries, with medians of \SI{0.88}{\second} and \SI{0.58}{\second}, respectively.

\begin{figure}[t]
	\centering
\includegraphics{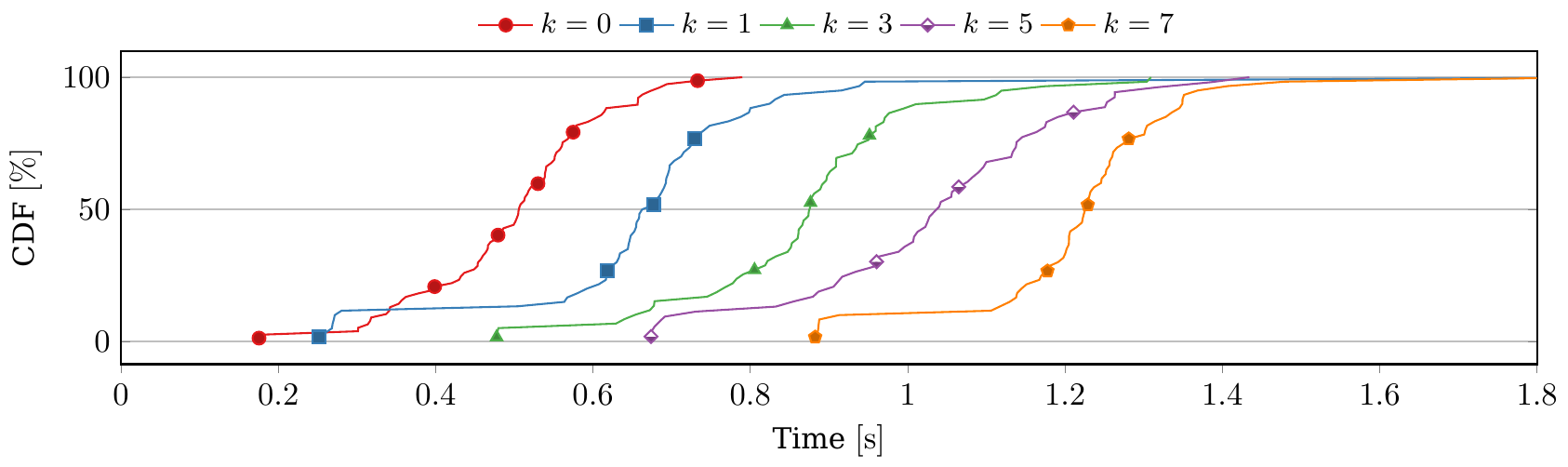}
\vskip 5mm
 	\caption{\label{fig:endtoend-latency-fakes}\cyclosa: Impact of k on latency.}
\end{figure}

In \Cref{fig:endtoend-latency-fakes}, we explore the impact of changing the number of fake queries.
Even in the worst case ($k=7$), the system still returns the results to clients in less than \SI{1.5}{\second}, with median latency at \SI{1.2}{\second}, which still allows for a usable web browsing experience.

\begin{figure}[tb]
	\centering
\includegraphics{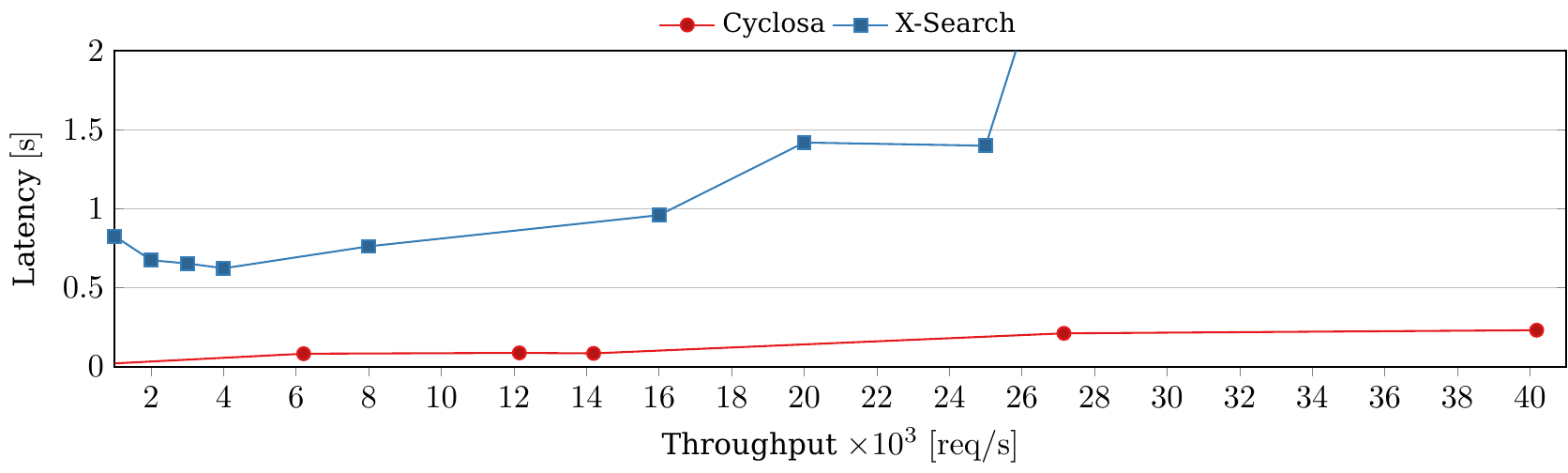}
\vskip 5mm
 	\caption{\label{fig:cyclxs:tput_lat}Throughput vs. latency of \cyclosa and \xsearch.}
\end{figure}

We also evaluate the capacity of a \cyclosa node to sustain high rates of requests in comparison to \xsearch.
\Cref{fig:cyclxs:tput_lat} presents the results.
We submit requests at increasingly high constant rates and measure the latency of a reply from the next hop (the \xsearch proxy or a \cyclosa relay), but without actually submitting the requests to the search engine.
\cyclosa is able to handle very high request rates with sub-seconds response delays.
In our evaluation, we achieved \SI{40,000}{\requests\per\second} with \SI{0.23}{\second} median response time while \xsearch starts straggling at \SI{30,000}{\requests\per\second}.

\begin{figure}[tb]
	\centering
\includegraphics{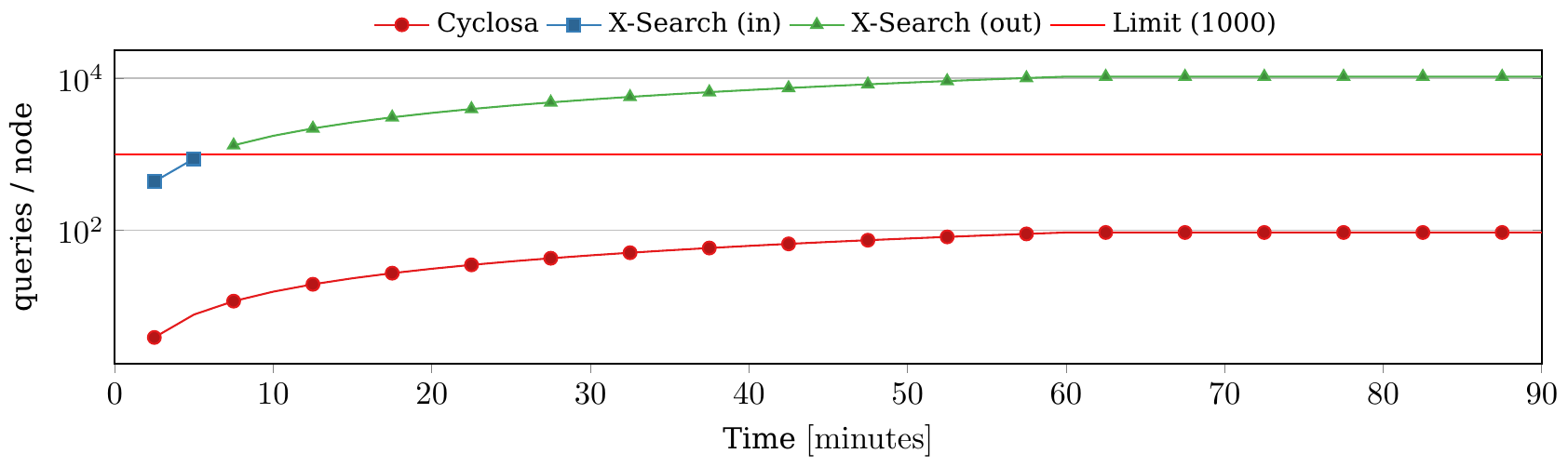}
\vskip 5mm 
 	\caption{\label{fig:load}Query protection vs. users blocked by search engine.}
\end{figure}

Such rates, although possible, cannot be observed in practice without being immediately blocked by a search engine's malicious user detection system.
In our experiments with Google search engine, this happened very soon.
Moreover, users submit searches at rates that are orders of magnitude slower: the 100 most active users from the \ac{AOL} dataset perform queries at a rate of \SI{31.23}{\requests\per\hour}.
Considering the same queries, \xsearch induces \SI{10,500}{\requests\per\hour} among real and fake ones when $k=3$. Therefore, it is eventually blocked by the search engine.
\cyclosa follows a more practical approach by spreading the load among the nodes with up to \SI{94}{\requests\per\hour} per node when $k=3$, as shown in the simulation-based results presented in \Cref{fig:load}, where \xsearch's curve is split by the blocking threshold after only \SI{5}{\second} of the testing set execution.

Our results show that \cyclosa efficiently detects sensitive queries while limiting the number of overprotected requests which are not sensitive.
We also show that it provides slightly better privacy protection than state-of-the-art solutions and drastically reduces the end-to-end latency without impact on accuracy.
Particularly, \cyclosa:
\begin{enumerate}[label={(\roman*)}]
	\item resists re-identification attacks better than alternatives with a low re-identification rate of \SI{4}{\%};
	\item enables sub-second response times, which is on average \num{13}$\times$ faster than using \tor;
	\item can sustain throughputs higher than \SI{40,000}{\requests\per\second}, enabling parallel users to securely browse the search engine; and 
	\item fairly balances the load between the participating nodes enabling all users to securely query the search engine without reaching its rate limitation.
\end{enumerate}
 
\section{Summary}

Major service providers track user behaviour, which is a serious privacy threat in today's Internet.
As one of the most widely used online services, search engines handle queries that may reveal sensitive information about individual users, such as sexual orientation, religion or political preferences.
Existing solutions for enabling users to access Web search engines in a privacy-preserving manner do not tolerate strong adversaries or have poor performance.
In this chapter, we explored the usage of \acp{TEE} to improve the capabilities of current solutions both in terms of privacy assurances and performance.

First, we proposed a novel architecture for privacy-preserving Web search which relies on \ac{SGX} to support stronger adversarial models.
\xsearch (\Cref{sec:xsearch}) operates as a proxy which stores and leverages user past queries within a protected \ac{SGX} enclave and generates obfuscated queries on the user's behalf.
It does so by aggregating random past queries in such a way that the search engine is not able to distinguish which one is the original query, yet providing relevant results back to the user.
Upon receiving a response from the search engine, \xsearch filters results so that  only those related to the initial query are forwarded.

We implemented a working prototype and evaluated it both analytically and experimentally using real-world datasets.
Our observations indicated that \xsearch can indeed provide accurate results without disclosing personal information about individuals. 
Most importantly, \xsearch did so with a throughput improvement that reached orders of magnitude when compared to its predecessors, \ie, \peas and \tor.

The centralised nature of \xsearch imposes some constraints though. 
The most notable being the easiness with which search engines could block requests originating from \xsearch proxies as part of anti-bot mechanisms.
To tackle that, we proposed \cyclosa (\Cref{sec:cyclosa}), a decentralised, private and accurate Web search solution that protects users against the risk of re-identification.
It provides adaptive protection to different query sensitivity levels by combining the analysis of linkability to its issuer's profile and query semantic.

\cyclosa follows a fully decentralised architecture for higher scalability, and is based on \ac{SGX} for preventing data leakage by relay nodes in its distributed setup.
At the same time, \cyclosa reaches perfect accuracy by leveraging distinct paths for real and fake queries employed as obfuscation.
In contrast to \xsearch, it therefore obviates the need for finer grained filtering of results returned by search engines.
Our implementation, evaluation and comparison to alternatives showed that \cyclosa is the most robust system against user re-identification and provided accurate responses to Web searches in an efficient and scalable way.
\chapter{Conclusion}
\label{chap:conclusion}
\acresetall

Designing secure distributed systems is complex.
Their dispersed nature multiplies the number of attack vectors.
Apart from that, with the growth of cloud computing such systems tend to be at least partially deployed in shared infrastructures.
This brings additional risks that stem from providers, their personnel or co-located tenants.
Existing countermeasures are either infeasible due to computational complexity or fail at protecting running code and data from powerful adversaries like the \ac{OS}.
Because of that, the design of secure systems can greatly benefit from \acp{TEE}.

In this thesis, we have proposed several distributed architectures that leverage \acp{TEE}.
We started in \textbf{\Cref{chap:clouds}} with systems that are likely to be deployed in cloud environments.
For each of them, we isolate the minimum processing unit that handles sensitive data and put that component inside the trusted environment.
In \Cref{sec:scbr} that is done with the matcher component of a \ac{CBR} middleware.
In this way, sensitive information within subscriptions and publications is safely matched inside enclaves before the publication payload is forwarded towards matching subscribers.
We compare this approach with a software-only scheme with equivalent guarantees and reach better performances by one order of magnitude.

As for the lightweight \mapreduce (\Cref{sec:lwmp}) and SecureStreams (\Cref{sec:sstreams}), we isolate the Lua scripting language's interpreter within enclaves.
In such manner, encrypted code and data may enter the trusted environment, be deciphered and computed upon in plaintext form. 
Encrypted results are provided back to be handled by the underlying untrusted framework until reaching their final destination.
This framework is what discerns each of the two systems.
\mapreduce runs atop of \ac{SCBR} for code and data dissemination and has distinct behaviour based on each phase of the \mapreduce execution flow.
SecureStreams, in turn, has a second Lua \ac{VM} running in untrusted mode for the execution of a distributed reactive programming pipeline.
With all these systems we were able to observe performance losses due to \ac{SGX}, which are mostly influenced by memory usage.

In \textbf{\Cref{chap:sharing}}, we adopted a different design strategy.
Instead of selecting subcomponents of frameworks to be placed inside secure enclaves, we take advantage of the isolation property they provide to design more efficient cryptographic protocols for group data sharing.
The core mechanism relies on the generation of a master key in the secure environment where it is safely kept.
Based on this premise, we can make use of simple and efficient cryptographic constructs, and occasionally simplify computationally complex phases in legacy protocols.

That is what we did both in \ibbesgx (\Cref{sec:ibbe}) and \asky (\Cref{sec:asky}).
\ibbesgx simplifies the encryption step of an \ac{IBBE} scheme.
Using the \ac{IBBE} public key for such operation has quadratic complexity in the number of users in a group, while an equivalent formula that uses components of the master key makes the same with linear complexity.
Since the master key is securely kept in the enclave, we are able to use the latter approach.
This suffices to make the \ac{IBBE} practical for demanding applications and to reap benefits such as considerably smaller ciphertexts and no need for \ac{PKI}.
We then tackled anonymity in group settings with \asky.
To guarantee the anonymity, \asky uses a similar approach to \ac{GPG} but using symmetric cryptography instead of asymmetric cryptography.
This is possible because keys are maintained by the trusted environment.
Like \ibbesgx, the advantages come both in performance improvements and shorter ciphertexts.

\textbf{\Cref{chap:privacy}} brought the attention to users and showed how systems that rely on \acp{TEE} can protect their privacy against established service providers.
\acp{TEE} help shielding user queries that cross by relays before reaching the provider's premises.
Our first proposal, \xsearch (\Cref{sec:xsearch}), act as an intermediary between users and the Web search engine by proxying their requests, so that the provider cannot directly tell the user identity derived from the requester's \ac{IP} address (unlinkability).
Besides, it stores past user queries to obfuscate the requests by sending compound requests, so that the provider cannot tell the difference between a real query and a fake one (indistinguishability).
With this approach, we offer better resistance against re-identification attacks while achieving better performance than alternatives.

Being a centralised system, \xsearch does not scale.
Because of that, we propose \cyclosa (\Cref{sec:cyclosa}), a \ac{P2P} network of users who are Web search clients and relays at the same time.
These users, who have \ac{SGX}-capable machines, issue queries intermediated by the \cyclosa enclave, which selects random peers to forward real and fake requests through distinct paths.
Thanks to this, \cyclosa shows perfect accuracy of results as it just discards the ones coming from fake queries.
Moreover, it adapts their number based on the real request's sensitivity, thus not overprotecting non-sensitive queries and generating less traffic.
All combined, our evaluation shows that \cyclosa---the last of our contributions---achieves good performance while being scalable and delivering accurate results. 

Looking back at our objectives enlisted in \Cref{sec:intro:context} in conjunction with the contributions of this work, we are able to draw some conclusions.
\begin{itemize}
\item How can \acp{TEE} help to achieve security and privacy in distributed systems?

\acp{TEE} provide hardware isolation to applications.
\ac{SGX} does that by means of memory confidentiality and integrity guarantees of a running process' partition.
The subcomponents of a system that handle sensitive data are good candidates to be part of this trusted partition that is able to crunch plaintext data.

We show that for \ac{SCBR} (matcher), \securestreams and Lightweight \mapreduce (Lua interpreter), \ibbesgx and \asky (cryptographic operations), \xsearch and \cyclosa (storage or forwarding of queries).
The respective sensitive data consisted of publications and subscriptions (\ac{SCBR}), code and processing data (\securestreams and Lightweight \mapreduce), keys (\ibbesgx and \asky), and user queries (\xsearch and \cyclosa).
Whenever these data leave the enclave, they must be encrypted to be handled by untrusted code.

\item What are the drawbacks? %

Since the \ac{OS} is not trusted in the \ac{SGX} threat model, enclaves cannot perform system calls.
Because of that, running legacy applications in \ac{SGX} enclaves might require considerable effort, which tends to be proportional to the amount of \ac{OS} services those applications use.
Although some runtime frameworks aiming to help with that were proposed (\Cref{sec:sgx:vulnerabilities}), we decided to design our systems from scratch in order to have control over the partitioning between trusted/untrusted code and to keep a minimised \ac{TCB}.

Because of the need to track the integrity of memory pages, \ac{SGX} has memory limitations (\Cref{sec:sgx:memory}).
There are two thresholds that once surpassed bring along performance overheads: the \ac{LLC} and the \ac{EPC} limits.
We saw these drawbacks in \ac{SCBR} both for oversubscribing the processor's cache (\Cref{fig:scbr:wloads}) and the \ac{SGX} protected memory (\Cref{fig:scbrepc}) .
In the Lightweight \mapreduce we observed an increase in cache misses as the size of input data grows (\Cref{fig:cachemisses}).
In the other systems, we measured overall overheads of \ac{SGX} in terms of latency and throughput for each solution.

\item What benefits may come from using \acp{TEE}?

The ability to securely process sensitive data in plaintext is key to achieve performance gains by replacing complex cryptographic constructs.
We showed this for \ac{SCBR} versus \ac{ASPE}, \ibbesgx versus \ac{IBBE} and \asky versus \ac{PGP}.
All of them achieved better performances by at least one order of magnitude in comparison to the equivalent cryptographic scheme with no \ac{TEE}.
Additionally, both \ibbesgx and \asky were able to produce a lot less metadata, entailing even further benefits in storage and network usage.

To possibly reduce infrastructure costs, the simple isolation of code and data enables the offloading of processing to untrusted environments, as we showed with the Lightweight \mapreduce and SecureStreams.
With \cyclosa and \xsearch such isolation was used to guard users' sensitive data in order to protect their privacy.
That was possible because the \ac{TEE} assurances allow users to trust relays in order to obfuscate and convey their queries, \ie, the \xsearch proxy or the \cyclosa's client-side \ac{P2P} network.

\end{itemize}

\section{The road ahead}

There are a number of interesting avenues for future work.
Specific to our contributions, there are a few improvements that could be done.
As we built a set of academic prototypes, many enhancements are required before turning them into production systems.
We briefly comment on potential future work for each of our proposals.

Although subscriptions and publications are encrypted in \ac{SCBR}, they are not integrity protected and could be replayed without detection.
This could be solved by the addition of \acp{MAC} and sequential numbers to each message.
Additionally, in order to reduce enclave transitions, \ac{SCBR} could benefit from processing publication batches instead of individual ones.
The source code of \ac{SCBR} is publicly available~\cite{pires:2016:scbrsource}.

The Lua interpreter that runs inside the enclave in the Lightweight \mapreduce and SecureStreams could be improved to provide additional security guarantees.
It could, for instance, check the scripts for memory violations and other malicious behaviour.
Just like \ac{SCBR}, messages are not integrity- and replay-protected.
The source code of both Lightweight \mapreduce and SecureStreams is publicly available~\cite{pires:2017:lwmrsource,havet:2017:sstreamssource}.

\ibbesgx could benefit from dynamically adapting the partition sizes based on the undergoing workload in order to balance speed optimisation of both administrator- and user-performed operations.
For availability and fault tolerance, both \ibbesgx and \asky would benefit from a distributed set of administrators.
Some sort of distributed locking would have to be employed.
Additionally, the possibility of auditing membership operations could be put in place with secure logs, possibly implemented with blockchains.
The source code of \asky is publicly available~\cite{pires:2018:askysource}.

\xsearch and \cyclosa were implemented to work with a single search engine: Bing and Google, respectively.
A more generic solution could be built by supporting more providers.
\cyclosa's peer-to-peer network was implemented with a library that only supports local area networks.
Obviously, real deployments should instead support world-wide deployments.
Besides, both \xsearch and \cyclosa would benefit from different browser integrations.

More generally, some security aspects could be further considered.
In general, our prototypes did not implement attestation and key provisioning.
When shared keys were not generated in enclaves, we assumed they were correctly provisioned, \ie, after attestation (see \Cref{sec:sgx:attestation}).
Key management systems are subject of active research, including involving \ac{SGX}~\cite{fortanix:2019,chakrabarti2017intel}.

Guaranteeing that a piece of persistent data is the most recent one is not an easy task, since stale data that is correctly encrypted and signed may be replayed by attackers.
The enclaves in \ac{SCBR}, \asky and \xsearch persistently store active subscriptions, keys \& membership information and user queries, respectively.
They all would benefit from effective rollback prevention.
Intel currently provides persistent monotonic counters for that, but they are known to be slow and vulnerable (see \Cref{sec:sgx:platservices}).

All in all, we believe that \acs{TEE} came to stay and further research will certainly help to advance this exciting area.
In this respect, the first commercially available machines with support for \ac{SGX}2 were recently released.
They have larger \ac{EPC}, allow dynamic loading of memory pages and permit executions in release mode without keys provisioned by Intel.
\Wrt attacks, side channels will continue to be a hot topic for a while.
We also believe that more applications that use \acp{TEE} will keep appearing, with special highlight to the domain of blockchains~\cite{kosba:2016:hawk} and secure computing maketplaces~\cite{dang:2019:marketplace}.

 \renewcommand{\chaptermark}[1]{\markboth{Appendix \thechapter.\ #1}{}}
\begin{appendices}
\chapter{Publications}

\section*{2016}

\begin{enumerate}
\item Rafael Pires, Marcelo Pasin, Pascal Felber, and Christof Fetzer.\\
\textbf{Secure content-based routing using intel software guard extensions}\\
in Proceedings of the 17th International Middleware Conference, Middleware ’16.\\
Trento, Italy: ACM, December 2016, pp. 10:1–10:10.\\
doi: \href{https://doi.org/10.1145/2988336.2988346}{10.1145/2988336.2988346}. %
\end{enumerate}

\section*{2017}

\begin{enumerate}
\setcounter{enumi}{1}
\item Florian Kelbert, Franz Gregor, Rafael Pires, Stefan K\"opsell, Marcelo Pasin, Aur\'elien Havet, Valerio Schiavoni, Pascal Felber, Christof Fetzer, and Peter Pietzuch.\\
\textbf{SecureCloud: Secure big data processing in untrusted clouds}\\
in Design, Automation Test in Europe Conference Exhibition (DATE), 2017.\\
Lausanne, Switzerland, March 2017, pp. 282–285.\\
doi: \href{https://doi.org/10.23919/DATE.2017.7926999}{10.23919/DATE.2017.7926999}. %

\vspace{5mm}
\item Rafael Pires, Daniel Gavril, Pascal Felber, Emanuel Onica, and Marcelo Pasin.\\
\textbf{A lightweight \mapreduce framework for secure processing with SGX}\\
in 2017 17th IEEE/ACM International Symposium on Cluster, Cloud and Grid Computing (CCGRID). International Workshop on Assured Cloud Computing and QoS aware Big Data (WACC'17).\\
Madrid, Spain, May 2017, pp. 1100–1107.\\
doi: \href{https://doi.org/10.1109/CCGRID.2017.129}{10.1109/CCGRID.2017.129}. %

\vspace{5mm}
\item Aur\'elien Havet, Rafael Pires, Pascal Felber, Marcelo Pasin, Romain Rouvoy, and Valerio Schiavoni\\
\textbf{SecureStreams: A reactive middleware framework for secure data stream processing}\\
in Proceedings of the 11th ACM International Conference on Distributed and Event-based Systems, DEBS ’17.\\
Barcelona, Spain: ACM, June 2017, pp. 124–133.\\
doi: \href{https://doi.org/10.1145/3093742.3093927}{10.1145/3093742.3093927}. %

\vspace{5mm}
\item Sonia Ben Mokhtar, Antoine Boutet, Pascal Felber, Marcelo Pasin, Rafael Pires, and Valerio Schiavoni\\
\textbf{X-search: Revisiting private web search using intel SGX}\\
in Proceedings of the 18th Middleware Conference, Middleware ’17.\\
Las Vegas, USA: ACM, December 2017, pp. 198–208.\\
doi: \href{https://doi.org/10.1145/3135974.3135987}{10.1145/3135974.3135987}. %
\end{enumerate}

\newpage
\section*{2018}

\begin{enumerate}
\setcounter{enumi}{5}
\item Stefan Contiu, Rafael Pires, S\'ebastien Vaucher, Marcelo Pasin, Pascal Felber, and Laurent R\'eveill\`ere.\\
\textbf{\ibbesgx: Cryptographic group access control using trusted execution environments}\\
in 2018 48th Annual IEEE/IFIP International Conference on Dependable Systems and Networks (DSN).\\
Luxembourg, June 2018, pp. 207–218.\\
doi: \href{https://doi.org/10.1109/DSN.2018.00032}{10.1109/DSN.2018.00032}. %

\vspace{5mm}
\item S\'ebastien Vaucher, Rafael Pires, Pascal Felber, Marcelo Pasin, Valerio Schiavoni, and Christof Fetzer.\\
\textbf{SGX-aware container orchestration for heterogeneous clusters}\\
in 2018 IEEE 38th International Conference on Distributed Computing Systems (ICDCS).\\
Vienna, Austria, July 2018, pp. 730– 741.\\
doi: \href{https://doi.org/10.1109/ICDCS.2018.00076}{10.1109/ICDCS.2018.00076}. %

\vspace{5mm}
\item Rafael Pires, David Goltzsche, Sonia Ben Mokhtar, Sara Bouchenak, Antoine Boutet, Pascal Felber, R\"udiger Kapitza, Marcelo Pasin, and Valerio Schiavoni.\\
\textbf{Cyclosa: Decentralizing private web search through \ac{SGX}-based browser extensions}\\
in 2018 IEEE 38th International Conference on Distributed Computing Systems (ICDCS).\\
Vienna, Austria, July 2018, pp. 467–477.\\
doi: \href{https://doi.org/10.1109/ICDCS.2018.00053}{10.1109/ICDCS.2018.00053}. %

\vspace{5mm}
\item Christian Göttel, Rafael Pires, Isabelly Rocha, Sébastien Vaucher, Pascal Felber, Marcelo Pasin, and Valerio Schiavoni.\\
\textbf{Security, performance and energy trade-off of hardware-assisted memory protection mechanisms}\\
in 2018 IEEE 37th International Symposium on Reliable Distributed Systems (SRDS).\\
Salvador, Brazil, October 2018, pp. 133-142.\\
doi: \href{https://doi.org/10.1109/SRDS.2018.00024}{10.1109/SRDS.2018.00024}. %
\end{enumerate}

\section*{2019}

\begin{enumerate}
\setcounter{enumi}{9}
\item Andrei Mogage, Rafael Pires, Cr\u{a}ciun Vlad, Emanuel Onica, and Pascal Felber.\\
\textbf{Supply chain malware targets \ac{SGX}: Take care of what you sign}\\
in 2019 IEEE 38th International Symposium on Reliable Distributed Systems (SRDS).\\
Lyon, France, October 2019, pp. 52-60.\\
doi: \href{https://doi.org/10.1109/SRDS.2019.00016}{10.1109/SRDS.2019.00016} %

\vspace{5mm}
\item Stefan Contiu, S\'ebastien Vaucher, Rafael Pires, Marcelo Pasin, Pascal Felber, and Laurent R\'eveill\`ere.\\
\textbf{Anonymous and confidential file sharing over untrusted clouds}\\
in 2019 IEEE 38th International Symposium on Reliable Distributed Systems (SRDS).\\
Lyon, France, October 2019, pp. 21-31.\\
doi: \href{https://doi.org/10.1109/SRDS.2019.00013}{10.1109/SRDS.2019.00013} %

\end{enumerate}

\end{appendices}
 \newpage
{
\addcontentsline{toc}{chapter}{Bibliography}
\renewcommand*{\UrlFont}{\sffamily\small\relax}
\printbibliography
}
\end{document}